%% file: QCMPaper_Main.tex
\documentclass[10pt, a4paper, xcolor=dvipsnames]{article}

\usepackage{dsfont}
\usepackage{rotating}
\usepackage{floatpag}
\rotfloatpagestyle{empty}

\usepackage{amsmath,amsthm,amssymb}
\usepackage{xspace,enumerate,color,epsfig}
\usepackage{graphicx}
\usepackage{marvosym}
\usepackage{tikzfig}
\usepackage{stmaryrd}
\usepackage{docmute}
\usepackage{keycommand}
\usepackage{enumitem}

\input{tikzstyles.tex}

\usepackage{makeidx}
\makeindex

\theoremstyle{plain}
\newtheorem*{main theorem}{Main Theorem}
\newtheorem{theorem}{Theorem}[section]

\newtheorem{lemma}[theorem]{Lemma}
\newtheorem{proposition}[theorem]{Proposition}

\newtheorem{definition}[theorem]{Definition}

\newtheorem{example*}[theorem]{Example*}
\newtheorem{examples*}[theorem]{Examples*}
\newtheorem{remark}[theorem]{Remark}
\newtheorem{remark*}[theorem]{Remark*}

\newtheorem*{search problem}{Search Problem}

\usepackage{color}

\usepackage[left=2.5cm,right=2.5cm,top=2.0cm,bottom=2.0cm]{geometry}

\usepackage[utf8]{inputenc} 
\usepackage[english]{babel}  
\usepackage{verbatim} 
\usepackage{dsfont}
\usepackage[T1]{fontenc}  
\usepackage{amstext} 
\usepackage{ae}               

\usepackage{amsmath}
\allowdisplaybreaks		
\usepackage{mathtools,cancel}
\usepackage{amsfonts}
\usepackage{amssymb}
\usepackage{slashed}
\usepackage{xfrac} 	
\usepackage{empheq}  
\usepackage{braket}

\usepackage{graphicx}
\usepackage{epstopdf}
\usepackage[font=small]{caption}
\usepackage{subcaption}
\usepackage{xcolor}
\usepackage{lscape}
\usepackage{tikz}  
\usepackage[rightcaption]{sidecap} 

\usepackage{wrapfig} 
\usepackage{lipsum}  
\usepackage{mdframed} 
\usepackage{tabularx}
\usepackage{url}
\usepackage{hhline} 		
\usepackage{fancyhdr}       
\usepackage{varwidth}   
\usepackage[absolute,overlay]{textpos}
\usepackage{enumitem}
\usepackage{array}           
\usepackage{float}			
\usepackage{booktabs}
\usepackage{dashbox}
\usepackage{soul}   
\usepackage{breqn}    		

\usepackage[noadjust]{cite}
\usepackage{mcite}
\usepackage{mciteplus}

\usepackage[colorinlistoftodos,prependcaption,textsize=tiny,textwidth=2cm,linecolor=black, backgroundcolor=white]{todonotes}
\usepackage{pdfpages}         
\usepackage[toc,page]{appendix}
\usepackage{hyperref}

\definecolor{darkgreen}{rgb}{0.01, 0.75, 0.24}

\newcommand{\key}[1]{\textbf{#1}}
\newcommand{\Scale}{2.6}
\newcommand{\PRLsep}{\noindent\makebox[\linewidth]{\resizebox{0.5\linewidth}{1pt}{$\bullet$}}\bigskip}
\newcommand{\etal}{\textit{et al.} }

\newcommand{\overbar}[1]{\mkern 1.5mu\overline{\mkern-1.5mu#1\mkern-1.5mu}\mkern 1.5mu}
\newcommand{\xdownarrow}[1]{%
  {\left\downarrow\vbox to #1{}\right.\kern-\nulldelimiterspace}
}

\newcommand{\bigCI}{\mathrel{\text{\scalebox{1.07}{$\perp\mkern-10mu\perp$}}}}
\newcommand{\clop}{\kappa}         
\newcommand{\Trace}{\textnormal{Tr}}
\newcommand{\NestAbbr}[1]{\overleftarrow{#1}}
\newcommand{\nbigCI}{\cancel{\mathrel{\text{\scalebox{1.07}{$\perp\mkern-10mu\perp$}}}}}
\newcommand{\nosig}{\nrightarrow^{s}}

\author{Jonathan Barrett, Ognyan Oreshkov, Robin Lorenz}
\sidecaptionvpos{figure}{c}
\numberwithin{equation}{section}

\begin{document}

\begin{center}
\begin{Large}
\key{Quantum Causal Models} \\[0.5cm]
\end{Large}
Jonathan Barrett$^1$, Robin Lorenz$^1$, Ognyan Oreshkov$^{1,2}$ \\[0.2cm]
\small{\textit{$ ^1$Department of Computer Science, University of Oxford, 
Wolfson Building, Parks Road, Oxford OX1 3QD, UK \\
$ ^2$ QuIC, Ecole polytechnique de Bruxelles, C.P. 165, Universit\'e libre de Bruxelles, 1050 Brussels, Belgium
}}	
\end{center}

{\small
It is known that the classical framework of causal models is not general enough to allow for causal reasoning about quantum systems. The classical framework has been generalized in a variety of different ways to the quantum case, for example by allowing multiple quantum sources and analysing the correlations between the outcomes of measurements performed on the quantum systems. Much of this work, however, leaves open the question of whether causal concepts are fundamental to quantum theory, or whether such concepts only find application at an emergent level of classical devices and measurement outcomes. Here, we present a framework of quantum causal models, with causal relations defined in terms intrinsic to quantum theory, and the central object of study being the quantum process itself. Following Allen~\emph{et al.}, Phys. Rev. X {\bf 7}, 031021 (2017), the approach defines quantum causal relations in terms of evolution described by a unitary circuit, where the circuit admits the possibility of intervening upon some of the systems carried by the wires. This is in analogy with an approach to classical causal models that assumes underlying determinism and situates causal relations in functional dependences between variables. We show that any unitary quantum circuit has a causal structure corresponding to a directed acyclic graph, and that when marginalising over local noise sources, the resulting quantum process satisfies a Markov condition with respect to the graph. We also prove a converse to this statement. We introduce an intrinsically quantum notion that plays a role analogous to the conditional independence of classical variables, and (generalizing a central theorem of the classical framework) show that d-separation is sound and complete for it in the quantum case. We present generalizations of the three rules of the classical \emph{do-calculus}, in each case relating a property of the causal structure to a formal property of the quantum process, and to an operational statement concerning the outcomes of interventions. In addition to the results concerning quantum causal models, we introduce and derive similar results for \emph{classical split-node causal models}, which are more closely analogous to quantum causal models than the classical causal models that are usually studied. 
}
\vspace*{0.3cm}
		
\vspace*{0.3cm}
\PRLsep	
\vspace*{-0.8cm}
\setcounter{tocdepth}{2}
\tableofcontents	
\vspace*{0.3cm}
\PRLsep

\section{Introduction}

Reasoning in causal terms is omnipresent, from fundamental physics to medicine, social sciences and economics, and in everyday life. The engine starts whenever the car key is turned, and we believe that this is because turning the key causes the engine to start. However, as the mantra  `correlation does not imply causation' suggests, the connections between observed data and causal relations between variables are not always obvious. The framework of causal models, most prominently developed by Judea Pearl \cite{Pearl_Causality, Pearl_2018_TheBookOfWhy} and by Spirtes \etal \cite{SpirtesEtAL_2000_BookCausationPredictionSearch}, makes these connections precise. This enables, for example, the development of `causal discovery algorithms', which spell out in a principled way the assumptions under which causal structure can indeed be inferred from observational data. The framework also allows predictions about the consequences of intervening upon (rather than merely observing) variables, and provides a method of evaluating counterfactual claims.

When it comes to quantum systems, however, the framework of classical causal models is not able to give satisfactory causal explanations of all possible correlations. Bell's theorem \cite{Bell_2004_SpeakableAndUnspeakable} rules out an explanation of Bell-correlations in terms of free choices of measurements in space-like separated regions and a variable in the common past. An account of Bell-correlations within the framework of classical causal models is still possible, for instance through the strategies of invoking retrocausal explanations, superdeterminism or superluminal signalling. But any such strategy requires some form of fine-tuning \cite{WoodEtAl_2014_CausalExplanationAndBellInequalities, Cavalcanti_2018_ClassicalCausalModelsFineTunedForNonlocality, PearlAndCavalcanti_2019_ClassicalCausalModelsFinedTunedForNonLocality}, with the consequence that the \emph{lack} of retrocausality, or superdeterminism, or superluminal signalling at the everyday operational level becomes mysterious.

In recent years, many works, taking inspiration from classical causal modelling, have generalized the basic Bell scenario, to situations in which measurements, or sequences thereof, are performed on systems produced by multiple independent sources. Here, the problem of characterizing the correlations that can be produced, assuming either quantum or classical sources, is technically more difficult than in the standard Bell scenario, but a lot of progress has been made \cite{BranciardEtAl_2010_CharacterizingNonLocalCorrelations, Fritz_2012_BeyondBellsTheorem,  HensonEtAl_2014_TheoryIndependentLimitsGeneralisedBayesianNetworks, ChavesEtAl_2015_InformationImplicationsOfQuantumCausalStructures, Fritz_2016_BeyondBellsTheorem, WeilenmannEtAl_2016_InabilityOfEntropyApproach,  Pienaar_2017_CausalQuantumClassicalGap, WeilenmannEtAl_2017_AnalysingCausalStructuresWithEntropy, MiklinEtAl_2017_EntropicApproachToCausalCorrelations, FraserEtAl_2018_CausalCompatibilityInequalities, WeilenmannEtAl_2018_NonShannonInequalitiesInEntropyApproach, VanHimbeeck_2019_QquantumViolationsInstrumentalScenario, PozasEtAl_2019_BoundingSetsOfCorrelationsInNetworks, WolfeEtAl_2019_InflationTechniqueForCausalInference, NavascuesEtAl_2020_InflationTechniqueSolvesCompatibilityProblem}. One motivation for doing so is for application to device-independent protocols: if quantum devices can produce correlated measurement outcomes that could not be produced by classical devices in a similar arrangement, then there is the possibility of cryptographic protocols whose security, by exploiting this fact, does not rely on any particular quantum description of the systems involved \cite{BarrettEtAl_2005_NoSignallingAndQKD, PironioEtAl_2010_RandomNumbersCertifiedByBell, ColbeckEtAl_2012_FreeRandomnessCanBeAmplified, LeeEtAl_2018_TowardsDeviceIndependentInfoProcessingInNetworks, VaziraniEtAl_2019_FullyDeviceIndependentQKD}.

This line of research, however, in which the principle object of study is the set of correlations that can be produced by measurements on quantum systems, leaves open many fundamental questions that can be asked concerning causality in a quantum universe. These questions include the following. If A is a cause of B, then what sort of thing are A and B? What does it mean to assert a causal relation between A and B? Do causal concepts involve, in an essential way, interventions by agents? Are causal relations directed in time? What, in the overall picture, is ontic (i.e., factual and independent of the agent), and what is epistemic (i.e., relative to an agent's knowledge or beliefs)? Are causal concepts fundamental to quantum theory, or do such concepts only find application at an emergent level of classical devices and measurement outcomes? From a quantum treatment, can the classical framework of causal models be recovered as a limiting case? In addition to the work generalizing Bell's theorem described above, the last two decades have seen much work on the study of causal reasoning in quantum theory and the development of quantum versions of causal modelling in which at least some causal concepts are defined in intrinsically quantum terms \cite{Tucci_1995_QuantumBayesianNets, 
Tucci_2007_FactorisationOfDensityMatricesAccordingToNetworks,  
Leifer_2006_QuantumDynamicsAndConditionalProbability, 
Laskey_2007_QuantumCausalNetworks,
LeiferEtAl_2008QuantumGraphicalModels,
LeiferEtAl_2013_QTAsBayesianInference,  
FitzsimonsEtAl_2015_QuantumCorrelationsWhichImplyCausation, 
PienaarEtAl_2015_GraphSeparationTheoremForQCMs, 
RiedEtAl_2015_QuantumAdvantageForCausalInference,
CostaEtAl_2016_QuantumCausalModeling,
AllenEtAl_2016_QCM, 
MilzEtAl_2017KolmogorovExtensionTheorem, 
PollockEtAl_2018_OperationalMarkovConditionForQuantum, 
Pienaar_2018_QuantumCausalModelsViaQuantumBayesianism, 
Pienaar_2018_QBistQCMs_ShortVersion, 
Pienaar_2019_TimeReversibleQCM, 
SchmidEtAl_2019_CausalPerspectiveOnCPMaps}.
Any such work involves, either explicitly or implicitly, answers to at least some of the questions above, but as might be expected, there are big differences between these approaches with respect to the answers that are given. 

We present in this work a framework of quantum causal models, based on the initial proposal of Ref.~\cite{AllenEtAl_2016_QCM}. In common with various other works, we take the relata of quantum causal relationships to be, not quantum systems or states, but \emph{loci of possible intervention} (what we later refer to as \emph{quantum nodes}). Such a locus corresponds to two Hilbert spaces, which can be thought of as representing an incoming and an outgoing system. A distinctive feature of the approach of this work is that it takes causal relations to be relations of influence between quantum nodes mediated by unitary evolution. This is in analogy with an approach to classical causal modelling that assumes underlying determinism and defines causal relations in terms of functional dependences between variables. In the classical case, probabilities then arise from an agent's uncertainty about unobserved systems. In the quantum case, while causal relations are defined by unitary evolution, the object of study is the (in general mixed) quantum process that similarly arises from an agent's uncertainty about unobserved systems. We show that given an ordinary, unitary quantum circuit, with the possibility of intervening upon the systems at some of the wires, the causal structure can be represented by a directed acyclic graph and that when marginalising over local noise sources, the resulting quantum process satisfies a Markov condition with respect to the graph, hence defines a quantum causal model. We also provide a converse to this result: given any directed acyclic graph and a quantum process that satisfies the corresponding Markov condition, there exists a circuit with appropriate causal structure such that the quantum process is returned when marginalising over unobserved systems. 

In classical causal modelling, of central importance is the idea that the causal relations between a set of random variables impose constraints on a probability distribution over the variables in the form of conditional independences. A simple example of this is that asserted by Reichenbach's principle \cite{Reichenbach_1991_DirectionOfTime, CavalcantiEtAl_2014_OnModificationsOfRP, AllenEtAl_2016_QCM}, wherein if correlation between variables $Y$ and $Z$ is to be explained by the hypothesis that a variable $X$ is the complete common cause of $Y$ and $Z$, then the probability distribution over all three should satisfy the constraint that $Y$ and $Z$ are conditionally independent given $X$. 
For an arbitrary causal structure, represented by a directed acyclic graph, Reichenbach’s principle is generalized by the constraint that a joint distribution over the variables should satisfy a Markov condition with respect to the graph. A central theorem then states that the graphical condition of d-separation is sound and complete for conditional independence in any probability distribution that is Markov for the graph. 
In order to express the constraints that causal structure places upon the quantum process, we introduce an intrinsically quantum notion, referred to below as \emph{strong relative independence}, which plays a role analogous to that of the conditional independence of two classical variables given a third. We relate this to quantum causal structure via a d-separation theorem. (This notion of quantum relative independence can also be seen as generalizing the property of vanishing quantum conditional mutual information of tripartite quantum states, and will, we believe, be of independent interest regardless of the connection to causal structure.) We then extend this result, presenting three theorems that generalize the three rules of Pearl's do-calculus \cite{Pearl_Causality}. In each case, causal structure is shown to impose constraints on the form of the marginal quantum process defined over a subset of the nodes, and these constraints are in turn shown to have operational consequences. Finally, in addition to the main results, which concern quantum causal models, we establish similar results for \emph{classical split-node causal models} (defined below), which are more closely analogous to the quantum formalism than the classical causal models of  Refs.~\cite{Pearl_Causality, Pearl_2018_TheBookOfWhy, SpirtesEtAL_2000_BookCausationPredictionSearch}, and which we have not seen before in the literature. 

The motivations for the work are both fundamental and practical. The first fundamental motivation is to establish an account of causality in quantum theory's own terms, without assuming a separate realm of classical systems or measurement outcomes. The account provides its own answer to many of the basic questions concerning causality in a quantum universe that are listed above. We do not discuss these at length, as this paper is mostly devoted to the technical development, but we expand a little in the Discussion section below.  A second fundamental motivation is provided by the idea of indefinite causal order, defined and explored in, e.g.,   
Refs.~\cite{Hardy_2005_ProbabilityTheoriesWithDynamicCausalStructure,
OreshkovEtAl_CorrelationsWithoutCausalOrder,
Chiribella_2012_PerfectDiscriminationOfChannelsViaSuperpositionCS,
ChiribellaEtAl_2013_QuantumCompWithoutDefCausalStructure,
AraujoEtAl_2014_ComputationalAdvantage,
AraujoEtAl_2015_WitnessingCausalNonSeparability,
OreshkovEtAl_2016_CausallySeparableProcesses,
GuerinEtAl_2016_ExponentialCommunicationComplexityAdvantage,
BranciardEtAl_2015_SimplestCausalInequalitiesAndViolations,
Baumeler_EtAl_2016_SpaceOfLOgicallyConsistentClassicalProcesses,
SilvaEtAl_2017_ConnectingIndefiniteWithMultiTime,
AbbottEtAl_2017_GenuinelyMultipartiteNoncausality,
Oreshkov_2019_TimeDelocalisedSystems, 
SalekEtAl_2018_QuantumCommunication}, wherein quantum (or classical) processes can be written down that are not compatible with any particular fixed causal order, nor even with any probabilistic mixture thereof. Indefinite causal order might be expected to arise, for example, in a future theory of both quantum systems and gravity, at least if causal orders themselves are subject to quantum uncertainty. Progress in understanding quantum indefinite causality will be facilitated by first properly understanding definite causal structure from a causal model perspective, in such a way that this can then be extended to the indefinite case.  

Practical motivation stems from the fact that as quantum technologies develop, quantum engineers will increasingly be concerned with medium to large scale quantum devices and networks that manipulate large numbers of quantum systems (say, tens of thousands rather than three or four as in laboratory experiments of the past). The device or network in principle realises some quantum process, but full tomography of that process is impractical, both because full tomography requires a number of measurement readings that is exponential in the number of systems, and because intervention at some of the nodes of the process may be expensive or difficult. There may be hard-to-characterize sources of noise on which direct intervention is impossible, or some parts of the process may be in the hands of an adversary, with the adversary's actions unknown. Thus a very generic problem is to make predictions concerning future instantiations of a quantum process, perhaps with varying parameters or new interventions, on the basis of existing data that offers only very partial tomography of the process as a whole. This is the realm, for example, of the subfield of quantum verification, on which there is by now a sizeable literature (see \cite{EisertEtAl_2020_QuantumCertificationAndBenchmarking} for a review). Our work offers the possibility of a novel and distinctly causal approach to this generic problem. 

The paper is written assuming standard background in quantum information science, but not in classical causal modelling, hence aims to be somewhat pedagogical with respect to the latter. The structure is as follows. Section~\ref{Sec_CCMIntro} summarizes the framework of classical causal models. Section~\ref{Sec_QCMIntro} provides necessary preliminaries, in particular the definition of a \emph{process operator} and the definition of a quantum causal model. At that stage, the definition of a quantum causal model will have simply been presented without much explanation or justification. Section~\ref{Sec_QCMsAndUnitaryDynamics} presents the fundamental link between unitary dynamics and quantum causal models, showing in particular that unitary evolution of quantum systems has a causal structure that can be represented by a directed graph, and that when marginalising over local noise sources, the resulting process is Markov for that graph, hence defines a quantum causal model. Section~\ref{Sec_ClassicalSplitNodeFormalism} introduces \emph{classical split-node causal models}, and gives an overview on how quantum and the different kinds of classical causal models are related to one other. Sections~\ref{Sec_IndependenceNotions} and \ref{Sec_DSeparationTheorem} present notions of conditional and relative independence, and the quantum d-separation theorem. Sections~\ref{Sec_DoCalculus_CCM} and \ref{Sec_QuantumDoCalculus} introduce the classical do-calculus, and generalize its rules to the quantum framework. Section~\ref{Sec_NoteCausalDiscovery} contains some first observations on the problem of causal inference. Sections~\ref{Sec_RelatedWork} and \ref{Sec_Discussion} give a brief account of related work and conclude.

\section{Classical causal models \label{Sec_CCMIntro}} 

\subsection{The definition \label{SubSec_DefinitionCCM}}

A \emph{classical causal model} specifies a finite set of random variables, which in this work are always assumed to take values in a finite set. A \emph{causal structure} for the random variables corresponds to a directed acyclic graph (DAG)\footnote{Directed graphs, which may have directed cycles, play some role in classical causal modeling, but will be ignored in the present work.}, with the variables as nodes, and with an arrow from a node $X$ to a node $Y$ representing the relation that $X$ is a direct cause of $Y$. If there is a directed path from $X$ to $Y$, but no arrow from $X$ to $Y$, then $X$ is an indirect cause of $Y$, meaning that $X$ should still be thought of as a cause of $Y$, but with the causal relation mediated by other variables in the model.\footnote{Note that the distinction between direct and indirect causes is dependent on the choice of variables.} In the following, the terminology of kinship will be useful to describe relationships between nodes of the graph. If there is an arrow from a node $X$ to a node $Y$, then $X$ is a \emph{parent} of $Y$, and $Y$ is a \emph{child} of $X$. If there is a directed path from a node $X$ to a node $Y$, then $X$ is an \emph{ancestor} of $Y$, and $Y$ is a \emph{descendant} of $X$. For a node $X$, let $Pa(X)$ denote the set of parents of $X$, and let $Nd(X)$ denote the set of all nodes that are not descendants of $X$, excluding $X$ itself.

\begin{figure}[H]
\centering
\vspace*{0.3cm}
\input{Figures/Fig_ExampleDAG.tex}
\caption{Example of a classical causal model.  \label{Fig_ExampleDAG}}
\end{figure}
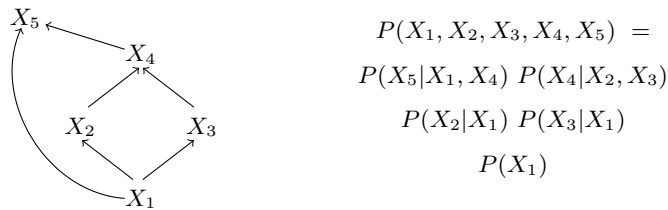

A classical causal model is then defined as follows.
\begin{definition} \label{Def_CCM} \textnormal{(Classical causal model):} A classical causal model is given by
\begin{enumerate}[label=\textnormal{(\arabic*)}, leftmargin=1.0cm]
\item a causal structure represented by a DAG $G$ with vertices corresponding to random variables $X_1 , \allowbreak \ldots  , \allowbreak X_n$, 
\item for each $X_i$, a classical channel $P(X_i | Pa(X_i))$.
\end{enumerate}
The classical causal model defines a probability distribution over $X_1 , \ldots ,X_n$, given by
\begin{equation}\label{Eq_Factorizationcondition}
P(X_1 ,... ,X_n) = \prod_i P(X_i | Pa(X_i) ).
\end{equation}
\end{definition}
Fig.~\ref{Fig_ExampleDAG} shows an example of a classical causal model.
Central to classical causal modelling is the idea that if a joint probability distribution $P(X_1 , ... , X_n)$ represents observed data, perhaps gathered over a large number of repeated trials, then a classical causal model that recovers $P(X_1 , ... , X_n)$ can be regarded as a candidate \emph{explanation} for the observations. Furthermore, as explained in Section~\ref{SubSec_dointerventions} below, this explanation is useful because it allows further kinds of predictions to be made. Verifying these predictions constitutes a test of the classical causal model.

In virtue of Eq.~(\ref{Eq_Factorizationcondition}), a given causal structure, represented by a DAG $G$, imposes a constraint on any probability distribution that can be explained by a classical causal model with DAG $G$. This is captured by the following definition.
\begin{definition} \textnormal{(Markov condition):} \label{Def_ClassicalMarkov}
Given a DAG $G$, with nodes corresponding to random variables $X_1 , \ldots , X_n$, a joint probability distribution $P(X_1, \ldots , X_n)$ is \emph{Markov} for $G$ if and only if there exists for each $i$ a classical channel $P( X_i | Pa(X_i) )$ such that  
\begin{equation}\label{Eq_Markovcondition}
P(X_1, \ldots , X_n) = \prod_{i=1}^n P\big( X_i|Pa(X_i) \big) \ .
\end{equation}
\end{definition}

\subsection{Why Markovianity? \label{SubSec_WhyMarkovianity}}

Given that a set of random variables has a causal structure corresponding to a DAG, it is reasonable to ask why a probability distribution over the variables \emph{should} be Markov for the DAG. If causal relations are facts in the world, and probabilities the degrees of belief of a rational agent, then the question becomes: why should a rational agent arrange her beliefs so as to be Markov for the causal structure? Alternatively, assuming that probabilities have at least some connection with observed relative frequencies in experimental trials, then why should these frequencies have any connection to the Markov condition?

One approach sees the Markov condition as a normative principle, which guides us in finding an appropriate set of variables for causal reasoning (see, e.g., \cite{SpirtesEtAL_2000_BookCausationPredictionSearch}, \cite{SEP_ProbabilisticCausation}). Pearl, on the other hand, obtains the Markov condition as a theorem, following an assumption of underlying determinism \cite{Pearl_Causality}. In this work (following \cite{AllenEtAl_2016_QCM}), we also take this approach. The idea is that direct causal relations in the model will be understood in terms of functional dependences, such that with additional (perhaps unobserved) variables $\lambda_1,\ldots, \lambda_n$, each variable $X_i$ of the original causal model can be expressed as a function of its parents and of the additional variable $\lambda_i$. The $\lambda_i$ are assumed to be statistically independent, i.e., $P(\lambda_1,...,\lambda_n)=\prod_i P(\lambda_i)$ -- a substantial but common assumption, which we will not further justify in this work (also see the discussion in Section~\ref{Sec_Discussion}). Seeing as each of the $\lambda_i$ is only a `local disturbance', that is it only influences one variable $X_i$, the $\lambda_i$ do not constitute additional common causes. Hence the causal structure of the variables $X_1,\ldots, X_n$, as represented by the DAG $G$, can be seen as a representation of functional dependences amongst the variables, modulo additional local noise sources. Probability distributions on this view arise due an agent's lack of knowledge of, or control over, the $\lambda_i$. 

If a distribution $P(X_1, \ldots , X_n)$ can be seen to arise from such a situation through marginalization over the $\lambda_i$, then we will say that it is \emph{compatible} with the causal structure expressed by the DAG $G$. More formally,
\begin{definition} \label{Def_ClassCompDAG} \textnormal{(Classical compatibility with a DAG)}: Given a DAG $G$ with nodes $X_1, \ldots, X_n$, a joint probability distribution $P(X_1, \ldots , X_n)$ is \emph{compatible} with $G$ if and only if there exist $n$ additional variables $\lambda_1,..., \lambda_n$ and functions $f_i : Pa(X_i) \times \lambda_i \rightarrow X_i$, along with distributions $P(\lambda_i)$ such that
\begin{equation}
P(X_1, \ldots , X_n) = \sum_{\lambda_1,..., \lambda_n} \left[ \ \prod_{i=1}^n \delta \Big( X_i , \ f_i(Pa(X_i) , \lambda_i) \ \Big) \ P(\lambda_i) \ \right]  \ . \label{Eq_Def_ClassicalMC}
\end{equation} 
\end{definition}
The connection with the Markov condition is then as follows.
\begin{theorem} \label{Thm_ClassicalEquivalence} \textnormal{(Equivalence of classical compatibility and Markovianity \cite{Pearl_Causality}):}
Given a distribution $P(X_1, \allowbreak \ldots, \allowbreak X_n)$ and a DAG $G$ with nodes $X_1, \ldots , X_n$, the following are equivalent:
\begin{enumerate}[label=\textnormal{(\arabic*)}, leftmargin=3cm]
\item $P(X_1, \ldots , X_n)$ is compatible with $G$.
\item $P(X_1, \ldots , X_n)$ is Markov for $G$. 
\end{enumerate}
\end{theorem}
The $(1)\rightarrow (2)$ direction can be seen as justifying the stipulation that a joint probability distribution $P(X_1 , \allowbreak ... , \allowbreak X_n)$, representing observed data, should be explained by a classical causal model with a DAG $G$ such that $P(X_1 , ... , X_n)$ is Markov for $G$. The $(2)\rightarrow (1)$ direction shows that, given a DAG $G$, any joint probability distribution that is Markov for $G$ admits an explanation in terms of underlying functional relationships among variables, such that the variables have the causal structure of $G$.

Fig.~\ref{Fig_Examplefunctionalmodel} shows the classical causal model of Fig.~\ref{Fig_ExampleDAG}, dilated to a functional model.
\begin{figure}[H]
\centering
\vspace*{0.3cm}
\small
\input{Figures/Fig_Examplefunctionalmodel.tex}
\caption{A functional model. \label{Fig_Examplefunctionalmodel}}
\end{figure}

\subsection{Do-interventions}\label{SubSec_dointerventions}

We may have observations of the weather conditions and readings of a barometer, but what is the probability of rain if we have built an airtight box around the barometer and forced it to show a particular pressure? A joint probability distribution allows, via Bayes' rule, the calculation of any conditional distribution $P(X_j | X_i)$, that is a set of values for the probability of finding $X_j=x_j$, given that one has observed $X_i=x_i$. However, the joint distribution is not in general sufficient to calculate probabilities for $X_j$ in an alternative scenario in which one intervenes on $X_i$, i.e., externally `reaches in', and changes the value of $X_i$ rather than simply observing it. A classical causal model allows probabilities to be computed in such alternative scenarios. 

A particularly important kind of intervention in the framework of classical causal models is a \emph{do-intervention}, wherein a variable $X_i$ is forced to take the value $x_i$, overriding the mechanisms which would have otherwise fixed the value of $X_i$. More generally, one may consider do-interventions that fix a subset $S$ of the variables to have a value $s$, where $s$ labels a tuple of values for the variables in $S$. In this case, the quantity of interest is the resulting joint distribution over the remaining variables. Importantly, it is assumed that the arrows in a causal structure $G$ represent stable and autonomous mechanisms: when do-intervening on $S$, the mechanisms by which variables $\notin S$ take values are unaltered. 
\begin{definition} \textnormal{(Do-conditional distribution):} \label{Def_DoConditionalProbability}
Consider a classical causal model given by a DAG $G$ with nodes $X_1 , \ldots , X_n$, and for each $i$ a classical channel $P(X_i | Pa(X_i))$. Let $S \subset \{X_1,...,X_n\}$ and let $T := \{X_1,...,X_n\} \setminus S$. The \emph{do-conditional distribution} for $T$, given a do-intervention on $S$, is given by
\begin{equation}
P( T | do(S) ) := \prod_{X_i \in T} P(X_i | Pa(X_i)). \label{Def_TruncatedFactorization}
\end{equation}
If $s$ is a particular value of $S$, i.e., a tuple containing a value for each variable $X_i \in S$, then
\begin{equation}
P( T | do(S=s) ) := \left. \left( \prod_{X_i \in T} P(X_i | Pa(X_i)) \right) \right|_{S=s}.\label{Def_TruncatedFactorizationwiths}
\end{equation}
\end{definition}
Given a do-conditional distribution $P( T | do(S) )$, a marginal for a variable $X_i\in T$ is obtained in the ordinary way as $P(X_i | do(S)):=\sum_{X_j\in T, j \neq i} P(T | do(S))$. Eq.~\ref{Def_TruncatedFactorization} is sometimes referred to as the \textit{truncated factorization formula}, as it results from dropping all those factors corresponding to variables in $S$ in the right hand side of the expression
\[
P(X_1,\ldots , X_n) = \prod_{i=1}^n P(X_i | Pa(X_i)).
\]

The quantity $P( T | do(S=s) )$ is to be interpreted as the resulting probability distribution over the variables in $T$, when a do-intervention sets the variables in $S$ to have the value $s$. Much as with the Markov condition, the definition of a do-conditional distribution can be justified in terms of underlying functional models. In terms of an underlying functional model, when a do-intervention is performed on a set of variables $S$, fixing a joint value $s$, the stability of mechanisms means that for any variable $X_j \notin S$, the function $f_j (Pa(X_j) , \lambda_j)$, which determines $X_j$ in terms of the parents of $X_j$ and a local disturbance $\lambda_j$, is assumed to be unaltered by the intervention upon $S$. Any variable $X_i \in S$, on the other hand, is no longer given as $X_i = f_i (Pa(X_i) , \lambda_i )$, but is simply fixed to have the value determined by $s$. It is also assumed that the distributions $P(\lambda_i)$ are unchanged by the intervention. It is then a simple matter \cite{Pearl_Causality} to show that, marginalizing over the local noise sources, the resulting joint distribution over the remaining variables $T$ is given by Eq.~\ref{Def_TruncatedFactorizationwiths}.

\begin{figure}[H]
\centering
\vspace*{0.3cm}
\small
\input{Figures/Fig_Exampledoconditional.tex}
\caption{Evaluating a do-conditional. A do-intervention at $X_4$ overrides the mechanisms that would normally fix $X_4$ in terms of its parents, $X_2$ and $X_3$, and a local disturbance $\lambda_4$. The quantity $P(X_1,X_2,X_3,X_5 | do(X_4=x))$ can be seen as arising from a mutilated DAG, in which arrows towards the node $X_4$ have been removed.  \label{Fig_Exampledoconditional}}
\end{figure}
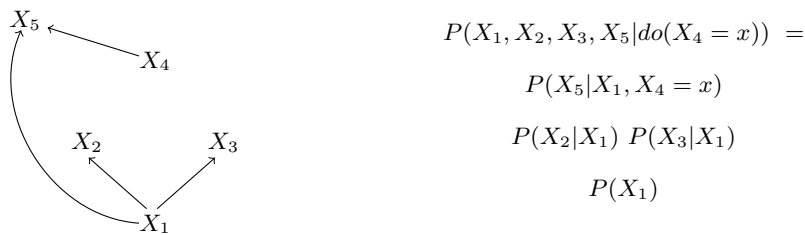

Fig.~\ref{Fig_Exampledoconditional} shows an example of a do-conditional, evaluated with the classical causal model of Fig.~\ref{Fig_ExampleDAG}.

\subsection{Causal influence, signalling, fine tuning, and causal inference \label{SubSec_RemarkClassicalCausalRelation}} 

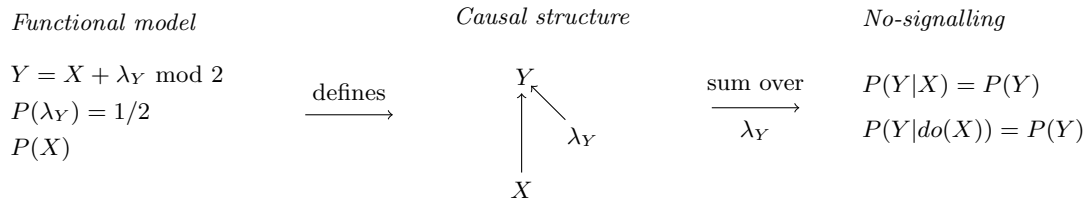
\begin{figure}[H]
\centering
\vspace*{0.3cm}
\small
\input{Figures/Fig_Examplefinetuning.tex}
\caption{Illustrating the distinction between causal influence and the possibility of signalling. \label{Fig_Examplefinetuning}}
\end{figure}

One particular feature of classical causal modelling is worth emphasising, as it will be important for the quantum case developed below. Consider the example of Fig.~\ref{Fig_Examplefinetuning}, where a binary variable $Y$ is given by a function of a binary variable $X$ and a binary local disturbance $\lambda_Y$, such that $Y = f_Y(X,\lambda_Y) = X + \lambda_Y \mod 2$. The joint distribution is given by $P(X,Y,\lambda_Y) = P(Y|X,\lambda_Y) P(X) P(\lambda_Y)$, where $P(Y|X,\lambda_Y) = \delta(Y , f_Y(X,\lambda_Y))$. 
Observe that if $P(\lambda_Y=0) = P(\lambda_Y=1) = 1/2$, then $P(Y|X) = P(Y)$, that is $Y$ is statistically independent of $X$. Furthermore, the do-conditional $P(Y | do(X))$ similarly satisfies $P(Y | do(X)) = P(Y)$. It is therefore impossible to send signals from $X$ to $Y$ by means of deliberately varying the value of $X$. Nonetheless, in this classical causal model, $X$ is a direct cause of $Y$, since in the function $f_Y$, the variable $Y$ does indeed have a dependence on $X$. Suppose the variables $X$ and $Y$ are accessible, while $\lambda_Y$ is inaccessible, but happens to be described by the probability distribution $P(\lambda_Y)$. This example then shows that the possibility of signalling from $X$ to $Y$ via $P(Y|do(X))$,  while sufficient, is not a necessary condition for $X$ to be a cause of $Y$. It is easy to see that with $f_Y$ fixed, any other distribution on $\lambda_Y$ would indeed lead to signalling from $X$ to $Y$. The absence of signalling only holds for the specific choice $P(\lambda_Y=0) = P(\lambda_Y=1) = 1/2$. In general, this phenomenon -- where a certain independence holds only for a specific distribution on other variables, and is not forced by the causal structure -- is known as \emph{fine tuning}. 

More generally, consider a classical causal model with DAG $G$ and distribution $P(X_1,\ldots , X_n)$ that is Markov for $G$, equivalently compatible with $G$. In the functional model that is asserted to exist by the definition of compatibility (Def.~\ref{Def_ClassCompDAG}), suppose that the distributions $P(\lambda_i)$ are modified, to become $P'(\lambda_i)$, while keeping the functions $f_i$ unchanged. Marginalizing over the $\lambda_i$, one may calculate a new distribution $P'(X_1,\ldots , X_n)$, in general different from  $P(X_1,\ldots , X_n)$. However, the distribution  $P'(X_1,\ldots , X_n)$ will be (by definition) compatible with $G$, hence (by Theorem~\ref{Thm_ClassicalEquivalence}) Markov for $G$. This means that  $P'(X_1,\ldots , X_n)$ will also satisfy all conditional independences that are forced by Markovianity for $G$ -- we may distinguish these conditional independences, which arise from the causal structure itself, from any accidental conditional independences that arise in virtue of a specific, fine tuned, choice of distributions on the $\lambda_i$. 

In classical causal modelling, the problem of causal inference is that of producing a classical causal model, given observed data summarized by a joint distribution $P(X_1,\ldots , X_n)$, such that the causal model can be seen as an explanation of the data. It is usually regarded as a desideratum that any conditional independences satisfied by $P(X_1,\ldots , X_n)$ are forced by the causal structure of the model, that is do not need to be obtained via a fine-tuned choice of distribution on unobserved variables $\lambda_i$.

\section{Introducing quantum causal models \label{Sec_QCMIntro}}

\subsection{Quantum definitions and notation 1 \label{SubSec_QuantumPreliminaries}} 

This work considers, for simplicity, only finite-dimensional quantum systems. Let $\mathcal{H}_A$ denote the Hilbert space of system $A$ and $d_A$ the dimension of $\mathcal{H}_A$. (For better readability, the proofs in the Appendices will sometimes denote the Hilbert space of system $A$, simply as $A$, rather than $\mathcal{H}_A$.) Let $\mathcal{L}(\mathcal{H})$ denote the vector space of linear operators on $\mathcal{H}$. A completely positive (CP) map $\mathcal{E}: \mathcal{L}(\mathcal{H}_A) \rightarrow \mathcal{L}(\mathcal{H}_B)$ can be represented as an operator via the Choi-Jamio\l kowski (CJ) isomorphism \cite{Jamolkowski_1972, Choi_1975}. We will use a variant of the CJ isomorphism such that a CP map $\mathcal{E}: \mathcal{L}(\mathcal{H}_A) \rightarrow \mathcal{L}(\mathcal{H}_B)$ is represented as a positive semi-definite operator on $\mathcal{H}_B \otimes \mathcal{H}_A^*$, where $\mathcal{H}_A^*$ is the dual space of $\mathcal{H}_A$:
\begin{equation}
	\rho_{B|A} := \sum_{i,j} \ \mathcal{E}(\ket{i}_A\bra{j}) \otimes  \ket{i}_{A^*}\bra{j}.
\end{equation}
Here, $\left\{ \ket{i}_A \right\}$ is an orthonormal basis of $\mathcal{H}_A$, and $\left\{ \ket{i}_{A^*} \right\}$ the corresponding dual basis. Under this definition, the CJ representation of a CP map is both positive semi-definite and basis-independent\footnote{A basis-dependent version of the CJ representation with a physical interpretation, where the choice of basis is related to the transformation of time reversal, was proposed in \cite{OreshkovEtAl_2016_OperationalQuantumTheoryWithoutTime}. Here, we will not be concerned with this time-neutral approach but will regard the CJ isomorphism simply as a mathematical tool that provides an equivalent representation of standard quantum theory (hence the basis-independent representation).}. 
The converse direction is then given by 
$\mathcal{E}(\rho_A)=\Trace_{AA^*} \big[ \tau^{\text{id}}_{A}\  \rho_{B|A} \ \rho_A \big]$ for $\rho_A\in \mathcal{L}(\mathcal{H}_A)$, where the operator $\tau^{\text{id}}_A$, 
is given by
\begin{equation}
	\tau^{\text{id}}_{A} := \sum_{i,j} \ \ket{i}_{A^*} \bra{j} \otimes  \ket{i}_{A}\bra{j} \ . \label{Eq_DefLinkOperator}
\end{equation}

The most general deterministic transformation of the state of a quantum system that can result from an interaction of the system with an environment, when the state of the system is initially uncorrelated from the state of the environment, is described by a trace-preserving CP (CPTP) map.
The CJ representation of a CPTP map $\rho_{B|A}$ satisfies $\Trace_B [\rho_{B|A}]=\mathds{1}_{A^*}$, where $\mathds{1}_{A}$ denotes the identity operator on $\mathcal{H}_A$. At times it will be useful to consider the operator on $\mathcal{H}_B \otimes \mathcal{H}_A^*$ associated with the representation $\rho_{B|A}$ of a channel, but normalised to have unit trace, which we write as $\hat{\rho}_{B|A} := (1/d_A)\rho_{B|A}$. 

The most general intervention upon a quantum system (initially uncorrelated with the agent) corresponds to a \emph{quantum instrument}, that is a set of trace-non-increasing CP maps $\left\{ \mathcal{E}^k : \mathcal{L}(\mathcal{H}_{A}) \rightarrow \mathcal{L}(\mathcal{H}_{B})\right\}$, where $k$ labels the classical outcome of the intervention, and where $\mathcal{E} = \sum_k \mathcal{E}^k$, is a CPTP map. The probability of obtaining outcome $k$ is $P(k) = \Trace ( \mathcal{E}^k (\rho_A) )$, and the state after the intervention, conditioned upon obtaining outcome $k$, is $\mathcal{E}^k (\rho_A) / P(k)$.

Whenever we write products of the form $\rho_{B|DA} \ \rho_{C|AE}$, `padding' with identity operators on complementary Hilbert spaces is to be understood implicitly, i.e., the expression $\rho_{B|DA} \ \rho_{C|AE}$ is short for $(\rho_{B|DA}\otimes \mathds{1}_{E^*} \otimes \mathds{1}_{C})(\rho_{C|AE} \otimes \mathds{1}_{B} \otimes \mathds{1}_{D^*})$.

\subsection{Quantum process operators \label{SubSec_Processoperators}} 

There is a rich literature studying quantum correlations between causally related systems using formalisms that are closely related to one another. A common ingredient in many of them is that a local quantum system is described by two Hilbert spaces, one associated with the causal past of the system, and one with the causal future. The two Hilbert spaces together can be interpreted as a locus at which an agent may or may not decide to intervene upon the system. If an intervention is performed, then the intervention corresponds to a quantum instrument, in general with a classical outcome, that defines an additional evolution of the system, mediating the causal past and the causal future. This approach was first introduced as the \emph{multi-time formalism} \cite{AharonovEtAl_2008_TimeInQM,AharonovEtAl_2009_MultitimeStates,SilvaEtAl_2014_PreAndPostSelectedQuantumStates, Aharonovetal_2014_EachInstant, SilvaEtAl_2017_ConnectingIndefiniteWithMultiTime}, and was later generalized under the terminology of \emph{quantum combs}  \cite{ChirbiellaEtAl_2009_QuantumNetworkFramework, ChiribellaEtAl_2013_QuantumCompWithoutDefCausalStructure}, and \emph{process matrices} \cite{OreshkovEtAl_CorrelationsWithoutCausalOrder, OreshkovEtAl_2016_OperationalQuantumTheoryWithoutTime, OreshkovEtAl_2016_CausallySeparableProcesses, AraujoEtAl_2015_WitnessingCausalNonSeparability}. The papers that appear under these different banners tend to address different questions, and use different conventions, but the basic objects under study are similar.

The approach taken by this work to describe causally related quantum systems is again similar, with conventions tailored to our purposes that follow those of Ref.~\cite{AllenEtAl_2016_QCM}. A \emph{quantum node} is a pair of Hilbert spaces, one associated with the causal past of the system, or an incoming system, and one with the causal future, or an outgoing system. For a quantum node $A$, associated with a system with Hilbert space $\mathcal{H}_A$, the incoming Hilbert space is denoted $\mathcal{H}_{A^{\text{in} }} \equiv \mathcal{H}_A$. An intervention at the node is associated with a quantum instrument that maps states on the incoming Hilbert space to states on the outgoing Hilbert space. In the literature on quantum combs and process matrices, it is common to allow that the two Hilbert spaces are of different dimension, which is consistent as long as an intervention at the node is assumed that changes the dimension of the quantum system. The present work does not adopt this extra generality, for the reason that we wish to allow always for the possibility that there is no intervention at a given node, hence the outgoing Hilbert space has the same dimension as, and may be regarded as a copy of, the incoming Hilbert space. 

Our particular choice of CJ representation means that channels representing evolution between nodes are represented with operators that act on incoming Hilbert spaces and the duals of outgoing Hilbert spaces. Formulae become cluttered when the $^*$-symbol needs to be used frequently to denote the dual of a Hilbert space, so let $\mathcal{H}_{A^{\text{out}}}$ represent the dual of the outgoing Hilbert space, such that $\mathcal{H}_{A^{\text{out}}} \cong \mathcal{H}_A^*$. 

An intervention at a quantum node is a quantum instrument of the form $\left\{ \mathcal{E}^{k_A} : \mathcal{L}(\mathcal{H}_{A}) \rightarrow \mathcal{L}(\mathcal{H}_{A})\right\}$, where each of these maps has an associated operator $\rho^{\mathcal{E}^{k_A}}_{A|A} = \sum_{i,j} \ \mathcal{E}^{k_A}_A(\ket{i}_{A} \bra{j}) \otimes  \ket{i}_{A^*}\bra{j}$ in the CJ representation. When it is interventions at quantum nodes that are considered, rather than evolution between quantum nodes, it is a useful convention to represent the maps through the transposes of the CJ operators, which we denote $\tau^{k_A}_A$:
\begin{equation}
\tau^{k_A}_A := \left( \rho^{\mathcal{E}^{k_A}}_{A|A} \right)^T  \ . \label{Eq_Def_GeneralIntervention}
\end{equation}
A special case of an intervention at a quantum node $A$ is that of a quantum channel, corresponding to a CPTP map, which in general will be denoted $\tau_A$. There is, in this case, no classical outcome of the intervention (or, equivalently, the classical outcome takes a single value that occurs with probability $1$). The operator $\tau_A^{\text{id}}$ from Eq.~\ref{Eq_DefLinkOperator} is the representation of the identity channel, which can be interpreted as `no external intervention'. 

The evolution of systems outside the nodes is described by a \emph{quantum process operator}:
\begin{definition} \textnormal{(Quantum process operator):} \label{Def_ProcessOperator}
A \emph{quantum process operator} (\emph{process operator} for short) over $n$ quantum nodes $A_1 , \ldots , A_n$  is a positive semi-definite operator $\sigma_{A_1 ... A_n} \in \mathcal{L}( \bigotimes_i \mathcal{H}_{A^{\text{in}}_i} \otimes \mathcal{H}_{A^{\text{out}}_i})$, for which 
\[
\Trace [ \sigma_{A_1 ... A_n} \ \tau_{A_1} \otimes \cdots \otimes \tau_{A_n} ] = 1,
\] 
for any set of quantum channels $\{\tau_{A_i}\}$ at the $n$ nodes. 
\end{definition}

Given a process operator $\sigma_{A_1 ... A_n}$, suppose that an intervention is performed at each node, with the intervention at the $i$th node corresponding to $\{ \tau^{k_{A_i}}_{A_i} \}$. The joint probability of obtaining outcomes $k_{A_1},\ldots, k_{A_n}$ is given by\footnote{Here we see the reason for the convention of representing interventions at quantum nodes by the operators $\tau^{k_{A_i}}_{A_i}$ of Eq.~(\ref{Eq_Def_GeneralIntervention}): it enables outcome probabilities to be returned via a simple trace rule.}
\begin{equation}\label{Eq_quantumprocesstracerule}
P(k_{A_1}, \ldots , k_{A_n}) = \Trace \Big[ \ \sigma_{A_1 ... A_n} \ \tau^{k_{A_1}}_{A_1} \otimes \cdots \otimes \tau^{k_{A_n}}_{A_n}  \Big] \ .
\end{equation} 
\begin{remark}\label{Remark_Processoperators}
Given a process operator $\sigma_{A_1 ... A_n}$, it follows from Def.~\ref{Def_ProcessOperator} that
\[
\Trace_{A_1^{\text{in}} ... A_n^{\text{in}} } (\sigma_{A_1 ... A_n}) = \mathds{1}_{A_1^{\text{out}}} \otimes \cdots \otimes \mathds{1}_{A_n^{\text{out}}}.
\]
Hence, given the choice of CJ representation of Section~\ref{SubSec_QuantumPreliminaries}, a process operator over quantum nodes  $A_1 , \ldots , A_n$ defines formally a CPTP map $\mathcal{L}(\mathcal{H}_{A_1} \otimes \cdots \otimes \mathcal{H}_{A_n}) \rightarrow \mathcal{L}(\mathcal{H}_{A_1} \otimes \cdots \otimes \mathcal{H}_{A_n})$, from the Hilbert spaces representing outgoing systems at the quantum nodes to the Hilbert spaces representing incoming systems. However, it is not the case that any CPTP map $\mathcal{L}(\mathcal{H}_{A_1} \otimes \cdots \otimes \mathcal{H}_{A_n}) \rightarrow \mathcal{L}(\mathcal{H}_{A_1} \otimes \cdots \otimes \mathcal{H}_{A_n})$ corresponds to a valid process operator. For alternative characterizations of the set of valid process operators (equivalent, up to minor differences in conventions), see Refs.~\cite{OreshkovEtAl_CorrelationsWithoutCausalOrder,AraujoEtAl_2015_WitnessingCausalNonSeparability, OreshkovEtAl_2016_CausallySeparableProcesses}. 
\end{remark}

\subsection{Quantum definitions and notation 2}

A few more definitions and notational conventions will be useful in the following. If $S$ is a set of quantum nodes, let $\mathcal{H}_S := \bigotimes_{A\in S} \mathcal{H}_A$, let $\mathcal{H}_{S^{\text{in}}} := \bigotimes_{A\in S} \mathcal{H}_{A^{\text{in}}} = \mathcal{H}_S$, and let $\mathcal{H}_{S^{\text{out}}} := \bigotimes_{A\in S} \mathcal{H}_{A^{\text{out}}} \cong \mathcal{H}_S^*$. In a slight abuse of notation, if $A$ is a quantum node, let
\[
\Trace_A (\cdots ) := \Trace_{A^{\text{in}} A^{\text{out}}}(\cdots ),
\]
and similarly $\Trace_S (\cdots ) =  \Trace_{S^{\text{in}} S^{\text{out}}}(\cdots )$ for $S$ a set of quantum nodes.

Given a process operator $\sigma_V$, defined over a set $V$ of quantum nodes, let $S$ be a subset of $V$. A \emph{local intervention} at $S$ consists of an intervention performed separately at each quantum node in $S$. If, at each node $A\in S$, the intervention $\{ \tau^{k_A}_A \}$ is performed, then the joint outcome $k_S$ is a tuple specifying the value of $k_A$ for each $A\in S$, and corresponds to the product operator $\tau^{k_S}_S = \bigotimes_{A: A\in S} \tau_A^{k_A}$. This work mostly considers local interventions. On the occasions when an intervention on a set of nodes that is not a local intervention needs to be considered, it will be referred to as a \emph{global intervention}, and described explicitly.

Given a set $V$ of quantum nodes, and a subset $S\subset V$, let $T:= V \backslash S$, and let $\sigma_{ST}$ be a process operator defined over the quantum nodes $V$. The process operator defined over $S$ nodes alone, which returns the correct predictions for interventions at $S$, will in general depend on which, if any, interventions are performed at $T$ nodes. For $\tau_T$ a (local or global, trace-preserving) intervention at $T$, the resulting marginal process operator over nodes in $S$ is written 
\[
\sigma_S^{\tau_T} := \Trace_T ( \sigma_{ST} \,  \tau_T ).
\]
If there are no interventions at $T$ nodes, write simply\footnote{This notation is in keeping with an attitude adopted in this work, which is that no-intervention is the default assumption, relative to which interventions are only considered when explicitly specified.} 
\[
\sigma_S := \sigma_S^{\tau^{\text{id}}_T} = \Trace_T ( \sigma_{ST}  \,  \tau^{\text{id}}_T ) 
\]
It is convenient to introduce the short-hand notation 
\[
\overline{\Trace}_T[ X_{ST} ] := \Trace_T \big[X_{ST} \,  \tau^{\text{id}}_{T} \big],
\]
where $X_{ST}$ is an arbitrary operator on $\mathcal{H}_{S^{\text{in}}} \otimes \mathcal{H}_{S^{\text{out}}} \otimes \mathcal{H}_{T^{\text{in}}} \otimes \mathcal{H}_{T^{\text{out}}}$ . Using this notation, if $\sigma_{ST}$ is a process operator on sets of nodes $S$ and $T$, then the marginal at $S$ assuming no interventions at $T$ is given by
\[
\sigma_S =  \overline{\Trace}_T (\sigma_{ST}) .
\]

\subsection{Quantum causal models: definition \label{SubSec_QuantumCausationAndQCMs}} 

\begin{figure}[H]
\centering
\vspace*{0.3cm}
\small
\input{Figures/Fig_Exampleqcm.tex}
\caption{A quantum causal model.  \label{Fig_Exampleqcm}}
\end{figure}
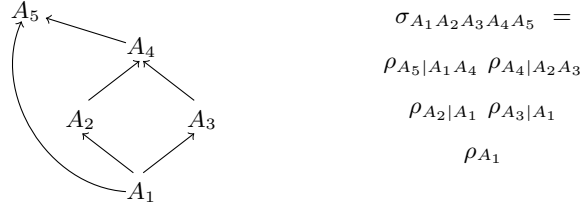

The framework of quantum causal models shares with classical causal models the graphical language of DAGs to represent causal structure. The following first appeared in Ref.~\cite{AllenEtAl_2016_QCM}.
\begin{definition}  \label{Def_QCM} \textnormal{(Quantum causal model):}  A quantum causal model is given by:
\begin{enumerate}[label=\textnormal{(\arabic*)}, leftmargin=2cm]
\item a causal structure represented by a DAG $G$ with vertices corresponding to quantum nodes $A_1, \ldots , A_n$, 
\item for each $A_i$, a quantum channel $\rho_{A_i | Pa(A_i)} \in \mathcal{L}(\mathcal{H}_{A_i} \otimes \mathcal{H}_{Pa(A_i)}^*)$, such that $[\rho_{A_i | Pa(A_i)} \ , \ \allowbreak \rho_{A_j | Pa(A_j)} ]=0$ for all $i,j$. 
\end{enumerate}
The quantum causal model defines a process operator over $A_1 , \ldots , A_n$, given by
\[
\sigma_{A_1 ... A_n} = \prod_i \rho_{A_i | Pa(A_i)}.
\]
\end{definition}
Fig.~\ref{Fig_Exampleqcm} shows an example of a quantum causal model, with the same causal structure as the classical causal model of Fig.~\ref{Fig_ExampleDAG}.
The following is then the analogue of Def.~\ref{Def_ClassicalMarkov}.
\begin{definition} \textnormal{(Quantum Markov condition)}:  \label{Def_QuantumMC}
Given a DAG $G$, with quantum nodes $A_1,...,A_n$, a process operator $\sigma_{A_1 ... A_n}$ is \emph{Markov} for $G$ if and only if there exists for each $i$ a quantum channel $\rho_{A_i | Pa(A_i)}  \in \mathcal{L}(\mathcal{H}_{A_i} \otimes \mathcal{H}_{Pa(A_i)}^*)$, such that $[ \rho_{A_i | Pa(A_i)} \ , \ \rho_{A_j | Pa(A_j) } ]=0$ for all $i,j$, and 
\begin{equation}
\sigma_{A_1 ... A_n} = \prod_{i=1}^n \rho_{A_i | Pa(A_i) }. \label{Eq_DefQuantumMarkov}
\end{equation}
\end{definition}

Defs.~\ref{Def_QCM} and \ref{Def_QuantumMC} have been presented in this section without any explanation or justification. This is provided in the next section, which defines quantum causal relations, shows that they can be represented by a directed graph, and explains why it is that the quantum process of a quantum causal model is Markov for the graph. 

\section{Quantum causal models and unitary dynamics \label{Sec_QCMsAndUnitaryDynamics}}

The purpose of this section is to analyse quantum causal structure in a manner similar to what was done in the classical case, but with unitary evolution replacing functions. We will establish that unitary evolution of quantum systems has a natural causal structure that can be represented by a DAG. 
Then we motivate and justify Def.~\ref{Def_QCM} with a quantum analogue of Theorem~\ref{Thm_ClassicalEquivalence}. 

\subsection{Motivation: why unitaries? \label{SubSec_quantumcausalmotivation}} 

\begin{figure}[H]
\centering
\small
\input{Figures/Fig_Example_BrokenCiruit.tex}
\caption{A broken cirucit with quantum nodes $A$, $B$, $C$, $D$ and $E$. \label{Fig_GeneralCPmapBrokenCircuit}}
\end{figure}
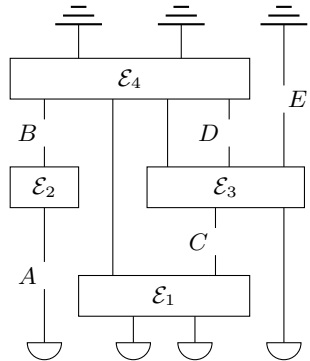

Before introducing unitary processes, let us consider a more general quantum evolution, represented by a circuit defined over multiple time steps, with gates corresponding to CPTP maps. Suppose that there is the possibility for an agent to intervene at some of the locations corresponding to wires in the circuit. Such a location corresponds to a quantum node. We draw these quantum nodes in a circuit diagram as broken wires, with the interpretation that the lower part of the wire carries the incoming Hilbert space and the upper part of the wire the outgoing Hilbert space, and refer to the resulting diagram as a broken circuit. (This is essentially a quantum network, as introduced in, e.g., Ref.~\cite{ChirbiellaEtAl_2009_QuantumNetworkFramework}.) An example is shown in Fig.~\ref{Fig_GeneralCPmapBrokenCircuit}.
The circuit defines outcome probabilities for any quantum interventions performed at the broken wires, hence defines a process operator $\sigma_{ABCDE}$ on the quantum nodes corresponding to the broken wires. This process operator is a quantum comb \cite{ChirbiellaEtAl_2009_QuantumNetworkFramework}.

Any such broken circuit defines a partial order over the quantum nodes, such that node $N$ precedes node $N'$ in the partial order if and only if there is a path from $N$ to $N'$ along the (broken or unbroken) wires of the circuit. Let us call this the \emph{causal order}. 
One may also discuss the signalling structure of a given process operator $\sigma$, which defines whether a subset of nodes $S$ can signal to another subset of nodes $T$, for some choice of interventions at the nodes not in $S$ or $T$. Here, `signal' means that agents stationed at $S$, by varying their choices of which interventions to perform, can vary the outcome probabilities for interventions performed by agents at $T$. If a process operator $\sigma_{ABCDE}$ is derived from a broken circuit, like that of Fig.~\ref{Fig_GeneralCPmapBrokenCircuit}, then the causal order of the circuit imposes constraints on the process operator, which in turn constrain signalling. For example, node $A$ can never signal to node $D$ in $\sigma_{ABCDE}$, since node $A$ does not precede node $D$ in the causal order defined by the circuit. 

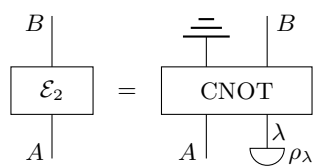
\begin{figure}[H]
\centering
\small

\renewcommand{\Scale}{1.8}
\newcommand{\xScale}{2.0}
\begin{tikzpicture}
	\node at (0.0*\Scale,0.0*\Scale) {$\mathcal{E}_2$};
	\draw (-0.6*\Scale, -0.35*\Scale) rectangle (0.6*\Scale, 0.35*\Scale);
	\node at (-0.25*\Scale, 1.0*\Scale) {$B$};
	\draw (0.0*\Scale, 0.35*\Scale) to (0.0*\Scale,1.1*\Scale);
	\node at (-0.25*\Scale, -0.9*\Scale) {$A$};
	\draw (0.0*\Scale, -0.35*\Scale) to (0.0*\Scale,-1.0*\Scale);
\end{tikzpicture}
\hspace*{0.1cm}
$=$
\hspace*{0.1cm}
\begin{tikzpicture}
	\node at (0.0*\Scale,0.0*\Scale) {$\text{CNOT}$};
	\draw (-1.1*\Scale, -0.35*\Scale) rectangle (1.1*\Scale, 0.35*\Scale);
	\node at (0.65*\xScale,1.0*\Scale) {$B$};
	\draw (0.4*\xScale, 0.35*\Scale) to (0.4*\xScale,1.1*\Scale);
	\node [style=upground] at (-0.4*\xScale,0.925*\Scale) {};
	\draw (-0.4*\xScale, 0.35*\Scale) to (-0.4*\xScale,0.8*\Scale);
	\node at (-0.65*\xScale,-0.9*\Scale) {$A$};
	\draw (-0.4*\xScale, -0.35*\Scale) to (-0.4*\xScale,-1.0*\Scale);
	\node [style=smallcpoint] at (0.4*\xScale, -0.96*\Scale) {};
	\node at (0.55*\xScale,-0.6*\Scale) {$\lambda$};
	\node at (0.85*\xScale,-0.97*\Scale) {\small{$\rho_{\lambda}$}};
	\draw (0.4*\xScale, -0.35*\Scale) to (0.4*\xScale,-0.85*\Scale);
\end{tikzpicture}

\caption{Realizsation of the channel $\rho^{\mathcal{E}_2}_{B|A} = \mathds{1}_{B^{\text{in}}}/2 \otimes \mathds{1}_{A^{\text{out}}}$ through the quantum Controlled-Not (CNOT) gate with control on $A$ and the maximally mixed state $\rho_{\lambda}=\mathds{1}_{\lambda}/2$. \label{Fig_Quantumfinetuningexample}}
\end{figure}

Importantly, causal influence should not simply be defined by the possibilities for signalling in a generic process operator. This is because there might always be additional unobserved variables such that there is causal influence even though there is no apparent signalling. This phenomenon was already seen in the classical case (with random variables instead of quantum nodes) in the example of Section~\ref{SubSec_RemarkClassicalCausalRelation}. For a quantum example, suppose that all systems in the circuit of Fig.~\ref{Fig_GeneralCPmapBrokenCircuit} are $2$-dimensional, and that the gate $\mathcal{E}_2$ is the CP map with CJ operator $\rho^{\mathcal{E}_2}_{B|A} = \mathds{1}_{B^{\text{in}}}/2 \otimes \mathds{1}_{A^{\text{out}}}$. Then there is no signalling from $A$ to $B$ in $\sigma_{ABCDE}$ since $\mathcal{E}_2$ simply ignores its input and prepares a maximally mixed state on $B^{\text{in}}$ (and the interventions at nodes $C,D,E$ are irrelevant). But a possible physical realization of this channel is as in Fig.~\ref{Fig_Quantumfinetuningexample}, where there is a maximally mixed state on an unobserved system $\lambda$, and a quantum Controlled-NOT gate acting on the output system at the $A$ node and on $\lambda$. Suppose that the preparation method for the $\lambda$ system is to toss a fair coin, and then to prepare $|0\rangle\langle 0|$ on heads and $|1\rangle\langle 1 |$ on tails. Agents not knowing the value of the coin toss will assign $\rho^{\mathcal{E}_2}_{B|A} = \mathds{1}_{B^{\text{in}}}/2 \otimes \mathds{1}_{A^{\text{out}}}$ and will not be able to signal from $A$ to $B$. But agents knowing the value of the coin toss will be able to send signals from $A$ to $B$, hence there must be causal influence from $A$ to $B$.

In a quantum universe, then, what is a suitable notion of quantum causal influence, distinct from the possibilities that a process operator $\sigma$ affords for agents to send signals to one another? We argue that as in the classical case, while an absence of apparent signalling can be due to fine-tuned unobserved systems, causal relations \emph{can} be defined in terms of signalling once the description of the scenario is expanded such that all relevant systems are included. In this case, we are dealing with a closed system. In the classical case, assuming determinism, relations between variables in a closed system are functional, and causal influence can be defined in terms of functional dependence, leading to the treatment of Sec.~\ref{Sec_CCMIntro}. In the quantum case, at least in standard quantum theory, evolution of a closed system is unitary. Hence our approach is to define quantum causal influence in terms of unitary evolution.

\subsection{Causal influence in unitary channels \label{SubSec_causalinfluenceunitary}} 

The task, then, is to analyze quantum causal structure in terms of unitary evolution, and to understand the constraints that causal structure imposes on more general quantum processes that are obtained as marginals when some systems are left out of the description (i.e., traced over). Before considering more complicated processes, we begin by analyzing the causal structure of an individual unitary transformation, in terms of an output having or not having a dependence on a given input. 
\begin{definition} \textnormal{(No-influence condition)} \label{Def_NoInfluenceCondition}
Let $U: \mathcal{H}_{A} \otimes \mathcal{H}_{B} \rightarrow \mathcal{H}_{C} \otimes \mathcal{H}_{D}$ be a unitary transformation, and let $\rho^U_{CD|AB}$ be the CJ representation of the corresponding channel. Write $A \nrightarrow D$ (`$A$ does not influence $D$') if and only if there exists a quantum channel $\mathcal{M}: \mathcal{L}(\mathcal{H}_B)\rightarrow \mathcal{L}(\mathcal{H}_D)$, with CJ representation $\rho^{\mathcal{M}}_{D|B}$ such that $\Trace_{C} [ \rho^U_{CD|AB} ] = \rho^{\mathcal{M}}_{D|B} \otimes \mathds{1}_{A^*}$.
\end{definition}
This condition was studied, for example, in Ref.~\cite{Schumacher_2005_LocalityNoInfluenceConditions}, where it is shown that it is equivalent to the following statement: for any choice of $\rho_B$, if the product state $\rho_A\otimes \rho_B$ is the input to the unitary channel, then $\rho_D = \Trace_C (U \rho_A \otimes \rho_B U^\dagger)$, the marginal state for system $D$ at the output, is independent of the choice of $\rho_A$. In other words: for any choice of $\rho_B$, it is not possible to send signals from $A$ to $D$ by varying the input state at $A$. Similar properties of channels were studied in Refs.~\cite{BeckmannEtAl_2011_CausalAndLocalisableQuantumOperations, EggelingEtAl_2002_SemicausalOperations}.
\begin{definition} \textnormal{(Direct cause):} \label{Def_CausalRelation}
Given a unitary channel $\rho^U_{CD|AB}$, say that $A$ is a direct cause of $D$ if and only if $A$ can influence $D$, i.e. $\neg (A \nrightarrow D)$ holds.
\end{definition}
This notion of direct cause in a unitary channel matches the classical notion of direct cause in terms of the output of a function having a dependence on a particular input. 

The following theorem is now an immediate corollary of Theorem~4 of Ref.~\cite{AllenEtAl_2016_QCM}. It will be useful in proving many of the results that follow and is interesting in its own right.
\begin{theorem} \textnormal{(Factorization of a unitary channel from no-influence conditions):} \label{Thm_FactorizationOfUnitary} 
Let $\rho^U_{B_1...B_k|A_1....A_n}$ be the CJ representation of a unitary channel with $n$ input and $k$ output systems. Let $S_i \subseteq \left\{ A_1,...,A_n \right\}$, $i=1,...,k$, be $k$ subsets of input systems such that there is no influence from the complementary sets to $B_i$, i.e., $A_j \nrightarrow B_i$ for all $A_j \notin S_i$. Then the operator factorizes in the following way
\begin{eqnarray}
\rho^U_{B_1...B_k|A_1....A_n} = \prod_{i=1}^k \rho_{B_i | S_i} \ ,
\end{eqnarray} 
where the marginal channels commute pairwise, $[\rho_{B_i | S_i} \ , \ \rho_{B_j | S_j} ]=0$ for all $i,j$.
\end{theorem}

This theorem has an important consequence, which we label as a Remark for future reference.
\begin{remark}\label{Remark_Causalinfluenceunitary}
Let $U$ be a unitary transformation, with inputs $A_1 , \ldots, A_n$, and outputs $B_1 , \ldots , B_k$, and suppose that $A_i$ does not influence $B_j$ and $A_i$ does not influence $B_{j'}$. A consequence of Theorem~\ref{Thm_FactorizationOfUnitary} is that if $B_j$ and $B_{j'}$ are grouped together to form a single system $B' = B_j B_{j'}$, then it is also the case that $A_i$ does not influence $B'$. In particular, it is not possible to signal from $A_i$ to the outputs $B_j$, $B_{j'}$, even if an agent stationed at the outputs performs joint measurements on $B_j B_{j'}$. It follows that the causal structure of a unitary transformation is completely specified by giving, for each output $B_j$, the subset of the inputs that are direct causes of $B_j$. The causal structure of a unitary transformation can thus be represented by drawing an arrow to each output $B_j$ from each input that is a direct cause of $B_j$. A similar remark holds trivially for classical functions with $n$ inputs and $k$ outputs. Note that a similar remark does not hold for the more general cases of non-unitary quantum channels, or non-deterministic classical channels. Here, full specification of the possibilities for signalling through the channel requires consideration of arbitrary subsets of inputs and outputs, and cannot be fully represented by drawing a graph with arrows from individual inputs to individual outputs. 
\end{remark}

 \subsection{Causal influence in unitary circuits \label{SubSec_causalinfluenceunitaryprocess}} 

Section~\ref{SubSec_causalinfluenceunitary} defined notions of no-influence and direct cause that apply to the inputs and outputs of a unitary channel. But the approach of this work is ultimately that the relata of quantum causal relations are quantum nodes, not the inputs and outputs of channels. This subsection therefore considers processes consisting of unitary evolution between quantum nodes, and by building on Section~\ref{SubSec_causalinfluenceunitary}, defines notions of no-influence and direct cause as relations between the nodes. 

\begin{figure}[H]
\centering
\small
\input{Figures/Fig_GenericBrokenUnitaryCiruit.tex}
\caption{A broken unitary circuit. \label{Fig_GeneralBrokenCircuit}}
\end{figure}
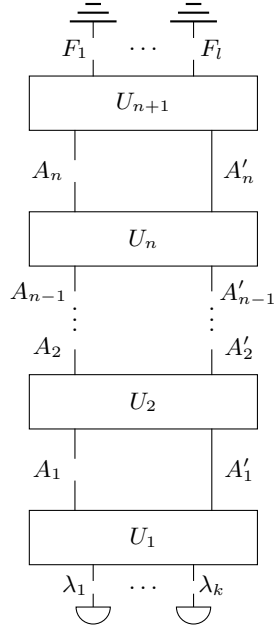

Consider a unitary circuit with some wires broken, as in the discussion of Section~\ref{SubSec_quantumcausalmotivation}, with independent initial preparations $\lambda_1 , \ldots , \lambda_k$, and with final output systems $F_1 , \ldots , F_l$. Any such circuit can be written in the form of that in Fig.~\ref{Fig_GeneralBrokenCircuit}, where the wires corresponding to inputs and outputs have also been broken, hence correspond to additional quantum nodes. Once the circuit is written in this form, the causal order defined by the original circuit structure is no longer apparent: for example, it may be that in the original circuit, $A_1$ and $A_2$ were parallel nodes, neither preceding the other. This does not matter, since causal structure will not be defined in terms of the diagrammatic representation of the circuit on the page.
The circuit defines a process operator on the quantum nodes corresponding to the broken wires. For the circuit of Fig.~\ref{Fig_GeneralBrokenCircuit}, the process operator is defined over quantum nodes $A_1 , \ldots , A_n , \lambda_1 , \ldots , \lambda_{k}, F_1, \ldots , F_l$, and is of the form
\[
\sigma_{A_1 ... A_n \lambda_1 ... \lambda_{k} F_1 ... F_l} = \rho_{A_1 ... A_n F_1 ... F_l | A_1 ... A_n \lambda_1 ... \lambda_k } \ \rho_{\lambda_1} \otimes\cdots\otimes \rho_{\lambda_{k}},
\]	
where 
\[
\rho_{A_1 ... A_n F_1 ... F_l | A_1 ... A_n \lambda_1 ... \lambda_{k}} = \overline{\Trace}_{A'_1...A'_n} \left[\rho^{U_{n+1}}_{F_1 ... F_l | A_n A'_n } \left(\prod_{i=2}^n \rho^{U_i}_{A_i A'_i | A_{i-1} A'_{i-1} } \right) \rho^{U_1}_{A_1 A'_1 | \lambda_1 ... \lambda_k }\right].
\]
It is easy to verify that the operator $\rho_{A_1 ... A_n F_1 ... F_l | A_1 ... A_n \lambda_1 ... \lambda_{k} }$ is the CJ representation of a unitary transformation. It is useful to define a term for process operators that have this form, i.e., consist of a unitary transformation multiplied by an initial product input state.
\begin{definition}\textnormal{(Unitary quantum process operator with inputs):} \label{Def_UnitaryProcess}
A quantum process operator $\sigma_{V}$ is a \textnormal{unitary quantum process operator with inputs}  (\emph{unitary process with inputs} for short)  if the nodes can be relabelled and $V$ partitioned as $V = P \cup S \cup R$, where $P = \{\lambda_1 , \ldots , \lambda_k \}$ , $S = \{ F_1 , \ldots , F_l \}$, and $R:= V \setminus (P \cup S$), such that $\sigma_{V} = \rho^{U}_{RS|RP} \ \rho_{\lambda_1}\otimes \cdots \otimes \rho_{\lambda_k}$, for some unitary transformation $U : \bigotimes_{N \in R\cup P} \mathcal{H}_{N} \rightarrow  \bigotimes_{N \in  R\cup S} \mathcal{H}_{N}$ and some states $\rho_{\lambda_i} \in \mathcal{L}(\mathcal{H}_{\lambda_i})$.
\end{definition}

The idea is now simple: a node (one of the $A$s or $\lambda$s) will be defined to be a direct cause of another node (one of the $A$s or $F$s) if the output space of the former is a direct cause of the input space of the latter in the unitary channel $\rho_{A_1 ... A_n F_1 ... F_l | A_1 ... A_n \lambda_1 ... \lambda_k }$ , according to Def.~\ref{Def_CausalRelation}. The causal structure is given by a directed graph defined over the quantum nodes, with an arrow between two nodes if the first is a direct cause of the second. This is made precise in the following definitions.

\begin{definition}\textnormal{(No-influence and direct cause in a unitary quantum process operator with inputs):}  \label{Def_DirectCauseInUnitaryProcess}
Consider a unitary process with inputs $\sigma_{RPS}$, where $R = \{ A_1 , \ldots , A_n \}$, $P = \{ \lambda_1 , \ldots , \lambda_k \}$, $S = \{ F_1 , \ldots , F_l \}$, and 
\[
\sigma_{RPS} =  \rho^{U}_{A_1 \ldots A_n F_1 \ldots F_l | A_1 \ldots A_n \lambda_1 \ldots \lambda_k}  \rho_{\lambda_1} \otimes \cdots \otimes \rho_{\lambda_k}.
\]
Write $N \nrightarrow N'$ (`$N$ does not influence $N'$') if and only if any of the following conditions hold:
\begin{itemize}
\item $N \in R\cup P$, $N' \in R\cup S$, and the output space of $N$ does not influence the input space of ${N'}$ in the unitary channel $\rho^{U}_{A_1 \ldots A_n F_1 \ldots F_l | A_1 \ldots A_n \lambda_1 \ldots \lambda_k} $.
\item $N \in S$.
\item $N' \in P$.
\end{itemize}
Node $N$ is a \textnormal{direct cause} of node $N'$, written $N \rightarrow N'$ if and only if $\neg (N \nrightarrow N')$. 
\end{definition}
\begin{definition}\textnormal{(Causal structure of a unitary quantum process operator with inputs):} \label{Def_Causalstructureunitaryprocess}
Consider a unitary process with inputs, $\sigma_{RPS}$. The \emph{causal structure} of $\sigma_{RPS}$ is a directed graph $G$, with nodes $R\cup P \cup S$, constructed such that for any nodes $N$, $N'$, there is an arrow in $G$ from $N$ to $N'$ if and only if $N$ is a direct cause of $N'$ in $\sigma_{RPS}$.
\end{definition}

In the case that the unitary process with inputs arises from an ordinary quantum circuit, as in Fig.~\ref{Fig_GeneralBrokenCircuit}, it is easy to verify that the graph is acyclic. Given a broken unitary circuit of the form of Fig.~\ref{Fig_GeneralBrokenCircuit}, let its causal structure be denoted by the DAG $G$. Then it is immediate from Theorem~\ref{Thm_FactorizationOfUnitary} that the process operator $\sigma_{A_1 ... A_n \lambda_1 ... \lambda_{k} F_1 ... F_l} $ is Markov for $G$. The next subsection turns to consider the marginal process operator on the nodes $A_1 , \ldots , A_n$.

\subsection{Equivalence of compatibility and Markovianity \label{SubSec_JustificationMC}}

We follow the treatment of the classical case in Section~\ref{Sec_CCMIntro}. There, causal structure was analyzed in terms of an underlying functional model governing the evolution of the variables. Given such a functional model, it is straightforward that causal structure can be represented by a DAG, with an arrow from $X$ to $Y$ if $Y$ has a direct dependency on $X$ in the functional description. The $(1)\rightarrow (2)$ direction of Theorem~\ref{Thm_ClassicalEquivalence} shows that if a probability distribution over observed variables is obtained from such a functional model by marginalizing over local noise sources, then the distribution is Markov for the DAG obtained by removing local noise sources. (For example, marginalizing over the $\lambda$s in Fig.~\ref{Fig_Examplefunctionalmodel} yields a distribution over $X$s that is Markov for the DAG of Fig.~\ref{Fig_ExampleDAG}.) The $(2)\rightarrow (1)$ direction shows that if a given probability distribution is Markov for a particular DAG, then there exists a functional model with additional local noise sources, where each variable is a function of its parents in the DAG and a local noise source, such that the probability distribution is recovered by marginalizing over the local noise sources. Taken together, these results motivate and justify Def.~\ref{Def_CCM}.

We motivate and justify Def.~\ref{Def_QCM} in a similar way with Theorem~\ref{Thm_QuantumEquivalence} below, which is the quantum analogue of Theorem~\ref{Thm_ClassicalEquivalence}. The idea is that, given a circuit like that of Fig.~\ref{Fig_GeneralBrokenCircuit}, with causal structure $G$, if each $\lambda_i$ is a direct cause of at most one of the $A_j$ (and any number of the $F$s), then with the $F$s ignored, each $\lambda_i$ constitutes a local noise source. Such $\lambda_i$ are not acting as common causes for the $A$ nodes, and are not responsible for correlations amongst the $A$ nodes. In this case, as the $(1)\rightarrow (2)$ direction of Theorem~\ref{Thm_QuantumEquivalence} shows, marginalizing over the $\lambda$s and the $F$s yields a process operator over $A_1 , \ldots , A_n$ that is Markov for the DAG $G'$ that is obtained from $G$ by removing the $\lambda$s and the $F$s. (This is the quantum analogue for the situation in which marginalizing over the $\lambda$s in Fig.~\ref{Fig_Examplefunctionalmodel} yields a distribution over $X$s that is Markov for the DAG of Fig.~\ref{Fig_ExampleDAG}.) The DAG $G'$ and the marginal process $\sigma_{A_1 ... A_n}$ are precisely what a quantum causal model describes.

One can also ask if any quantum causal model can be obtained in this manner: that is, given some DAG $G'$ with nodes $A_1 , \ldots , A_n$, and a process operator $\sigma_{A_1 ... A_n}$ that is Markov for $G'$, does there always exist a broken unitary circuit with additional local noise sources and output systems, such that each $A_i$ has as direct causes only its parents in $G'$ and a local noise source, and such that $\sigma_{A_1 ... A_n}$ is recovered by marginalizing over the local noise sources and output systems? If there does exist such a circuit, then we say that $\sigma$ is \emph{compatible} with $G$.  The $(2)\rightarrow (1)$ direction of Theorem~\ref{Thm_QuantumEquivalence} shows that such a circuit always exists.

The precise definition of compatibility is as follows, and generalizes Def.~\ref{Def_ClassCompDAG} to the quantum case.\footnote{Quantum compatibility with a DAG $G$, as we define it here, only requires the existence of a unitary process with inputs that satisfies the no-influence conditions dictated by $G$. That is, the \emph{absence} of causal relations must be respected, but there is no requirement that the nodes of the unitary process operator with inputs \emph{have} a direct cause relation whenever there is an arrow in $G$. The observant reader may have already noticed that the classical notion of compatibility given in Def.~\ref{Def_ClassCompDAG} is similar: in the functions that are asserted to exist by Def.~\ref{Def_ClassCompDAG}, there is no requirement that the value of a function actually have a non-trivial dependence on each of its arguments.} 
\begin{definition} \textnormal{(Quantum compatibility with a DAG)}: \label{Def_QuantumCompatibility}
Given a DAG $G$ with nodes $A_1,...,A_n$, a quantum process operator $\sigma_{A_1 ... A_n}$ is \emph{compatible} with $G$ if and only if there exists a unitary process with inputs, $\sigma_{A_1 \ldots A_n \lambda_1  \ldots \lambda_n F} =\rho^U_{A_1 ...A_nF|A_1...A_n\lambda_1...\lambda_n} \  \rho_{\lambda_1} \otimes\cdots\otimes\rho_{\lambda_n}$, such that 
\begin{enumerate}
\item The quantum process operator $\sigma_{A_1 ... A_n}$ is recovered as the marginal process on nodes $A_1 , \ldots , A_n$:
\begin{equation}
\sigma_{A_1 ... A_n} \ = \ \overline{\Trace}_{\lambda_1...\lambda_nF} \left[ \rho^U_{A_1...A_nF|A_1...A_n\lambda_1...\lambda_n} \ \rho_{\lambda_1} \otimes\cdots\otimes \rho_{\lambda_n} \right], \label{Eq_UnitaryDilationForSigma}
\end{equation}
\item The unitary process with inputs satisfies the following no-influence conditions (with $Pa(A_i)$ referring to $G$):
\begin{equation}
\left\{ A_j \nrightarrow A_i \right\}_{A_j \notin Pa(A_i)} \ , \ \left\{ \lambda_j \nrightarrow A_i \right\}_{j \neq i}. \label{Eq_CompatibilityNoInfluenceCond} 
\end{equation}
\end{enumerate}		
\end{definition}

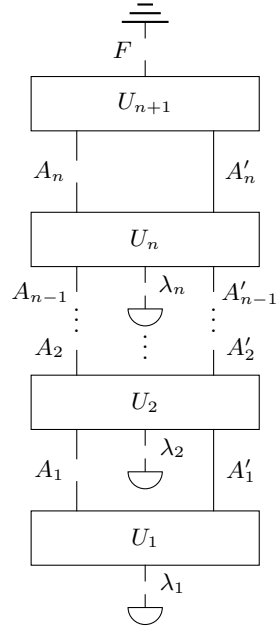
\begin{figure}[H]
\centering
\small
\input{Figures/Fig_JustificationCircuit_NewNotation_2.tex}
\caption{A broken unitary circuit as implied to exist by compatibility of a quantum process operator with a DAG. Each $\lambda_i$ is a direct cause of no other nodes, except possibly $A_i$ and $F$, hence the circuit can always be written in this form, rather than as in Fig.~\ref{Fig_GeneralBrokenCircuit}. There is no loss of generality in assuming a single output system $F$, since multiple outputs can be assumed collected together into a single system labelled $F$. \label{Fig_JustificationCircuit}}
\end{figure}

Importantly, if a quantum process operator $\sigma_{A_1 ... A_n}$ is compatible with a DAG $G$, then the unitary process with inputs, $\rho^U_{A_1...A_nF|A_1...A_n\lambda_1...\lambda_n} \ \rho_{\lambda_1}\otimes\cdots\otimes\rho_{\lambda_n}$, that this implies can always be realized as a broken unitary circuit. This is not completely obvious, hence we state it in the form of a theorem.  
\begin{theorem}\label{Thm_Compatibilitywithbroken}\textnormal{(Circuit form for underlying unitary transformation}):
Consider a DAG $G$ with nodes $A_1 , \ldots , A_n$, labelled such that the total order $A_1 < \cdots < A_n$ is compatible with the partial order defined by $G$. Suppose that a process operator $\sigma_{A_1 ... A_n}$ is compatible with $G$. Let $\sigma_{A_1...A_n  \lambda_1 \ldots \lambda_n F}$ be the unitary process with inputs, whose existence is asserted by the definition of compatibility, such that  
\[
\sigma_{A_1 ... A_n} \ = \ \overline{\Trace}_{\lambda_1...\lambda_n F} \left[ \sigma_{A_1...A_n  \lambda_1 \ldots \lambda_n F} \right],
\]
and such that the no-influence conditions~\ref{Eq_CompatibilityNoInfluenceCond} hold. Then there exists a broken unitary circuit $\mathcal{C}$, of the form of Fig.~\ref{Fig_JustificationCircuit}, such that $\mathcal{C}$ is a realization of $\sigma_{A_1...A_n  \lambda_1 \ldots \lambda_n F}$, i.e., 
\[
\sigma_{A_1...A_n  \lambda_1 \ldots \lambda_n F} = \overline{\Trace}_{A'_1...A'_n} \left[\rho^{U_{n+1}}_{F | A_n A'_n} \left(\prod_{i=2}^n \rho^{U_i}_{A_i A'_i | A_{i-1} A'_{i-1} \lambda_i } \right) \rho^{U_1}_{A_1 A'_1 | \lambda_1 }\right] \left( \rho_{\lambda_1}\otimes\cdots\otimes\rho_{\lambda_n} \right).
\]
\end{theorem}
\noindent {\bf Proof.} See Appendix~\ref{SubSec_Proof_Thm_Compatibilitywithbroken}. \hfill $\square$

Finally, the analogue of Theorem~\ref{Thm_ClassicalEquivalence} for quantum causal models is:
\begin{theorem} \textnormal{(Equivalence of quantum compatibility and Markovianity)}: \label{Thm_QuantumEquivalence}
Given a DAG $G$ with nodes $A_1, \ldots ,A_n$, and a process operator $\sigma_{A_1 ... A_n}$ , the following are equivalent:
\begin{enumerate}
\item $\sigma_{A_1 ... A_n}$ is compatible with $G$.
\item $\sigma_{A_1 ... A_n}$ is Markov for $G$.
\end{enumerate}
\end{theorem}

\noindent {\bf Proof: $(1)\rightarrow (2)$.} By Theorem~\ref{Thm_FactorizationOfUnitary}, the no-influence conditions specified in Eq.~\ref{Eq_CompatibilityNoInfluenceCond} of the definition of compatibility imply
\begin{eqnarray}
	\rho^U_{A_1...A_nF|A_1...A_n\lambda_1...\lambda_n} = \rho_{F|A_1...A_n\lambda_1...\lambda_n} \Big( \prod_i \ \rho_{A_i|Pa(A_i) \lambda_i} \Big) \ . \label{Eq_CompatibilityFactorization}
\end{eqnarray}
Marginalizing over the $\lambda_i$ and $F$ gives
\begin{eqnarray}
	\sigma_{A_1 ... A_n} &=& \overline{\Trace}_{\lambda_1...\lambda_n} \Trace_{F} \left[ \rho^U_{A_1...A_nF|A_1...A_n\lambda_1...\lambda_n} \ \rho_{\lambda_1} \otimes\cdots\otimes \rho_{\lambda_n} \right] \nonumber \\
			&=& \Trace_{F} \big[ \rho_{F|A_1...A_n\lambda_1...\lambda_n} \big] \
		 \Big( \prod_i \ \overline{\Trace}_{\lambda_i} \big[ \ \rho_{A_i|Pa(A_i)\lambda_i} \ \rho_{\lambda_i} \ \big] \Big) \label{Eq_MarkovStability} \\
			&=& \prod_{i} \rho_{A_i|Pa(A_i)} \ , 
\end{eqnarray}
where $[ \rho_{A_i | Pa(A_i)},  \rho_{A_j | Pa(A_j)}] = 0$ for all $i,j$, since $[\rho_{A_i | Pa(A_i)\lambda_i} ,\rho_{A_j | Pa(A_j)\lambda_j} ] = 0$ for all $i,j$. Hence $\sigma_{A_1 ... A_n}$ is Markov for $G$.  \hfill $\square$
\\ \\
\noindent {\bf Proof: $(2)\rightarrow (1)$.} See Appendix~\ref{Subsec_Proof_Thm_QuantumEquivalence21}. \hfill $\square$

\subsection{Quantum causal influence, signalling, fine tuning, and causal inference}

We end this section with some remarks concerning quantum causal relations, which we have defined for a broken unitary circuit with a product state input, and which can be represented as a DAG. Note first that the fact that quantum causal relations can be represented by a DAG \emph{at all} (as opposed to some other kind of mathematical object, say a partial order, or some kind of hyper-graph) is not assumed in this work. It is, rather, a derived result, a consequence of Theorem~\ref{Thm_FactorizationOfUnitary}, and in particular of Remark~\ref{Remark_Causalinfluenceunitary}, from which it can be argued that the graph is a complete representation of the causal structure. 

Note that given a particular broken circuit, the arrangement of gates itself defines a DAG over the quantum nodes, with an arrow from node $A_i$ to node $A_j$ if and only if there is a path from $A_i$ to $A_j$ along the unbroken wires of the circuit. Let us call this the \emph{connectivity DAG} associated with the broken circuit. Of course, the gate arrangement is no longer visible when the circuit is drawn in the form of Fig.~\ref{Fig_GeneralBrokenCircuit}. But the connectivity DAG can be taken as given by the original, more general, form of the circuit. Importantly, the DAG representing the quantum causal structure should be distinguished from the connectivity DAG, and is in general a subgraph of the connectivity DAG. This is because causal structure is not defined simply by the presence or absence of pathways in the circuit: the unitary transformations that appear as gates can have, individually and in concert, non-trivial no-influence relations between inputs and outputs that are not captured by information about the gate arrangement alone.\footnote{Any DAG defines a partial order such that node $N$ precedes node $N’$ in the partial order if and only if there is a path from $N$ to $N’$ along the arrows of the DAG. In the case of the connectivity DAG, the associated partial order is the causal order defined in Section~\ref{SubSec_quantumcausalmotivation}. The DAG representing causal structure also defines a partial order, in general stricter than the causal order.}

As discussed in Section~\ref{SubSec_quantumcausalmotivation}, the quantum causal structure should also be distinguished from the possibilities for signalling that a process operator affords. 
Just as in the classical case, there can be situations where a node $A_i$ is a direct cause of $A_j$, yet signalling from $A_i$ to $A_j$ is not possible in the marginal process $\sigma_{A_1 ... A_n}$ because it is `washed out' by a fine-tuned value of the state on an unobserved system. Furthermore this idea generalizes just as it does in the classical case (see Section~\ref{SubSec_RemarkClassicalCausalRelation}): the framework of quantum causal models allows a distinction to be made between features of a process operator that are forced by causal structure, and features that are merely accidental and arise as a result of fine tuning. Consider a quantum causal model with DAG $G$ and a process operator $\sigma_{A_1 ... A_n}$ that is Markov for $G$, hence compatible with $G$. In the broken unitary circuit that is asserted to exist by Theorems~\ref{Thm_Compatibilitywithbroken} and \ref{Thm_QuantumEquivalence}, suppose that the input states $\rho_{\lambda_i}$ are modified, to become $\rho'_{\lambda_i}$, while keeping the unitary transformations of the circuit unchanged. Marginalizing over the $\rho'_{\lambda_i}$, one may calculate a new process operator $\sigma'_{A_1 ... A_n}$, in general different from $\sigma_{A_1 ... A_n}$. However, $\sigma'_{A_1 ... A_n}$ will be (by definition) compatible with $G$, hence (by Theorem~\ref{Thm_QuantumEquivalence}) Markov for $G$. This means that $\sigma'_{A_1 ... A_n}$ will also satisfy all constraints that are forced by Markovianity for $G$ -- these constraints, which arise from the causal structure itself, should be distinguished from any accidental properties of $\sigma_{A_1 ... A_n}$ that arise in virtue of a specific, fine tuned, choice of distributions on the $\rho_{\lambda_i}$. 

Any time quantum causal structure is being inferred from limited data (where the limited data might be, say, the classical measurement outcomes that have been obtained from interventions that constitute only partial tomography on only a subset of the quantum nodes), we suggest a desideratum matching the one that Section~\ref{SubSec_RemarkClassicalCausalRelation} describes for the classical case: a model is to be preferred if features of a quantum process corresponding to equalities satisfied by the process operator are seen as forced by causal structure, rather than as arising due to finely tuned states on unobserved systems. 

The equality constraints satisfied by marginal quantum processes that follow from Markovianity of the overall quantum process, i.e., are forced by causal structure, are investigated in detail in Sections~\ref{Sec_IndependenceNotions}, \ref{Sec_DSeparationTheorem}, \ref{Sec_DoCalculus_CCM} and \ref{Sec_QuantumDoCalculus}.

\section{Classical split nodes, classical process maps, and classical split-node causal models \label{Sec_ClassicalSplitNodeFormalism}}

This section introduces the notion of a \emph{classical split-node causal model}\footnote{In Ref.~\cite{AllenEtAl_2016_QCM} called `classical interventional model'.} , which is a closer analogue to a quantum causal model than the standard classical causal models described in Section~\ref{Sec_CCMIntro}, and is essential to the discussions of independence and the do-calculus in Sections~\ref{Sec_IndependenceNotions}, \ref{Sec_DSeparationTheorem} and \ref{Sec_QuantumDoCalculus} below.

\subsection{Classical split nodes and classical process maps}

By analogy with a quantum node, which is a pair of Hilbert spaces, let a \emph{classical split node} $X$ consist of two `copies' of a random variable,  the input variable $X^{\text{in}}$ and the output variable $X^{\text{out}}$, where the output variable ranges over the same set of values as the input variable. The interpretation is much the same as that of a quantum node: the classical split node represents a locus at which an agent may intervene, with the possibility that the intervention has a classical outcome of its own. An intervention at node $X$ corresponds to a \emph{classical instrument} $P(k_X,X^{\text{out}}|X^{\text{in}})$, where a separate variable $k_X$ records the outcome of the intervention. A special case (analogous to a quantum channel) is that of a classical channel, i.e., an intervention with no outcome of its own corresponding to $P(X^{\text{out}}|X^{\text{in}})$. A noise-free, non-disturbing measurement of the variable $X$ corresponds to the intervention $P(k_X,X^{\text{out}} | X^{\text{in}})\linebreak[1]=\delta(k_X,X^{\text{in}})\delta(X^{\text{out}},X^{\text{in}})$. No intervention corresponds to the classical channel $P(X^{\text{out}}|X^{\text{in}})=\delta(X^{\text{out}},X^{\text{in}})$, in analogy with the identity channel $\tau^{\text{id}}_A$ for a quantum node. 

A framework to study the most general, logically consistent classical processes connecting classical split nodes has been introduced by Baumeler, Feix and Wolf in \cite{BaumelerEtAl_2014_MaximalIncompatibilityClassicalBehaviourGlobalCausalOrder} \footnote{Therein classical split nodes are called `local laboratories'.} and studied extensively in \cite{BaumelerEtAl_2014_MaximalIncompatibilityClassicalBehaviourGlobalCausalOrder} and \cite{Baumeler_EtAl_2016_SpaceOfLOgicallyConsistentClassicalProcesses}. 
The following defines these processes using our terminology\footnote{What we call a classical process map is called an `environment' in \cite{Baumeler_EtAl_2016_SpaceOfLOgicallyConsistentClassicalProcesses}.}, and mirrors the introduction of quantum process operators in Section~\ref{Sec_QCMIntro}.
\begin{definition} \textnormal{(Classical process map):} \label{Def_Classicalprocess}
A \emph{classical process map} $\clop_{X_1 ... X_n}$ over classical split nodes $X_1 , \ldots , X_n$  is a map 
\[
\clop_{X_1 ... X_n} : X_1^{\text{in}}\times X_1^{\text{out}} \times \cdots\times X_n^{\text{in}}\times X_n^{\text{out}} \rightarrow [0,1],
\]
such that
\[
\sum_{X_1^{\text{in}}, X_1^{\text{out}},...,X_n^{\text{in}}, X_n^{\text{out}}} \left(\clop_{X_1 ... X_n} \prod_i P(X_i^{\text{out}} | X_i^{\text{in}})\right) = 1,
\]
for any set of classical channels $\{ P(X_i^{\text{out}} | X_i^{\text{in}})  \}$. 
\end{definition}
Given a classical process map, and an intervention at each node, the joint probability of obtaining outcomes $k_1 , \ldots , k_n$ is given by the analogue of Eq.~(\ref{Eq_quantumprocesstracerule}):
\begin{equation}
P(k_1,...,k_n)=\sum_{X_1^{\text{in}}, X_1^{\text{out}},...,X_n^{\text{in}}, X_n^{\text{out}}} \Big[ \ \clop_{X_1 ... X_n} \ \Big( \prod_i P(k_i,X_i^{\text{out}}|X_i^{\text{in}}) \Big) \ \Big] \ . \label{Eq_CIMOutcomeProbs}
\end{equation}
The definition of a classical process map ensures that for any set of interventions, the joint outcome probabilities satisfy $0 \leq P(k_1,...,k_n) \leq 1$, and $\sum_{k_1...k_n} P(k_1,...,k_n) = 1$.\footnote{The study of classical process maps in the literature has focused largely on the causally indefinite case, that on classical process maps that are not compatible with any particular causal ordering of the nodes, nor even with a probabilistic mixture of different causal orderings. The first such classical process map was presented in \cite{BaumelerEtAl_2014_MaximalIncompatibilityClassicalBehaviourGlobalCausalOrder}. The present work is concerned only with causally definite quantum process operators and classical process maps, for which the causal structure can be described with a DAG.}
\begin{remark}\label{Remark_Classicalprocesses}
(cf Remark~\ref{Remark_Processoperators} for the quantum case.) Given a classical process map $\clop_{X_1 ... X_n}$, it follows from Def.~\ref{Def_Classicalprocess} that
\[
\sum_{X_1^{\text{in}} ... X_n^{\text{in}} } (\clop_{X_1 ... X_n})\big|_{X_1^{\text{out}} = x_1 , \ldots ,  X_n^{\text{out}} = x_n} = 1 \qquad \forall x_1,\ldots x_n.
\]
Hence a classical process map over classical split nodes  $X_1 , \ldots , X_n$ defines formally a classical channel $P(X_1^{\text{in}}\ldots \allowbreak X_n^{\text{in}} \allowbreak | \allowbreak X_1^{\text{out}} \ldots X_n^{\text{out}})$. However, it is not the case that any classical channel $P(X_1^{\text{in}} \ldots X_n^{\text{in}} | \allowbreak X_1^{\text{out}} \ldots \allowbreak X_n^{\text{out}})$ corresponds to a valid classical process map \cite{Baumeler_EtAl_2016_SpaceOfLOgicallyConsistentClassicalProcesses}.
\end{remark}

\subsection{Classical definitions and notation}

The classical definitions and notation introduced in this section closely match those of Section~\ref{Sec_QCMIntro}. If $S$ is a set of classical split nodes, let $S^{\text{in}}$ denote the collection of random variables consisting of $X^{\text{in}}$ for each $X\in S$, and similarly $S^{\text{out}}$. In a slight abuse of notation, if $X$ is a classical split node, let
\[
\sum_X (\cdots ) := \sum_{X^{\text{in}} X^{\text{out}}}(\cdots ),
\]
and similarly $\sum_S (\cdots ) =  \sum_{S^{\text{in}} S^{\text{out}}}(\cdots )$ for $S$ a set of classical split nodes.

Given an outcome $k_X$ of an intervention at a classical split node $X$, the corresponding $P(k_X, X^{\text{out}} | X^{\text{in}})$ (seen as a map $X^{\text{out}}\times X^{\text{in}}\rightarrow [0,1]$) is written $\tau^{k_X}_X$. This notation does not distinguish classical from quantum interventions, but the meaning will always be obvious in context. If the intervention is a classical channel (i.e., has no outcomes of its own), then it is written $\tau_X$. Similarly, if $S$ is a set of classical split nodes, write $\tau^{k_S}_S$ for the maps corresponding to outcomes of an intervention at $S$, and $\tau_S$ for an intervention at $S$ with no outcome. A local intervention at $S$ is defined in a manner similar to the quantum case. When it appears in a classical split-node context, let $\tau^{\text{id}}_S = \delta (S^{\text{out}} , S^{\text{in}})$.

The notation for marginals is also defined in a similar manner to the quantum case. Consider a classical process map $\clop_{ST}$, defined over sets of classical split nodes $S$ and $T$. For $\tau_T$ a (local or global) intervention at $T$, the resulting marginal classical process map over $S$ is written 
\[
\clop_S^{\tau_T} := \sum_T ( \clop_{ST} \,  \tau_T ).
\]
If there are no interventions at $T$ nodes, write simply
\[
\clop_S := \clop_S^{\tau^{\text{id}}_T} = \sum_T ( \clop_{ST}  \,  \tau^{\text{id}}_T ). 
\]

\subsection{Classical split-node causal models}

\begin{definition} \label{Def_CSM} \textnormal{(Classical split-node causal model):} A classical split-node causal model is given by
\begin{enumerate}[label=\textnormal{(\arabic*)}, leftmargin=2cm]
\item a causal structure represented by a DAG $G$ with vertices corresponding to classical split nodes $X_1, \ldots ,X_n$
\item for each $X_i$, a classical channel $P(X_i^{\text{in}} | Pa(X_i)^{\text{out}} )$.
\end{enumerate}
The classical split-node causal model defines a classical process map $\clop_{X_1 ... X_n}$, given by
\[
\clop_{X_1...X_n} :=\prod_i P(X_i^{\text{in}}|Pa(X_i)^{\text{out}}).
\]
\end{definition}
\begin{definition}\textnormal{(Classical split-node Markov condition):}
Given a DAG $G$, with classical split nodes $X_1 , \ldots , X_n$, a classical process map $\clop_{X_1 ... X_n}$ is \emph{Markov} for $G$ if and only if there exists for each $i$ a classical channel $P(X_i^{\text{in}}|Pa(X_i)^{\text{out}})$, such that
\begin{equation}
\clop_{X_1 ... X_n} = \prod_i P(X_i^{\text{in}}|Pa(X_i)^{\text{out}}). \label{Eq_DefClassicalsplitnodeMarkov}
\end{equation}
\end{definition}
It is a straightforward matter to define a notion of compatibility with a DAG for classical process maps, in analogy with Defs.~\ref{Def_ClassCompDAG} and \ref{Def_QuantumCompatibility}, and to derive a theorem to the effect that compatibility is equivalent to Markovianity. These aspects will not be essential in the following, hence we omit the details.

\subsection{Overview of classical and quantum processes and causal models}\label{Sec_ClassicalQauntumComparison}

The following relationships between the different types of processes and causal models will be useful. They are labelled for ease of reference.
\begin{itemize}[leftmargin=1.9cm]
\item[{\bf $I_{\sigma\rightarrow\kappa}$}] Given a quantum process operator $\sigma_{A_1...A_n}$, suppose that there exists an orthonormal basis at each node (that is, an orthonormal basis for $\mathcal{H}_{A_i^{\text{in}}}$, along with the dual basis for $\mathcal{H}_{A_i^{\text{out}}}$), such that $\sigma_{A_1...A_n}$ is diagonal with respect to the product of these bases. Then the quantum process operator defines a classical process map, with in and out variables $X_i^{\text{in}}$, $X_i^{\text{out}}$ at the $i$th node labelling the basis elements of $\mathcal{H}_{A_i^{\text{in}}}$ and $\mathcal{H}_{A_i^{\text{out}}}$, and the diagonal entries of $\sigma_{A_1...A_n}$ interpreted as $P(X_1^{\text{in}} , \ldots , X_n^{\text{in}} | X_1^{\text{out}} , \ldots , X_n^{\text{out}})$. If the quantum process operator is Markov for a particular DAG over $A_1,\ldots ,A_n$, then the induced classical process map is Markov for the equivalent DAG over $X_1,\ldots ,X_n$.
\item[{\bf $I_{\kappa\rightarrow\sigma}$}] A classical process map straightforwardly induces a quantum process operator by interpreting the variables $X_i^{\text{in}}$ and $X_i^{\text{out}}$ as labelling the elements of an orthonormal basis of $\mathcal{H}_{A_i^{\text{in}}}$, and the dual basis of $\mathcal{H}_{A_i^{\text{out}}}$, respectively, and by encoding the conditional probabilities $P(X_1^{\text{in}} , \allowbreak \ldots , \allowbreak X_n^{\text{in}} | X_1^{\text{out}}, \allowbreak \ldots , X_n^{\text{out}})$ as the diagonal elements of a matrix, which is then interpreted as a quantum process operator $\sigma_{A_1...A_n}$. If the classical process map is Markov for a particular DAG over $X_1, \ldots , X_n$, then the induced quantum process operator is Markov for the equivalent DAG over $A_1 , \ldots , A_n$.
\item[{\bf $I_{\kappa\rightarrow P}$}] A classical process map $\kappa_{X_1...X_n}$ straightforwardly induces a classical probability distribution $P(X_1 , \allowbreak \ldots , \allowbreak X_n)$ by identifying input with output variables, and marginalizing over input variables, to obtain
\[
P(X_1^{\text{out}}, \ldots , X_n^{\text{out}}) = \sum_{X_1^{\text{in}}...X_n^{\text{in}}} \left(\clop_{X_1 ... X_n} \prod_i \delta(X_i^{\text{in}} , X_i^{\text{out}})\right),
\]
and then identifying each variable $X_i^{\text{out}}$ with a single variable $X_i$ such that  $P(X_1, \allowbreak \ldots , \allowbreak X_n) = P(X_1^{\text{out}} , \allowbreak \ldots , \allowbreak X_n^{\text{out}})$. If the classical process map is Markov for a particular DAG over classical split nodes $X_1, \ldots , X_n$, then the probability distribution $P(X_1 , \ldots , X_n)$ is Markov for the equivalent DAG over random variables $X_1 , \ldots , X_n$.
\item[{\bf $I_{CCM\rightarrow CSM}$}] Given a classical causal model, with DAG $G$ and channels $P(X_i | Pa(X_i))$, a classical split-node causal model is straightforwardly induced by replacing each variable $X_i$ with the pair $X_i^{\text{in}}, X_i^{\text{out}}$, and by replacing the channels $P(X_i | Pa(X_i) )$ with $P(X_i^{\text{in}} | Pa(X_i)^{\text{out}} )$. 
\end{itemize}
Observe that the first three of these inductions can be applied to the objects $\sigma_{A_1...A_n}$ and $\kappa_{X_1...X_n}$ without specification of any particular DAG (or even the assumption that there exists a DAG for which they are Markov). The induction $I_{CCM\rightarrow CSM}$, on the other hand, takes one type of causal model to another type of causal model, with specification of the DAG required. The reason for this is that, even in the absence of a DAG, classical and quantum processes encode causal information in a way that a joint probability distribution does not. 

\begin{figure}[H]
\centering
\small
\input{Figures/Fig_SchematicalOverviewCausalModels.tex}
\caption{Summary of relationships between the notions of quantum causal model (QCM), classical split-node causal model (CSM) and classical causal model (CCM).
\label{Fig_Schematicoverviewcausalmodels}
}
\end{figure}
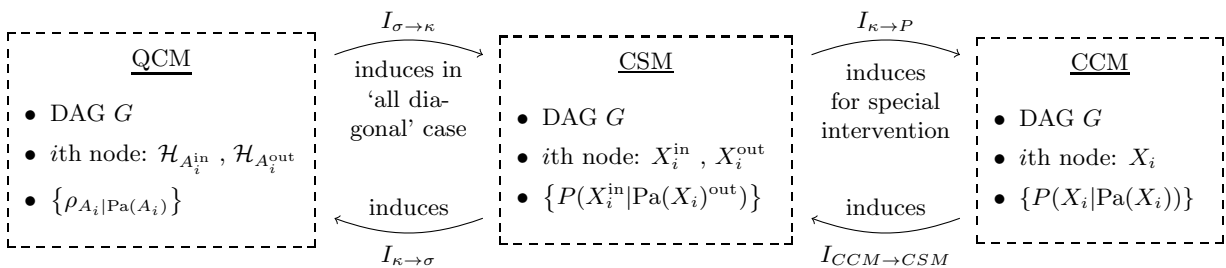

When these inductions are applied to processes that do result from causal models, they yield relationships between the different types of causal models, as summarized in Fig.~\ref{Fig_Schematicoverviewcausalmodels}. Note in particular that classical causal models and classical split-node causal models are in 1-to-1-correspondence.

\section{Quantum and classical split-node do-interventions \label{Sec_QuantumDoInterventions}} 

The do-interventions of Section~\ref{SubSec_dointerventions} have a natural formulation in quantum and classical split-node causal models, wherein a do-intervention at a node (or set of nodes) fixes an output state, independently of the input state. This corresponds to overriding the mechanisms that would normally fix the output to be identical with the input, with the input state determined via the outputs of parental nodes.
\begin{definition} \textnormal{(do-conditional process operator):} \label{Def_QuantumDoCond}
Consider a set of quantum nodes $V$, with $S\subset V$ and $T=V\setminus S$, and let $\sigma_{ST}$ be a process operator over the nodes in $V$. The \emph{do-conditional process operator} for a do-intervention on $S$ is given by
\begin{equation}
\sigma_{T do(S)} \ := \ \Trace_{S^{\text{in}}} (\sigma ). \label{Eq_QuantumDoCondDef}
\end{equation}
\end{definition}
\noindent A do-conditional process operator
$\sigma_{T do(S)}$ is an operator on $T^{\text{in}} \otimes T^{\text{out}} \otimes S^{\text{out}} $, such that if a fixed state $\rho_S \in \mathcal{L}(\mathcal{H}_S)$ is prepared and released to the outputs of $S$, the process operator on the remaining nodes $T$ is given by
\[
\sigma_{T do(S = \rho_S)} = \overline{\Trace}_S \left( \sigma_{T do(S)} \rho_S \right).
\]
If $\sigma_{ST}$ is Markov for $G$, then $\sigma_{T do(S)}=\prod_{A_i\not \in S} \rho_{A_i|Pa(A_i)}$, in analogy with the classical truncated factorization formula Eq.~(\ref{Def_TruncatedFactorization}). \footnote{An important part of the notion of mechanisms for classical causal models is their stability and autonomy. While the tensor product structure in quantum theory ensures that local interventions by definition do not affect the channels associated with other nodes, that one can actually in practice implement a quantum instrument without interacting with any other node constitutes an equally substantial assumption as in the classical case.}

In the case of a classical process map, similarly defined over nodes $V$, with $S\subset V$ and $T=V\setminus S$, the \emph{do-conditional process map} $\clop_{T do(S)}$ is given by
\[
\clop_{T do(S)} = \sum_{S^{\text{in}}} (\clop_{ST} ),
\]
which is consistent with Eq.~\ref{Eq_QuantumDoCondDef} if $\clop_{ST}$ is interpreted as a diagonal process operator. This definition also reproduces Eq.~\ref{Def_TruncatedFactorization} for $P(T | do(S))$ if the $T$ nodes are reduced to the single variables of a classical causal model via $I_{\kappa\rightarrow P}$.

\section{Notions of independence \label{Sec_IndependenceNotions}} 

An important aspect of classical causal modelling is that -- via the Markov condition -- the structure of a causal DAG imposes constraints on a probability distribution in the form of conditional independences that must hold. For example, given a DAG $G$, with two nodes $X$ and $Y$, if neither $X$ nor $Y$ is a descendant of the other, and they have no common ancestors, then it follows from the Markov condition that the variables must be statistically independent, i.e., $P(X,Y) = P(X) P(Y)$. This section presents notions of `unconditional' and `conditional' independence for classical process maps and process operators, and relates them to the usual notion of statistical independence in a probability distribution. Sections~\ref{Sec_DSeparationTheorem}, \ref{Sec_DoCalculus_CCM} and \ref{Sec_QuantumDoCalculus} will develop a series of theorems relating the structure of a causal DAG to independences in classical process maps and quantum process operators.

\subsection{2-place independence relations \label{SubSec_Independence}} 

\subsubsection{Classical probability distributions} 

Given a probability distribution $P(Y,Z)$ over two (sets of) variables $Y$ and $Z$, the (sets of) variables $Y$ and $Z$ are (statistically) independent if $P(Y,Z)=P(Y)P(Z)$. In this case, write $(Y \bigCI Z)_P$.

\subsubsection{Quantum states}

Given a quantum state $\rho_{AB}$ of two (sets of) systems $A$ and $B$, the (sets of) systems $A$ and $B$ are independent\footnote{In more common terminology, \emph{uncorrelated}.} if $\rho_{AB} = \rho_A \otimes \rho_B$. In this case, write $(A \bigCI B)_{\rho_{AB}}$.

\subsubsection{Classical process maps}

When the object under study is a classical process map, rather than a joint probability distribution, there are at least two interesting notions of independence, one of which is logically stronger than the other.
\begin{definition} \textnormal{(Classical strong independence):} \label{Def_StrongIndependence}
Given a classical process map $\clop_{YZ}$, the sets $Y$ and $Z$ are \emph{strongly independent}, written $(Y \bigCI Z)_{\clop_{YZ}}$, if and only if $\clop_{YZ}=\clop_{Y} \clop_{Z}$.
\end{definition}
The following proposition provides an operational statement, that is a statement in terms of the outcomes of interventions, that is equivalent to strong independence.
\begin{proposition}\label{Prop_ClassicalindependenceOS1}
Given a classical process map $\clop_{YZ}$, the condition $(Y \bigCI Z)_{\clop_{YZ}}$ holds if and only if for all local interventions at $Y$ and $Z$, with outcomes $k_Y$ and $k_Z$, respectively, the probability distribution $P(k_Y , k_Z)$ satisfies $(k_Y \bigCI k_Z)_P$.
\end{proposition}
The proof is omitted, as it is essentially the same as the proof of Proposition~\ref{Prop_QuantumIndependecneOS1} below, restricted to diagonal matrices.

The weaker notion of independence is given by:
\begin{definition} \textnormal{(Classical weak independence):} \label{Def_WeakIndependence}
Given a classical process map $\clop_{YZ}$, let $P(Y,Z)$ be the probability distribution over the sets of single variables $Y$ and $Z$, obtained via $I_{\kappa\rightarrow P}$. The sets of classical split nodes $Y$ and $Z$ are \emph{weakly independent} if and only if $P(Y,Z)=P(Y)P(Z)$.
\end{definition}

As may be seen from Proposition~\ref{Prop_ClassicalindependenceOS1}, strong independence can be seen as a statement about the outcomes of all possible interventions that agents may perform. Weak independence, on the other hand, may be seen as a statement about the statistics of the variables themselves, in the event that there are no interventions. In order to see that $(Y \bigCI Z)_{\clop_{YZ}}$ is logically stronger than weak independence, consider the simple example of Fig.~\ref{Fig_strongvsweakindependence}, with binary variables $Y$ and $Z$. Strong independence fails -- in particular, an agent stationed at $Y$ can send signals to an agent stationed at $Z$ -- but the distribution $P(Y,Z)$ defined by the induced classical causal model is the point distribution $P(Y=y, Z=z) = 1 \mathrm{\ iff\ } y=z=0$, hence $P(Y,Z) = P(Y)P(Z)$.
\begin{figure}
\begin{center}
\small
\begin{minipage}{10cm}
\begin{minipage}{1cm}
\begin{center}
\begin{tikzpicture}
\node [style=none]  at (0, 0) {$Y$};
\node [style=none]  at (0, 2.5) {$Z$};
\node [style=none] (1) at (0, 0.5) {};
\node [style=none] (2) at (0, 2) {};
\draw [->] (1) to (2);
\end{tikzpicture}
\end{center}
\end{minipage}
\hfill
\begin{minipage}{9cm}
\begin{equation}
Z^{\text{in}} = Y^{\text{out}} , \qquad P(Y^{\text{in}} = 0) = 1.
\end{equation}
\end{minipage}
\end{minipage}
\end{center}
\caption{ Example of a scenario in which weak independence holds but strong independence fails. \label{Fig_strongvsweakindependence}}
\end{figure}

\subsubsection{Quantum process operators}

The quantum analogue of strong independence is obvious.
\begin{definition} \textnormal{(Quantum strong independence):} \label{Def_QuantumIndependence}
Given a process operator $\sigma_{YZ}$, the sets $Y$ and $Z$ are \emph{strongly independent}, written $(Y \bigCI Z)_{\sigma_{YZ}}$, if and only if $\sigma_{YZ}=\sigma_{Y} \sigma_{Z}$.
\end{definition} 
Just as with classical split nodes, the notion of quantum strong independence has an operational interpretation.
\begin{proposition} \label{Prop_QuantumIndependecneOS1}
Given a process operator $\sigma_{YZ}$, the condition $(Y \bigCI Z)_{\sigma_{YZ}}$ holds if and only if for all local interventions at $Y$ and $Z$, with outcomes $k_Y$ and $k_Z$ respectively, the probability distribution $P(k_Y , k_Z)$ satisfies $(k_Y \bigCI k_Z)_P$.
\end{proposition}
\noindent {\bf Proof.} See Appendix~\ref{Subsubsec_Prop_QuantumIndependecneOS1}. \hfill $\square$

Under $I_{\sigma \rightarrow \kappa}$ quantum strong independence obviously implies classical strong independence.
Whether there is a quantum analogue of weak independence is a question that we take to be of considerable interest, but do not address in this work. For further discussion, and a proposal for the quantum analogue, see Ref.~\cite{BarrettEtAl_2019_WeakMeasurements}.
	
\subsection{3-place independence relations \label{SubSec_CondIndependence}} 

\subsubsection{Classical probability distributions} 

Given random variables $Y,Z,W$, and a joint distribution $P(Y,Z,W)$, a common definition states that $Y$ is independent from $Z$ conditioned on $W$ if 
\begin{equation}\label{Eq_classicalconditionalindep}
P(Y,Z|W) = P(Y|W) \ P(Z|W).
\end{equation}
This way of stating it has a minor deficiency, which is that  $P(Y,Z|W=w)$ is undefined for values $w$ such that $P(W=w)=0$. Let us, therefore, say that \emph{$Y$ and $Z$ are independent conditioned on $W$}, and write $(Y \bigCI Z | W)_P$, if $P(Y,Z|W=w) = P(Y|W=w) \ P(Z|W=w)$ for all values $w$ such that $P(W=w) > 0$. The intuitive meaning of this is that if an agent already knows $W$, then learning $Z$ provides no further information about $Y$, and vice versa.
\begin{proposition} 
\label{Prop_EquivalentStatementsClassicalCI}
Given (sets of) variables $Y,Z,W$, and a probability distribution $P(Y,Z,W)$, the following are equivalent:
\begin{enumerate}
\item $(Y \bigCI Z | W)_P$
\item $P(Y,Z,W) \ P(W) =  P(Y,W) \ P(Z,W)$
\item There exist real functions $\alpha : Y \times W \rightarrow \mathbb{R}$ and $ \beta : Z \times W \rightarrow \mathbb{R}$,  such that $P(Y,Z,W) =  \alpha(Y,W) \ \beta(Z,W) $ 
\item The conditional mutual information satisfies $I(Y : Z | W) = 0$.
\end{enumerate}
\end{proposition}
The proofs of these equivalences are straightforward, and omitted.

\subsubsection{Quantum states}

Given a quantum state $\rho_{ABC}$ of three (sets of) systems $A,B,C$, let us say that $A$ and $B$ are independent relative to $C$, and write $(A \bigCI B | C)_{\rho_{ABC}}$,  if the quantum conditional mutual information between $A$ and $B$ given $C$ vanishes: $I(A:B|C)=0$.\footnote{Here, and in the definitions below for classical process maps and process operators, we write `relative to' rather than `conditioned on' so as to not invoke associations of conditioning `on the value of a variable', which is a concept absent in quantum theory, and is in any case potentially misleading when the object under study is a (classical or quantum) process.}

As with classical conditional independence, various equivalent formulations can be given. In order to state one of these, define first the following product of operators \cite{CerfEtAl_1997_NegativeEntropy, WarmuthEtAL_2006_BayesRuleForDensityMatrices, LeiferEtAl_2008QuantumGraphicalModels}:
\begin{eqnarray}
	\star &:& \mathcal{L}(\mathcal{H}) \times \mathcal{L}(\mathcal{H}) \rightarrow \mathcal{L}(\mathcal{H}) \nonumber \\
	(A,B) &\mapsto& A \star B := \lim_{n \rightarrow \infty} \left( A^{1/n} \ B^{1/n} \right)^n \label{Def_StarProduct_1}
\end{eqnarray}
This product has some useful properties. It is associative and commutative, and reduces to the ordinary product $AB$ if $[A,B]=0$. For the special case of strictly positive definite operators, 
\begin{equation}
	A \star B = \text{exp}(\text{log}(A) + \text{log}(B)) \ .	\label{Def_StarProduct_2}
\end{equation}
In the case of positive semi-definite operators $A$ and $B$, the above equation can be extended in the sense that $\text{log}( A \star B) = \text{log}(A) + \text{log}(B)$ holds, where the logarithms are restricted to the respective supports of the operators \cite{WarmuthEtAL_2006_BayesRuleForDensityMatrices}.  We extend our convention of suppressing identities in products, writing $\sigma_{XY} \star \sigma_{YZ}$ as short-hand for $(\sigma_{XY}\otimes \mathds{1}_{Z}) \star (\mathds{1}_{X} \otimes \sigma_{YZ})$. 

\begin{proposition} 
\label{Prop_EquivalentStatementsQuantumCI}
Given (sets of) systems $A,B,C$, and a quantum state $\rho_{ABC}$, the following are equivalent:
\begin{enumerate}[label=\textnormal{(\arabic*)}, leftmargin=1cm]
\item $(A \bigCI B | C)_{\rho_{ABC}}$
\item $\rho_{ABC} \star \rho_C = \rho_{AC} \star \rho_{BC}$
\item There exist Hermitian operators $\alpha_{AC}$ and $ \beta_{BC}$,  such that $\rho_{ABC} =  \alpha_{AC} \ \beta_{BC}$. 
\item The Hilbert space $\mathcal{H}_C$ decomposes as $\mathcal{H}_C = \bigoplus_m \mathcal{H}_{C_m^1}\otimes \mathcal{H}_{C_m^2}$, such that $\rho_{ABC} = \sum_m q_m \allowbreak \big(\rho_{A C_m^1} \otimes \rho_{B C_m^2} \big)$, where $0\leq q_m \leq 1$, $\sum_m q_m = 1$, and for each $m$, $\rho_{A C_m^1}$ and $\rho_{B C_m^2}$ are density operators on the Hilbert spaces indicated by the subscripts.
\end{enumerate}
If (3) holds, then $[\alpha_{AC}, \beta_{BC}] = 0$.
\end{proposition}
\noindent {\bf Proof.} For `$(1) \Leftrightarrow (2)$', see Refs.~\cite{Ruskai_2002_InequalitiesforQuantumEntropy,LeiferEtAl_2008QuantumGraphicalModels}. For `$(1) \Leftrightarrow (4)$', see Ref.~\cite{HaydenEtAl_2003_StructureOfStatesAubadditivity}. It is immediate that `$(4) \Rightarrow (3)$' and straightforward to verify that `$(3) \Rightarrow (2)$'. If $\rho_{ABC} =  \alpha_{AC} \ \beta_{BC}$, for Hermitian $\alpha_{AC}$ and $\beta_{BC}$, then taking the Hermitian conjugate of both sides of the equation yields $[\alpha_{AC}, \beta_{BC}] = 0$. \hfill $\square$

\subsubsection{Classical process maps}

As with the $2$-place relation of independence, when the object under study is a classical process map, rather than a classical probability distribution, there are at least two interesting notions that correspond to a $3$-place relation of relative independence. One is logically stronger than the other.
\begin{definition} \textnormal{(Classical strong relative independence):} \label{Def_StrongClassicalCI}
Given a classical process map $\clop_{YZW}$, say that \emph{$Y$ and $Z$ are strongly independent relative to $W$}, written $(Y \bigCI Z|W)_{\clop_{YZW}}$, if and only if there exist real functions $\alpha_{YW} : Y^{\text{in}} \times Y^{\text{out}} \times W^{\text{in}} \times W^{\text{out}} \rightarrow \mathds{R}$ and $\beta_{ZW}: Z^{\text{in}} \times Z^{\text{out}} \times W^{\text{in}} \times W^{\text{out}} \rightarrow \mathds{R}$, such that $\clop_{YZW}=\alpha_{YW} \ \beta_{ZW}$. 
\end{definition}
Observe the formal similarity with Part~(3) of Prop.~\ref{Prop_EquivalentStatementsClassicalCI}. Observe also that if $\clop_{YZW}$ has the form $\clop_{YZW} = I_{Y^{\text{out}}} \times I_{Z^{\text{out}}} \times I_{W^{\text{out}}} \times P(Y^{\text{in}} , Z^{\text{in}} , W^{\text{in}})$ for  a probability distribution $P(Y^{\text{in}} , Z^{\text{in}} , W^{\text{in}})$, as it must, e.g., in the case of 3 nodes none of which is a causal descendant of another, then $(Y \bigCI Z|W)_{\clop_{YZW}}$ reduces to the ordinary notion of conditional independence in the probability distribution $P(Y^{\text{in}} , Z^{\text{in}} , W^{\text{in}})$.

Def.~\ref{Def_StrongClassicalCI} may be understood in terms of an equivalent operational formulation that concerns the outcomes of interventions at the nodes. In order to state the operational formulation, define first the following subclass of classical interventions:
\begin{definition}\textnormal{(Maximally informative intervention):} \label{Def_MaxInfMeas} 
A \emph{maximally informative intervention} at a node $W$, with outcome $k_W$, is a classical intervention such that $W^{\text{in}}$ and $W^{\text{out}}$ can each be inferred from $k_W$. A necessary and sufficient condition is that: 
\begin{equation}
P(k_W,W^{\text{out}}|W^{\text{in}}) = \delta \big( g^{\text{in}}(k_W), W^{\text{in}} \big) \ \delta \big( g^{\text{out}}(k_W), W^{\text{out}} \big) \ P(k_W,W^{\text{out}}|W^{\text{in}}) \ ,
\end{equation}
where $g^{\text{in}}$ is a surjective function and $g^{\text{out}}$ an arbitrary function. For $W$ a set of nodes, a maximally informative local intervention at $W$ is a product of maximally informative interventions at each node in $W$.
\end{definition}
The idea is that an agent who performs a maximally informative intervention at a node $W$, and records the outcome, knows the values of both $W^{\text{in}}$ and $W^{\text{out}}$. Note that this is more general than the non-disturbing measurement, which simply records the value of $W^{\text{in}}$, and fixes $W^{\text{out}} = W^{\text{in}}$. Maximally informative interventions include disturbing interventions, such as that which records the value of $W^{\text{in}}$, and fixes $W^{\text{out}} = w$ for a constant value $w$.
\begin{proposition} \label{Prop_CSCI_EquivalentOperationalStatements}
Given a classical process map $\clop_{YZW}$, the condition $(Y \bigCI Z|W)_{\clop_{YZW}}$ holds if and only if for any choice of maximally informative local intervention at $W$ with outcome $k_W$, and any local interventions at $Y$ and $Z$, with outcomes $k_Y$ and $k_Z$ respectively, the joint probability distribution $P(k_Y , k_Z , k_W)$ satisfies $(k_Y \bigCI k_Z | k_W)_P$.
\end{proposition}
\noindent {\bf Proof.} See Appendix~\ref{Subsubsec_Proof_CSCI_EquivalentOperationalStatements}.  \hfill $\square$

Intuitively speaking, and as expected for a classical notion of relative independence, once one knows everything about $W$ -- the only `thing through which $Y$ and $Z$ interact' -- $Y$ is independent from $Z$.  
\begin{remark}
One may be tempted to formulate Prop.~\ref{Prop_CSCI_EquivalentOperationalStatements} in terms of perfect, non-disturbing measurements at $W$ only -- that is interventions that record the value of $W^{\text{in}}$ and fix $W^{\text{out}} = W^{\text{in}}$ -- rather than quantifying over all maximally informative interventions at $W$. In fact, the resulting statement is inequivalent, hence not equivalent to classical strong relative independence. The reader is invited to construct an example that demonstrates the inequivalence.
\end{remark}

The following weaker notion of relative independence is relevant to the case in which no interventions are performed at nodes.
\begin{definition} \textnormal{(Classical weak relative independence):} \label{Def_WeakClassicalCI}
Given a classical process map $\clop_{YZW}$, let $P(Y,Z,W)$ be the joint probability distribution over sets of single variables $Y,Z,W$ obtained via $I_{\kappa\rightarrow P}$. For the sets of classical split nodes, \emph{$Y$ and $Z$ are weakly independent relative to $W$} if and only if $(Y \bigCI Z | W)_P$.
\end{definition}

Strong implies weak relative independence, as can easily be seen by applying $I_{\kappa\rightarrow P}$ to a classical process map $\clop_{YZW}$ for which $(Y \bigCI Z|W)_{\clop_{YZW}}$, and bearing in mind the equivalences of Prop.~\ref{Prop_EquivalentStatementsClassicalCI}. That the converse does not hold is inherited from the special case $W=\emptyset$.

\subsubsection{Quantum process operators}

The quantum definition for strong relative independence is the obvious analogue of Def.~\ref{Def_StrongClassicalCI} and condition~(3) of Prop.~\ref{Prop_EquivalentStatementsQuantumCI}.
\begin{definition} \textnormal{(Quantum strong relative independence):} \label{Def_QuantumCI}
Given a process operator $\sigma_{YZW}$, say that \emph{$Y$ is strongly independent from $Z$ relative to $W$}, and write $(Y \bigCI Z|W)_{\sigma_{YZW}}$, if and only if there exist Hermitian operators $\alpha_{YW}$ and $\beta_{ZW}$ such that
\begin{equation}
\sigma_{YZW} \ = \ \alpha_{YW} \ \beta_{ZW}.
\end{equation}
\end{definition}
Observe that quantum strong relative independence reduces to quantum strong independence $(Y \bigCI Z)_{\sigma_{YZ}}$ when $W=\emptyset$\footnote{The only way that $\sigma_{YZ} = \alpha_Y \beta_Z$ can be true is if $\alpha_Y$ and $\beta_Z$ coincide with the marginal operators $\sigma_Y$ and $\sigma_Z$.}, and reduces to classical strong relative independence in the case that $\sigma_{YZW}$ is diagonal with respect to a product basis.\footnote{\label{Footnote_RelInd_SigmaToKappa} That $\alpha_{YW}$ and $\beta_{ZW}$ can be chosen to be diagonal with respect to the same preferred basis that is assumed to exist in virtue of $I_{\sigma \rightarrow \kappa}$,  is obvious when $\sigma_{YZW}$ is non-degenerate and slightly less obvious when $\sigma_{YZW}$ is degenerate. The claim follows, e.g., from Statement~2 of Prop.~\ref{Prop_EquivalentStatementsCQCI}, along with Remark \ref{Rem_EquivalentStatementsRelIndCSM}.} Observe also that if $\sigma_{YZW}$ has the form $\sigma_{YZW} = \mathds{1}_{Y^{\text{out}}} \otimes \mathds{1}_{Z^{\text{out}}} \otimes \mathds{1}_{W^{\text{out}}} \otimes \rho_{Y^{\text{in}}Z^{\text{in}}W^{\text{in}}}$ for a density operator $\rho_{Y^{\text{in}}Z^{\text{in}}W^{\text{in}}}$, as it must, e.g., in the case of 3 nodes none of which is a causal descendant of another, then $(Y \bigCI Z|W)_{\sigma_{YZW}}$ reduces to the ordinary notion of quantum relative independence on the state $\rho_{Y^{\text{in}}Z^{\text{in}}W^{\text{in}}}$.

In the case of classical strong relative independence, we gave an equivalent operational statement in terms of maximally informative interventions at the $W$ nodes. There is no obvious quantum analogue of a maximally informative intervention that would yield an operational statement equivalent to $(Y \bigCI Z|W)_{\sigma_{YZW}}$. The following proposition gives an operational statement that is implied by (but not equivalent to) $(Y \bigCI Z|W)_{\sigma_{YZW}}$.
\begin{proposition} \label{Prop_Implication_QOS1a} 
Consider a process operator $\sigma_{YZW}$. If $(Y \bigCI Z|W)_{\sigma_{YZW}}$, then there exists a global intervention at the $W$ nodes, with outcome $k_W$, such that for all local interventions at $Y$, $Z$, with joint outcomes $k_Y$, $k_Z$ respectively, the joint probability distribution $P(k_Y , k_Z , k_W)$ satisfies $(k_Y \bigCI k_Z | k_W)_P$.
\end{proposition}
\noindent {\bf Proof.} See Appendix~\ref{Subsubsec_Proof_Implication_QOS1a}. \hfill $\square$
\begin{remark} 
The required intervention at the $W$ nodes, which gives $(k_Y \bigCI k_Z | k_W)_P$ for all interventions at $Y$ and $Z$ can always be assumed to have the following form: let an agent be stationed at an additional locus $E$, such that for each node $N \in W$, the quantum system at $N^{\text{in}}$ is sent to $E$, one of a maximally entangled pair of systems is fed into $N^{\text{out}}$, and the other one is sent to $E$. The agent at $E$ then performs an appropriately chosen joint von Neumann measurement on those systems that are incoming at $E$.
\end{remark}
\begin{remark}
That there is no obvious quantum analogue of a maximally informative intervention corresponds to the fact that it is not possible in quantum theory to measure all observables at once: for example, measurement of the $z$-spin of a spin-$1/2$ particle precludes measurement of $x$-spin. The content of Prop.~\ref{Prop_Implication_QOS1a}, roughly speaking, is that there always exists a suitable joint measurement on the $W$ systems, which reveals the value of the `correct' observable to render measurements at $Y$ and $Z$ independent. Other measurements at $W$ would reveal the values of other observables, at the expense of learning the value of the `correct' observable, hence conditioned on their outcome, measurements at $Y$ and $Z$ will not in general be rendered independent.
\end{remark}

In order to see that the converse direction of Prop.~\ref{Prop_Implication_QOS1a} does not hold, it suffices to consider a process operator on $3$ quantum nodes, $\sigma_{YZW} = \rho_{W|YZ} \rho_Y \rho_Z$ such that $(Y \bigCI Z | W)_{\sigma_{YZW}}$ fails. Given any trace-preserving intervention at $W$, with a trivial, i.e., single-valued outcome $k_W$ -- it holds that conditioned on $k_W$, the outcomes of any interventions at $Y$ and $Z$ are independent. For a more interesting case in which the converse direction of Prop.~\ref{Prop_Implication_QOS1a} fails, consider $\sigma_{YZW} = \rho^{\text{copy}}_{YZ | W} \rho_W$, where $Y,Z,W$ are each $2$-dimensional systems, and $\rho^{\text{copy}}_{YZ|W}$ is the CJ representation of the following isometry (a coherent copy operation in the computational basis):
\begin{eqnarray*}
|0\rangle_W &\rightarrow& |0\rangle_Y \otimes |0\rangle_Z \\
|1\rangle_W &\rightarrow& |1\rangle_Y \otimes |1\rangle_Z.
\end{eqnarray*}
The consequent of Prop.~\ref{Prop_Implication_QOS1a} holds, with the required intervention at $W$ being a von Neumann measurement on the computational basis. However, it is false that $(Y \bigCI Z|W)_{\sigma_{YZW}}$, since $ \rho^{\text{copy}}_{YZ | W} \ne \rho^{\text{copy}}_{Y | W} \rho^{\text{copy}}_{Z | W}$ \cite{AllenEtAl_2016_QCM}.

Finally, see Ref.~\cite{BarrettEtAl_2019_WeakMeasurements} for discussion of, and a proposal for, a quantum analogue of weak relative independence.

\subsection{Mathematically equivalent expressions}

Equivalent formulations of quantum strong relative independence can be given as follows.
\begin{proposition} 
\label{Prop_EquivalentStatementsCQCI}
Let $\sigma_{YZW}$ be a process operator. The following are equivalent:
\begin{center}
\begin{minipage}{14cm}
\begin{enumerate}[label=\textnormal{(\arabic*)}, leftmargin=1cm]
\item $(Y \bigCI Z|W)_{\sigma_{YZW}}$
\item $\sigma_{YZW} \ \star \ \sigma_{W}^{\tau_Y \tau_Z} \ = \ \sigma_{YW}^{\tau_Z} \  \star \ \sigma_{ZW}^{\tau_Y}$ \hspace{0.5cm} $\forall \ \ \mathrm{local \ interventions\ }\tau_Y, \ \tau_Z$.
\item $I(Y:Z|W)=0$, evaluated on $\hat{\sigma}_{YZW}$. 
\end{enumerate}
\end{minipage}
\end{center}
\end{proposition}
\noindent {\bf Proof.} The proposition is a special case of Prop.~\ref{Prop_EquivalentStatementsRule1}, which is proven in Appendix~\ref{Subsubsec_Proof_Prop_EquivalentStatementsRule1}.  \hfill $\square$
\begin{remark} \label{Rem_EquivalentStatementsRelIndCSM}
Very similar statements can be derived for classical process maps. Replacing $\sigma$ with $\clop$ throughout, the star products in (2) become ordinary multiplication, and the quantum conditional mutual information in (3) becomes the classical conditional mutual information. These statements can be established via the quantum proofs restricted to diagonal matrices.
\end{remark}

\section{d-separation and independence \label{Sec_DSeparationTheorem}}

\subsection{The d-separation theorem for classical causal models \label{Subsec_ClassicalDSeparationTheorem}}

Given a joint probability distribution over random variables, perhaps estimated from observational data, it is always possible in principle (if computationally difficult) to determine the conditional independence relations that the distribution satisfies. But what is the link with causal relations described by a DAG? This link is made precise by one of the core theorems of the framework of classical causal models. The theorem involves a graphical property of DAGs, known as \emph{d-separation}, which is a statement about the connectedness of subsets of nodes. 
\begin{definition} \textnormal{(Blocked paths and d-separation \cite{Pearl_1988_Probabilistic}):}  \label{Def_DSeparation}
Given a DAG $G$, a path between nodes $y$ and $z$ is \emph{blocked} by the set of nodes $W$ if the path contains either 
\begin{enumerate}[leftmargin=2cm]
\item a chain $a \rightarrow w \rightarrow  c$ or a fork  $a \leftarrow w \rightarrow  c$ with the middle node $w\in W$ 
\item a collider $a \rightarrow r \leftarrow  c$ such that neither $r$ nor any descendant of $r$ lies in $W$.
\end{enumerate}
For subsets of nodes $Y$, $Z$ and $W$, say that $Y$ and $Z$ are \emph{d-separated} by $W$, and write $(Y \bigCI Z | W)_G$, if for every $y\in Y$ and $z\in Z$, every path between $y$ and $z$ is blocked by $W$.
\end{definition}

\begin{theorem} \textnormal{(d-separation theorem \cite{Verma&Pearl_1990_CausalNetworks, GeigerEtAl_1990_IdentifyingIndependence}, see also Ref.~\cite{Pearl_Causality}):} \label{Thr_ClassicalDSeparation}
Consider a DAG $G$, with nodes $X_1, \ldots , X_n$, and disjoint subsets of nodes $Y$, $Z$, and $W$.
\begin{enumerate}[label=(\alph*)]
\item (Soundness): if $(Y \bigCI Z | W)_G$, then any distribution $P(X_1, \ldots , X_n)$ that is Markov for $G$ satisfies $(Y \bigCI Z | W)_P$.
\item (Completeness): if $(Y \bigCI Z | W)_G$ does not hold, then there exists a probability distribution $P(X_1, \ldots , \allowbreak X_n)$ such that $P(X_1, \ldots , X_n)$ is Markov for $G$ and $(Y \bigCI Z | W)_P$ does not hold. 
\end{enumerate}
\end{theorem}

\subsubsection{Example}

The concept of d-separation, along with Theorem~\ref{Thr_ClassicalDSeparation}, is illustrated by the DAG $G$ of Fig.~\ref{Fig_Example_DAGForRules}.
\begin{figure}[H]
\centering
\small
\input{Figures/Fig_Example_DAGForRules.tex}
\caption{\label{Fig_Example_DAGForRules} }
\end{figure}
Suppose that $G$ represents the causal structure of a classical causal model, with joint distribution 
$P(N_1, \allowbreak N_2, \allowbreak N_3, \allowbreak N_4, \allowbreak N_5)$ 
that is Markov for $G$. Then $N_2$ and $N_4$ are in general correlated, both in virtue of the common cause $N_3$, and in virtue of the causal pathway $N_2\rightarrow N_5 \rightarrow N_4$. However, $\{N_2\}$ is d-separated from $\{N_4\}$ by $\{N_3,N_5\}$. This captures the idea that if $N_1$ is ignored, then once the causal influences from the common cause $N_3$, and through the chain involving $N_5$, have been `picked up' by conditioning on $N_3$ and $N_5$, there are no further mechanisms by which $N_2$ and $N_4$ can be correlated, hence by Theorem~\ref{Thr_ClassicalDSeparation}, $(\{ N_2\} \bigCI \{ N_4 \} | \{N_3 , N_5\})_P$. 

On the other hand, $\{N_2\}$ is not d-separated from $\{N_4\}$ by $\{N_1,N_3,N_5\}$. The reason for this is, in terms of Def.~\ref{Def_DSeparation}, is that $(N_2 , N_1 , N_4)$ is an unblocked path, featuring as it does a collider at the node $N_1$, which is contained in the set $\{N_1,N_3,N_5\}$. The significance is that even when conditioning on $N_3$, and $N_5$, conditioning also on $N_1$ tends to induce correlations between $N_2$ and $N_4$. The completeness part of Theorem~\ref{Thr_ClassicalDSeparation} ensures that there exists a joint distribution $P(N_1 , N_2 , N_3 , N_4 , N_5)$, Markov for $G$, for which this is indeed the case. (In fact it will be the case for almost all joint distributions $P(N_1 , N_2 , N_3 , N_4 , N_5)$, the exceptions having measure zero within the set of joint distributions Markov for $G$.)

\subsection{A d-separation theorem for quantum causal models \label{Subsec_QuantumDSeparationTheorem}}

Consider a set of quantum nodes $V$, with $Y,Z,W$ disjoint subsets of $V$, and $R:=V \setminus ( Y \cup Z \cup W)$. Suppose that a process operator $\sigma_{YZWR}$ over the nodes in $V$ is given. The marginal process operator over $Y \cup Z \cup W$ in general depends on any interventions that are performed at the $R$ nodes, and is written $\sigma_{YZW}^{\tau_R}$. Hence whether a statement like \emph{$Y$ and $Z$ are strongly independent relative to $W$} holds also depends, in general, on any interventions that are performed at $R$ nodes.\footnote{The same is of course true in classical split-node causal models. Note that the same is \emph{also} true in classical causal models when evaluating marginals of the distribution $P(X_1 , ... , X_n)$ and statements like $(Y \bigCI Z | W )_{P}$; however, in the framework of classical causal models, it is understood implicitly that there is no intervention upon a variable unless stated explicitly, hence marginals of $P(X_1 ,..., X_n)$ are computed in the ordinary way.} The following theorem is as strong as it could be, in the sense that the soundness part holds for arbitrary local interventions at $R$ nodes, and the completeness part holds assuming no intervention at $R$ nodes. 
\begin{theorem} \textnormal{(Quantum d-separation theorem):} \label{Thm_QuantumDSeparationTheorem}
Consider a DAG $G$, with a set $V$ of quantum nodes, and disjoint subsets of nodes $Y$, $Z$, and $W$, with $R:=V \setminus ( Y \cup Z \cup W)$.
\begin{enumerate}[label=(\alph*)]
\item (Soundness): If $(Y \bigCI Z | W)_G$, then for any quantum process operator $\sigma_{YZWR}$ that is Markov for $G$, and any local intervention $\tau_R$ at $R$, the marginal $\sigma_{YZW}^{\tau_R}$ satisfies $(Y \bigCI Z | W)_{\sigma_{YZW}^{\tau_R}}$.
\item (Completeness): If $(Y \bigCI Z | W)_G$ does not hold, then there exists a process operator $\sigma_{YZWR}$ that is Markov for $G$, such that with no interventions at $R$ nodes, the marginal $\sigma_{YZW}$ does not satisfy $(Y \bigCI Z | W)_{\sigma_{YZW}}$.
\end{enumerate}
\end{theorem} 
\noindent {\bf Proof.} Consider, quite generally, a set $V$, with a ternary relation $S$ defined on the subsets of $V$. The \emph{semi-graphoid axioms}, for arbitrary disjoint subsets $Y,Z,W,X \subseteq V$, are as follows  \cite{Pearl&Paz_1985_Graphoids, GeigerEtAl_1990_IdentifyingIndependence}:
\begin{eqnarray}
	& \text{symmetry} \hspace*{0.5cm} & S(Y,Z;W) \hspace*{0.3cm} \Leftrightarrow \hspace*{0.3cm}  S(Z,Y;W) \\
	& \text{decomposition} \hspace*{0.5cm} & S(Y,XZ;W)  \hspace*{0.3cm}  \Rightarrow  \hspace*{0.3cm}  S(Y,Z;W) \\
	& \text{weak union} \hspace*{0.5cm} & S(Y,XZ;W)  \hspace*{0.3cm}  \Rightarrow  \hspace*{0.3cm}  S(Y,Z;XW) \\
	& \text{contraction} \hspace*{0.5cm} & S(Y,Z;W) \ \wedge \ S(Y,X;ZW)  \hspace*{0.3cm}  \Rightarrow  \hspace*{0.3cm}  S(Y,ZX;W).
\end{eqnarray}
A relation $S$ that satisfies these axioms is known as a \emph{semi-graphoid}. Given a DAG $G$, with nodes $V$, if a semi-graphoid $S$ additionally satisfies the \emph{local Markov condition},
\[
S(\{X\},Nd(X)\setminus Pa(X);Pa(X)) \qquad \forall X\in V,
\]
then it follows that d-separation is sound for $S$ \cite{Verma&Pearl_1990_CausalNetworks, Lauritzen_2011_DirectedMarkovProperties}, i.e.,
\[
(Y \bigCI Z | W)_G \ \Rightarrow \ S(Y,Z;W).
\]

Given a set of quantum nodes $V$, then, consider a process operator $\sigma_V$, and define a ternary relation $T$ on the subsets of $V$ such that
\begin{equation}\label{Eq_DefofTrelation}
T(Y,Z;W) \qquad \mathrm{iff} \qquad \forall \mathrm{\ local\ interventions\ }\tau_R, (Y \bigCI Z | W)_{\sigma_{YZW}^{\tau_R}}.
\end{equation}
The soundness part of Theorem~\ref{Thm_QuantumDSeparationTheorem} follows from the following two lemmas.
\begin{lemma}\label{Lemma_quantumdsepsemigraphoid}
Given a process operator $\sigma_{YZWR}$, the relation $T$ defined by Eq.~(\ref{Eq_DefofTrelation}) satisfies the semi-graphoid axioms.
\end{lemma}
\noindent {\bf Proof.} See Appendix~\ref{Subsubsec_Proof_Thm_QuantumDSeparationTheorem} \hfill $\square$
\begin{lemma}\label{Lemma_quantumdseplocalmarkov}
Consider a DAG $G$, with nodes $V$, and a process operator $\sigma_V$ that is Markov for $G$. The relation $T$ defined by Eq.~(\ref{Eq_DefofTrelation}) satisfies the local Markov condition.
\end{lemma}
\noindent {\bf Proof.} See Appendix~\ref{Subsubsec_Proof_Thm_QuantumDSeparationTheorem} \hfill $\square$

Finally, for the completeness part of Theorem~\ref{Thm_QuantumDSeparationTheorem}, consider a DAG $G$ with quantum nodes $V$, and suppose that $(Y \nbigCI Z | W)_G$ for disjoint subsets of nodes $Y$, $Z$, and $W$. Let $R = V \backslash (Y \cup Z \cup W)$. Associate a classical random variable with each node, ranging over a set of values whose cardinality is the same as the dimension of the quantum node. By virtue of the completeness part of Theorem~\ref{Thr_ClassicalDSeparation}, there exists a joint probability distribution $P(..)$ over these random variables, Markov for $G$, for which $( Y \nbigCI Z | W )_{P}$. This classical causal model induces a classical split-node causal model with the same DAG via $I_{CCM\rightarrow CSM}$ of Section~\ref{Sec_ClassicalQauntumComparison}, for which $( Y \nbigCI Z | W )_{\clop^{\tau_R^{\text{id}}}_{YZW}}$. This in turn induces a quantum causal model with the same DAG via $I_{\kappa\rightarrow\sigma}$ of Section\ref{Sec_ClassicalQauntumComparison}, for which $( Y \nbigCI Z | W )_{\sigma^{\tau_R^{\text{id}}}_{YZW}}$.
\hfill $\square$
\begin{remark} 
Theorem~\ref{Thm_QuantumDSeparationTheorem} holds, \emph{mutatis mutandis}, with process operators replaced by classical process maps. The proof is the same as the proof of Theorem~\ref{Thm_QuantumDSeparationTheorem}, with operators restricted to the diagonal case. Theorem~\ref{Thm_QuantumDSeparationTheorem} also holds, \emph{mutatis mutandis}, with process operators replaced by classical process maps, and strong relative independence replaced by weak relative independence. In this case, the proof of the soundness part follows from the fact that for classical process maps, strong relative independence implies weak relative independence. Completeness follows from the completeness part of Theorem~\ref{Thr_ClassicalDSeparation}, along with the definition of weak relative independence.
\end{remark}

\subsubsection{Example}

Consider again the DAG of Fig.~\ref{Fig_Example_DAGForRules}, and suppose that $G$ is the causal structure for a quantum causal model, with $N_1 , \ldots , N_5$ quantum nodes, and a process operator $\sigma_{N_1 .... N_5}$ that is Markov for $G$. The fact that $\{N_2\}$ is d-separated from $\{N_4\}$ by $\{N_3,N_5\}$ ensures that for any trace-preserving intervention $\tau_{N_1}$ at $N_1$, the resulting $\sigma^{\tau_{N_1}}_{N_2 ... N_5}$ satisfies $(\{N_2\} \bigCI \{N_4\} | \{N_3 , N_5\})_{\sigma^{\tau_{N_1}}_{N_2 ... N_5}}$. It follows from Prop.~\ref{Prop_Implication_QOS1a} that there is an intervention at $N_3 N_5$, such that, conditioned on its outcome, the outcomes of any interventions at $N_2$ and $N_4$ are independent. 

The fact that $\{N_2\}$ is not d-separated from $\{N_4\}$ by $\{N_1,N_3,N_5\}$ corresponds to the fact that no matter what intervention is performed at $N_3 N_5$, conditioning on the outcome of an intervention at $N_1$ will tend to correlate $N_2$ and $N_4$. Note, however, that the operational statement of Prop.~\ref{Prop_Implication_QOS1a} does not actually fail here. There is always the possibility of a trace-preserving intervention at $N_1$ -- with trivial, or single-valued, outcome -- along with a suitable intervention at $N_3 N_5$, such that conditioned on the outcomes of both, the nodes $N_2$ and $N_4$ become strongly independent.

\section{The do-calculus for classical causal models \label{Sec_DoCalculus_CCM}}

There are many situations, ranging through all disciplines from the natural sciences and medicine to the social sciences and policy making, where one would like to know the \textit{causal effect} of $X$ on $Y$, as Pearl calls the marginal post-intervention distribution $P(Y|do(X))$ of two (disjoint subsets of) variables $X$ and $Y$. However, obtaining experimental data for the situation in which one controls $X$ is often not feasible. This could be due to practical limitations that simply prevent the control of a certain variable, or it could be that controlling the variable is unethical. An example of the latter that is often given concerns the tobacco industry's genotype theory to explain the correlation between smoking and lung cancer through the confounding common cause of a carcinogenic genotype which both increases the probability for suffering from lung cancer and leads to a craving for nicotine. People cannot be forced to smoke for the sake of the study. 

In Section~\ref{Sec_CCMIntro} it was pointed out that one of the reasons why the framework of classical causal models is so useful is because it allows predictions for post-intervention distributions. Given a classical causal model, the causal effect $P(Y | do(X))$ can always be calculated using the truncated factorization formula Eq.~\ref{Def_TruncatedFactorization}. In many practical situations, though, it may be that a causal structure in the form of a DAG is posited, but that not all variables are observed. This leads to the problem of \emph{causal effect identifiability}  \cite{Pearl_Causality}: given a causal structure, and a joint probability distribution over the observed variables only, which in general correspond to a proper subset of the nodes of the DAG, when is it possible to calculate $P(Y|do(X))$ unambiguously? 

This question is answered via the \textit{do-calculus} \cite{Pearl_Causality}, a set of inference rules that relate interventional and observational statements to one another in the case that certain graphical properties of the DAG hold. The rules are stated in terms of mutilated versions of the causal DAG $G$, that is DAGs obtained by removing some of the arrows from $G$. We use the same notation as Ref.~\cite{Pearl_Causality}. Given a DAG $G$, let $S$ be a subset of the nodes. Write $G_{\overline{S}}$ for the DAG identical with $G$ except that all arrows incident upon elements of $S$ have been removed, and write $G_{\underline{S}}$ for the DAG identical with $G$ except that all outgoing arrows from elements of $S$ have been removed. 
\begin{theorem} \label{Thm_ClassicalDoCalculus} 
\textnormal{(Rules of the do-calculus for classical causal models \cite{Pearl_Causality}):} 
Let a classical causal model be given by a DAG $G$ and a probability distribution $P(...)$.\footnote{As remarked elsewhere, when conditional probabilities are used, as they are in this theorem, these are undefined if there are values of the variable being condition on that have zero probability. In each case, the claim should be understood as asserting that the equality holds for all values of conditioned-on variables that have positive probability.}
Let $X, Y, Z$, and $W$ be disjoint subsets of the variables.	
	\begin{center}
		\begin{minipage}{14.5cm}
			\key{Rule 1} (insertion/deletion of observations): \\[0.2cm]
			\noindent \hspace*{1.2cm} $(Y \bigCI Z|X,W)_{G_{\overline{X}}} \hspace{1cm} \Rightarrow \hspace{1cm} P(Y|do(X),Z,W)=P(Y|do(X),W)$ \\[0.2cm]
			\key{Rule 2} (exchange of observations and interventions):\\[0.2cm]
			\noindent \hspace*{1.2cm}
				$(Y \bigCI Z|X,W)_{G_{\overline{X}\underline{Z}}}  \hspace{1cm} \Rightarrow \hspace{1cm}  P(Y|do(X),do(Z),W)=P(Y|do(X),Z,W)$  \\[0.2cm]
			\key{Rule 3} (insertion/deletion of interventions):\\[0.2cm]
			\noindent \hspace*{1.2cm}
				$(Y \bigCI Z|X,W)_{G_{\overline{X},\overline{Z(W)}}}  \hspace{1cm} \Rightarrow \hspace{1cm}  P(Y|do(X),do(Z),W)=P(Y|do(X),W) $  \\[0.1cm]
			\noindent \hspace*{1.2cm} where $Z(W)$ denotes the set of nodes in $Z$ that are not ancestors of $W$ in $G_{\overline{X}}$.
		\end{minipage}
	\end{center}
\end{theorem}

Note that the soundness part of Theorem~\ref{Thr_ClassicalDSeparation} is a special case of Rule~1, corresponding to $X=\emptyset$. The do-calculus is complete for the problem of causal effect identifiability in the following sense \cite{Pearl_Causality}: whenever a causal effect is identifiable, then it can be computed in a finite sequence of steps, each employing one of the rules of the do-calculus, such that $P(Y|do(X))$ is identified with a final expression written entirely in terms of (conditional) probabilities over observed variables.

\section{Constraints on processes from causal structure \label{Sec_QuantumDoCalculus}} 

This section presents three pairs of theorems, each of which can be regarded as a generalization of one of the three rules of Theorem~\ref{Thm_ClassicalDoCalculus} (the do-calculus), first to classical split-node causal models, and then to quantum causal models. Each of the pairs of theorems involves the same graphical antecedent as the corresponding rule of the do-calculus. The theorems are generalizations of the rules of the do-calculus in the sense that they reduce to the latter via the inductions $I_{\sigma\rightarrow\kappa}$ and $I_{\kappa\rightarrow P}$ of Section~\ref{Sec_ClassicalQauntumComparison}. Of course, when the object under study is a (classical or quantum) process, rather than a probability distribution, the problem of causal effect identifiability does not arise. Even if a causal model involves some nodes that are regarded as unobserved, if a (classical or quantum) process is given over the observed nodes only, there is no ambiguity concerning the outcomes of interventions at the observed nodes -- probabilities for the outcomes of such interventions are by definition given by the (classical or quantum) process. The theorems that we present in this section, therefore, do not play the role of a calculus that is complete for a certain problem. The significance of the theorems lies, rather, in the fact that causal structure expressed by a DAG can be related to properties that any (classical or quantum) process must have if it is Markov for that DAG. Furthermore, as this section also shows, these properties are equivalent to (in the classical case) or imply (in the quantum case) operational statements, in the form of independences that must hold when considering the outcomes of interventions. The results of Section~\ref{Sec_QCMsAndUnitaryDynamics}, along with the fact that quantum causal models admit a theorem corresponding to each rule of the do-calculus, establish Def.~\ref{Def_QCM} as the most appropriate quantum generalization of classical causal models.

\subsection{Generalizations of rule 1 \label{SubSubsec_Rule1}} 

The first pair of theorems are slight generalizations of the soundness part of Theorem~\ref{Thm_QuantumDSeparationTheorem}, and its classical counterpart, to include the case in which an additional subset of nodes is subject to a do-intervention. This matches the sense in which Rule~1 of the classical do-calculus is a slight generalization of the soundness part of Theorem~\ref{Thr_ClassicalDSeparation}.

\subsubsection{Classical process maps}

\begin{definition} \textnormal{(Classical strong relative independence with a do-intervention):} \label{Def_StrongClassicalCIdo}
Given a classical process map $\clop_{YZWX}$, say that \emph{$Y$ and $Z$ are strongly independent relative to ($W$, $do(X)$)}, and write $(Y \bigCI Z | W do(X) )_{\clop_{YZWX}}$, if and only if there exist real functions $\alpha_{YWX^{\text{out}}} : Y^{\text{in}} \times Y^{\text{out}} \times W^{\text{in}} \times W^{\text{out}} \times X^{\text{out}} \rightarrow \mathds{R}$ and $\beta_{ZWX^{\text{out}}}: Z^{\text{in}} \times Z^{\text{out}} \times W^{\text{in}} \times W^{\text{out}} \times X^{\text{out}} \rightarrow \mathds{R}$, such that $\clop_{YZW do(X)}=\alpha_{YW X^{\text{out}}} \ \beta_{ZW X^{\text{out}}}$. 
\end{definition}
Roughly speaking, the interpretation of this definition is that with a do-intervention at $X$ nodes, i.e., with input variables at $X$ ignored, and output variables fixed to have a certain value, and with full knowledge of both inputs and outputs at $W$ nodes, learning the outcome of an intervention at $Y$ tells an agent nothing about the outcome of an intervention at $Z$. This interpretation is captured more precisely by the following proposition, which provides an operational formulation of Def.~\ref{Def_StrongClassicalCIdo}.
\begin{proposition} \label{Prop_Equivalence_COS1}
Given a classical process map $\clop_{YZWX}$, the condition $(Y \bigCI Z|Wdo(X))_{\clop_{YZWX}}$ holds if and only if:
\begin{center}
\begin{minipage}{15cm}
\begin{enumerate}[label=\textnormal{(COS{\arabic*})}, leftmargin=1cm]
\setcounter{enumi}{0}
\item For all values $X^{\text{out}}=x$ of a do-intervention at $X$, any maximally informative intervention at $W$ with outcome $k_W$, and any local interventions at $Y$ and $Z$ with outcomes $k_Y$ and $k_Z$, respectively, the joint probability distribution $P(k_Y , k_Z , k_W)$ satisfies $(k_Y \bigCI k_Z | k_W)_P$.
\label{COS1}
\end{enumerate}
\end{minipage}
\end{center}
\end{proposition}
\noindent {\bf Proof.} Observe that the condition $(Y \bigCI Z|Wdo(X))_{\clop_{YZWX}}$ is equivalent to the statement that  $(Y \bigCI Z| \allowbreak W)_{\clop_{YZWdo(X=x)}}$ holds for all values $x$ of a do-intervention at $X$. The result then follows from Prop.~\ref{Prop_CSCI_EquivalentOperationalStatements}. \hfill $\square$

\begin{theorem} \textnormal{(Rule 1 analogue for classical process maps):} \label{Thm_Rule1_cl}
Consider a DAG $G$, with a set $V$ of classical split nodes, and disjoint subsets of nodes $Y$, $Z$, $W$ and $X$, with $R:=V \setminus ( Y \cup Z \cup W \cup X)$. For any classical process map $\clop_{YZWXR}$ that is Markov for $G$,
\[
(Y \bigCI Z|X,W)_{G_{\overline{X}}} \quad \Rightarrow \quad \forall \tau_R \ \  (Y \bigCI Z | W do(X) )_{\clop_{YZWX}^{\tau_R}}.
\]
\end{theorem} 
\noindent {\bf Proof.} The proof is essentially the same as that of Theorem~\ref{Thm_Rule1_qu} below, with classical process maps replacing quantum process operators, classical channels replacing quantum channels, and classical interventions replacing quantum interventions. \hfill $\square$

Note that under $I_{\kappa \rightarrow P}$, the condition $(Y \bigCI Z | W do(X) )_{\kappa_{YZWX}}$ reduces to the consequent of Rule~1 of Theorem~\ref{Thm_ClassicalDoCalculus}, that is $P(Y|do(X),Z,W)=P(Y|do(X),W)$.

\subsubsection{Quantum process operators}

The quantum analogue of Def.~\ref{Def_StrongClassicalCIdo} is the following.
\begin{definition} \textnormal{(Quantum strong relative independence with a do-intervention):} \label{Def_StrongQuantumCIdo}
Given a process operator $\sigma_{YZWX}$, say that \emph{$Y$ and $Z$ are strongly independent relative to ($W$, $do(X)$)}, and write $(Y \bigCI Z | W do(X) )_{\sigma_{YZWX}}$, if and only if there exist Hermitian operators $\alpha_{YWX^{\text{out}}}$ and $\beta_{ZWX^{\text{out}}}$ such that \allowbreak $\sigma_{YZW do(X)} \allowbreak = \allowbreak \alpha_{YW X^{\text{out}}} \ \allowbreak \beta_{ZW X^{\text{out}}}$. 
\end{definition}
As in the discussion of strong relative independence in Section~\ref{SubSec_CondIndependence}, there is no obvious quantum analogue of the notion of a maximally informative intervention. For this reason, the following proposition provides an operational implication of Def.~\ref{Def_StrongQuantumCIdo}, rather than an equivalence.
\begin{proposition}\label{Prop_Implication_QOS1}
Given a process operator $\sigma_{YZWX}$, if $(Y \bigCI Z|Wdo(X))_{\sigma_{YZWX}}$, then:
\begin{center}
\begin{minipage}{15cm}
\begin{enumerate}[label=\textnormal{(QOS{\arabic*})}, leftmargin=1cm]
\setcounter{enumi}{0}
\item There exists a global intervention at $W X^{\text{out}}$, with outcome $k_{W X^{\text{out}}}$, such that for all local interventions at $Y$, $Z$, with outcomes $k_Y$, $k_Z$ respectively, the joint probability distribution $P(k_Y , k_Z , k_{WX^{\text{out}}})$, satisfies $(k_Y \bigCI k_Z | k_{WX^{\text{out}}})_P$.
\label{QOS1}
\end{enumerate}
\end{minipage}
\end{center}
\end{proposition}
\noindent {\bf Proof.} See Appendix~\ref{Subsec_Proof_Prop_Implication_QOS1}. \hfill $\square$
\begin{remark}\label{formofglobalinterventiononwxout}
The required intervention at $W X^{\text{out}}$, which gives $(k_Y \bigCI k_Z | k_{WX^{\text{out}}})_P$ for all interventions at $Y$ and $Z$, can be assumed to have the following form: let an agent be stationed at an additional locus $E$, such that for each node $N \in W$, the quantum system at $N^{\text{in}}$ is sent to $E$, one of a maximally entangled pair of systems is fed into $N^{\text{out}}$, and the other one is sent to $E$. Similarly, for each node $N$ in $X$, the system at $N^{\text{in}}$ is ignored, one of a maximally entangled pair of systems is fed into $N^{\text{out}}$, and the other one is sent to $E$. The agent at $E$ then performs an appropriately chosen joint von Neumann measurement on those systems that are incoming at $E$.
\end{remark}

\begin{theorem} \textnormal{(Analogue of rule 1 for process operators):} \label{Thm_Rule1_qu}
Consider a DAG $G$, with a set $V$ of quantum nodes, and disjoint subsets of nodes $Y$, $Z$, $W$ and $X$, with $R:=V \setminus ( Y \cup Z \cup W \cup X)$. For any process operator $\sigma_{YZWXR}$ that is Markov for $G$,
\[
(Y \bigCI Z|X,W)_{G_{\overline{X}}} \quad \Rightarrow \quad \forall \tau_R \ \  (Y \bigCI Z | W do(X) )_{\sigma_{YZWX}^{\tau_R}}.
\]
\end{theorem} 
\noindent {\bf Proof.} See Appendix~\ref{Subsubsec_Proof_Rule1}. \hfill $\square$

Note that under $I_{\sigma \rightarrow \kappa}$, the condition $(Y \bigCI Z | W do(X) )_{\sigma_{YZWX}}$ reduces to $(Y \bigCI Z | W do(X) )_{\kappa_{YZWX}}$ for the induced classical process map\footnote{An analogous comment to that of footnote~\ref{Footnote_RelInd_SigmaToKappa} applies, now relying on Prop.~\ref{Prop_EquivalentStatementsRule1} and Remark~\ref{Rem_EquivalentStatements_Kappa_Rule1}.}.

\subsubsection{Example}

Consider again the DAG $G$ of Fig.~\ref{Fig_Example_DAGForRules}, for convenience reproduced here as Fig.~\ref{Fig_Example_DAGForRules2}. Suppose that $G$ is the causal structure for a quantum causal model, with quantum nodes $N_1 , \ldots , N_5$, and a process operator $\sigma_{N_1 ... N_5}$ that is Markov for $G$.
\begin{figure}[H]
\centering
\small
\input{Figures/Fig_Example_DAGForRules.tex}
\caption{\label{Fig_Example_DAGForRules2} }
\end{figure}
In order to illustrate Theorem~\ref{Thm_Rule1_qu},  consider the subsets of nodes defined in Fig.~\ref{Fig_Example_Rule1_a}, along with the corresponding mutilated DAG. 
\begin{figure}[H]
\small
\begin{minipage}{16.5cm}
\begin{minipage}{8cm}
\centering
\begin{subfigure}{\textwidth}
\centering
\input{Figures/Fig_Example_Rule1_a.tex}
\caption{\label{Fig_Example_Rule1_a}} 
\end{subfigure}
\end{minipage}
\hfill
\begin{minipage}{8cm}
\centering
\begin{subfigure}{\textwidth}
\centering
\input{Figures/Fig_Example_Rule1_b.tex}
\caption{\label{Fig_Example_Rule1_b} }
\end{subfigure}
\end{minipage}
\end{minipage}
\caption[]{The mutilated DAGs $G_{\overline{X}}$ and $G_{\overline{X'}}$ for different choices of subsets of nodes of the DAG $G$ in Fig. \ref{Fig_Example_DAGForRules2}. 
The condition $(Y \bigCI Z | X,W)_{G_{\overline{X}}}$ holds, but the condition $(Y \bigCI Z | X',W')_{G_{\overline{X'}}}$ does not hold.}
\end{figure}
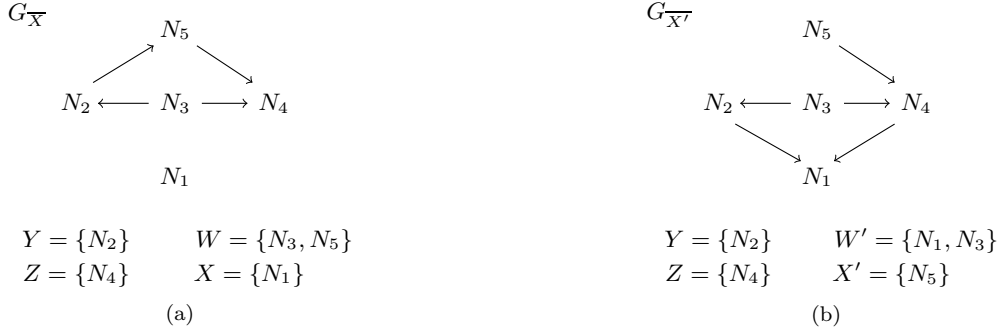
Here, there is a do-intervention at $N_1$, hence the incoming arrows at $N_1$ have been removed, and the condition $(Y \bigCI Z | W X)_{G_{\overline{X}}}$ holds. This captures the idea that a do-intervention does not reveal any information about $N_1$ that would tend to correlate $N_2$ and $N_4$. Hence, much as when $N_1$ is simply ignored, once the causal influences from the common cause $N_3$, and through the chain involving $N_5$, have been `picked up' by conditioning on the outcome of an appropriate intervention at $N_3 N_5$, the outcomes of arbitrary interventions at $N_2$ and $N_4$ are independent.

In contrast, if considering a do-intervention at $N_5$ instead of $N_1$, the mutilated graph of Fig.~\ref{Fig_Example_Rule1_b} fails to satisfy a corresponding d-separation condition, that is, the condition $(\{N_2\} \bigCI \{N_4\} | \{N_1, N_3, N_5\})_{G_{\overline{\{N_5\}}}}$ does not hold. This means that in general, $\sigma_{N_1 ... N_4 do(N_5)}$ does not satisfy the condition $\sigma_{N_1 ... N_4 do(N_5)} = \alpha_{N_2 N_1 N_3 N_5^{\text{out}}} \beta_{N_4 N_1 N_3 N_5^{\text{out}}}$. The statement (QOS1) does in fact hold, since with a trace-preserving intervention at $N_1$ (with trivial, or single-valued, outcome), there exists an intervention at $N_3 N_5^{\text{out}}$ such that conditioned on the outcome, the outcomes of any interventions at $N_2$ and $N_4$ are independent.

\subsubsection{Mathematically equivalent expressions}

\begin{proposition}  
\label{Prop_EquivalentStatementsRule1}
Let $\sigma_{YZWX}$ be a process operator over the disjoint sets of nodes $Y, Z, W$ and $X$. The following statements are equivalent:
\begin{center}
\begin{minipage}{15cm}
\begin{enumerate}[label=\textnormal{(\arabic*)}, leftmargin=1cm]
\item $(Y \bigCI Z | W do(X) )_{\sigma_{YZWX}}$
\item $\sigma_{YZWdo(X)} \ \star \ \sigma_{Wdo(X)}^{\tau_Y, \tau_Z} \ = \ \sigma_{YWdo(X)}^{ \tau_Z} \ \star \ \sigma_{ZWdo(X)}^{\tau_Y}$ \hspace{0.5cm} $\forall \ \  \mathrm{local \ interventions\ }\tau_Y, \ \tau_Z$.
\item $I(Y:Z|WX^{\text{out}})=0$, evaluated on $\hat{\sigma}_{YZWdo(X)}$.
\end{enumerate}
\end{minipage}
\end{center}
\end{proposition}

\noindent {\bf Proof.} See Appendix~\ref{Subsubsec_Proof_Prop_EquivalentStatementsRule1}.  \hfill $\square$
\begin{remark} \label{Rem_EquivalentStatements_Kappa_Rule1}
Very similar statements can be derived for classical process maps. Replacing $\sigma$ with $\clop$ throughout, the star products in (2) become ordinary multiplication, and the quantum conditional mutual information in (3) becomes the classical conditional mutual information. The proofs are essentially the same, with classical process maps replacing quantum process operators.
\end{remark}

\subsection{Generalizations of rule 2 \label{Subsubsec_Rule2}} 

\subsubsection{Classical process maps}

\begin{definition} \textnormal{(Classical strong independence from broken nodes):} \label{Def_ClassicalRIYZin}
Given a classical process map $\clop_{YZWX}$, say that \emph{$Y$ is strongly independent from $Z^{\text{in}}$ relative to ($W$, $do(X)$, $Z^{\text{out}}$)}, and write $(Y \bigCI Z^{\text{in}} | W do(X) Z^{\text{out}} )_{\clop_{YZWX}}$, if and only if there exist real functions $\alpha_{YWX^{\text{out}}Z^{\text{out}}} : Y^{\text{in}} \times Y^{\text{out}} \times W^{\text{in}} \times W^{\text{out}} \times X^{\text{out}} \times Z^{\text{out}} \rightarrow \mathds{R}$ and $\beta_{ZWX^{\text{out}}}: Z^{\text{in}} \times Z^{\text{out}} \times W^{\text{in}} \times W^{\text{out}} \times X^{\text{out}} \rightarrow \mathds{R}$, such that $\clop_{YZW do(X)}=\alpha_{YW X^{\text{out}} Z^{\text{out}}} \ \beta_{ZW X^{\text{out}}}$. 
\end{definition}
Roughly speaking, the interpretation of this definition is that with a do-intervention at $X$ nodes, i.e., with input variables at $X$ ignored, and output variables fixed to have a certain value, with full knowledge of both inputs and outputs at $W$ nodes, and full knowledge of outputs at $Z$ nodes, gaining information about the values of the inputs at $Z$ nodes tells an agent nothing about the outcome of an intervention at $Y$. 

In order to capture this interpretation more precisely with an equivalent operational formulation, define first the following subclass of interventions.
\begin{definition} \textnormal{(Classical breaking intervention):}\label{Def_BreakingDefinitioncl}
A \emph{breaking intervention} at a node $Z$ consists of a measurement of $Z^{\text{in}}$, giving outcome $k_Z$, and the preparation of a fixed value $z$ of $Z^{\text{out}}$. A necessary and sufficient condition is that: 
\begin{equation}
P(k_Z,Z^{\text{out}}|Z^{\text{in}}) = P(k_Z | Z^{\text{in}}) \delta (Z^{\text{out}} , z).
\end{equation}
For $Z$ a set of nodes, a breaking local intervention at $Z$ is a product of breaking interventions at each node in $Z$.
\end{definition}
A breaking intervention is a generalization of a do-intervention in which the value of the input variable is not ignored, but is allowed to influence an outcome $k_Z$. The following proposition then gives the equivalent operational formulation of Def.~\ref{Def_ClassicalRIYZin}.
\begin{proposition} \label{Prop_Rule2_CSMOperationalStatements}
Given a classical process map $\clop_{YZWX}$, the condition $(Y \bigCI Z^{\text{in}} | W do(X) Z^{\text{out}} )_{\clop_{YZWX}}$ holds if and only if:	
\begin{center}
\begin{minipage}{14cm}
\begin{enumerate}[label=\textnormal{(COS\arabic*)}, leftmargin=1cm]
\setcounter{enumi}{1}
\item For all values $X^{\text{out}}=x$ of a do-intervention at $X$, all maximally informative local interventions at $W$ with outcome $k_W$, all local interventions at $Y$ with outcome $k_Y$, and all breaking local interventions at $Z$ with outcome $k_Z$, the joint probability distribution $P(k_Y , k_Z , k_W)$ satisfies $(k_Y \bigCI k_Z | k_W)_P$.
\label{COS2}
\end{enumerate}
\end{minipage}
\end{center}
\end{proposition}

\noindent {\bf Proof.} See Appendix~\ref{Subsubsec_Proof_Prop_Rule2_CSMOperationalStatements}.  \hfill $\square$

\begin{theorem} \textnormal{(Rule 2 analogue for classical process maps):} \label{Thm_Rule2_cl}
Consider a DAG $G$, with a set $V$ of classical split nodes, and disjoint subsets of nodes $Y$, $Z$, $W$ and $X$, with $R:=V \setminus ( Y \cup Z \cup W \cup X)$. For any classical process map $\clop_{YZWXR}$ that is Markov for $G$,
\[
(Y \bigCI Z|X,W)_{G_{\overline{X}\underline{Z}}} \quad \Rightarrow \quad \forall \tau_R \ \  (Y \bigCI Z^{\text{in}} | W do(X) Z^{\text{out}} )_{\clop_{YZWX}^{\tau_R}}.
\]
\end{theorem} 
\noindent {\bf Proof.} The proof is essentially the same as that of Theorem~\ref{Thm_Rule2_qu} below, with classical process maps replacing quantum process operators, classical channels replacing quantum channels, and classical interventions replacing quantum interventions. \hfill $\square$

Note that under $I_{\kappa \rightarrow P}$, the condition $(Y \bigCI Z^{\text{in}} | W do(X) Z^{\text{out}} )_{\clop_{YZWX}}$ reduces to the consequent of Rule~2 of Theorem~\ref{Thm_ClassicalDoCalculus}, that is $P(Y|do(X),do(Z),W)=P(Y|do(X),Z,W)$.

\subsubsection{Quantum process operators}

\begin{definition} \textnormal{(Quantum strong independence from broken nodes):} \label{Def_QuantumRIYZin}
Given a process operator $\sigma_{YZWX}$, say that \emph{$Y$ is strongly independent from $Z^{\text{in}}$, relative to ($W$, $do(X)$, $Z^{\text{out}}$ )}, and write $(Y \bigCI Z^{\text{in}} | \allowbreak W do(X) \allowbreak Z^{\text{out}} )_{\sigma_{YZWX}}$, if and only if there exist Hermitian operators $\alpha_{YWX^{\text{out}}Z^{\text{out}}}$ and $\beta_{ZWX^{\text{out}}}$ such that $\sigma_{YZW do(X)}= \allowbreak\alpha_{YW X^{\text{out}} Z^{\text{out}}} \ \allowbreak \beta_{ZW X^{\text{out}}}$. 
\end{definition}
An operational implication of this notion can be given in terms of the existence of a global intervention at $W X^{\text{out}} Z^{\text{out}}$. Conditioned on its outcome, measurements of $Z^{\text{in}}$ give no information about the outcome of an intervention at $Y$.
\begin{proposition} \label{Prop_Implication_QOS2}
Consider a process operator $\sigma_{YZWX}$. If $(Y \bigCI Z^{\text{in}} | W do(X) Z^{\text{out}} )_{\sigma_{YZWX}}$, then:
\begin{center}
\begin{minipage}{15cm}
\begin{enumerate}[label=\textnormal{(QOS{\arabic*})}, leftmargin=1cm]
\setcounter{enumi}{1}
\item There exists a global intervention at $W X^{\text{out}} Z^{\text{out}} $, with outcome $k_{W X^{\text{out}} Z^{\text{out}} }$, such that for all interventions at $Y$, with outcome $k_Y$, and all measurements of $Z^{\text{in}}$ with outcome $k_{Z^{\text{in}}}$, the joint probability distribution $P(k_Y , k_{Z^{\text{in}}} , k_{W X^{\text{out}} Z^{\text{out}} })$ satisfies $(k_Y \bigCI k_{Z^{\text{in}}} | k_{W X^{\text{out}} Z^{\text{out}} } )_P$.
\label{QOS2}
\end{enumerate}
\end{minipage}
\end{center}
\end{proposition}
\noindent {\bf Proof.} See Appendix~\ref{Subsec_Proof_ImpliedQOS2}.  \hfill $\square$
\begin{remark}
The global intervention at $W X^{\text{out}} Z^{\text{out}} $ has a similar form to that of Remark~\ref{formofglobalinterventiononwxout}, except that one of a maximally entangled pair of systems is additionally sent to $Z^{\text{out}}$, with the other one going to $E$. 
\end{remark}

\begin{theorem} \textnormal{(Rule 2 analogue for process operators):} \label{Thm_Rule2_qu}
Consider a DAG $G$, with a set $V$ of quantum nodes, and disjoint subsets of nodes $Y$, $Z$, $W$ and $X$, with $R:=V \setminus ( Y \cup Z \cup W \cup X)$. For any process operator $\sigma_{YZWXR}$ that is Markov for $G$,
\[
(Y \bigCI Z|X,W)_{G_{\overline{X}\underline{Z}}} \quad \Rightarrow \quad \forall \tau_R \ \  (Y \bigCI Z^{\text{in}} | W do(X) Z^{\text{out}} )_{\sigma^{\tau_R}_{YZWX}}.
\]
\end{theorem} 
\noindent {\bf Proof.} See Appendix~\ref{Subsec_Proof_Rule_2}. \hfill $\square$

Note that  under $I_{\sigma \rightarrow \kappa}$, the condition $(Y \bigCI Z^{\text{in}} | W do(X) Z^{\text{out}} )_{\sigma_{YZWX}}$ reduces to $(Y \bigCI Z^{\text{in}} | \allowbreak W do(X) \allowbreak Z^{\text{out}} )_{\kappa_{YZWX}}$ for the induced classical process map\footnote{Again, an analogous comment to that of footnote~\ref{Footnote_RelInd_SigmaToKappa} applies, now relying on Prop.~\ref{Prop_EquivalentStatementsRule2} and Remark~\ref{Rem_EquivalentStatements_Kappa_Rule2}.}.

\subsubsection{Example}

Consider again the DAG $G$ of Fig.~\ref{Fig_Example_DAGForRules}, for convenience reproduced here as Fig.~\ref{Fig_Example_DAGForRules3}. Suppose that $G$ is the causal structure for a quantum causal model, with quantum nodes $N_1 , \ldots , N_5$, and a process operator $\sigma_{N_1 ... N_5}$ that is Markov for $G$.
\begin{figure}[H]
\centering
\small
\input{Figures/Fig_Example_DAGForRules.tex}
\caption{\label{Fig_Example_DAGForRules3} }
\end{figure}
In order to illustrate Theorem~\ref{Thm_Rule2_qu},  consider the subsets of nodes defined in Fig.~\ref{Fig_Example_Rule2_a}, along with the corresponding mutilated DAG. 
\begin{center}
\begin{figure}[H]
\small
\begin{minipage}{16.cm}
\begin{minipage}{6cm}
\centering
\begin{subfigure}{\textwidth}
\centering
\input{Figures/Fig_Example_Rule2_a.tex}
\caption{\label{Fig_Example_Rule2_a}} 
\end{subfigure}
\end{minipage}
\hfill
\begin{minipage}{3cm}
\begin{eqnarray}
Y&=&\{N_4,N_5\} \nonumber \\
Z&=&\{N_2\} \nonumber \\
W&=&\{N_3\} \nonumber \\
X&=&\{N_1\}	\nonumber 
\end{eqnarray}	
\vspace*{0.5cm}				
\end{minipage}
\hfill
\begin{minipage}{6cm}
\centering
\begin{subfigure}{\textwidth}
\centering
\input{Figures/Fig_Example_Rule2_b.tex}
\caption{\label{Fig_Example_Rule2_b} }
\end{subfigure}
\end{minipage}
\end{minipage}
\caption[]{The mutilated DAGs $G_{\overline{X},\underline{Z}}$ and $G_{\overline{X},\underline{Y}}$ for the depicted choice of subsets of nodes of the DAG $G$ in Fig.~\ref{Fig_Example_DAGForRules3}. The condition $(Y \bigCI Z | X,W)_{G_{\overline{X},\underline{Z}}}$ holds, but the condition $(Z \bigCI Y | X,W)_{G_{\overline{X},\underline{Y}}}$ does not hold. } 
\end{figure}
\end{center}

Observe that in the DAG $G$ of Fig.~\ref{Fig_Example_DAGForRules3},  $\{N_2\}$ is not d-separated from $\{ N_4 , N_5 \} $ by $\{ N_1 , N_3\} $, since $N_2$ is a direct cause of $N_5$. The same observation holds in the DAG $G_{\overline{\{N_1\}}}$. Hence even with a do-intervention at $N_1$, the outcome of an intervention at $N_2$ is in general correlated with those at $N_4$ and $N_5$.

However, the condition $(Y \bigCI Z | X,W)_{G_{\overline{X},\underline{Z}}}$ does hold. This captures the idea that -- relative to $N_3$ -- there is no backwards inferential pathway from $N_2$ to $\{N_4 , N_5\}$, hence -- relative to $N_3$ -- all correlation between $N_2$ and $\{ N_4 , N_5 \}$ is due to the causal pathway from $N_2$ to $N_5$. Theorem~\ref{Thm_Rule2_qu}, together with Prop.~\ref{Prop_Implication_QOS2}, implies (QOS2): with a suitable intervention at $N_3 N_1^{\text{out}} N_2^{\text{out}}$ -- in particular one that accounts for the common cause $N_3$ -- the outcome of a measurement of $N_2^{\text{in}}$ is independent of the outcome of any intervention at $N_4 N_5$.

On the other hand, with the roles of $Y$ and $Z$ exchanged, as in Fig.~\ref{Fig_Example_Rule2_b}, the condition $(Z \bigCI Y | X,W)_{G_{\overline{X},\underline{Y}}}$ does not hold. Here, it will in general be the case that for any intervention at $N_3 N_1^{\text{out}} N_4^{\text{out}} \allowbreak N_5^{\text{out}}$, even conditioned on its outcome, a measurement of (say) $N_5^{\text{in}}$ can be correlated with an intervention at $N_2$.

\subsubsection{Mathematically equivalent expressions}

\begin{proposition}\label{Prop_EquivalentStatementsRule2}
Given a process operator $\sigma_{YZWX}$, the following statements are equivalent:
\begin{center}
\begin{minipage}{15cm}
\begin{enumerate}[leftmargin=1cm]
\item $(Y \bigCI Z^{\text{in}} | W do(X) Z^{\text{out}} )_{\sigma_{YZWX}}$
\item $\sigma_{YZWdo(X)} \star \sigma_{Wdo(X)do(Z)}^{\tau_Y} \ = \ \sigma_{ZWdo(X)}^{\tau_Y}  \star  \sigma_{YWdo(X)do(Z)} \ $ \hspace{0.5cm}  $\forall \ \ \mathrm{local\ interventions\ }\tau_Y$.
\item $I(Y:Z^{\text{in}} |WX^{\text{out}}Z^{\text{out}} )=0$, evaluated on $\hat{\sigma}_{YZWdo(X)}$.
\end{enumerate}
\end{minipage}
\end{center}
\end{proposition}
\noindent {\bf Proof.} See Appendix~\ref{Subsec_Proof_EquivalentStRule_2}.  \hfill $\square$
\begin{remark} \label{Rem_EquivalentStatements_Kappa_Rule2}
Very similar statements can be derived for classical process maps. Replacing $\sigma$ with $\clop$ throughout, the star products in (2) become ordinary multiplication, and the quantum conditional mutual information in (3) becomes the classical conditional mutual information. The proofs are essentially the same, with classical process maps replacing quantum process operators.
\end{remark}

\subsection{Generalizations of rule 3 \label{Subsec_Rule3}} 

\subsubsection{Classical process maps}

\begin{definition} \textnormal{(Classical strong independence from settings):} \label{Def_ClassicalRIYZset}
Given a classical process map $\clop_{YZWX}$, say that \emph{$Y$ is strongly independent from the setting at $Z$, relative to ($W$, $do(X)$)}, and write $(Y \bigCI Set(Z) | W do(X) )_{\clop_{YZWX}}$, if and only if there is a real-valued function $\eta_{YWX^{\text{out}}} : Y^{\text{in}} \times Y^{\text{out}} \times W^{\text{in}} \times W^{\text{out}}\times X^{\text{out}} \rightarrow  \mathds{R}$ such that for all local interventions $\tau_Z$ at $Z$, there is a real valued function $\xi^{\tau_Z}_{W X^{\text{out}}} : W^{\text{in}}\times W^{\text{out}} \times X^{\text{out}} \rightarrow  \mathds{R}$, such that 
\[
\clop_{YW do(X)}^{\tau_Z} = \eta_{YW X^{\text{out}}} \xi^{\tau_Z}_{W X^{\text{out}}}.
\]
\end{definition}
At first sight, this definition looks involved, but the intuitive meaning is fairly simple. Consider first the case where $X=W=\emptyset$. The definition then simply expresses that $\clop_{Y}^{\tau_Z}$ is independent of the choice of $\tau_Z$, hence the outcome of any intervention at $Y$ is independent of the choice of $\tau_Z$: in other words, there is no signalling from $Z$ to $Y$. In the general case of non-empty $X$ and $W$, the definition captures a similar idea: from the perspective of an agent who knows the value of a do-intervention at $X$, and has full knowledge of both inputs and outputs at $W$ nodes, learning the value of $\tau_Z$, that is the choice of intervention at $Z$, tells the agent nothing about the outcome of an intervention at $Y$. This is captured more precisely by the following proposition, which gives an operational statement equivalent to Def.~\ref{Def_ClassicalRIYZset}.
\begin{proposition} \label{Prop_Rule3_CSMOperationalStatements}
Given a classical process map $\clop_{YZWX}$, the condition $(Y \bigCI Set(Z) | W do(X) )_{\clop_{YZWX}}$ holds if and only if:
\begin{center}
\begin{minipage}{14cm}
\begin{enumerate}[label=\textnormal{(COS\arabic*)}, leftmargin=1cm]
\setcounter{enumi}{2}
\item For all values $X^{\text{out}}=x$ of a do-intervention at $X$, all maximally informative local interventions at $W$ with outcome $k_W$, and all local interventions at $Y$ with outcome $k_Y$, the conditional probability $P(k_Y|k_W)$ is independent of the choice of local intervention at $Z$.
\label{COS3}
\end{enumerate}
\end{minipage}
\end{center}
\end{proposition}
\noindent {\bf Proof.} See Appendix~\ref{Subsec_Proof_Rule3_CSMOperationalStatements}.  \hfill $\square$

\begin{theorem} \textnormal{(Rule 3 analogue for classical process maps):} \label{Thm_Rule3_cl}
Consider a DAG $G$, with a set $V$ of classical split nodes, and disjoint subsets of nodes $Y$, $Z$, $W$ and $X$, with $R:=V \setminus ( Y \cup Z \cup W \cup X)$. For any classical process map $\clop_{YZWXR}$ that is Markov for $G$,
\[
(Y \bigCI Z|X,W)_{G_{\overline{X},\overline{Z(W)}}} \quad \Rightarrow \quad \forall \tau_R \ \ (Y \bigCI Set(Z) | W do(X) )_{\clop^{\tau_R}_{YZWX}}.
\]
\end{theorem} 
\noindent {\bf Proof.} The proof is essentially the same as that of Theorem~\ref{Thm_Rule3_qu} below, with classical process maps replacing quantum process operators, classical channels replacing quantum channels, and classical interventions replacing quantum interventions.  \hfill $\square$

Note that under $I_{\kappa \rightarrow P}$, the condition $(Y \bigCI Set(Z) | W do(X) )_{\clop_{YZWX}}$ reduces to the consequent of Rule~3 of Theorem~\ref{Thm_ClassicalDoCalculus}, that is $P(Y|do(X),do(Z),W)=P(Y|do(X),W)$.

\subsubsection{Quantum process operators}

\begin{definition} \textnormal{(Quantum strong independence from settings):} \label{Def_QuantumRIYZset}
Given a process operator $\sigma_{YZWX}$, say that \emph{$Y$ is strongly independent from the setting at $Z$, relative to ($W$, $do(X)$)}, and write $(Y \bigCI Set(Z) | W do(X) )_{\sigma_{YZWX}}$, if and only if there is a Hermitian operator $\eta_{YWX^{\text{out}}}$ such that for all local interventions $\tau_Z$ at $Z$, there is a Hermitian operator $\xi^{\tau_Z}_{W X^{\text{out}}}$, such that 
\[
\sigma_{YW do(X)}^{\tau_Z} = \eta_{YW X^{\text{out}}} \xi^{\tau_Z}_{W X^{\text{out}}}.
\]
\end{definition}
The interpretation of this definition is similar to that of Def.~\ref{Def_ClassicalRIYZset}, with the usual wrinkle that there is no quantum equivalent of an agent who has full knowledge of both $W^{\text{in}}$ and $W^{\text{out}}$. The following proposition gives an operational statement that is implied by, but not equivalent to, Def.~\ref{Def_QuantumRIYZset}.
\begin{proposition} \label{Prop_Implication_QOS3}
Consider a process operator $\sigma_{YZWX}$. If $(Y \bigCI Set(Z) | W do(X) )_{\sigma_{YZWX}}$, then:
\begin{center}
\begin{minipage}{15cm}
\begin{enumerate}[label=\textnormal{(QOS{\arabic*})}, leftmargin=1cm]
\setcounter{enumi}{2}
\item There exists a global intervention at $W X^{\text{out}}$, with outcome $k_{W X^{\text{out}}}$, such that for all local interventions at $Y$, with outcome $k_Y$, the conditional probability $P(k_Y | k_{WX^{\text{out}}})$ is independent of the choice of local intervention at $Z$.
\label{QOS3}
\end{enumerate}
\end{minipage}
\end{center}
\end{proposition}
\noindent {\bf Proof.} See Appendix~\ref{Subsec_Proof_Prop_Implication_QOS3}. \hfill $\square$
\begin{remark}
The global intervention at $W X^{\text{out}}$ has the same form as that of Remark~\ref{formofglobalinterventiononwxout}.
\end{remark}

\begin{theorem} \textnormal{(Rule 3 analogue for quantum processes):} \label{Thm_Rule3_qu}
Consider a DAG $G$, with a set $V$ of quantum nodes, and disjoint subsets of nodes $Y$, $Z$, $W$ and $X$, with $R:=V \setminus ( Y \cup Z \cup W \cup X)$. For any process operator $\sigma_{YZWXR}$ that is Markov for $G$,
\[
(Y \bigCI Z|X,W)_{G_{\overline{X},\overline{Z(W)}}} \quad \Rightarrow \quad \forall \tau_R \ \ (Y \bigCI Set(Z) | W do(X) )_{\sigma^{\tau_R}_{YZWX}}
\]
\end{theorem} 
\noindent {\bf Proof.} See Appendix~\ref{SubSec_Appendix_Proof_Rule3}. \hfill $\square$

Note that under $I_{\sigma \rightarrow \kappa}$, the condition $(Y \bigCI Set(Z) | W do(X) )_{\sigma_{YZWX}}$ reduces to $(Y \bigCI Set(Z) | \allowbreak W \allowbreak do(X) )_{\kappa_{YZWX}}$ for the induced classical process map\footnote{An analogous comment to that of footnote~\ref{Footnote_RelInd_SigmaToKappa} applies. In order to see that $\eta_{YWX^{out}}$ and $\xi_{WX^{out}}^{\tau_Z}$ can be chosen to be diagonal with respect to the same preferred basis in which $\sigma_{YWdo(X)}^{\tau_Z}$ is diagonal, observe the following.
In the proof of Prop. \ref{Prop_Implication_QOS3} it was shown that, if $(Y \bigCI Set(Z) | W do(X) )_{\sigma_{YZWX}}$, then there exists a decomposition into orthogonal subspaces of the form $WX^{\text{out}}=\bigoplus_i F_i^L \otimes F_i^R$ such that $\sigma_{YWdo(X)}^{\tau_Z}  = \sum_i \eta_{YF_i^L} \otimes \xi_{F_i^R}^{\tau_Z}$  for all $\tau_Z$. Hence, if the left-hand side is diagonal in the preferred basis, so are $\eta_{YWX^{out}}$ and $\xi_{WX^{out}}^{\tau_Z}$.}.

\subsubsection{Example}

Consider again the DAG $G$ of Fig.~\ref{Fig_Example_DAGForRules}, for convenience reproduced here as Fig.~\ref{Fig_Example_DAGForRules4}. Suppose that $G$ is the causal structure for a quantum causal model, with quantum nodes $N_1 , \ldots , N_5$, and a process operator $\sigma_{N_1 ... N_5}$ that is Markov for $G$. 
\begin{figure}[H]
\centering
\small
\input{Figures/Fig_Example_DAGForRules.tex}
\caption{\label{Fig_Example_DAGForRules4} }
\end{figure}
In order to illustrate Theorem~\ref{Thm_Rule3_qu},  consider the subsets of nodes defined in Fig.~\ref{Fig_Example_Rule3_a}, along with the corresponding mutilated DAG. 
\begin{center}
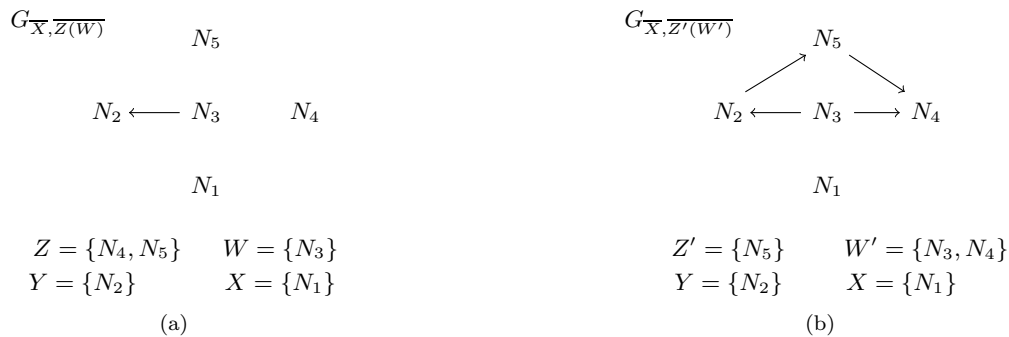
\begin{figure}[H]
\small
\begin{minipage}{16.5cm}
\begin{minipage}{8cm}
\centering
\begin{subfigure}{\textwidth}
\centering
\input{Figures/Fig_Example_Rule3_a.tex}
\caption{\label{Fig_Example_Rule3_a}} 
\end{subfigure}
\end{minipage}
\hfill
\begin{minipage}{8cm}
\centering
\begin{subfigure}{\textwidth}
\centering
\input{Figures/Fig_Example_Rule3_b.tex}
\caption{\label{Fig_Example_Rule3_b} }
\end{subfigure}
\end{minipage}
\end{minipage}
\caption[]{The mutilated DAGs $G_{\overline{X},\overline{Z(W)}}$ and $G_{\overline{X},\overline{Z'(W')}}$ for the specified choices of subsets of nodes of the DAG $G$ in Fig.~\ref{Fig_Example_DAGForRules4}. The condition $(Y \bigCI Z | X,W)_{G_{\overline{X},\overline{Z(W)}}}$ holds, but the condition $(Y \bigCI Z' | X,W')_{G_{\overline{X},\overline{Z'(W')}}}$ does not hold. }
\end{figure}
\end{center}

The condition $(Y \bigCI Z | X,W)_{G_{\overline{X},\overline{Z(W)}}}$ holds. Hence Theorem~\ref{Thm_Rule3_qu}, together with Prop.~\ref{Prop_Implication_QOS3}, implies (QOS3): for a suitable intervention at $N_3 N_1^{\text{out}}$, then conditioned on its outcome there is no signalling from $\{ N_4 N_5 \}$ to $N_2$. (In fact, in this case, the statement holds for any intervention at $N_3 N_1^{\text{out}}$.)

On the other hand, with $Z'=\{N_5\}$ and $W'=\{N_3,N_4\}$, as in Fig.~\ref{Fig_Example_Rule3_b}, the condition $(Y \bigCI Z' | X,W')_{G_{\overline{X},\overline{Z'(W')}}}$ fails. The failure of the condition follows from the fact that $N_5$ is an ancestor of $N_4$, and corresponds to the idea that conditioning on the outcome of an intervention at $N_4$ in general induces a correlation between the outcome of an intervention at $N_2$ and the choice of intervention at $N_5$. That is: in a scenario involving the postselection of an outcome at $N_4$, the `backwards signalling' from $N_5$ to $N_2$ appears possible.  Again, however, the operational statement (QOS3) is weak: it still holds in this scenario, for the simple reason that the intervention at $W' X^{\text{out}}$ can be chosen to be a trace-preserving intervention with trivial, or single-valued, outcome.

\subsection{Overview}

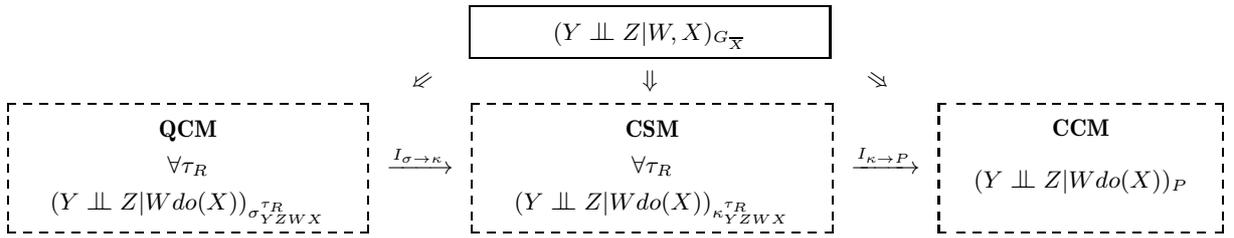
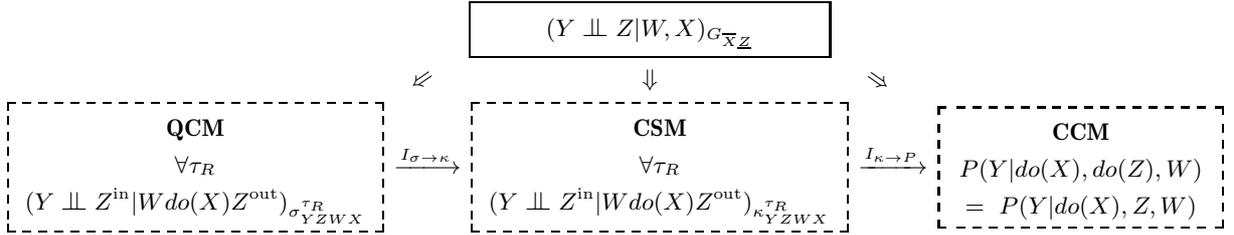
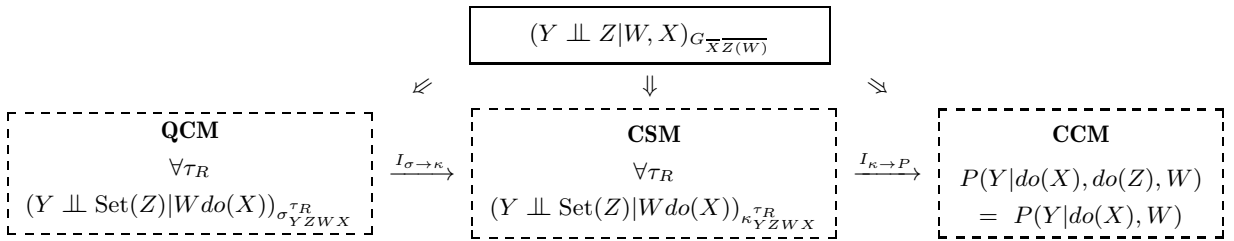
\begin{figure}[H]
\small
\begin{subfigure}{\textwidth}
\begin{center}
	\begin{minipage}{16.5cm}
		\begin{center}
			\hspace*{0.3cm}
			\fbox{
				\begin{minipage}{4.3cm}
					\vspace*{0.1cm}
					\centering
					$( Y \bigCI Z | W,X)_{G_{\overbar{X}}}$	
				\end{minipage}
			}\\[0.1cm]
			\hspace*{0.3cm}
			\renewcommand{\Scale}{6.0}
			\begin{tikzpicture}
				\node [style=none, rotate=-55] at (-1.0*\Scale, 0) {$\Downarrow$};
				\node [style=none] at (0,0) {$\Downarrow$};
				\node [style=none, rotate=55] at (1.0*\Scale, 0) {$\Downarrow$};
			\end{tikzpicture}
			\vspace*{0.1cm}
		\end{center}
	\end{minipage}
	\begin{minipage}{16.0cm}
		\dbox{
			\begin{minipage}{4.3cm}
				\vspace*{0.1cm}
				\centering
				\key{QCM}\\[0.1cm]
				$\forall \tau_R$ \\[0.1cm]
				$( Y \bigCI Z | W do(X) )_{\sigma_{YZWX}^{\tau_R}}$
			\end{minipage}
		}
		\hfill
		$\xrightarrow{I_{\sigma\rightarrow\kappa}}$
		\hfill
		\dbox{
			\begin{minipage}{4.3cm}
				\vspace*{0.1cm}
				\centering
				\key{CSM}\\[0.1cm]
				$\forall \tau_R$ \\[0.1cm]
				$( Y \bigCI Z | W do(X) )_{\clop_{YZWX}^{\tau_R}}$ 
			\end{minipage}
		}
		\hfill
		$\xrightarrow{I_{\kappa\rightarrow P}}$
		\hfill
		\dbox{
			\begin{minipage}{3.3cm}
				\vspace*{0.1cm}
				\centering
				\key{CCM} \\[0.3cm]
				$( Y \bigCI Z | W do(X) )_{P}$ \\[0.1cm]
				\vspace*{0.3cm}
			\end{minipage}
		}
	\end{minipage}
\end{center}
\caption{Generalization of rule 1.}
\end{subfigure}

\vspace*{0.7cm}
\begin{subfigure}{\textwidth}
\begin{center}
	\begin{minipage}{16.5cm}
		\begin{center}
			\hspace*{0.3cm}
			\fbox{
				\begin{minipage}{4.3cm}
					\vspace*{0.1cm}
					\centering
					$( Y \bigCI Z |W,X )_{G_{\overbar{X}\underline{Z}}}$	
				\end{minipage}
			}\\[0.1cm]
			\hspace*{0.3cm}
			\renewcommand{\Scale}{6.0}
			\begin{tikzpicture}
				\node [style=none, rotate=-55] at (-1.0*\Scale, 0) {$\Downarrow$};
				\node [style=none] at (0,0) {$\Downarrow$};
				\node [style=none, rotate=55] at (1.0*\Scale, 0) {$\Downarrow$};
			\end{tikzpicture}
			\vspace*{0.1cm}
		\end{center}
	\end{minipage}
	\begin{minipage}{16.0cm}
		\dbox{
			\begin{minipage}{4.5cm}
				\vspace*{0.1cm}
				\centering
				\key{QCM}\\[0.1cm]
				$\forall \tau_R$ \\[0.1cm]
				\small{$(Y \bigCI Z^{\text{in}} | W do(X) Z^{\text{out}} )_{\sigma^{\scriptscriptstyle \tau_R}_{\scriptscriptstyle YZWX}}$}
			\end{minipage}
		}
		\hfill
		$\xrightarrow{I_{\sigma\rightarrow\kappa}}$
		\hfill
		\dbox{
			\begin{minipage}{4.5cm}
				\vspace*{0.1cm}
				\centering
				\key{CSM}\\[0.1cm]
				$\forall \tau_R$ \\[0.1cm]
				\small{$(Y \bigCI Z^{\text{in}} | W do(X) Z^{\text{out}} )_{\clop^{\scriptscriptstyle \tau_R}_{\scriptscriptstyle YZWX}}$}
			\end{minipage}
		}
		\hfill
		$\xrightarrow{I_{\kappa\rightarrow P}}$
		\hfill
		\dbox{
			\begin{minipage}{3.3cm}
				\vspace*{0.1cm}
				\begin{center}
					\key{CCM} \\[0.1cm]
					$P(Y|do(X),do(Z),W)$ \\[0.1cm]
					$= \ P(Y|do(X),Z,W)$
				\end{center}
			\end{minipage}
		}
	\end{minipage}
\end{center}
\caption{Generalization of rule 2.}
\end{subfigure}

\vspace*{0.7cm}
\begin{subfigure}{\textwidth}
\begin{center}
	\begin{minipage}{16.5cm}
		\begin{center}
			\hspace*{0.3cm}
			\fbox{
				\begin{minipage}{4.3cm}
					\vspace*{0.1cm}
					\centering
					$( Y \bigCI Z | W,X )_{G_{\overbar{X}\overbar{Z(W)}}}$	
				\end{minipage}
			}\\[0.1cm]
			\hspace*{0.3cm}
			\renewcommand{\Scale}{6.0}
			\begin{tikzpicture}
				\node [style=none, rotate=-55] at (-1.0*\Scale, 0) {$\Downarrow$};
				\node [style=none] at (0,0) {$\Downarrow$};
				\node [style=none, rotate=55] at (1.0*\Scale, 0) {$\Downarrow$};
			\end{tikzpicture}
			\vspace*{0.1cm}
		\end{center}
	\end{minipage}
	\begin{minipage}{16.0cm}
		\dbox{
			\begin{minipage}{4.35cm}
				\centering
				\key{QCM}\\[0.1cm]
				$\forall \tau_R$ \\[0.1cm]
				\small{$(Y \bigCI \text{Set}(Z) | W do(X) )_{\sigma^{\scriptscriptstyle \tau_R}_{\scriptscriptstyle YZWX}}$}
			\end{minipage}
		}
		\hfill
		$\xrightarrow{I_{\sigma\rightarrow\kappa}}$
		\hfill
		\dbox{
			\begin{minipage}{4.35cm}
				\vspace*{0.1cm}
				\centering
				\key{CSM}\\[0.1cm]
				$\forall \tau_R$ \\[0.1cm]
				\small{$(Y \bigCI \text{Set}(Z) | W do(X) )_{\clop^{\scriptscriptstyle \tau_R}_{\scriptscriptstyle YZWX}}$} 
			\end{minipage}
		}
		\hfill
		$\xrightarrow{I_{\kappa\rightarrow P}}$
		\hfill
		\dbox{
			\begin{minipage}{3.3cm}
				\vspace*{0.1cm}
				\begin{center}
					\key{CCM} \\[0.2cm]
					$P(Y|do(X),do(Z),W)$ \\[0.1cm]
					$= \ P(Y|do(X),W)$
				\end{center}
			\end{minipage}
		}
	\end{minipage}
\end{center}
\caption{Generalization of rule 3.}
\end{subfigure}

\caption{\label{Fig_robinsbigdiagram} Overview of the three pairs of theorems, along with the corresponding rules of the classical do-calculus, and the relationships between them.}
\end{figure}

Fig.~\ref{Fig_robinsbigdiagram} summarizes the generalizations of the three rules, as applied to quantum causal models, classical split-node causal models, and classical causal models.

\section{Note on causal inference \label{Sec_NoteCausalDiscovery}} 

In previous sections, the causal structure of a set $V$ of quantum nodes is mostly assumed given, and the questions that are answered are to do with the constraints that the causal structure imposes on marginal process operators on subsets of the nodes. The problem of causal inference concerns a situation in which the causal structure is not known: given at least some information about the process operator $\sigma_V$, what can we say about the causal structure of the nodes in $V$? We do not address this problem in any kind of generality. In this section, we discuss the simplest case, which is that the process operator $\sigma_V$ is in fact completely specified. In case this seems trivial, note that even given  $\sigma_V$, there is substantial causal knowledge to be discovered: is there a causal order of the nodes at all? Is $V$ a causally complete set or are common causes missing? Does there exist a quantum causal model, which would serve as a candidate causal explanation of $\sigma_V$? 

Giarmatzi and Costa consider some of these questions in Ref.~\cite{GiarmatziEtAl_2018_CausalDiscoveryAlgorithm}, presenting a quantum causal discovery algorithm. However, as presented, the algorithm is based on a different definition of the quantum Markov condition (see Section~\ref{Sec_RelatedWork}). This section presents similar ideas to those of Ref.~\cite{GiarmatziEtAl_2018_CausalDiscoveryAlgorithm}, but adapted for the framework we have introduced.

\begin{definition} \textnormal{(`No-signalling relation for a channel')} \label{Def_NoSig}	
	Let $\mathcal{C} : \mathcal{L}(\mathcal{H}_{A}) \otimes \mathcal{L}(\mathcal{H}_{B}) \rightarrow \mathcal{L}(\mathcal{H}_{C}) \otimes \mathcal{L}(\mathcal{H}_{D})$ be a channel, and let $\rho^{\mathcal{C}}_{CD|AB}$ be the CJ representation of it. Write $A \nosig D$ (`$A$ does not signal to $D$') if and only if there exists a quantum channel $\mathcal{M}: \mathcal{L}(\mathcal{H}_B)\rightarrow \mathcal{L}(\mathcal{H}_D)$, with CJ representation $\rho^{\mathcal{M}}_{D|B}$ such that $\Trace_{C} [ \rho^{\mathcal{C}}_{CD|AB} ] = \rho^{\mathcal{M}}_{D|B} \otimes \mathds{1}_{A^*}$.
\end{definition}
Recall that in a generic channel $\rho_{CD|AB}$, it is possible that $A \nosig C$ and $A \nosig D$, while $A$ can signal to the composite $CD$ (see Remark~\ref{Remark_Causalinfluenceunitary}).
Based on this notion of a no-signalling relation, observe that any process operator $\sigma_V$, seen as the representation of a CPTP map $\mathcal{L}(\mathcal{H}_{A_1} \otimes \cdots \otimes \mathcal{H}_{A_n}) \rightarrow \mathcal{L}(\mathcal{H}_{A_1} \otimes \cdots \otimes \mathcal{H}_{A_n})$, defines a graph encoding the relations of direct signalling to single nodes:
\begin{definition} \textnormal{(Simple induced graph):}
Consider a process operator $\sigma_V$ over the set $V=\{A_1,...,A_n\}$ of quantum nodes, viewed as a CPTP map $\mathcal{L}(\mathcal{H}_{A_1} \otimes \cdots \otimes \mathcal{H}_{A_n}) \rightarrow \mathcal{L}(\mathcal{H}_{A_1} \otimes \cdots \otimes \mathcal{H}_{A_n})$. Let the \textnormal{simple induced graph}, written $G^{S}_{\sigma}$, be the graph with nodes $V$, and an arrow $A_j \rightarrow A_i$ if and only if $A_j$ can signal to $A_i$, in the sense of Def.~\ref{Def_NoSig}, in $\sigma_V$.
\end{definition}
Note that such an induced graph $G^{S}_{\sigma}$ only encodes the no-signalling relations for single systems, but not those also considering composite systems, and therefore does in general  not capture the causal structure of a generic process operator $\sigma$. However, in the special case of $\sigma$ being Markov for its graph $G^{S}_{\sigma}$, the latter does capture the causal structure. Moreover, a simple algorithm reveals whether this is the case. 

The algorithm takes a process operator $\sigma_V$ as input. The first step is to determine the graph $G^{S}_{\sigma}$. This involves checking, for each $i \ne j$, whether  
\begin{equation}
\frac{1}{d_{A_j}} \mathds{1}_{A_j^{\text{out}}} \ \otimes \ \Trace_{A_j^{\text{out}}} \Big[ \ \rho_{A_i|A_1...A_n}  \ \Big] \ = \ \rho_{A_i|A_1...A_n}, 
\end{equation}
where $\rho_{A_i|A_1...A_n} = \Trace_{\{A_k^{\text{in}}\}, k \neq i}[\sigma_V]$ (a total of $n(n-1)$ linear constraints).
Note that it holds by definition that $\rho_{A_i|A_1...A_n} = \rho_{A_i|Pa(A_i)}$. The algorithm then checks whether $G^{S}_{\sigma}$ is a DAG or not. If $G^{S}_{\sigma}$ is a DAG, the algorithm checks whether $\sigma_V$ is Markov for $G^{S}_{\sigma}$: this involves checking whether the $n$ operators $\rho_{A_i|Pa(A_i)}$ commute pairwise, and if so, whether their product gives back $\sigma_V$. In case the algorithm outputs that $G^{S}_{\sigma}$ is a DAG and that $\sigma_V$ is Markov for $G^{S}_{\sigma}$, the pair $(G^{S}_{\sigma}, \sigma)$ constitutes a quantum causal model and $G^{S}_{\sigma}$ is a candidate causal explanation of $\sigma_V$ revealed by these steps. In case $\sigma_V$ is not Markov for the DAG $G^{S}_{\sigma}$ or else $G^{S}_{\sigma}$ is not a DAG, i.e., contains directed cycles, then a causal discovery algorithm has more work to do. In fact, a plethora of interesting questions can then be asked concerning the causal structure, and we leave the study of these to future work.

This section concludes with the following observation.
\begin{remark} \label{Rem_Markovianity} 
Suppose that a process operator $\sigma_V$ is Markov for the simple induced graph $G^S_{\sigma}$. Then $\sigma_V$ is also Markov for any DAG of which $G^S_{\sigma}$ is a subgraph, since adding arrows that do not represent direct signalling relations in $\sigma_V$ does not destroy the Markov property. 
That $\sigma_V$ cannot be Markov for a DAG with fewer arrows than $G^S_{\sigma}$ is obvious as that would contradict the definition of $G^S_{\sigma}$\footnote{Note that the channels $\rho_{A_i|Pa(A_i)}$, which are asserted to exist by virtue of $\sigma_V$ being Markov for some DAG $G$, have to have non-trivial action at least on the output spaces of all parent nodes  in $G^S_{\sigma}$.}.
Hence, there is a unique sense in which a process operator is either Markov or not, independently from consideration of any particular DAG. 	
\end{remark}

\section{Related work \label{Sec_RelatedWork}}


Many works are explicitly concerned with causal structure, but not to the end of a quantum generalization of causal models. These include many works already cited, as well as, for example, Refs.~\cite{ChiribellaEtAl_2010_Purification,
CoeckeEtAl_2011_CausalCategories,
ArrighiEtAl_2011_UnitarityPlusCausalityImpliesLocalisability,
KissingerEtAl_2017_EquivalenceRelCausalStructureAndTerminality,
Portmann_2017_CausalBoxes,
ArrighiEtAL_2017_QuantumCausalGraphDynamics, 
CotlerEtAl_2018_QuantumCausalInfluence}. We do not present any further discussion of these here. 

Other works that have generalized at least some of the ideas of classical causal models to the quantum case include, for example, Refs.~\cite{BranciardEtAl_2010_CharacterizingNonLocalCorrelations, Fritz_2012_BeyondBellsTheorem, ChavesEtAl_2015_InformationImplicationsOfQuantumCausalStructures, HensonEtAl_2014_TheoryIndependentLimitsGeneralisedBayesianNetworks,  Fritz_2016_BeyondBellsTheorem, Pienaar_2017_CausalQuantumClassicalGap, FraserEtAl_2018_CausalCompatibilityInequalities, PozasEtAl_2019_BoundingSetsOfCorrelationsInNetworks, VanHimbeeck_2019_QquantumViolationsInstrumentalScenario, WeilenmannEtAl_2017_AnalysingCausalStructuresWithEntropy, MiklinEtAl_2017_EntropicApproachToCausalCorrelations, WeilenmannEtAl_2018_NonShannonInequalitiesInEntropyApproach, WeilenmannEtAl_2016_InabilityOfEntropyApproach}. In these works, causal structure is represented by a DAG, with nodes of the DAG corresponding to quantum processes with classical outcomes, and edges to the passing of physical systems between these processes. The mathematical object constrained by causal structure is taken to be a probability distribution over classical variables -- either choices of settings or outcomes of measurements. This yields a framework that allows the derivation and study of Bell-type inequalities in arbitrary causal scenarios, and the study of conditional independence relations between the classical variables. Our approach differs, again in that we take causal relations to be defined by the notion of influence in a unitary transformation, and in that we introduce intrinsically quantum notions of relative independence, rather than restrict the study to conditional independences amongst classical random variables.

Early work by Tucci \cite{Tucci_1995_QuantumBayesianNets, Tucci_2007_FactorisationOfDensityMatricesAccordingToNetworks} describes a quantum generalization of classical Bayesian networks and causal models, obtained by associating a single Hilbert space (rather than split in- and output spaces) and a complex conditional probability amplitude (rather than a channel) with each node of a DAG. The resulting formalism is very different from that which we have presented here. More closely related to our work is that of Leifer and Poulin \cite{LeiferEtAl_2008QuantumGraphicalModels}, which presents (amongst other things) an approach to quantum Bayesian networks, wherein a quantum state is associated with a DAG, and must satisfy independence relationships formalized by the quantum mutual information, given by the structure of the DAG. The results of Ref.~\cite{LeiferEtAl_2008QuantumGraphicalModels} have at various times been used in our proofs. Leifer and Spekkens \cite{LeiferEtAl_2013_QTAsBayesianInference} adapt the ideas of Ref.~\cite{LeiferEtAl_2008QuantumGraphicalModels} to quantum causal models, using a particular definition of a quantum conditional state. Our approach differs from that of Ref.~\cite{LeiferEtAl_2013_QTAsBayesianInference} in taking influence in unitary transformations as defining of causal relations, in its use of the process operator formalism, and in the fact that we don't use quantum conditional states.

Apart from Ref.~\cite{AllenEtAl_2016_QCM}, which contains the core ideas that our work builds on, the nearest precursor is perhaps the work of Costa and Shrapnel \cite{CostaEtAl_2016_QuantumCausalModeling}. Ref.~\cite{CostaEtAl_2016_QuantumCausalModeling} presents a fully quantum notion of causal model, in which the central object under study is a process operator, with a Markov condition expressed in terms of a factorization of the process operator.
The main difference is that in the approach of Ref.~\cite{CostaEtAl_2016_QuantumCausalModeling}, the output Hilbert space of a node factorizes into as many subsystems as the node has children. This is the natural approach if direct causal influence relations, represented by edges in the DAG, are understood in terms of distinct systems travelling between local laboratories at nodes, rather than in terms of causal influence in unitary transformations as we have done.

The rough sketch of an algorithm for causal inference in Section~\ref{Sec_NoteCausalDiscovery} is very similar to the quantum causal discovery algorithm of Giarmatzi and Costa in Ref.~\cite{GiarmatziEtAl_2018_CausalDiscoveryAlgorithm}, where a detailed discussion and concrete implementation are presented. The main differences are due to the algorithm being adapted for the different approach to quantum causal models presented in this work. In particular: the algorithm we presented takes as input a process operator, and does not require additional information about Hilbert space decompositions and dimensions of factors.
Ref.~\cite{GiarmatziEtAl_2018_CausalDiscoveryAlgorithm} does discuss adaptation of the quantum causal discovery algorithm to the quantum causal models of Ref.~\cite{AllenEtAl_2016_QCM} (hence also the present work), but suggests that this would result in an exponential scaling (in the number of nodes) vs the original algorithm's quadratic scaling. The reason for this is that the adaptation suggested looks for children of nodes, rather than parents, which requires consideration of the power set of nodes. The algorithm sketched in Section~\ref{Sec_NoteCausalDiscovery}, by contrast, has the same quadratic scaling behaviour as that of Ref.~\cite{GiarmatziEtAl_2018_CausalDiscoveryAlgorithm}.

Finally, more recent work \cite{Pienaar_2018_QuantumCausalModelsViaQuantumBayesianism, Pienaar_2018_QBistQCMs_ShortVersion, Pienaar_2019_TimeReversibleQCM} develops an approach to quantum causal models, inspired by the QBist approach to quantum theory \cite{FuchsandStacey_2016_QuantumTheoryasHeroHandbook}, and based on symmetric informationally-complete postive operator-valued measurements (SIC POVMs). This leads to a quantum Markov condition in the form of a constraint on the outcome distribution for SIC-POVMs carried out at nodes, and has the interesting feature of being compatible with a reversal of time, i.e., the quantum Markov property is preserved under reversal of all arrows of the DAG.

\section{Discussion \label{Sec_Discussion}}

\subsection{Summary}

The main aim of this work is to provide an account of causal relations in quantum theory, and the constraints that they place on a quantum process. In the approach that we have developed, causal relations are not defined by factors that are either external to or emergent in quantum theory, such as the arrangement of classical devices, nor by spacetime configurations. They are instead defined by structural properties of unitary transformations, i.e., in the terms of the theory itself. Similarly, the relata of causal relations are not emergent classical variables such as measurement outcomes, but are instead quantum nodes, that is particular loci within an overall unitary process. The object that is constrained by causal relations is not a probability distribution over classical variables (at least not directly), but the quantum process itself. This leads to a framework of quantum causal models, with the existing classical framework recovered as a special case when operators are diagonal. 

The first part of the paper shows that when quantum systems evolve according to a unitary circuit, with some wires broken to represent places where an agent has an option of intervening, a causal structure can be defined and represented with a DAG. Theorem~\ref{Thm_QuantumEquivalence} states that when marginalizing over local noise sources and any final output systems that are not of interest, the resulting quantum process operator satisfies a condition that matches the well known causal Markov condition of the classical case: the process operator is Markov for the DAG. A quantum causal model, by specifying the DAG and a process operator that is Markov for the DAG, describes just such a situation. The assumption that only local noise sources are missing matches the spirit of classical causal modelling, wherein local noise sources are irrelevant to the explanation of correlation between observed variables, hence can be marginalized over, but where common causes can generate correlation, hence should be included in the model. Theorem~\ref{Thm_QuantumEquivalence} also states a converse result, which is that for any quantum causal model, the process operator can be recovered from an underlying unitary circuit, without postulating additional causal links or common causes that are not in the model. In other words, the set of quantum causal models is not redundant, since there is no quantum causal model that does not represent a valid scenario. These results generalize to arbitrary DAGs those of Ref.~\cite{AllenEtAl_2016_QCM}, where similar things are shown for the very simple case of three nodes, in which node $A$ is a common cause of nodes $B$ and $C$, and the results interpreted as a quantum version of Reichenbach's principle. 

Sections~\ref{Sec_IndependenceNotions}-\ref{Sec_QuantumDoCalculus} derive constraints that the marginal quantum process operator on a subset of nodes has to satisfy, given the overall causal structure. In the classical case, a core theorem states that if sets of variables $Y$ and $Z$ are d-separated by a third set $W$, where d-separation is a condition involving paths through the DAG, then $Y$ and $Z$ must be conditionally independent given $W$. Prior to this work, whether a generalization of this theorem holds in the quantum case was not obvious because there is no standard notion of quantum systems being conditionally independent, when some of those systems are earlier in time, some later, and they are causally related to one another. Section~\ref{Sec_IndependenceNotions} therefore introduced new notions of independence of quantum systems, including the one we call \emph{quantum strong relative independence}, which we expect will be of interest in their own right. 
In the quantum case, Theorem~\ref{Thm_QuantumDSeparationTheorem} shows that no matter what interventions are performed at other nodes, if the subsets of nodes $Y$ and $Z$ are d-separated by a third subset $W$, then the marginal process on $YZW$ satisfies quantum strong relative independence, thus generalizing the classical theorem. Section~\ref{Sec_QuantumDoCalculus} introduces other notions of independence, and states similar theorems that express constraints on marginal quantum process operators. When all operators are suitably diagonal, the classical case is recovered, with these theorems reducing to the rules of Pearl's do-calculus. 

Finally, Section~\ref{Sec_NoteCausalDiscovery} presented in outline a causal inference algorithm for the case in which full tomography is assumed and a process operator completely specified.

\subsection{Practical application and technical open questions}

As briefly discussed in the Introduction, the most obvious practical application of this work is to situations in which a process operator over a large number of quantum nodes is not fully known or characterised. This is because full tomography of the process requires a number of readings exponential in the number of quantum nodes, hence is impractical, and because intervening at some nodes may be difficult or impossible, either due to technological constraints or because those nodes are controlled by an adversary. Partial data may be gathered in the form of the classical outcomes from limited interventions carried out on a subset of nodes, over some number of trials. The goal is to make predictions about what will happen in subsequent trials, possibly with some parameters changed, or with interventions carried out at new nodes that haven't been tested before, or with an adversary doing something different. Such predictions will involve inferences about the causal structure that can be made from the partial data. (Note that this scenario can be distinguished from that addressed in Section~\ref{Sec_NoteCausalDiscovery}, where it is assumed that a full process operator is given, and inferences are made concerning the causal structures that it is compatible with.) This very general problem will be at least somewhat easier in any situation in which time-stamps can be associated with each node since later nodes cannot be causes of earlier nodes. But in some applications, it may be that reliable time-stamps for all nodes are not available (perhaps because the time at which some difficult-to-perform measurement is carried out is significantly uncertain relative to the time-scale of the process itself \cite{Shrapnel_2019_DiscoveringQuantumCausalModels}).

The practical application of classical causal modelling is similar, with the distinction that the partial data is usually assumed to arise from non-disturbing observations of the variable at each of a subset of the nodes. Hence a major problem in classical causal modelling is this: given a classical causal structure (a DAG $G$) with the set of variables $V$ partitioned into a set $S$ of observed variables and a set $R$ of unobserved variables, characterize the set of probability distributions over $S$ that can be realized as the marginal of a distribution over $V$ that satisfies the Markov condition with respect to $G$ \cite{TianEtAl_2002_TestableImplicationsOfCausalModelsWithHiddenVariables, Richardson_2003_MarkovPropertiesForMixedDAGs, Wermuth_2011_ProbabilityDIstributionsAndSummaryGraphs, 
Fritz_2012_BeyondBellsTheorem, 
EvansEtAl_2014_MarkovianADMGs, 
WeilenmannEtAl_2017_AnalysingCausalStructuresWithEntropy, 
FraserEtAl_2018_CausalCompatibilityInequalities, 
WeilenmannEtAl_2018_NonShannonInequalitiesInEntropyApproach, 
WolfeEtAl_2019_InflationTechniqueForCausalInference, 
VanHimbeeck_2019_QquantumViolationsInstrumentalScenario, 
NavascuesEtAl_2020_InflationTechniqueSolvesCompatibilityProblem}. This problem can be addressed both with known cardinalities of the variables at unobserved nodes \cite{LeeEtAl_2017_CausalInferenceViaAlgebraicGeometry}, or with unknown and unrestricted cardinalities. In each case, the set of realizable probability distributions over observed nodes is characterized both by equalities and by inequalities that must be satisfied. The realizable marginal distributions are contained in the \emph{nested Markov set} \cite{ShpitserEtAl_2014_IntroductionNestedMarkovModels, RichardsonEtAl_2017_NestedMarkovPropertiesMixedGraphs, Evans_2018_MarginsOfDiscreteBayesianNetworks, EvansEtAl_2019_SmoothIdentifiableSupermodels}, which is the set of probability distributions that need satisfy only the equality constraints.

The corresponding problem in the context of quantum causal models is this: given a quantum causal structure $G$, with nodes $V$ partitioned into an observed set $S$ and an unobserved set $R$, characterize the set of process operators over $S$ that can be realized as the marginal of a process operator over $V$ that is Markov for $G$, assuming arbitrary interventions at the nodes in $R$. As in the classical case, the problem can be stated assuming either known dimensions of unobserved nodes, or unknown and unconstrained dimensions. In each case, the set will be defined both by equalities and by inequalities. The set of realizable process operators over $S$ is contained in the quantum analogue of the nested Markov set, which is the set of valid process operators over $S$ satisfying only the equality constraints.

Our Theorems~\ref{Thm_Rule1_qu}, \ref{Thm_Rule2_qu} and \ref{Thm_Rule3_qu} address this problem, stating equalities that are satisfied by the marginal process operator on the subset $S = Y \cup Z \cup W \cup X$, where the subset $R$ that appears in the statements of the theorems can be thought of as the set of unobserved nodes. As in the classical case, the problem can be stated assuming either known dimensions of unobserved nodes, or unknown dimensions. Our theorems hold independently of the dimension of unobserved nodes (except to the extent that our analysis assumes finite dimension throughout), hence can still be applied even in the case it is finite but unknown. We have not attempted to derive inequalities that constrain the set of marginal process operators. Nor have we attempted to derive the full set of equality constraints that would characterize the quantum analogue of the nested Markov set. We hypothesize, but have not shown, that the constraints implied by Theorems~\ref{Thm_Rule1_qu}, \ref{Thm_Rule2_qu} and \ref{Thm_Rule3_qu} are complete, in the sense that given a quantum causal structure $G$, if a process operator satisfies these constraints for all possible choices of $Y,Z,W,X,R$, then it is Markov for $G$. If this hypothesis holds, then a characterization of the quantum analogue of the nested Markov set could perhaps be obtained from a similar approach to the classical case, splitting the observed nodes of the DAG into subsets called districts \cite{ShpitserEtAl_2014_IntroductionNestedMarkovModels, RichardsonEtAl_2017_NestedMarkovPropertiesMixedGraphs, Evans_2018_MarginsOfDiscreteBayesianNetworks, EvansEtAl_2019_SmoothIdentifiableSupermodels}, and applying a series of nested conditions based on Theorems~\ref{Thm_Rule1_qu}, \ref{Thm_Rule2_qu} and \ref{Thm_Rule3_qu}, depending on the districts.

Of course, experimental data is in the form of the classical outcomes of interventions, so the constraints on the joint probability distribution over outcomes of interventions at a subset of the nodes must also be studied. As discussed in Section~\ref{Sec_RelatedWork}, previous works such as Refs.~\cite{BranciardEtAl_2010_CharacterizingNonLocalCorrelations, 
Fritz_2012_BeyondBellsTheorem,  HensonEtAl_2014_TheoryIndependentLimitsGeneralisedBayesianNetworks, ChavesEtAl_2015_InformationImplicationsOfQuantumCausalStructures, Fritz_2016_BeyondBellsTheorem, WeilenmannEtAl_2016_InabilityOfEntropyApproach,  Pienaar_2017_CausalQuantumClassicalGap, WeilenmannEtAl_2017_AnalysingCausalStructuresWithEntropy, WeilenmannEtAl_2018_NonShannonInequalitiesInEntropyApproach, MiklinEtAl_2017_EntropicApproachToCausalCorrelations, FraserEtAl_2018_CausalCompatibilityInequalities, VanHimbeeck_2019_QquantumViolationsInstrumentalScenario, PozasEtAl_2019_BoundingSetsOfCorrelationsInNetworks} have studied the achievable joint distributions using formalisms involving quantum systems and DAGs, but not with Def.~\ref{Def_QCM}, and not with an interpretation of the DAG as the causal structure of a unitary circuit. The frameworks that these earlier results rely on can be recovered as special cases of the framework we have presented, but our framework additionally allows novel scenarios to be considered involving novel sorts of constraints. As a very simple example, it would be easy to derive the set of correlations that can be obtained from measuring a bipartite system for which a single qubit is the complete common cause \cite{AllenEtAl_2016_QCM}, a possibility that previous frameworks do not admit.

Future work will develop techniques for characterizing both the set of achievable marginal process operators on subsets of nodes, and the set of achievable probability distributions over outcomes of interventions. Another avenue for future work is the development of causal discovery algorithms, for situations in which the causal structure must be inferred from only limited data.

 \subsection{Some remarks concerning time symmetry}

The bulk of this paper has focused on the technical development of quantum causal models. The technical results, along with practical applications of them, do not require any particular commitment to conceptual claims concerning the nature of causality in a quantum universe. Nonetheless, as mentioned in the Introduction, the results do suggest answers to various conceptual questions. Some of these are mentioned in the Summary above. We end with a brief discussion of one more of these, which concerns the status of time symmetry. 

As shown in subsequent work \cite{LorenzEtAl_2020_CausalAndCompositionalStructure} (see also \cite{ArrighiEtAl_2011_UnitarityPlusCausalityImpliesLocalisability}), the causal structure of a unitary transformation is reversible in a strong sense, which goes beyond the obvious fact that unitaries are reversible transformations: if the causal structure of a unitary transformation $U$ is represented by drawing arrows from inputs to outputs (as in Remark~\ref{Remark_Causalinfluenceunitary}), then the causal structure of $U^{-1}$ is given by inverting the arrows. In case this seems obvious, note that a similar thing is not true for classical reversible functions (see Ref.~\cite{LorenzEtAl_2020_CausalAndCompositionalStructure}: a counterexample is the classical Controlled-NOT gate). This remark about reversibility of a unitary transformation may be lifted to the case of a unitary process with inputs, corresponding to a circuit like that of Fig.~\ref{Fig_GeneralBrokenCircuit}. In the inverse process obtained by reading the figure from top to bottom, replacing each $U_i$ with $U_i^{-1}$, and swapping traces with initial preparations, the causal structure is given by inverting the arrows of the causal structure of the original process.

Any temporal asymmetry, then, in the framework of quantum causal models does not arise in the specification of the causal relations themselves. One source of asymmetry lies in the differing treatment of $\lambda$ nodes and $F$ nodes in the transition from a broken circuit like that of Fig.~\ref{Fig_GeneralBrokenCircuit} to a quantum causal model over the nodes $A_1 , \ldots , A_n$. The marginal process $\sigma_{A_1 ... A_n}$ is only guaranteed to satisfy the (asymmetric) Markov condition if the $\lambda$s are local noise sources, meaning that each $\lambda$ is a direct cause of at most one of the $A$ nodes, and that a product input state over the $\lambda$s is assumed. There are no similar conditions applied to the $F$ nodes. Another source of asymmetry is that interventions at nodes by agents can influence causal descendants of those nodes but not causal ancestors. Whether these assumptions that lead to the time asymmetric aspects of quantum causal models can be further justified from less ad hoc ones is an open question, but one which may well be hard and inseparable from understanding the role of boundary conditions in quantum physics \cite{OreshkovEtAl_2015_OperationalFormulationTimeReversal}. For interesting discussion of time symmetry within the context of a different approach to quantum causal modelling, see Refs.~\cite{Pienaar_2018_QuantumCausalModelsViaQuantumBayesianism, Pienaar_2018_QBistQCMs_ShortVersion, Pienaar_2019_TimeReversibleQCM}.

Concerning quantum indefinite causality (briefly discussed in the Introduction), the quantum causal models of this work can be generalized by allowing directed cycles, such that a quantum causal model is defined by a directed graph $G$ (not necessarily acyclic) and a process operator that satisfies our Markov condition with respect to $G$. This is carried out in subsequent work of ours \cite{BarrettEtAl_2020_CyclicQCMs}, and the approach used to obtain novel results concerning the structure of causally indefinite processes. We leave open the question of how the possibility of cyclic causal structures intersects with questions of time symmetry.


\section{Acknowledgements} 

We would like to thank Matt Pusey, Matty Hoban, Sina Salek, Rob Spekkens and David Schmid for helpful discussions. This work was supported by the EPSRC National Quantum Technology Hub in Networked Quantum Information Technologies, the Wiener-Anspach Foundation, and by the Perimeter Institute for Theoretical Physics. Research at Perimeter Institute is supported by the Government of Canada through the Department of Innovation, Science and Economic Development Canada and by the Province of Ontario through the Ministry of Research, Innovation and Science. This project/publication was made possible through the support of a grant from the John Templeton Foundation. The opinions expressed in this publication are those of the author(s) and do not necessarily reflect the views of the John Templeton Foundation. This work was supported by the Program of Concerted Research Actions (ARC) of the Universit\'{e} Libre de Bruxelles. O. O. is a Research Associate of the Fonds de la Recherche Scientifique (F.R.S.--FNRS).

\phantomsection
\renewcommand\bibname{References}
\addcontentsline{toc}{section}{\bibname}
\bibliographystyle{utphys}

\providecommand{\href}[2]{#2}\begingroup\raggedright\endgroup


\appendix

\section{Appendix part I}

\subsection{Useful tools I \label{SubSec_UsefulTools}} 

The following statements, which will feature in subsequent proofs, are presented separately with proofs here for better readability.

\begin{lemma} \textnormal{(Splitting from commutation relations I):} \label{Lem_SplitFromCommutation_I} 
Let $\rho_{A|CD}$ and $\rho_{B|DE}$ be CJ representations of channels. If they commute $[ \rho_{A|CD} \ , \ \rho_{B|DE} ] \ = \ 0$, then there exists a decomposition of the Hilbert space on which the domains of the channels overlap, here denoted as $D$, into orthogonal subspaces
\begin{equation}
\mathcal{H}_{D} = \bigoplus_i \mathcal{H}_{(D)_i^L} \ \otimes \ \mathcal{H}_{(D)_i^R}, \label{Eq_BasicSplitting}
\end{equation}
and channels $\rho_{A|CD_i^L}$ and $\rho_{B|D_i^RE}$, such that
\begin{eqnarray}
\rho_{A|CD} &=& \sum_i \rho_{A|CD_i^L} \ \otimes \ \mathds{1}_{D_i^R} \label{Eq_SplitChannel_1} \nonumber \\
\rho_{B|DE} &=& \sum_i \mathds{1}_{D_i^L} \ \otimes \ \rho_{B|D_i^RE}   \label{Eq_SplitChannel_2} \ . \label{Eq_splittingequations}
\end{eqnarray}	 
\end{lemma}

\noindent (NB Here, and on a few occasions below, the same notation is used for an operator defined on a subspace, and the same operator extended to the whole Hilbert space via the stipulation that it is zero on the orthogonal subspace. This enables, for example, Eqs.~(\ref{Eq_splittingequations}) to be written with ordinary summation, rather than a direct sum.)\\

\noindent {\bf Proof.} The proof appears in Ref.~\cite{AllenEtAl_2016_QCM}, but is included here for completeness. Define the channel $\rho_{AB|CDE}:=\rho_{A|CD} \rho_{B|DE}$ which yields $\rho_{A|CD}$ and $\rho_{B|DE}$ as the marginal channels into $A$ and $B$ respectively. 
One can directly verify that the quantum conditional mutual information $I(AC:BE|D)$ vanishes, if evaluated on the trace-1 quantum state $\hat{\rho}_{AB|CDE} = (1/(d_Cd_Dd_E)) \rho_{AB|CDE}$.
Theorem~6 of Ref.~\cite{HaydenEtAl_2003_StructureOfStatesAubadditivity} then implies that there is a decomposition of $\mathcal{H}_D$ into orthogonal subspaces of the form of Eq.~(\ref{Eq_BasicSplitting}), along with a probability distribution $\left\{ p_i \right\}$, such that
\begin{equation}
\hat{\rho}_{AB|CDE} = \sum_i p_i \big( \hat{\rho}_{A|CD_i^L} \ \otimes \ \hat{\rho}_{B|D_i^RE} \big) \ ,
\end{equation}
where $\hat{\rho}_{A|CD_i^L}$ and $\hat{\rho}_{B|D_i^RE}$ are (trace-1) quantum states on the indicated Hilbert spaces. The normalization condition for the  CJ representation of a channel, $\Trace_{AB} (\rho_{AB|CDE} ) = \mathds{1}_{CDE} $, fixes the $p_i$ such that
\begin{eqnarray}
\rho_{AB|CDE} = \sum_i \big[ \ \rho_{A|CD_i^L} \ \otimes \  \rho_{B|D_i^RE} \ \big] \ ,
\end{eqnarray}
where now each operator on the RHS is normalized as CJ operator of a channel, that is for each $i$, $\Trace_{A}(\rho_{A|CD_i^L})= \mathds{1}_{CD_i^L}$, and $\Trace_{B}(\rho_{B|D_i^R E})= \mathds{1}_{D_i^R E}$. The orthogonality of the subspaces means that the marginals $\rho_{A|CD} $ and $\rho_{B|DE}$ can indeed be written as claimed in eqs. \ref{Eq_SplitChannel_1}-\ref{Eq_SplitChannel_2}. This completes the proof. \hfill $\square$

\begin{lemma} \textnormal{(Splitting from commutation relations II - nesting):} \label{Lem_SplitFromCommutation_II}
Suppose that the three channels $\rho_{A|CD}$, $\rho_{B|DEF}$ and $\rho_{H|DEG}$ commute pairwise. Then there exists a `nested' decomposition of the $DE$ Hilbert space into orthogonal subspaces of the form
	\begin{eqnarray}
		\mathcal{H}_{D} \otimes \mathcal{H}_{E} 
			\ = \ \Big( \bigoplus_i  \mathcal{H}_{D_i^L} \otimes \mathcal{H}_{D_i^R} \Big) \otimes  \mathcal{H}_{E}  \ = \ \bigoplus_{i} \mathcal{H}_{D_i^L} \otimes   \Big( \bigoplus_{j_i} \ \mathcal{H}_{(D_i^RE)_{j_i}^L} \otimes \mathcal{H}_{(D_i^RE)_{j_i}^R} \ \Big) \label{Eq_NestedSplitting}
	\end{eqnarray}
such that the given channels are block diagonal with respect to this decomposition in the following way:
	\begin{eqnarray}
		\rho_{A|CD} &=& \sum_i \ \rho_{A|CD_i^L} \ \otimes \ \mathds{1}_{D_i^R} \label{Eq_NestedChannel_1} \\
		\rho_{B|DEF} &=& \sum_{i,j_i} \ \mathds{1}_{D_i^L} \ \otimes \ \rho_{B|(D_i^RE)_{j_i}^LF} \ \otimes \ \mathds{1}_{((D)_i^RE)_{j_i}^R}   \label{Eq_NestedChannel_2} \\ 
		\rho_{H|DEG} &=& \sum_{i,j_i} \ \mathds{1}_{D_i^L} \ \otimes \ \mathds{1}_{(D_i^RE)_{j_i}^L} \ \otimes \ \rho_{H|((D)_i^RE)_{j_i}^RG} \ . \label{Eq_NestedChannel_3}
	\end{eqnarray}
\end{lemma}

\noindent {\bf Proof.} This is a mere iteration of the previous proof. Consider the operator $\rho_{ABH|CDEFG}:= \rho_{A|CD} \ \allowbreak \rho_{B|DEF} \ \allowbreak \rho_{H|DEG}$ and the commutation relation $[\rho_{A|CD} \ , \  \rho_{BH|DEFG} ] = 0$, where $\rho_{BH|DEFG} = \rho_{B|DEF} \ \allowbreak \rho_{H|DEG}$. Lemma~\ref{Lem_SplitFromCommutation_I} implies a decomposition of $D$ such that
\begin{eqnarray}
	\rho_{A|CD} = \sum_i \ \rho_{A|CD_i^L} \ \otimes \ \mathds{1}_{D_i^R}  \hspace{1cm} \text{and} \hspace{1cm}
	\rho_{BH|DEFG} = \sum_i \ \mathds{1}_{D_i^L} \ \otimes \ \rho_{BH|D_i^REFG} \ . \label{Eq_BasicNesting_Step1_Channel2} 
\end{eqnarray}	
Second, the commutation relation $[ \rho_{B|DEF} \ , \  \rho_{H|DEG} ] = 0$, together with the orthogonality of the distinct subspaces, implies that for each $i$, 
\begin{equation}
	[ \ \rho_{B|D_i^REF} \ , \  \rho_{H|D_i^REG} \ ] \ = \ 0  \ , \label{Eq_BasicNesting_Comm_3}
\end{equation}
where $\rho_{B|D_i^REF}$ and $\rho_{H|D_i^REG}$ are marginals of the operator $\rho_{BH|D_i^REFG}$. Now, Lemma~\ref{Lem_SplitFromCommutation_I} applies again and induces a splitting of $D_i^RE$, as a direct sum over orthogonal subspaces $j_i$, and a corresponding block-diagonal form of the operators in Eq.~\ref{Eq_BasicNesting_Comm_3}. This yields the claim of Eqs.~(\ref{Eq_NestedSplitting}- \ref{Eq_NestedChannel_3}). \hfill $\square$

\begin{remark}\label{Rem_SplittingForUnitary}
In the special case 
in which $ \rho_{A|CD} \rho_{B|DE} = \rho^U_{AB|CDE}$ represents a unitary channel, the purity of the CJ operator $\rho^U_{AB|CDE}$, along with Lemma~\ref{Lem_SplitFromCommutation_I}, implies the global factorization $\mathcal{H}_D = \mathcal{H}_{D^L}\otimes \mathcal{H}_{D^R}$, with $\rho^U_{AB|CDE} = \rho^U_{A|CD^L} \otimes \rho^U_{B|D^R E}$.
A similar statement applies to the `nested' case in Lemma~\ref{Lem_SplitFromCommutation_II}.
\end{remark}

The above also enables the following useful result to be established.
\begin{definition} \textnormal{(Reduced unitary\footnote{Called an autonomous CPTP map in Ref.~\cite{Schumacher_2005_LocalityNoInfluenceConditions} .}):} \label{Def_ReducedUnitary}
A channel $\rho_{B|A}$ is a \emph{reduced unitary} if and only if there exists a unitary channel $\rho^U_{BF|A}$ such that $\rho_{B|A}= \Trace_F [\rho^U_{BF|A}]$.
\end{definition}
\begin{lemma} \textnormal{(Products of reduced unitaries):} \label{Lem_ReducedUnitaryLemma} Suppose $\rho_{B|A}$ and $\rho_{C|A}$ are reduced unitaries such that $[\rho_{B|A},\rho_{C|A}]=0$. Then $\rho_{BC|A} := \rho_{B|A} \rho_{C|A}$ is also a reduced unitary.
\end{lemma}

\noindent {\bf Proof.} That $\rho_{B|A}$ is a reduced unitary by definition means that there exists a unitary channel $\rho^V_{BF|A}$ such that $\Trace_{F}[\rho^V_{BF|A}]=\rho_{B|A}$. By Theorem~\ref{Thm_FactorizationOfUnitary}, $\rho^V_{BF|A}=\rho_{B|A} \rho_{F|A}$, hence by Remark~\ref{Rem_SplittingForUnitary} there is a global factorization $A=A_b^L \otimes A_b^R$, with respect to which $\rho^V_{BF|A} = \rho_{B|A_b^L} \otimes \rho_{F|A_b^R}$. Similarly, there exists a unitary channel $\rho^W_{CG|A}$ such that $\Trace_{G}[\rho^W_{CG|A}]=\rho_{C|A}$ and a factorization $A = A_c^L \otimes A_c^R$ such that $\rho^W_{CG|A} = \rho_{G|A_c^L} \otimes \rho_{C|A_c^R}$. 

A priori, the relation between the factorizations $A_b^L \otimes A_b^R$ and $A_c^L \otimes A_c^R$ is unknown. However, by assumption it is also true that $[\rho_{B|A} , \rho_{C|A}]=0$. The operators $\rho_{B|A_b^L}$, $\rho_{F|A_b^R}$, $\rho_{G|A_c^L}$ and $\rho_{C|A_c^R}$ all have to represent unitary channels (for dimensional reasons and due to the purity of the operators), hence, up to normalization, can be seen as maximally entangled states. It is then straightforward to check that due to the commutation of $\rho_{B|A_b^L}$ with $\rho_{C|A_c^R}$, the operator $\rho_{C|A_c^R}$ acts trivially on $A_b^L$, and conversely that the operator $\rho_{B|A_b^L}$ acts trivially on $A_c^R$. Therefore, there exists a factorization $A= A_b^L \otimes A' \otimes A_c^R$ such that $\rho_{BC|A} = \rho_{B|A_b^L} \otimes \mathds{1}_{A'} \otimes \rho_{C|A_c^R}$. This establishes the claim. \hfill $\square$

\subsection{Proof of Theorem~\ref{Thm_Compatibilitywithbroken} \label{SubSec_Proof_Thm_Compatibilitywithbroken}} 

Consider a unitary transformation $U$, with CJ representation $\rho^U_{A_1...A_nF|A_1...A_n\lambda_1...\lambda_n}$ such that 
$U$ respects the no-influence conditions of Eq.~(\ref{Eq_CompatibilityNoInfluenceCond}). The following establishes the claim that this unitary channel has a realization as a circuit of the form of Fig.~\ref{Fig_JustificationCircuit}. 

Due to the no-influence conditions (and Theorem~\ref{Thm_FactorizationOfUnitary}), the unitary channel has to factorize as
\begin{eqnarray}
\rho^U_{A_1...A_nF|A_1...A_n\lambda_1...\lambda_n} \ = \ \rho_{F|A_1...A_n\lambda_1...\lambda_n} \ \Big( \prod_{i=1}^n  \rho_{A_i|Pa(A_i) \lambda_i} \Big) \ . \label{Eq_Factorization}
\end{eqnarray}
Let the enumeration of the $A_i$ be compatible with the partial order of $G$. This total order can be uniquely extended to the $2n+1$ nodes: $\lambda_1 < A_1 < \lambda_2 < A_2 \linebreak[1] < ... < \lambda_n \linebreak[1]< A_n < F$. Now consider the following data:
\begin{center}
\begin{minipage}{13.5cm}
\begin{minipage}{4cm}
\begin{figure}[H]
\centering
\small
\input{Figures/Fig_DAGProofJustification.tex}
\vspace*{0.2cm}
\caption{ DAG $G'$ \label{Fig_DAGProofJustification}}
\end{figure}
\end{minipage}
\newcommand{\Spacing}{0.14cm}
\hfill
\begin{minipage}{4cm}
\begin{eqnarray}
&&\rho_{F|\lambda_1 A_1 ...\lambda_n A_n}  \nonumber \\[\Spacing]
&&\rho_{A_n|Pa(A_n)\lambda_n}  \nonumber \\[\Spacing]
&&\rho_{\lambda_n}  \nonumber \\[\Spacing]
&& \hspace{0.5cm} \vdots  \nonumber \\[\Spacing]
&&\rho_{A_2|Pa(A_2)\lambda_2}  \nonumber \\[\Spacing]
&&\rho_{\lambda_2}  \nonumber \\[\Spacing]
&&\rho_{A_1|\lambda_1}  \nonumber \\[\Spacing]
&&\rho_{\lambda_1}  \nonumber \\[0.2cm]
&& \underbrace{\hspace{2.5cm}}_{\textnormal{\textit{Set A}}} \nonumber \\
\vspace*{0.2cm} \nonumber
\end{eqnarray}
\end{minipage}
\hfill
\begin{minipage}{0.3cm}
\rule{0.15mm}{5.75cm}
\end{minipage}
\hfill
\begin{minipage}{4cm}
\begin{eqnarray}
&&\rho_{F|\lambda_1 A_1 ...\lambda_n A_n}  \nonumber \\[\Spacing]
&&\rho_{A_n|\lambda_1 A_1 ...\lambda_{n-1} A_{n-1} \lambda_n }  \nonumber \\[\Spacing]
&&\rho_{\lambda_n}  \nonumber \\[\Spacing]
&& \hspace{0.5cm} \vdots  \nonumber \\[\Spacing]
&&\rho_{A_2|\lambda_1 A_1 \lambda_2}  \nonumber \\[\Spacing]
&&\rho_{\lambda_2}  \nonumber \\[\Spacing]
&&\rho_{A_1|\lambda_1}  \nonumber \\[\Spacing]
&&\rho_{\lambda_1}  \nonumber \\[0.2cm]
&& \underbrace{\hspace{2.5cm}}_{\textnormal{\textit{Set B}}} \nonumber \\
\vspace*{0.2cm} \nonumber
\end{eqnarray}
\end{minipage}
\end{minipage}
\end{center}

The DAG $G'$ is obtained from drawing an arrow from $\lambda_i$ to $A_i$ for all $i$, drawing an arrow into $F$ from all preceding nodes and drawing arrows between nodes $A_i$ as given by the original DAG $G$. The product of channels in \textit{Set A}, where $Pa(A_i)$ refers to the parental structure of $G$, is by construction Markov for $G'$. Focusing on the channels in \textit{Set B}, these are obtained by padding those in Eq.~\ref{Eq_Factorization} with identities:
\begin{equation}
\rho_{A_i|\lambda_1 A_1 ...\lambda_{i-1} A_{i-1} \lambda_i } \ := \ \rho_{A_i|Pa(A_i) \lambda_i} \otimes \mathds{1}_{\overline{(Pa(A_i) \lambda_i)}} \ , \label{Eq_PaddedChannels}
\end{equation}
where $\overline{(Pa(A_i) \lambda_i)}$ denotes the relative complement of $Pa(A_i) \cup \{ \lambda_i \}$ in the set $\left\{ \lambda_1 , A_1 , \ldots , \lambda_{i-1} , A_{i-1} , \lambda_i \right\}$. It follows from the pairwise commutation of the operators defined in Eq.~\ref{Eq_PaddedChannels} that
\begin{eqnarray}
[ \rho_{A_1|\lambda_1} \ , \ \rho_{A_2...A_n F|\lambda_1 A_1...\lambda_n A_n}  ] \ = \ 0 \ .
\end{eqnarray}
The unitarity of $U$, along with Lemma~\ref{Lem_SplitFromCommutation_I} and Remark~\ref{Rem_SplittingForUnitary} then implies a factorization $\lambda_1 = (\lambda_1)^L \otimes (\lambda_1)^R$ such that
\begin{eqnarray}
\rho_{A_1|\lambda_1} &=& \rho_{A_1|(\lambda_1)^L} \otimes \mathds{1}_{(\lambda_1)^R} \label{Eq_Channel1Step1} \\
\rho_{A_2...A_n F | \lambda_1 A_1...\lambda_n A_n} &=& \rho_{A_2...A_n F | (\lambda_1)^R A_1...\lambda_n A_n} \otimes \mathds{1}_{(\lambda_1)^L} \ . \label{Eq_Channel2Step1}
\end{eqnarray}
The proof now proceeds via iterative use of Lemma~\ref{Lem_SplitFromCommutation_II}. 
In the second step, the fact that
\begin{equation}
	[ \rho_{A_2 | \lambda_1 A_1 \lambda_2} \ , \ \rho_{A_3...A_n F | \lambda_1 A_1 \lambda_2 A_2...\lambda_n A_n}  ] = 0 \ ,
\end{equation}
along with Lemma~\ref{Lem_SplitFromCommutation_II} and Remark~\ref{Rem_SplittingForUnitary}, implies a factorization
$(\lambda_1)^R A_1 \lambda_2 = ((\lambda_1)^R A_1 \lambda_2)^L \otimes ((\lambda_1)^R A_1 \lambda_2)^R$
such that
\begin{eqnarray}
	\rho_{A_2 | \lambda_1 A_1 \lambda_2} &=& \rho_{A_2 | ((\lambda_1)^R A_1 \lambda_2)^L} \otimes \mathds{1}_{((\lambda_1)^R A_1 \lambda_2)^R} \otimes \mathds{1}_{(\lambda_1)^L} \label{Eq_Channel1Step2} \\
	\rho_{A_3...A_n F | \lambda_1 A_1 \lambda_2 A_2...\lambda_n A_n} &=&  \rho_{A_3...A_n F | ((\lambda_1)^R A_1 \lambda_2)^R A_2...\lambda_n A_n} \otimes \mathds{1}_{((\lambda_1)^R A_1 \lambda_2)^L} \otimes \mathds{1}_{(\lambda_1)^L} \ . \ \ 
\end{eqnarray}
By using the short-hand notation
\begin{equation}
(\NestAbbr{A_i\lambda_{i+1}}) \ := \ (\cdots (((\lambda_1)^R A_1 \lambda_2)^R A_2 \lambda_3)^R \cdots A_{i}\lambda_{i+1})^R \ ,
\end{equation}
the iteration of the above step yields the factorization
\begin{eqnarray}
	\lambda_1 A_1 \cdots \lambda_n  A_n &=& (\lambda_1)^L \ \otimes \ ((\lambda_1)^R A_1 \lambda_2)^L 
					\ \otimes \ (((\lambda_1)^R A_1 \lambda_2)^R A_2 \lambda_3)^L  \label{Eq_OverallFactorization} \\[0.1cm]
					&& \hspace{0.2cm} \otimes \  ((\NestAbbr{A_{2}\lambda_{3}})  A_{3}\lambda_4)^L \ \otimes \  ... \ \otimes  \ ((\NestAbbr{A_{n-2}\lambda_{n-1}})  A_{n-1}\lambda_n)^L \nonumber \\[0.1cm]
					&& \hspace{0.2cm}  \otimes  \  ((\NestAbbr{A_{n-2}\lambda_{n-1}})  A_{n-1}\lambda_n)^R \ \otimes \ A_n \ , \nonumber
\end{eqnarray}
along with channels on the respective factors such that
\begin{eqnarray}
	\rho^U_{A_1...A_nF|A_1...A_n\lambda_1...\lambda_n}  &=&  
									\rho_{A_1|(\lambda_1)^L} \ \ \rho_{A_2|((\lambda_1)^R A_1 \lambda_2)^L} \label{Eq_UnitaryFactorizationIntoUnitaries} \\[0.1cm]
								&& \rho_{A_3|((\NestAbbr{A_{1}\lambda_{2}})  A_{2}\lambda_3)^L} \ \ ... \ \ \rho_{A_n|((\NestAbbr{A_{n-2}\lambda_{n-1}})  A_{n-1}\lambda_n)^L}  \nonumber \\[0.1cm]
								&&  \rho_{F|((\NestAbbr{A_{n-2}\lambda_{n-1}})  A_{n-1}\lambda_n)^R A_n} 	\nonumber			
\end{eqnarray}
Note that all operators appearing on the right hand side of Eq.~\ref{Eq_UnitaryFactorizationIntoUnitaries} act on distinct spaces. The product of the $n+1$ operators (recalling the convention of suppressing identities) is therefore identical with the tensor product of the same $n+1$ operators: $\rho_{A_1|(\lambda_1)^L}\ \otimes \cdots \otimes \ \rho_{F|((\NestAbbr{A_{n-2}\lambda_{n-1}}) \ A_{n-1}\lambda_n)^R A_n}$. From the fact that the operator on the left hand side of Eq.~\ref{Eq_UnitaryFactorizationIntoUnitaries} is rank 1, it follows that each of the terms on the right hand side is rank 1, hence each of the terms on the right hand side is the CJ operator for an isometry. The fact that the left hand side is a unitary channel implies that $d_f \prod_i d_{A_i} = \prod_i d_{A_i} d_{\lambda_i}$, which can only be satisfied if each term on the right hand side is also a unitary channel. \\

Systems $A_i'$ ($i=1,...,n$), corresponding to the unbroken wires in the circuit of Fig.~\ref{Fig_JustificationCircuit}, are now defined using Eq.~\ref{Eq_OverallFactorization} as follows:
\begin{eqnarray}
	\mathcal{H}_{A_1'} &\cong & \mathcal{H}_{(\lambda_1)^R} \nonumber \\
	\mathcal{H}_{A_2'} &\cong & \mathcal{H}_{((\lambda_1)^R A_1 \lambda_2)^R} \nonumber \\
			&\vdots & \nonumber \\
	\mathcal{H}_{A_n'} &\cong & \mathcal{H}_{((\NestAbbr{A_{n-2}\lambda_{n-1}}) \ A_{n-1}\lambda_n)^R} . \nonumber 
\end{eqnarray}
Write 
\begin{equation}
\rho^{I_i}_{A_i'|((\NestAbbr{A_{i-2}\lambda_{i-1}}) \ A_{i-1}\lambda_i)^R}
\end{equation}
for the CJ operator corresponding to a unitary map $I_i$ between $A_i'$ and the indicated `right factor' -- such an $I_i$ exists by definition of the primed systems. Then define the following channels:
\begin{eqnarray}
\rho^{U_1}_{A_1A_1'|\lambda_1} &:=& \rho_{A_1|(\lambda_1)^L} \ \otimes \ \rho^{I_1}_{A_1'|(\lambda_1)^R} \label{Eq_NewChnnels_1} \\
\rho^{U_2}_{A_2A_2'|A_1'A_1\lambda_2} &:=& \rho_{A_2|(A_1'A_1 \lambda_2)^L} \ \otimes \ \rho^{I_2}_{A_2'|(A_1' A_1 \lambda_2)^R} \label{Eq_NewChnnels_2} \\
&\vdots & \nonumber \\
\rho^{U_n}_{A_nA_n'|A_{n-1}'A_{n-1}\lambda_n} &:=& \rho_{A_n|(A_{n-1}' A_{n-1}\lambda_n)^L} \ \otimes \  \rho^{I_n}_{A_n'|(A_{n-1}'A_{n-1}\lambda_n)^R} \\
\rho^{U_{n+1}}_{F|A_n'A_n} &:=& \rho_{F|A_n'A_n} , \label{Eq_NewChnnels_n}
\end{eqnarray}
which are by construction unitary channels. This notation is to be understood in the obvious way: $\rho_{A_2|(A_1'A_1 \lambda_2)^L}$ is short for $\rho_{A_2|((\lambda_1)^R A_1 \lambda_2)^L}$ post-composed with the unitary map $I_1$, and similarly for the other channels.
These channels, along with the states $\rho_{\lambda_i}$, define a circuit of the form of that of Fig.~\ref{Fig_JustificationCircuit}.  It remains to show that marginalizing over the primed systems to obtain
\begin{eqnarray}
\overline{\textnormal{Tr}}_{A_1',...,A_n'} \left[ \ \rho^{U_1}_{A_1A_1'|\lambda_1} \ \rho^{U_2}_{A_2A_2'|A_1'A_1\lambda_2} \ \ ... \ \ \rho^{U_n}_{A_nA_n'|A_{n-1}'A_{n-1}\lambda_n} \ \rho^{U_{n+1}}_{F|A_n'A_n} \ \right] \ 
\end{eqnarray}
returns the unitary channel $\rho^U_{A_1...A_nF|A_1...A_n\lambda_1...\lambda_n}$ of Eq.~(\ref{Eq_UnitaryFactorizationIntoUnitaries}). This is the case by construction, since
\begin{eqnarray}
&& \overline{\textnormal{Tr}}_{A_1'} \left[ \ \rho_{A_2|(A_1'A_1 \lambda_2)^L} \ \rho^{I_2}_{A_2'|(A_1' A_1 \lambda_2)^R} \ \rho^{I_1}_{A_1'|(\lambda_1)^R} \ \right] \nonumber \\ 
&& \hspace{3cm} = \ \textnormal{Tr}_{A_1'{A_1'}^*} \left[  \tau^{\text{id}}_{A_1'} \ \rho_{A_2|(A_1'A_1 \lambda_2)^L} \ \rho^{I_2}_{A_2'|(A_1' A_1 \lambda_2)^R} \ \rho^{I_1}_{A_1'|(\lambda_1)^R} \ \right] \nonumber \\  
&& \hspace{3cm} = \  \rho_{A_2|((\lambda_1)^R A_1 \lambda_2)^L} \ \rho^{I_2}_{A_2'|((\lambda_1)^R A_1 \lambda_2)^R} 	 \ ,
\end{eqnarray}
and similarly for $A_i'$, $i = 2 , \ldots , n$.
\hfill $\square$

\subsection{Proof of Theorem~\ref{Thm_QuantumEquivalence}: $(2)\rightarrow (1)$} \label{Subsec_Proof_Thm_QuantumEquivalence21}

Consider a DAG $G$, with nodes $A_1, \ldots , A_n$ and a process operator $\sigma_{A_1 ... A_n}$ that is Markov for $G$. For improved legibility of the following proof, write $P_i:= Pa(A_i)$. By the definition of Markovianity of a process operator (Def.~\ref{Def_QuantumMC}), there exist commuting channels  $\rho_{A_i|P_i}$ such that $\sigma_{A_1 ... A_n} = \prod_i \rho_{A_i|P_i}$. In particular, $[ \rho_{A_1|P_1} \ , \ \rho_{\overline{A_1}|P_1 \overline{P_1}} ] = 0$, where a bar denotes relative complement in the set $\{A_1, \ldots , A_n\}$, so that $\rho_{\overline{A_1}|P_1 \overline{P_1}} = \prod_{i\neq 1} \rho_{A_i|P_i}$. 
Lemma~\ref{Lem_SplitFromCommutation_I} implies that there exists a decomposition of $P_1$ into orthogonal subspaces
$P_1 = \bigoplus_j \ (P_1)_j^L \otimes (P_1)_j^R$ such that 
\begin{eqnarray}
\rho_{A_1|P_1} &=& \sum_j \ \rho_{A_1|(P_1)_j^L} \otimes \mathds{1}_{(P_1)_j^R} \\
\rho_{\overline{A_1}|P_1 \overline{P_1}} &=& \sum_j \ \mathds{1}_{(P_1)_j^L} \otimes \rho_{\overline{A_1}|(P_1)_j^R \overline{P_1}}   \ .
\end{eqnarray}
For each $j$, the channel represented by $\rho_{A_1|(P_1)_j^L}$ can be dilated to a unitary channel with unitary $U^j : (P_1)_j^L \otimes \lambda_1^j \allowbreak \rightarrow \allowbreak A_1 \otimes F^j$ and some appropriate state $\ket{0}_{\lambda_1^j}$.	Let $\lambda_1$ be a system of large enough dimension that these unitaries can be extended to $U^j : (P_1)_j^L\lambda_1 \rightarrow A_1F^j$ (for appropriate $F^j$) with a common auxiliary state $\ket{0}_{\lambda_1}$. Define the unitary $U_1: P_1 \otimes \lambda_1 \rightarrow A_1 \otimes F$ by setting $F:=\bigoplus_j F^j \otimes (P_1)_j^R$ and $U_1:= \sum_j \tilde{U}^j$, where  $\tilde{U}^j := U^j \otimes \mathds{1}_{(P_1)_j^R}$ acts only on the $j$th subspace and as zero everywhere else. By construction,
\begin{equation}
\rho_{A_1|P_1} \ = \ \Trace_{F} \overline{\Trace}_{\lambda_1} \Big[ \ \rho^{U_1}_{A_1F|P_1\lambda_1} \ \ket{0}_{\lambda_1} \bra{0} \ \Big] \ . 
\end{equation}
The marginal channel $\rho_{A_1|P_1\lambda_1}$ is by definition a reduced unitary, and it commutes with $\rho_{A_j|P_j}$ for all $j\neq1$. Next, consider $[ \rho_{A_2|P_2} \ , \ \rho_{\overline{A_2}|P_2 \overline{P_2}\lambda_1} ] = 0$, 
where $\rho_{\overline{A_2}|P_2 \overline{P_2}\lambda_1}:=\rho_{A_1|P_1 \lambda_1} \big( \prod_{i=3}^n \rho_{A_i|P_i} \big)$. A similar construction to the above yields a corresponding channel $\rho_{A_2|P_2\lambda_2}$. Iterating this procedure yields a set of pairwise commuting reduced unitaries, $\{ \rho_{A_i|P_i\lambda_i} \}$, such that for each $i$,  the reduced unitary $\rho_{A_i|P_i\lambda_i}$ returns the original channel $\rho_{A_i|P_i}$ given an input $\rho_{\lambda_i}$. Lemma~\ref{Lem_ReducedUnitaryLemma} then implies that $\prod_i \rho_{A_i|P_i\lambda_i}$ is a reduced unitary, hence there exists a unitary $U$ such that
\begin{equation}
\Trace_{F} \Big[ \ \rho^U_{A_1...A_nF|A_1...A_n\lambda_1...\lambda_n} \ \Big] = \prod_i \rho_{A_i|P_i\lambda_i} \ .
\end{equation}
By construction,
\begin{equation}
\sigma_{A_1 ... A_n} \ = \ \overline{\Trace}_{\lambda_1...\lambda_n} \Trace_{F} \Big[ \ \rho^U_{A_1...A_nF|A_1...A_n\lambda_1...\lambda_n} \ \rho_{\lambda_1} \otimes \cdots\otimes \rho_{\lambda_n} \ \Big],
\end{equation}
and the unitary channel $U$ satisfies the no-influence conditions:
\begin{equation}
\left\{ A_j \nrightarrow A_i \right\}_{A_j \notin Pa(A_i)} \ , \ \left\{ \lambda_j \nrightarrow A_i \right\}_{j \neq i} \hspace*{0.5cm} \forall i=1, \ldots ,n .
\end{equation}

It remains to show that the operator $\sigma_{A_1 ... A_n \lambda_1 ... \lambda_n F}  = \rho^U_{A_1...A_nF|A_1...A_n\lambda_1...\lambda_n} \ \rho_{\lambda_1} \otimes\cdots\otimes \rho_{\lambda_n} $ that has been constructed is a valid process operator, as required by Def.~\ref{Def_QuantumCompatibility}. This follows via an application of the proof of Theorem~\ref{Thm_Compatibilitywithbroken} in Appendix~\ref{SubSec_Proof_Thm_Compatibilitywithbroken}, to show that  $\sigma_{A_1 ... A_n \lambda_1 ... \lambda_n F}$
can be realized as a broken unitary circuit. Note in particular that the proof of Appendix~\ref{SubSec_Proof_Thm_Compatibilitywithbroken} does not need to assume that $\rho^U_{A_1...A_nF|A_1...A_n\lambda_1...\lambda_n} \ \rho_{\lambda_1} \otimes \cdots\otimes \rho_{\lambda_n}$ is a valid process operator. Rather, the fact that $\sigma_{A_1 ... A_n \lambda_1 ... \lambda_n F}$ is a valid process operator follows from the conclusion of the proof, since any broken circuit yields a valid process operator.
\hfill $\square$

\section{Appendix part II}

\subsection{Useful tools II \label{Subsec_Appendix_DSeparationSets}} 

The following definition and lemma, which will be helpful in several other proofs, are concerned solely with DAGs. 
\begin{definition}\label{Def_SR}\textnormal{(Relation $SR$ on the nodes of a DAG):}
Let $G$ be a DAG, with a set of nodes $V$. Let $Y$, $Z$ and $W$ denote arbitrary disjoint subsets of $V$, with $R:=V \setminus ( Y \cup Z \cup W)$. The 3-place relation $SR(Y,Z;W)$ holds if there exist partitions of $W$ and $R$
\begin{eqnarray}
R &=& R_Y \cup R_Z \cup R^c  \label{Eq_RPartition}\\
W &=& W_Y \cup W_Z  \label{Eq_WPartition},
\end{eqnarray}
such that: if $A$ is any of $Y$, $Z$, $R_Y$, $R_Z$ or $R^c$, and $B$ is any of $W_Y$, $W_Z$, $Y$, $Z$, $R_Y$, $R_Z$ or $R^c$, then the absence of an arrow  $A \rightarrow B$ in Fig.~\ref{Fig_DSepSubSets_Statement} implies that for any $a\in A, b\in B$, there is no arrow in $G$ from $a$ to $b$. (NB Nodes in $W_Y$ and $W_Z$ can have children in any other set, but these arrows are suppressed in Fig.~\ref{Fig_DSepSubSets_Statement} for better visibility.)
\begin{figure}[H]
\small
\begin{center}
\renewcommand{\Scale}{2.5}
\input{Figures/Fig_DSepSubSets_Statement.tex}
\caption{ \label{Fig_DSepSubSets_Statement}}
\end{center}
\end{figure}
\end{definition}
\begin{lemma} \label{Lem_DAGDSep}
Let $G$ be a DAG, with a set of nodes $V$, and let $Y$,$Z$, and $W$ be disjoint subsets of $V$. If $(Y \bigCI Z | W)_{G}$, then $SR(Y,Z;W)$.
\end{lemma}
\noindent {\bf Proof.} 
As observed in the proof of Theorem~\ref{Thm_QuantumDSeparationTheorem} in Section~\ref{Subsec_QuantumDSeparationTheorem} (see Refs.~\cite{Verma&Pearl_1990_CausalNetworks} and \cite{Lauritzen_2011_DirectedMarkovProperties}), the soundness of d-separation for a 3-place relation on subsets of nodes can be established by showing that the relation satisfies the \textit{local Markov condition} and the \textit{semi-graphoid axioms}. 

\begin{figure}[H]
\begin{center}
\small
\renewcommand{\Scale}{1.5}
\input{Figures/Fig_DSepSubSets_Proof_LMC.tex}
\caption{ $SR(\{X\},N;P)$ \label{Fig_DSepSubSets_Proof_LMC}}
\end{center}
\end{figure}
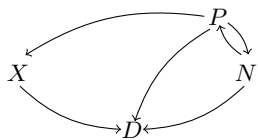

In order to see that the local Markov condition holds, let $X \in V$ and define $P:=Pa(X)$ and $N:=Nd(X) \setminus Pa(X)$. The nodes in $D = V \backslash (\{X\} \cup P \cup N)$ are the descendants of $X$. Without any further partitioning of sets, Fig.~\ref{Fig_DSepSubSets_Proof_LMC} is already of the required form, hence $SR(\{X\},N;P)$. 
Now consider the semi-graphoid axioms. The symmetry axiom $SR(Y,Z;W) \ \Leftrightarrow \  SR(Z,Y;W)$ is immediate. Concerning the decomposition axiom, $SR(Y,XZ;W)  \ \Rightarrow  \  SR(Y,Z;W)$, suppose that $SR(Y,XZ;W)$ holds, with corresponding decompositions $W = W_Y \cup W_{XZ}$ and $R = R_{Y} \cup R_{XZ} \cup R^c$. Defining $R_Z := R_{XZ} \cup X$ and $W_Z := W_{XZ}$, it follows immediately that $SR(Y,Z;W)$ holds, with corresponding decompositions $W = W_Y \cup W_Z$ and $R = R_Y \cup R_Z \cup R^c$. Concerning the weak union axiom, $SR(Y,XZ;W)  \ \Rightarrow  \  SR(Y,Z;XW)$, suppose that $SR(Y,XZ;W)$ holds, with corresponding decompositions $W = W_Y \cup W_{XZ}$ and $R = R_{Y} \cup R_{XZ} \cup R^c$. Defining $W_Z := W_{XZ} \cup X$ and $R_Z:=R_{XZ}$, it is immediate that $SR(Y,Z;XW)$ holds with corresponding decompositions $X\cup W = W_Y \cup W_Z$ and $R = R_Y \cup R_Z \cup R^c$. Finally, concerning the contraction axiom, suppose that $SR(Y,Z;W) \ \wedge \ SR(Y,X;ZW)$. Let the subsets implied by these two relations be labelled as in Figs.~\ref{Fig_DSepSubSets_Proof_Contraction_1} and \ref{Fig_DSepSubSets_Proof_Contraction_2}.\\
\begin{minipage}{16cm}
	\begin{minipage}{5cm}
		\begin{center}
			\begin{figure}[H]
				\begin{center}
					\small
					\renewcommand{\Scale}{2.3}
					\input{Figures/Fig_DSepSubSets_Proof_Contraction_1.tex}
					\caption{ $SR(Y,Z;W)$ \label{Fig_DSepSubSets_Proof_Contraction_1}}
				\end{center}
			\end{figure}
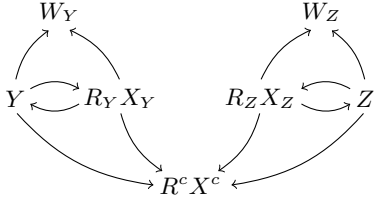
		\end{center}
	\end{minipage}
	\hfill
	\begin{minipage}{5cm}
		\begin{center}
			\begin{figure}[H]
				\begin{center}
					\small
					\renewcommand{\Scale}{2.3}
					\input{Figures/Fig_DSepSubSets_Proof_Contraction_2.tex}
					\caption{ $SR(Y,X;ZW)$ \label{Fig_DSepSubSets_Proof_Contraction_2}}
				\end{center}
			\end{figure}
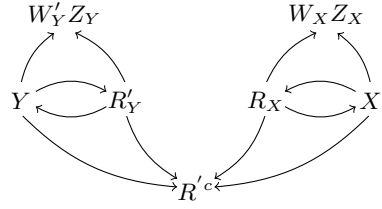
		\end{center}
	\end{minipage}
	\hfill
	\begin{minipage}{5cm}
		\begin{center}
			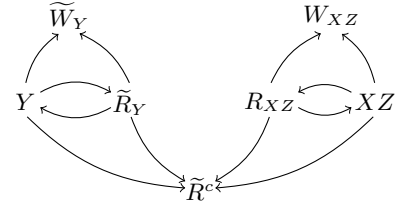
\begin{figure}[H]
				\begin{center}
					\small
					\renewcommand{\Scale}{2.3}
					\input{Figures/Fig_DSepSubSets_Proof_Contraction_3.tex}
					\caption{ $SR(Y,ZX;W)$ \label{Fig_DSepSubSets_Proof_Contraction_3}}
				\end{center}
			\end{figure}
		\end{center}
	\end{minipage}
\end{minipage}
\\[0.3cm]
Defining new sets as follows,
\begin{center}
	\begin{minipage}{13cm}
		\begin{minipage}{6cm}
			\begin{eqnarray}
				W_{XZ} &:=& W_X \cup W_Z \nonumber \\
				\widetilde{W}_Y &:=& W \setminus W_{XZ} \nonumber
			\end{eqnarray}
		\end{minipage}
		\hfill
		\begin{minipage}{6cm}
			\begin{eqnarray}
				R_{XZ} &:=&  R_X \cup R_Z \nonumber \\
				\widetilde{R}^c &:=& (R^c \cup R^{'c} ) \setminus R_{XZ} \nonumber \\
				\widetilde{R}_Y &:=& R \setminus ( R_{XZ} \cup \widetilde{R}^c) \ , \nonumber
			\end{eqnarray}
		\end{minipage}
	\end{minipage}
\end{center}
the diagram in Fig.~\ref{Fig_DSepSubSets_Proof_Contraction_3} correctly expresses which parent-child relations between those subsets are forbidden, i.e. there must not be any arrows from $R_{XZ}$, $X$ or $Z$ to the sets on the left and no arrows from $\widetilde{R}_Y$ or $Y$ to any sets on the right. This establishes $SR(Y,ZX;W)$. \hfill $\square$

\subsection{Proof of Prop.~\ref{Prop_QuantumIndependecneOS1} \label{Subsubsec_Prop_QuantumIndependecneOS1} } 

The only if direction is immediate. 

For the if direction, consider an informationally complete intervention at each node $X \in Y$: that is, an intervention corresponding to a quantum instrument $\{ \tau_{X}^{k_{X}} \}$ such that, varying over $k_X$, the operators $\tau_{X}^{k_{X}}$ span the real vector space of Hermitian operators on $\mathcal{H}_{X^{\text{in}}}\otimes \mathcal{H}_{X^{\text{out}}}$. 
Let $k_Y$ denote the joint outcome, and for each $k_Y$, let $\tau_Y^{k_Y}$ denote the corresponding tensor product of local operators. Varying over $k_Y$, the operators $\tau_Y^{k_Y}$ span the real vector space of Hermitian operators on $\mathcal{H}_{Y^{\text{in}}}\otimes \mathcal{H}_{Y^{\text{out}}}$, i.e., the local intervention at $Y$, consisting of a product of informationally complete interventions, is also informationally complete. Similarly, consider an informationally complete local intervention at $Z$, with joint outcome $k_Z$, and corresponding product operators $\tau_Z^{k_Z}$. 

Given $\sigma_{YZ}$, suppose that $P(k_Y , k_Z) = \Trace_{YZ} (\sigma_{YZ} \tau_Y^{k_Y}\otimes \tau_Z^{k_Z}) = P(k_Y)P(k_Z)$. Let $\alpha_Y$ be the operator such that $P(k_Y) = \Trace_Y (\alpha_Y \tau_Y^{k_Y})$ and $\beta_Z$ be the operator such that $P(k_Z) = \Trace_Z (\beta_Z \tau_Z^{k_Z})$. Since $\{ \tau_Y^{k_Y} \otimes \tau_Z^{k_Z} \}_{k_Y,k_Z}$ spans the tensor product space of operators, $\sigma_{YZ}$ and $\alpha_{Y} \otimes \beta_{Z}$ agree on a basis, hence $\sigma_{YZ} = \alpha_{Y} \otimes \beta_{Z} = \sigma_Y \otimes \sigma_Z$.  \hfill $\square$

\subsection{Proof of Prop.~\ref{Prop_CSCI_EquivalentOperationalStatements} \label{Subsubsec_Proof_CSCI_EquivalentOperationalStatements} } 

For the only if direction, assume that $(Y \bigCI Z|W)_{\clop}$ holds, hence $\clop_{YZW}$ can be written in the form $\clop_{YZW} = \alpha_{YW}\beta_{ZW}$. Consider arbitrary interventions at $Y$ and $Z$, and a maximally informative intervention at $W$. Then by Def.~\ref{Def_MaxInfMeas}, functions $g^{\text{in}}$ and $g^{\text{out}}$ exist such that
\begin{eqnarray}
P(k_Y,k_Z,k_W) &=& \sum_Y \  \sum_Z \  \sum_W \  \clop_{YZW}\ P(k_Y, Y^{\text{out}} | Y^{\text{in}}) \ P(k_Z, Z^{\text{out}} | Z^{\text{in}}) \nonumber \\[-0.3cm]
&& \hspace{2.0cm}  \ \delta ( g^{\text{in}}(k_W), W^{\text{in}} ) \ \delta ( g^{\text{out}}(k_W), W^{\text{out}} ) \ P(k_W, W^{\text{out}} | W^{\text{in}})   \nonumber \\[0.3cm]
&=& \sum_Y \ \alpha_{YW}(Y^{\text{in}}, Y^{\text{out}}, g^{\text{in}}(k_W), g^{\text{out}}(k_W)) \ P(k_Y, Y^{\text{out}} | Y^{\text{in}})\nonumber \\
&& \sum_Z \  \beta_{ZW}(Z^{\text{in}}, Z^{\text{out}}, g^{\text{in}}(k_W), g^{\text{out}}(k_W)) \ P(k_Z, Z^{\text{out}} | Z^{\text{in}})\ \ P(k_W, g^{\text{out}}(k_W) | g^{\text{in}}(k_W))  \ . \nonumber
\end{eqnarray}
Setting 
\[
 \alpha' (k_Y , k_W) =  \sum_Y \ \alpha_{YW}(Y^{\text{in}}, Y^{\text{out}}, g^{\text{in}}(k_W), g^{\text{out}}(k_W)) \ P(k_Y, Y^{\text{out}} | Y^{\text{in}}),
\]
and 
\[
\beta'(k_Z, k_W) = \sum_Z \  \beta_{ZW}(Z^{\text{in}}, Z^{\text{out}}, g^{\text{in}}(k_W), g^{\text{out}}(k_W)) \ P(k_Z, Z^{\text{out}} | Z^{\text{in}})\ \ P(k_W, g^{\text{out}}(k_W) | g^{\text{in}}(k_W)) 
\]
yields $P(k_Y,k_Z,k_W) = \alpha' (k_Y , k_W) \beta'(k_Z , k_W)$, hence $(k_Y \bigCI k_Z | k_W)_P$, by Prop.~\ref{Prop_EquivalentStatementsClassicalCI}.

For the if direction, assume that for any local interventions at $Y$ and $Z$, and any maximally informative local intervention at $W$, the joint outcome probabilities $P(k_Y,k_Z,k_W)$ satisfy $(k_Y \bigCI k_Z | k_W)_P$. Consider, in particular, an intervention at each node $X$ (where $X\in Y$, or $X\in Z$, or $X\in W$), such that the outcome $k_X$ is a pair  $k_X = (k_X^I , k_X^O)$, and 
\[
P(k_X, X^{\text{out}} | X^{\text{in}}) = \frac{1}{d_X} \delta(k_X^I , X^{\text{in}}) \delta(k_X^O , X^{\text{out}}),
\]
where $d_X$ is the cardinality of the set on which $X^{\text{in}}$ takes values. This corresponds to measuring the input variable, recording the value as $k_X^I$, choosing a value $k_X^O$ at random, and setting $X^{\text{out}} = k_X^O$. Assume that the local intervention at $Y$ is a product of interventions of this form, with joint outcome $k_Y = (k_Y^I , k_Y^O)$, where $k_Y^I$ is the tuple consisting of a value of $k_X^I$, for each $X\in Y$, and $k_Y^O$ is the tuple consisting of a value of $k_X^O$, for each $X\in Y$. Similarly $Z$ and $k_Z = (k_Z^I , k_Z^O)$, and $W$ and $k_W = (k_W^I , k_W^O)$. Observe that these interventions are maximally informative, in keeping with the assumption of a maximally informative intervention at $W$.

Given $(k_Y \bigCI k_Z | k_W)_P$, it follows from Prop.~\ref{Prop_EquivalentStatementsClassicalCI} that there exist $\alpha(k_Y , k_W)$ and $\beta(k_Z , k_W)$ such that $P(k_Y , k_Z , k_W) = \alpha(k_Y , k_W) \beta(k_Z , k_W)$. Define $\alpha'_{YW} : Y^{\text{in}} \times Y^{\text{out}} \times W^{\text{in}} \times W^{\text{out}} \rightarrow \mathbb{R}$ and $\beta'_{ZW} : Z^{\text{in}} \times Z^{\text{out}} \times W^{\text{in}} \times W^{\text{out}} \rightarrow \mathbb{R}$ such that
\begin{eqnarray}
\alpha'_{YW}(Y^{\text{in}} , Y^{\text{out}} , W^{\text{in}} , W^{\text{out}}) &=& d_Yd_W \ \alpha(k_Y , k_W)\big|_{k_Y = (Y^{\text{in}} , Y^{\text{out}}) , k_W = (W^{\text{in}} , W^{\text{out}})}      \nonumber\\
\beta'_{ZW}(Z^{\text{in}} , Z^{\text{out}} , W^{\text{in}} , W^{\text{out}}) &=& d_Z \ \beta(k_Z , k_W)\big|_{k_Z = (Z^{\text{in}} , Z^{\text{out}}) , k_W = (W^{\text{in}} , W^{\text{out}})}.     \nonumber
\end{eqnarray}
Observe that 
\begin{eqnarray*}
\alpha(k_Y , k_W) &=& \sum_{YW} \alpha'_{YW} P(k_Y, Y^{\text{out}} | Y^{\text{in}}) P(k_W, W^{\text{out}} | W^{\text{in}}),  \\
\beta(k_Z , k_W) &=& d_W \sum_{ZW} \beta'_{ZW} P(k_Z, Z^{\text{out}} | Z^{\text{in}}) P(k_W, W^{\text{out}} | W^{\text{in}}).
\end{eqnarray*}
Now,
\begin{eqnarray}\label{Eq_ClassicalprocessCIproof1}
P(k_Y , k_Z , k_W) &=& \alpha(k_Y , k_W) \beta(k_Z , k_W)\nonumber \\
&=& d_W \left(\sum_{YW} \alpha'_{YW} P(k_Y, Y^{\text{out}} | Y^{\text{in}}) P(k_W, W^{\text{out}} | W^{\text{in}})\right) \nonumber \\ 
&& \hspace*{0.6cm} \left(  \sum_{ZW} \beta'_{ZW} P(k_Z, Z^{\text{out}} | Z^{\text{in}}) P(k_W, W^{\text{out}} | W^{\text{in}})\right)\nonumber \\
&=& \sum_{YZW}  P(k_Y, Y^{\text{out}} | Y^{\text{in}}) P(k_Z, Z^{\text{out}} | Z^{\text{in}}) P(k_W, W^{\text{out}} | W^{\text{in}})  \alpha'_{YW} \beta'_{ZW},
\end{eqnarray}
where the third equality follows from the form of the $W$ intervention (in particular, the fact that the $W$ intervention is maximally informative).
From the definition of the classical process map,
\begin{equation}\label{Eq_ClassicalprocessCIproof2}
P(k_Y , k_Z , k_W) =  \sum_{YZW}  P(k_Y, Y^{\text{out}} | Y^{\text{in}}) P(k_Z, Z^{\text{out}} | Y^{\text{in}}) P(k_W, W^{\text{out}} | W^{\text{in}})  \clop_{YZW}.
\end{equation}
The measurements considered are informationally complete: that is, if for fixed $k_Y$, $k_Z$, $k_W$, the term $P(k_Y, Y^{\text{out}} | \allowbreak Y^{\text{in}}) \allowbreak P(k_Z, Z^{\text{out}} | \allowbreak Z^{\text{in}}) P(k_W, W^{\text{out}} | \allowbreak W^{\text{in}})$ is viewed as a real-valued function of $Y^{\text{in}}$, $Y^{\text{out}}$,  $Z^{\text{in}}$, $Z^{\text{out}}$, $W^{\text{in}}$, $W^{\text{out}}$, then, varying over $k_Y$, $k_Z$, $k_W$, these functions span the vector space of all real-valued functions of $Y^{\text{in}}$, $Y^{\text{out}}$,  $Z^{\text{in}}$, $Z^{\text{out}}$, $W^{\text{in}}$, $W^{\text{out}}$. Comparing Eqs.~(\ref{Eq_ClassicalprocessCIproof1}) and (\ref{Eq_ClassicalprocessCIproof2}) then gives $\clop_{YZW} =  \alpha'_{YW} \beta'_{ZW}$, that is $(Y \bigCI Z | W)_{\clop_{YZW}}$.
\hfill $\square$

\subsection{Proof of Prop.~\ref{Prop_Implication_QOS1a} \label{Subsubsec_Proof_Implication_QOS1a}}

Suppose that $\sigma_{YZW}=\alpha_{YW} \ \beta_{ZW}$. The commutation of $\alpha_{YW}$ and $\beta_{ZW}$ implies the existence of a decomposition $\mathcal{H}_{W^{\text{in}}} \otimes \mathcal{H}_{W^{\text{out}}} = \bigoplus_i \mathcal{H}_{F_i^L} \otimes \mathcal{H}_{F_i^R}$ into orthogonal subspaces such that $\sigma_{YZW} = \sum_i \alpha_{YF_i^L} \otimes \beta_{ZF_i^R}$ (see Lemma~\ref{Lem_SplitFromCommutation_I} and note that $\alpha$ and $\beta$ can be chosen to be positive\footnote{Strictly speaking, Lemma~\ref{Lem_SplitFromCommutation_I} is formulated for commuting operators which are correctly normalised as CJ representations of channels, while here $\alpha_{YW}$ and $\beta_{ZW}$ are only assumed to be positive. It is, however, easy to see that nothing in the proof of that lemma changes if one accounts for different normalization everywhere.}). 
Let $\{\ket{i,f_i^L}\ket{i,f_i^R}\}_{f_i^L,f_i^R}$ be a product orthonormal basis of the $i$th subspace $F_i^L \otimes F_i^R$. 

Consider the following global intervention at $W$. An agent is stationed at an additional locus $E$, such that for each node $N \in W$, the quantum system at $N^{\text{in}}$ is sent to $E$, one half of a maximally entangled state is fed into $N^{\text{out}}$, and the other half is sent to $E$. 
This defines a new process operator $\sigma_{YZE}$  over $Y,Z$ and $E$, with $E^{\text{in}}$ isomorphic to $\mathcal{H}_{W^{\text{in}}} \otimes \mathcal{H}_{W^{\text{out}}}$, such that $\sigma_{YZE}$ has a block-diagonal structure with respect to the induced decomposition. Let $\ket{i,f_i^L,f_i^R} := J^{-1} \ket{i,f_i^L}$ $\ket{i,f_i^R}$ label the induced orthonormal basis of $E^{\text{in}}$ (for a suitable isomorphism $J$). The agent performs the von Neumann measurement at $E$ corresponding to that basis. 
For a particular outcome $k_E$, corresponding to the basis state $\ket{i,f_i^L,f_i^R}$, define the operator
\begin{eqnarray}
	\Trace_{E^{\text{in}} E^{\text{out}}} \Big[ \sigma_{YZE} \ \ket{i,f_i^L,f_i^R} \bra{i,f_i^L,f_i^R} \Big]  
		&=&  \bra{i,f_i^L,f_i^R} \ J^{-1}  \Big(\sum_j \alpha_{YF_j^L} \otimes \beta_{ZF_j^R} \Big) J \ \ket{i,f_i^L,f_i^R} \nonumber \\
		&=& \bra{i,f_i^L} \alpha_{YF_i^L}\ket{i,f_i^L} \ \otimes \ \bra{i,f_i^R} \beta_{ZF_i^R}\ket{i,f_i^R} 
		\ =: \ \gamma_Y \otimes \eta_Z \nonumber 
\end{eqnarray}
The product form $\gamma_Y \otimes \eta_Z$ implies that the joint probability distribution for $k_E$ and outcomes $k_Y$ and $k_Z$ for arbitrary choices of interventions at $Y$ and $Z$ satisfies $P(k_Y,k_Z,k_E)=\phi(k_Y,k_E)\chi(k_Z,k_E)$ (for some functions $\phi$ and $\chi$). Recalling Prop.~\ref{Prop_EquivalentStatementsClassicalCI}, this establishes the claim. \hfill $\square$

\subsection{Proof of Lemmas~\ref{Lemma_quantumdsepsemigraphoid} and \ref{Lemma_quantumdseplocalmarkov} \label{Subsubsec_Proof_Thm_QuantumDSeparationTheorem}}

{\bf Proof of Lemma~\ref{Lemma_quantumdsepsemigraphoid}.} Let $\sigma_V$ be a process operator, 
and consider the relation $T$ of Eq.~(\ref{Eq_DefofTrelation}), defined over subsets of $V$. The symmetry axiom is immediate. For the decomposition axiom, suppose that $T(Y,XZ;W)$ holds, i.e., for all local interventions $\tau^R$, there exist $ \alpha_{YW}$ and $\beta_{XZW}$ such that $\sigma^{\tau_R}_{YXZW} = \alpha_{YW} \ \beta_{XZW}$. Then for any choice of local intervention $\tau_X$ at $X$, $\sigma^{\tau_R \tau_X}_{YZW} = \alpha_{YW} \ \Trace_{X}\big[ \tau_X \beta_{XZW} \big]$, hence $(Y \bigCI Z | W)_{\sigma^{\tau_R \tau_X}_{YZW}}$, hence $T(Y,Z;W)$ holds. The weak union axiom is immediate. 
Finally, for the contraction axiom, suppose that for all local interventions $\tau_R$ and $\tau_X$, $(Y \bigCI Z | W)_{\sigma^{\tau_R \tau_X}_{YZW}}$, and that for all local interventions $\tau_R$, $(Y \bigCI X | ZW)_{\sigma^{\tau_R}_{YXZW}}$. The first condition, along with Part~(3) of Prop.~\ref{Prop_EquivalentStatementsCQCI}, implies that for all local interventions $\tau_R$ and $\tau_X$, the quantum conditional mutual information $I(Y:Z|W)=0$ when evaluated on $\hat{\sigma}^{\tau_R \tau_X}_{YZW}$. Similarly, the second condition implies that for all local interventions $\tau_R$, $I(Y:X|ZW)=0$ when evaluated on $\hat{\sigma}^{\tau_R}_{YZWX}$.\footnote{The following steps are essentially the same as those used in Ref.~\cite{LeiferEtAl_2008QuantumGraphicalModels} to show that the condition $I(A:B|C)=0$ on ordinary quantum states satisfies the semi-graphoid axioms.}. Let $\tau_X$ be the intervention that, at each node in $X$, ignores the input and prepares a maximally mixed state on the output. 
This yields
\begin{eqnarray}
H(\Trace_{ZX}[\hat{\sigma}^{\tau_R}_{YZWX}])  + H(\Trace_{YX}[\hat{\sigma}^{\tau_R}_{YZWX}]) - H(\Trace_{YZX}[\hat{\sigma}^{\tau_R}_{YZWX}]) - H(\Trace_{X}[\hat{\sigma}^{\tau_R}_{YZWX}])  &=& 0 \\
H(\Trace_{X}[\hat{\sigma}^{\tau_R}_{YZWX}])  + H(\Trace_{Y}[\hat{\sigma}^{\tau_R}_{YZWX}]) - H(\Trace_{YX}[\hat{\sigma}^{\tau_R}_{YZWX}]) - H(\hat{\sigma}^{\tau_R}_{YZWX})  &=& 0
\end{eqnarray}
where $H(...)$ denotes the von Neumann entropy.  
Adding the two equations gives $I(Y:XZ|W)=0$, when evaluated on $\hat{\sigma}^{\tau_R}_{YZWX}$. Seeing as this holds for any local intervention $\tau_R$, Prop.~\ref{Prop_EquivalentStatementsCQCI} gives $T(Y,XZ;W)$.
 \hfill $\square$
\\ \\
{\bf Proof of Lemma~\ref{Lemma_quantumdseplocalmarkov}.} Consider a DAG $G$, with nodes $V$, and a process operator $\sigma_V$, such that $\sigma_V$ is Markov for $G$. Let $X\in V$ and set $P:=Pa(X)$, $N:=Nd(X) \setminus Pa(X)$ and $D:=V \setminus (X \cup P \cup N)$. The sets $\{X\}$, $P$, $N$ and $D$ constitute a partition of $V$, hence $\sigma_V = \rho_{X|Pa(X)} \ \rho_{P|Pa(P)} \ \rho_{N|Pa(N)} \ \rho_{D|Pa(D)}$. The set $D$ only contains descendants of $X$, hence $D$ cannot have children in any of the other sets. Given an arbitrary local intervention $\tau_D$, the marginal process operator over $XPN$ is therefore of the form $\sigma^{\tau_D}_{XPN} = \Trace_{D}[\tau_D \sigma_{XPND}] = \rho_{X|Pa(X)} \ \rho_{P|Pa(P)} \ \rho_{N|Pa(N)}$. By definition, $X \notin P$ and $N \cap P = \emptyset$, hence $\sigma^{\tau_D}_{XPN}$ is of the form $\alpha_{XP} \ \beta_{NP}$, with $\alpha_{XP} = \rho_{X|Pa(X)}$ and $\beta_{NP} = \rho_{P|Pa(P)} \ \rho_{N|Pa(N)}$. Therefore $T(X,N;P)$ holds.
\hfill $\square$

\subsection{Proof of Prop.~\ref{Prop_Implication_QOS1} \label{Subsec_Proof_Prop_Implication_QOS1}}
This is essentially the same as the proof of Prop.~\ref{Prop_Implication_QOS1a} in Appendix~\ref{Subsubsec_Proof_Implication_QOS1a}, except that with $\sigma_{YZWdo(X)} = \alpha_{YWX^{\text{out}}} \ \allowbreak \beta_{ZWX^{\text{out}}}$, the two factors have non-trivial action on the three Hilbert spaces $W^{\text{in}}$, $W^{\text{out}}$ and $X^{\text{out}}$. The relevant decomposition into orthogonal subspaces is therefore a decomposition of $W^{\text{in}} \otimes W^{\text{out}} \otimes X^{\text{out}}$. The proof proceeds with $W X^{\text{out}}$ replacing $W$.  \hfill $\square$

\subsection{Proof of Theorem~\ref{Thm_Rule1_qu} \label{Subsubsec_Proof_Rule1}}

Let $\sigma_{YZWXR}$ be a process operator that is Markov for $G$, hence $\sigma_{YZWRdo(X)} = \rho_{Y|Pa(Y)} \ \rho_{Z|Pa(Z)} \ \allowbreak \rho_{W|Pa(W)} \ \allowbreak \rho_{R|Pa(R)} \nonumber$, where all operators commute and may act non-trivially on $X^{\text{out}}$. Suppose that $( Y \bigCI Z | WX )_{G_{\overbar{X}}}$. Recalling Def.~\ref{Def_SR}, Lemma~\ref{Lem_DAGDSep} implies that $SR(Y,Z ; WX)$ holds in $G_{\overbar{X}}$. Therefore there exist partitions $R = R_Y \cup R_Z \cup R^c$ and $W \cup X  = W_Y \cup X_Y \cup W_Z \cup X_Z$, with
\[ 
\rho_{W | Pa(W)} =  \rho_{W_Y | Pa(W_Y)} \  \rho_{W_Z | Pa(W_Z)},    
\] 
and 
\[
\rho_{R | Pa(R)} = \rho_{R_Y | Pa(R_Y)} \ \rho_{R_Z | Pa(R_Z)} \ \rho_{R^c | Pa(R^c)},
\]
such that each of the operators $\rho_{Y|Pa(Y)}$,  $\rho_{W_Y|Pa(W_Y)}$,  $\rho_{R_Y|Pa(R_Y)}$ acts trivially on $Z\otimes R_Z \otimes R^c$, and each of the operators $\rho_{Z|Pa(Z)}$, $\rho_{W_Z|Pa(W_Z)}$ , $\rho_{R_Z|Pa(R_Z)}$ acts trivially on $Y\otimes R_Y \otimes R^c$. With $\tau_R = \tau_{R_Y} \otimes \tau_{R_Z} \otimes \tau_{R^c}$ an arbitrary local intervention at $R$, the term $\Trace_{R^c} [\tau_{R^c} \ \rho_{R^c | Pa(R^c)} ] = \mathds{1}_{Pa(R^c)}$, and it follows that 
\begin{eqnarray}
\sigma^{\tau_R}_{YZWdo(X)} \ = \Trace_{R} \left[ \ \tau_R\ \sigma_{YZWRdo(X)} \ \right] 
&=&  \Trace_{R_Y} \left[ \ \tau_{R_Y}\ \rho_{Y|Pa(Y)} \ \rho_{W_Y|Pa(W_Y)} \ \rho_{R_Y|Pa(R_Y)} \right]  \nonumber \\
&&  \Trace_{R_Z} \left[ \ \tau_{R_Z}\ \rho_{Z|Pa(Z)} \ \rho_{W_Z|Pa(W_Z)} \ \rho_{R_Z|Pa(R_Z)} \ \right] . \nonumber 
\end{eqnarray}
Setting
\[
 \alpha_{YWX^{\text{out}}} =  \Trace_{R_Y} \left[ \ \tau_{R_Y}\ \rho_{Y|Pa(Y)} \ \rho_{W_Y|Pa(W_Y)} \ \rho_{R_Y|Pa(R_Y)} \right], 
\]
and 
\[
 \beta_{ZWX^{\text{out}}} = \Trace_{R_Z} \left[ \ \tau_{R_Z}\ \rho_{Z|Pa(Z)} \ \rho_{W_Z|Pa(W_Z)} \ \rho_{R_Z|Pa(R_Z)} \ \right], 
\]
concludes the proof.

\hfill $\square$

\subsection{Proof of Prop.~\ref{Prop_EquivalentStatementsRule1} \label{Subsubsec_Proof_Prop_EquivalentStatementsRule1}}

\noindent {\bf $(1)\rightarrow (2)$} \\

Suppose that $\sigma_{YZWdo(X)} = \alpha_{YWX^{\text{out}}} \beta_{ZWX^{\text{out}}}$ for a pair of Hermitian operators $\alpha_{YWX^{\text{out}}}$ and $\beta_{ZWX^{\text{out}}}$. Taking the Hermitian conjugate of both sides of this equation establishes that $[  \alpha_{YWX^{\text{out}}} , \allowbreak \beta_{ZWX^{\text{out}}} ] = 0$. For arbitrary local interventions $\tau_Y$ at $Y$ and $\tau_Z$ at $Z$, let $\alpha^{\tau_Y}_{WX^{\text{out}}} = \Trace_Y \left[ \tau_Y \alpha_{YWX^{\text{out}}} \right]$ and $\beta^{\tau_Z}_{WX^{\text{out}}} = \Trace_Z \left[ \tau_Z \beta_{ZWX^{\text{out}}} \right]$. Observe that $[\alpha^{\tau_Y}_{WX^{\text{out}}} , \beta_{ZWX^{\text{out}}} ] = 0$ and $[\alpha_{YWX^{\text{out}}} \ , \ \beta^{\tau_Z}_{WX^{\text{out}}} ] = 0$

Using the commutativity and associativity of the `$\star$'-product, along with the fact that for arbitrary Hermitian operators $M$ and $N$, if $[M,N]=0$ then $M \star N = MN$,
\begin{eqnarray}
\sigma_{YZWdo(X)}  \star  \sigma^{\tau_Y ,\tau_Z}_{Wdo(X)} 
& = & \left( \alpha_{YWX^{\text{out}}} \  \beta_{ZWX^{\text{out}}}\right) \star  \left( \alpha^{\tau_Y}_{WX^{\text{out}}} \  \beta^{\tau_Z}_{WX^{\text{out}}}\right) \nonumber \\
& = &  \alpha_{YWX^{\text{out}}} \star  \beta_{ZWX^{\text{out}}} \star  \alpha^{\tau_Y}_{WX^{\text{out}}} \star  \beta^{\tau_Z}_{WX^{\text{out}}} \nonumber \\
& = &  \left(\alpha_{YWX^{\text{out}}} \ \beta^{\tau_Z}_{WX^{\text{out}}} \right) \star  \left( \alpha^{\tau_Y}_{WX^{\text{out}}} \ \beta_{ZWX^{\text{out}}} \right) \nonumber \\
&=& \sigma^{\tau_Z}_{YWdo(X)}  \star  \sigma^{\tau_Y}_{ZWdo(X)} \nonumber \ .
\end{eqnarray}

\noindent {\bf $(2)\rightarrow (3)$} \\

Assume (2), and consider local interventions $\tau_Y$ and $\tau_Z$ that consist of discarding the input, and preparing a maximally mixed state on the output, for each node in the corresponding set.
Dividing each side of (2) by $d_Y d_Z (d_W d_X)^2$, and taking logarithms, gives 
\begin{equation}
\text{log}\Big(\hat{\sigma}_{YZWdo(X)} \Big) + \text{log}\Big(\Trace_{YZ}(\hat{\sigma}_{YZWdo(X)}) \Big) \ = \ \text{log}\Big(\Trace_Z(\hat{\sigma}_{YZWdo(X)} ) \Big) + \text{log}\Big( \Trace_Y(\hat{\sigma}_{YZWdo(X)}) \Big),
\end{equation}
where the logarithms are restricted to the support of the respective operators. This implies (3). \\

\noindent {\bf $(3)\rightarrow (1)$} \\

Assume (3). By Theorem~6 of Ref.~\cite{HaydenEtAl_2003_StructureOfStatesAubadditivity}, there is a decomposition of the $WX^{\text{out}}$ Hilbert space of the form 
$\mathcal{H}_{WX^{\text{out}}} = \bigoplus_i \mathcal{H}_{ (WX^{\text{out}} )^L_i} \otimes \mathcal{H}_{ (WX^{\text{out}} )^R_i}$, and a probability distribution $\left\{ p_i \right\}$ such that $\hat{\sigma}_{YZWdo(X)} = \sum_i \ p_i \ \hat{\sigma}_{Y (WX^{\text{out}} )^L_i} \otimes \hat{\sigma}_{Z (WX^{\text{out}} )^R_i}$
for positive trace-1 operators $\hat{\sigma}_{Y (WX^{\text{out}} )^L_i}$ and $\hat{\sigma}_{Z (WX^{\text{out}} )^R_i}$. Define Hermitian operators
\begin{eqnarray}
	\alpha_{YWX^{\text{out}}} &:= & d_Y d_Z d_W d_X \sum_i \ p_i \ \hat{\sigma}_{Y (WX^{\text{out}} )^L_i} \otimes \mathds{1}_{Z(WX^{\text{out}} )^R_i} \\
	\beta_{ZWX^{\text{out}}}  &:= &   \sum_i \ \mathds{1}_{Y(WX^{\text{out}} )^L_i} \otimes \hat{\sigma}_{Z (WX^{\text{out}} )^R_i} \ .
\end{eqnarray}
Since the distinct subspaces are orthogonal, $\sigma_{YZWdo(X)} = \alpha_{YWX^{\text{out}}} \beta_{ZWX^{\text{out}}}$, which establishes the claim. \hfill $\square$

\subsection{Proof of Prop.~\ref{Prop_Rule2_CSMOperationalStatements} \label{Subsubsec_Proof_Prop_Rule2_CSMOperationalStatements}}

\noindent {\bf $(Y \bigCI Z^{\text{in}} | W do(X) Z^{\text{out}} )_{\clop_{YZWX}} \Rightarrow$ \ref{COS2}} \\

Let $\clop_{YZWX}$ represent a classical process map, and suppose that $\clop_{YZWdo(X)} = \alpha_{YWX^{\text{out}}Z^{\text{out}}} \ \beta_{ZWX^{\text{out}}}$, for suitable functions $\alpha_{YWX^{\text{out}}Z^{\text{out}}}$ and $\beta_{ZWX^{\text{out}}}$. Then, for an arbitrary local intervention at $Y$, a breaking local intervention at $Z$ that fixes $Z^{\text{out}} = z$, a maximally informative local intervention at $W$, and a do-intervention that fixes $X^{\text{out}} = x$,
\begin{eqnarray}
P(k_Y,k_Z,k_W) &=& \sum_Y \ \sum_Z \ \sum_W \ \sum_{X^{\text{out}}} \Big[ \ \clop_{YZWdo(X)}\ \delta(X^{\text{out}},x) \ \delta(Z^{\text{out}},z) \ P(k_Z|Z^{\text{in}}) \ P(k_Y,Y^{\text{out}}|Y^{\text{in}})  \nonumber \\[-0.4cm] 
&& \hspace{3cm} \delta \Big( g^{\text{in}}(k_W), W^{\text{in}} \Big) \ \delta \Big( g^{\text{out}}(k_W), W^{\text{out}} \Big) \ P(k_W, W^{\text{out}} | W^{\text{in}})  \ \Big] \nonumber \\[0.2cm]
&=& \left( \sum_Y \ P(k_Y,Y^{\text{out}}|Y^{\text{in}})  \, \alpha_{YWX^{\text{out}}Z^{\text{out}}}\left(Y^{\text{in}} , Y^{\text{out}} , g^I(k_W) , g^O(k_W) , x , z \right) \right) \nonumber \\
&& \left( \sum_{Z^{\text{in}}} P(k_Z|Z^{\text{in}}) P(k_W, g^{\text{out}}(k_W)|g^{\text{in}}(k_W)) \, \beta_{ZWX^{\text{out}}}\left(Z^{\text{in}} , z , g^I(k_W) , g^O(k_W) , x \right) \right).\nonumber
\end{eqnarray}
Setting
\[
\alpha'(k_Y , k_W) =  \sum_Y \ P(k_Y,Y^{\text{out}}|Y^{\text{in}}) \, \alpha_{YWX^{\text{out}}Z^{\text{out}}}\left(Y^{\text{in}} , Y^{\text{out}} , g^I(k_W) , g^O(k_W) , x , z \right),
\]
and 
\[
\beta'(k_Z,k_W) =   \sum_{Z^{\text{in}}} P(k_Z|Z^{\text{in}}) P(k_W, g^{\text{out}}(k_W)|g^{\text{in}}(k_W)) \,  \beta_{ZWX^{\text{out}}}\left(Z^{\text{in}} , z , g^I(k_W) , g^O(k_W) , x \right), 
\]
yields $P(k_Y,k_Z,k_W) = \alpha'(k_Y , k_W) \beta'(k_Z,k_W)$, hence $(k_Y \bigCI k_Z | k_W)_P$, by Prop.~\ref{Prop_EquivalentStatementsClassicalCI}.\\

\noindent {\bf \ref{COS2} $\Rightarrow (Y \bigCI Z^{\text{in}} | W do(X) Z^{\text{out}} )_{\clop_{YZWX}}$ }  \\

The converse direction proceeds by a similar argument to that of the proof of Prop.~\ref{Prop_CSCI_EquivalentOperationalStatements}.
Consider an intervention at each node $N$ (where $N\in Y$ or $N\in W$), such that the outcome $k_N$ is a pair  $k_N = (k_N^I , k_N^O)$, and 
\[
P(k_N, N^{\text{out}} | N^{\text{in}}) = \frac{1}{d_N} \delta(k_N^I , N^{\text{in}}) \delta(k_N^O , N^{\text{out}}),
\]
where $d_N$ is the cardinality of the set on which $N^{\text{in}}$ takes values. Assume that the local intervention at $Y$ is a product of interventions of this form, with joint outcome $k_Y = (k_Y^I , k_Y^O)$, where $k_Y^I$ is the tuple consisting of a value of $k_N^I$, for each $N\in Y$, and $k_Y^O$ is the tuple consisting of a value of $k_N^O$, for each $N\in Y$. Similarly $W$ and $k_W = (k_W^I , k_W^O)$. Consider a breaking local intervention at $Z$ that fixes $Z^{\text{out}} = z$, and returns $k_Z = Z^{\text{in}}$. Consider a do-intervention at $X$ that fixes $X^{\text{out}} = x$.

For each choice of $(x,z)$, let $P_{xz}(k_Y , k_Z , k_W)$ be the joint distribution over outcomes, and assume that the condition $(k_Y \bigCI k_Z | k_W)_{P_{xz}}$ holds. It follows from Prop.~\ref{Prop_EquivalentStatementsClassicalCI} that there exist $\alpha_{xz}(k_Y , k_W)$ and $\beta_{xz}(k_Z , k_W)$ such that $P_{xz}(k_Y , k_Z , k_W) = \alpha_{xz}(k_Y , k_W) \beta_{xz}(k_Z , k_W)$. Define $\alpha'_{YWX^{\text{out}}Z^{\text{out}}} : Y^{\text{in}} \times Y^{\text{out}} \times W^{\text{in}} \times W^{\text{out}} \times X^{\text{out}} \times Z^{\text{out}} \rightarrow \mathbb{R}$ and $\beta'_{ZWX^{\text{out}}} : Z^{\text{in}} \times Z^{\text{out}} \times W^{\text{in}} \times W^{\text{out}} \times X^{\text{out}} \rightarrow \mathbb{R}$ such that for each $x,z$,
\begin{eqnarray}
\alpha'_{YWX^{\text{out}}Z^{\text{out}}}(Y^{\text{in}} , Y^{\text{out}} , W^{\text{in}} , W^{\text{out}} , x , z) &=& d_Yd_W \ \alpha_{xz}(k_Y , k_W)\big|_{k_Y = (Y^{\text{in}} , Y^{\text{out}}) , k_W = (W^{\text{in}} , W^{\text{out}})}      \nonumber\\
\beta'_{ZWX^{\text{out}}}(Z^{\text{in}} , z , W^{\text{in}} , W^{\text{out}} , x) &=&  \beta_{xz}(k_Z , k_W)\big|_{k_Z = Z^{\text{in}} , k_W = (W^{\text{in}} , W^{\text{out}})}.     \nonumber
\end{eqnarray}
Observe that 
\begin{eqnarray*}
\alpha_{xz}(k_Y , k_W) &=& \sum_{YW} \alpha'_{YWX^{\text{out}}Z^{\text{out}}}( Y^{\text{in}} , Y^{\text{out}} , W^{\text{in}} , W^{\text{out}} , x , z ) \  P(k_Y, Y^{\text{out}} | Y^{\text{in}}) P(k_W, W^{\text{out}} | W^{\text{in}}),  \\
\beta_{xz}(k_Z , k_W) &=& d_W \sum_{Z^{\text{in}} W} \beta'_{ZWX^{\text{out}}} ( Z^{\text{in}} , z, W^{\text{in}} , W^{\text{out}} , x ) \ P(k_Z | Z^{\text{in}}) \ P(k_W, W^{\text{out}} | W^{\text{in}}).
\end{eqnarray*}
Now,
\begin{eqnarray}\label{Eq_ClassicalprocessCIproof3}
&& P_{xz}(k_Y , k_Z , k_W) = \alpha_{xz}(k_Y , k_W) \beta_{xz}(k_Z , k_W)\nonumber \\
&& \hspace*{0.3cm} = d_W \left(\sum_{YW}  \alpha'_{YWX^{\text{out}}Z^{\text{out}}}( Y^{\text{in}} , Y^{\text{out}} , W^{\text{in}} , W^{\text{out}} , x , z ) \ P(k_Y, Y^{\text{out}} | Y^{\text{in}}) P(k_W, W^{\text{out}} | W^{\text{in}})\right) \nonumber \\
&& \hspace*{0.8cm} \left(  \sum_{Z^{\text{in}} W} \beta'_{ZWX^{\text{out}}} ( Z^{\text{in}} , z, W^{\text{in}} , W^{\text{out}} , x )   \ P(k_Z | Z^{\text{in}}) P(k_W, W^{\text{out}} | W^{\text{in}})\right)\nonumber \\
&& \hspace*{0.3cm} = \sum_{YZ^{\text{in}}W}  P(k_Y, Y^{\text{out}} | Y^{\text{in}}) P(k_Z | Z^{\text{in}}) P(k_W, W^{\text{out}} | W^{\text{in}})  \left( \alpha'_{YWX^{\text{out}}Z^{\text{out}}} \beta'_{ZWX^{\text{out}}}\right)\big|_{X^{\text{out}}=x , Z^{\text{out}}=z}\nonumber\\
&& \hspace*{0.3cm} = \sum_{YZW} P(k_Y, Y^{\text{out}} | Y^{\text{in}}) P(k_Z , Z^{\text{out}} | Z^{\text{in}}) P(k_W, W^{\text{out}} | W^{\text{in}})  \left( \alpha'_{YWX^{\text{out}}Z^{\text{out}}} \beta'_{ZWX^{\text{out}}}\right)\big|_{X^{\text{out}}=x},\nonumber\\
\end{eqnarray}
where the third equality follows from the form of the $W$ intervention (in particular, the fact that the $W$ intervention is maximally informative), and where
\[
P(k_Z , Z^{\text{out}} | Z^{\text{in}}) = P(k_Z | Z^{\text{in}}) \delta(Z^{\text{out}} , z).
\]
From the definition of the do-conditional process map,
\begin{equation}\label{Eq_ClassicalprocessCIproof4}
P_{xz}(k_Y , k_Z , k_W) =  \sum_{YZW}  P(k_Y, Y^{\text{out}} | Y^{\text{in}}) P(k_Z, Z^{\text{out}} | Z^{\text{in}}) P(k_W, W^{\text{out}} | W^{\text{in}})  \clop_{YZWdo(X)}\big|_{X^{\text{out}}=x}.
\end{equation}

The measurements considered are informationally complete: that is, if for fixed $z$, $k_Y$, $k_Z$, $k_W$, the term $P(k_Y, Y^{\text{out}} | Y^{\text{in}}) P(k_Z, Z^{\text{out}} | Z^{\text{in}}) P(k_W, W^{\text{out}} | W^{\text{in}})$ is viewed as a real-valued function of $Y^{\text{in}}$, $Y^{\text{out}}$,  $Z^{\text{in}}$, $Z^{\text{out}}$, $W^{\text{in}}$, $W^{\text{out}}$, then, varying over $z$, $k_Y$, $k_Z$, $k_W$, these functions span the vector space of all real-valued functions of $Y^{\text{in}}$, $Y^{\text{out}}$,  $Z^{\text{in}}$, $Z^{\text{out}}$, $W^{\text{in}}$, $W^{\text{out}}$. Comparing Eqs.~(\ref{Eq_ClassicalprocessCIproof3}) and (\ref{Eq_ClassicalprocessCIproof4}) yields $\clop_{YZWdo(X)} =  \alpha'_{YWX^{\text{out}}} \beta'_{ZWX^{\text{out}}}$, that is $(Y \bigCI Z^{\text{in}} | W do(X) Z^{\text{out}} )_{\clop_{YZWX}}$.

\hfill $\square$

\subsection{Proof of Prop.~\ref{Prop_Implication_QOS2} \label{Subsec_Proof_ImpliedQOS2}}

The proof is similar to that of Prop.~\ref{Prop_Implication_QOS1a} (and Prop.~\ref{Prop_Implication_QOS1}). 

Assume $(Y \bigCI Z^{\text{in}} | W do(X) Z^{\text{out}} )_{\sigma_{YZWX}}$, hence there exist Hermitian operators $\alpha_{YWX^{\text{out}}Z^{\text{out}}}$ and $\beta_{ZWX^{\text{out}}}$ such that
$\sigma_{YZWdo(X)} = \alpha_{YWX^{\text{out}}Z^{\text{out}}} \ \beta_{ZWX^{\text{out}}}$. Lemma~\ref{Lem_SplitFromCommutation_I} implies that there is a decomposition $\mathcal{H}_{W^{\text{in}}} \otimes \mathcal{H}_{W^{\text{out}}} \otimes \mathcal{H}_{X^{\text{out}}} \otimes \mathcal{H}_{Z^{\text{out}}} = \bigoplus_i \mathcal{H}_{F_i^L} \otimes \mathcal{H}_{F_i^R}$ such that $\sigma_{YZWdo(X)} = \sum_i \alpha_{YF_i^L} \otimes \beta_{Z^{\text{in}} F_i^R}$, with $\alpha$ and $\beta$ positive. Let $\{\ket{i,f_i^L}\ket{i,f_i^R}\}_{f_i^L,f_i^R}$ be a product orthonormal basis of the $i$th subspace $F_i^L \otimes F_i^R$. 

Consider the following global intervention at $WX^{\text{out}} Z^{\text{out}}$. An agent is stationed at an additional locus $E$, such that for each node $N \in W$, the quantum system at $N^{\text{in}}$ is sent to $E$, one half of a maximally entangled state is fed into $N^{\text{out}}$, and the other half is sent to $E$. For each node $N \in Z$ or $N\in X$, one half of a maximally entangled state is fed into $N^{\text{out}}$, and the other half is sent to $E$. 
This defines an operator $\sigma_{YZ^{\text{in}}E}$  over $Y$, $Z^{\text{in}}$ and $E$, with $E^{\text{in}}$ isomorphic to $\mathcal{H}_{W^{\text{in}}} \otimes \mathcal{H}_{W^{\text{out}}} \otimes \mathcal{H}_{X^{\text{out}}} \otimes \mathcal{H}_{Z^{\text{out}}}$, such that $\sigma_{YZ^{\text{in}}E}$ has a block-diagonal structure with respect to the induced decomposition. Let $\ket{i,f_i^L,f_i^R} := J^{-1} \ket{i,f_i^L}\ket{i,f_i^R}$ label the induced orthonormal basis of $E^{\text{in}}$ (for a suitable isomorphism $J$). The agent performs the von Neumann measurement at $E$ corresponding to that basis. For a particular outcome $k_E$, corresponding to the basis state $\ket{i,f_i^L,f_i^R}$, define the operator
\begin{eqnarray}
	\Trace_{E^{\text{in}} E^{\text{out}}} \Big[ \sigma_{YZ^{\text{in}}E} \ \ket{i,f_i^L,f_i^R} \bra{i,f_i^L,f_i^R} \Big]  
		&=&  \bra{i,f_i^L,f_i^R} \ J^{-1}  \Big(\sum_j \alpha_{YF_j^L} \otimes \beta_{Z^{\text{in}}F_j^R} \Big) J \ \ket{i,f_i^L,f_i^R} \nonumber \\
		&=& \bra{i,f_i^L} \alpha_{YF_i^L}\ket{i,f_i^L} \ \otimes \ \bra{i,f_i^R} \beta_{Z^{\text{in}}F_i^R}\ket{i,f_i^R} 
		\ := \ \gamma_Y \otimes \eta_{Z^{\text{in}}} \nonumber 
\end{eqnarray}
The product form $\gamma_Y \otimes  \eta_{Z^{\text{in}}} $ implies that the joint probability distribution for $k_E$ and outcomes $k_Y$ and $k_Z$ for an arbitrary intervention at $Y$, and an arbitrary local measurement of $Z^{\text{in}}$, satisfies $P(k_Y,k_Z,k_E)=\phi(k_Y,k_E)\chi(k_Z,k_E)$ (for some functions $\phi$ and $\chi$). Recalling Prop.~\ref{Prop_EquivalentStatementsClassicalCI}, this establishes the claim. \hfill $\square$

\subsection{Proof of Thorem~\ref{Thm_Rule2_qu} \label{Subsec_Proof_Rule_2}}

\begin{figure}[H]
\begin{center}
\small
\renewcommand{\Scale}{2.2}
\input{Figures/Fig_Rule2SubsetRelation.tex}
\caption{The allowed parent-child relations in $G$, suppressing the set $X$, and suppressing arrows coming out of $W_Y$ and $W_Z$. Blue dashed arrows represent parent-child relationships that are allowed in $G$, but absent in $G_{\overbar{X}\underline{Z}}$. \label{Fig_Rule2SubsetRelation}}
\end{center}
\end{figure}
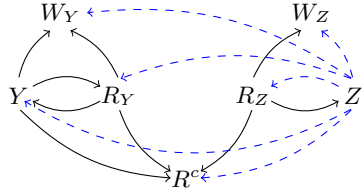

Suppose $(Y \bigCI Z | WX )_{G_{\overbar{X}\underline{Z}}}$. Then Lemma~\ref{Lem_DAGDSep} implies $SR(Y, Z ; WX)$ with respect to the mutilated DAG $G_{\overbar{X}\underline{Z}}$. Hence, with the set $X$ suppressed, there are partitions of the sets $W$ and $R$ such that allowed parent-child relationships are as shown in Fig.~\ref{Fig_Rule2SubsetRelation}.
Let $\sigma_V$ be a process operator that is Markov for $G$. It follows from the above that, with a local intervention $\tau_R = \tau_{R_Y} \otimes \tau_{R_Z} \otimes \tau_{R^c}$ at the $R$ nodes,
\begin{eqnarray}
\sigma^{\tau_R}_{YZWdo(X)} &=& \Trace_{R_Y} \left[ \ \tau_{R_Y}\  \rho_{Y|Pa(Y)} \ \rho_{W_Y|Pa(W_Y)} \ \rho_{R_Y|Pa(R_Y)} \right] \label{Eq_Rule2SigmaYZWDoX} \\
&&  \Trace_{R_Z} \left[ \ \tau_{R_Z} \ \rho_{Z|Pa(Z)} \ \rho_{W_Z|Pa(W_Z)} \ \rho_{R_Z|Pa(R_Z)} \ \right]  \nonumber \ , 
\end{eqnarray}
which is of the form $\sigma^{\tau_R}_{YZWdo(X)} = \alpha_{YWX^{\text{out}}Z^{\text{out}}} \ \beta_{ZWX^{\text{out}}}$. \\

\hfill $\square$

\subsection{Proof of Prop.~\ref{Prop_EquivalentStatementsRule2} \label{Subsec_Proof_EquivalentStRule_2}}

The proof is similar to that of Prop.~\ref{Prop_EquivalentStatementsRule1}. \\

\noindent {\bf $(1)\rightarrow (2)$}

Assume $\sigma_{YZWdo(X)} = \alpha_{Y W X^{\text{out}} Z^{\text{out} }} \beta_{Z W X^{\text{out}}}$. As in the proof of Prop.~\ref{Prop_EquivalentStatementsRule1}, use the associativity and commutativity of the star-product, and express the operators in Condition~(2) in terms of 
$\alpha^{\tau_Y}_{WX^{\text{out}}Z^{\text{out}}}:=\Trace_Y\big[ \tau_Y \alpha_{YWX^{\text{out}}Z^{\text{out}}}\big]$ and 
$\beta_{Z^{\text{out}}WX^{\text{out}}}:=\Trace_{Z^{\text{in}}} \big[ \beta_{ZWX^{\text{out}}} \big]$. \\

\noindent {\bf $(2)\rightarrow (3)$}

Consider the intervention at each node in $Y$ that discards the input, and prepares a maximally mixed state on the output. Condition~(2) then implies
\[
\hat{\sigma}_{YZWdo(X)} \star \Trace_{Y Z^{\text{in}}} (\hat{\sigma}_{YZWdo(X)}) = \Trace_Y (\hat{\sigma}_{YZWdo(X)})  \star  \Trace_{Z^{\text{in}}}(\hat{\sigma}_{YZWdo(X)}),
\]
which yields Condition~(3).\\

\noindent {\bf $(3)\rightarrow (1)$}

The proof is the same as that of the proof of the $(3)\rightarrow (1)$ direction of Prop.~\ref{Prop_EquivalentStatementsRule1}, except that $Z^{\text{in}}$ replaces $Z$, and $WX^{\text{out}}Z^{\text{out}}$ replaces $WX^{\text{out}}$.\\

\hfill $\square$

\subsection{Proof of Prop.~\ref{Prop_Rule3_CSMOperationalStatements} \label{Subsec_Proof_Rule3_CSMOperationalStatements}}

\noindent {\bf $(Y \bigCI Set(Z) | Wdo(X) )_{\clop_{YZWX}} \Rightarrow$ \ref{COS3}} \\

Given a classical do-conditional process $\clop_{YZWdo(X)}$, suppose that 
\begin{equation}\label{Eq_Proof_COS3_Assumption}
\clop_{YWdo(X)}^{\tau_Z} = \eta_{YWX^{\text{out}}} \ \xi_{WX^{\text{out}}}^{\tau_Z}  \hspace*{0.5cm}  \forall \ \mathrm{local\ interventions\ }\tau_Z .
\end{equation}
Consider an arbitrary local intervention at $Y$ given by $P(k_Y,Y^{\text{out}}|Y^{\text{in}})$, a maximally informative local intervention at $W$ given by
\[
P(k_W,W^{\text{out}}|W^{\text{in}}) = \delta \big( g^{\text{in}}(k_W), W^{\text{in}} \big) \ \delta \big( g^{\text{out}}(k_W), W^{\text{out}} \big) \ P(k_W,W^{\text{out}}|W^{\text{in}}),
\] 
and a do-intervention at $X$ that fixes $X^{\text{out}}=x$. For $\tau_Z$ a local intervention at $Z$, let $P^{\tau_Z}(k_Y , k_W)$ denote the resulting joint distribution over outcomes $k_Y$ and $k_W$ (where the notation suppresses dependence on $x$). This is given by
\begin{eqnarray}
P^{\tau_Z}(k_Y,k_W) &=&	\sum_Y \sum_W \sum_{X^{\text{out}}} \clop^{\tau_Z}_{YWdo(X)}\ P(k_Y,Y^{\text{out}}|Y^{\text{in}})\nonumber \\[-0.1cm]
&& \hspace{1.3cm} \delta \big( X^{\text{out}}, x) \ \delta \big( g^{\text{in}}(k_W), W^{\text{in}} \big) \ \delta \big( g^{\text{out}}(k_W), W^{\text{out}} \big) \ P(k_W,W^{\text{out}}|W^{\text{in}})  \nonumber \\[0.2cm]
&=&\sum_Y \eta_{YWX^{\text{out}}}(Y^{\text{in}},Y^{\text{out}},g^{\text{in}}(k_W),g^{\text{out}}(k_W),x) \ P(k_Y,Y^{\text{out}}|Y^{\text{in}}) \nonumber \\
&& \xi_{WX^{\text{out}}}^{\tau_Z}(g^{\text{in}}(k_W),g^{\text{out}}(k_W),x) \ P(k_W, g^{\text{out}}(k_W) | g^{\text{in}}(k_W)) . \nonumber
\end{eqnarray}
Setting
\[
\alpha(k_Y,k_W,x) = \sum_Y \eta_{YWX^{\text{out}}}(Y^{\text{in}},Y^{\text{out}},g^{\text{in}}(k_W),g^{\text{out}}(k_W),x) \ P(k_Y,Y^{\text{out}}|Y^{\text{in}}),
\]
and 
\[
\beta^{\tau_Z}(k_W,x) =  \xi_{WX^{\text{out}}}^{\tau_Z}(g^{\text{in}}(k_W),g^{\text{out}}(k_W),x) \ P(k_W, g^{\text{out}}(k_W) | g^{\text{in}}(k_W)), 
\]
yields $P^{\tau_Z}(k_Y,k_W) = \alpha(k_Y,k_W,x) \ \beta^{\tau_Z}(k_W,x)$, which implies that $k_Y$ is independent of the choice of intervention $\tau_Z$, conditioned on $k_W$.
This holds for all interventions at $Y$, all maximally informative interventions at $W$, and all $x$, hence \ref{COS3} follows.\\

\noindent {\bf  \ref{COS3} $\Rightarrow (Y \bigCI Set(Z) | Wdo(X) )_{\clop_{YZWX}}$ } \\

Consider a classical process map $\clop_{YZWX}$, and suppose that \ref{COS3} holds for $\clop_{YZWdo(X)}$. Consider local interventions at $Y$ and $W$ corresponding to an intervention at each node $N$ ($N\in Y$ or $N\in W$) of the form
\[
P(k_N, N^{\text{out}} | N^{\text{in}}) = \frac{1}{d_N} \delta(k_N^I , N^{\text{in}}) \delta(k_N^O , N^{\text{out}}),
\]
where $d_N$ is the cardinality of the set on which $N^{\text{in}}$ takes values. (Similar interventions were considered in Section~\ref{Subsubsec_Proof_CSCI_EquivalentOperationalStatements}.) Let the joint outcome of the intervention at $Y$ be $k_Y = (k_Y^I , k_Y^O)$, where $k_Y^I$ is the tuple consisting of a value of $k_N^I$, for each $N\in Y$, and $k_Y^O$ is the tuple consisting of a value of $k_N^O$, for each $N\in Y$. Similarly $W$ and $k_W = (k_W^I , k_W^O)$. Consider a do-intervention at $X$ that sets $X^{\text{out}} = x$, and an arbitrary local intervention $\tau_Z$ at $Z$.

As above, let $P^{\tau_Z}(k_Y , k_W)$ denote the resulting joint distribution over outcomes $k_Y$ and $k_W$, given a local intervention $\tau_Z$ at $Z$, where the dependence on $x$ is suppressed. The intervention at $W$ is maximally informative, hence by assumption, the probability of outcome $k_Y$ is independent of $\tau_Z$ when conditioned on $k_W$. This implies that there exist a function $\alpha(k_Y, k_W, x)$, and for each $\tau_Z$, a function $\beta^{\tau_Z}(k_W,x)$, such that
\begin{equation}
P^{\tau_Z}(k_Y, k_W) = \alpha(k_Y, k_W, x) \beta^{\tau_Z}(k_W, x).
\end{equation}

Define $\eta_{YWX^{\text{out}}}$ such that 
\[
\eta_{YWX^{\text{out}}}(Y^{\text{in}} , Y^{\text{out}} , W^{\text{in}} , W^{\text{out}} , x) = d_Yd_W \ \alpha(k_Y, k_W, x)\big|_{k_Y = (Y^{\text{in}} , Y^{\text{out}}) , k_W = (W^{\text{in}} , W^{\text{out}}) }, 
\]
and for each $\tau_Z$, a function $\xi^{\tau_Z}_{WX^{\text{out}}}$ such that
\[
\xi^{\tau_Z}_{WX^{\text{out}}}(W^{\text{in}} , W^{\text{out}} , x) = \beta^{\tau_Z}(k_W, x)\big|_{k_W = (W^{\text{in}} , W^{\text{out}})}. 
\]
Observe that
\begin{eqnarray*}
\alpha(k_Y, k_W, x) &=& \sum_{YW} \eta_{YWX^{\text{out}}}\big|_{X^{\text{out}}=x} P(k_Y , Y^{\text{out}} | Y^{\text{in}})  P(k_W , W^{\text{out}} | W^{\text{in}}) \\
\beta^{\tau_Z}(k_W, x) &=& d_W \sum_W \xi^{\tau_Z}_{WX^{\text{out}}}\big|_{X^{\text{out}}=x} \ P(k_W , W^{\text{out}} | W^{\text{in}}).
\end{eqnarray*}
Therefore,
\begin{eqnarray}
P^{\tau_Z}(k_Y, k_W) &=& \alpha(k_Y, k_W, x) \beta^{\tau_Z}(k_W, x)\nonumber\\
&=& d_W \  \left( \sum_{YW} \eta_{YWX^{\text{out}}}\big|_{X^{\text{out}}=x} P(k_Y , Y^{\text{out}} | Y^{\text{in}})  P(k_W , W^{\text{out}} | W^{\text{in}}) \right) \nonumber \\
&& \left(\sum_W \xi^{\tau_Z}_{WX^{\text{out}}}\big|_{X^{\text{out}}=x} \ P(k_W , W^{\text{out}} | W^{\text{in}})\right)\nonumber \\
&=& \sum_{YW} P(k_Y , Y^{\text{out}} | Y^{\text{in}})  P(k_W , W^{\text{out}} | W^{\text{in}}) \  \eta_{YWX^{\text{out}}}\big|_{X^{\text{out}}=x} \  \xi^{\tau_Z}_{WX^{\text{out}}}\big|_{X^{\text{out}}=x}.\label{Eq_lollylolly}
\end{eqnarray}
From the definition of a classical process map,
\begin{equation}\label{Eq_lollylolly2}
P^{\tau_Z}(k_Y, k_W) = \sum_{YW} P(k_Y , Y^{\text{out}} | Y^{\text{in}})  P(k_W , W^{\text{out}} | W^{\text{in}}) \clop^{\tau_Z}_{YWdo(X)}\big|_{X^{\text{out}}=x}.
\end{equation}
The interventions considered at $Y$ and $W$ are informationally complete: that is, if for fixed $k_Y$, $k_W$, the term $P(k_Y, Y^{\text{out}} | Y^{\text{in}}) P(k_W, W^{\text{out}} | W^{\text{in}})$ is viewed as a real-valued function of $Y^{\text{in}}$, $Y^{\text{out}}$, $W^{\text{in}}$, $W^{\text{out}}$, then, varying over $k_Y$, $k_W$, these functions span the vector space of all real-valued functions of $Y^{\text{in}}$, $Y^{\text{out}}$, $W^{\text{in}}$, $W^{\text{out}}$. Comparing Eqs.~(\ref{Eq_lollylolly}) and (\ref{Eq_lollylolly2}) then gives $\clop^{\tau_Z}_{YZWdo(X)} =  \eta_{YWX^{\text{out}}} \xi^{\tau_Z}_{WX^{\text{out}}}$, that is $(Y \bigCI Set(Z) | Wdo(X))_{\clop_{YZWX}}$.

\hfill $\square$

\subsection{Proof of Prop.~\ref{Prop_Implication_QOS3} \label{Subsec_Proof_Prop_Implication_QOS3}}

This proof is similar to that of Prop.~\ref{Prop_Implication_QOS1a} in Section~\ref{Subsubsec_Proof_Implication_QOS1a}. 

Consider a process operator $\sigma_{YZWX}$, and assume that there is a Hermitian operator $\eta_{YWX^{\text{out}}}$, and for each local intervention $\tau_Z$, a Hermitian operator $\xi_{WX^{\text{out}}}^{\tau_Z}$, such that 
\[
\sigma_{YWdo(X)}^{\tau_Z} = \eta_{YWX^{\text{out}}} \ \xi_{WX^{\text{out}}}^{\tau_Z}.
\]
Note that $\eta_{YWX^{\text{out}}}$ commutes with $\xi_{WX^{\text{out}}}^{\tau_Z}$ for each $\tau_Z$. The set $\{ \xi_{WX^{\text{out}}}^{\tau_Z} \}$ for varying $\tau_Z$ generates a *-subalgebra of the form $\mathcal{A} = \mathds{1}_{Y} \otimes \mathcal{A}_{WX^{\text{out}}}$, where $\mathcal{A}_{WX^{\text{out}}}$ is a *-subalgebra of 
$\mathcal{L}(\mathcal{H}_{W^{\text{in}}} \otimes \mathcal{H}_{W^{\text{out}}} \otimes \mathcal{H}_{X^{\text{out}}})$. A fundamental representation-theoretic result concerning finite-dimensional C*-algebras (see, e.g., Lemma~13 of Ref.~\cite{HaydenEtAl_2003_StructureOfStatesAubadditivity}) then implies that there exists a decomposition $\mathcal{H}_{W^{\text{in}}} \otimes \mathcal{H}_{W^{\text{out}}} \otimes \mathcal{H}_{X^{\text{out}}} = \bigoplus_i \mathcal{H}_{F_i^L} \otimes \mathcal{H}_{F_i^R}$ such that $\mathcal{A}_{WX^{\text{out}}} = \bigoplus_i \mathds{1}_{F_i^L} \otimes \mathcal{L}(\mathcal{H}_{F_i^R})$. The commutant $\mathcal{A}'$ of $\mathcal{A}$, that is, the subalgebra of operators that commute with all elements of $\mathcal{A}$, is of the form $\mathcal{A}' = \bigoplus_i \mathcal{L}( \mathcal{H}_{Y} \otimes \mathcal{H}_{F_i^L}) \otimes \mathds{1}_{F_i^R}$. 
Since $\eta_{YWX^{\text{out}}} \in \mathcal{A}'$, and the distinct subspaces labelled by $i$ are orthogonal, $\sigma_{YWdo(X)}^{\tau_Z}$ can be written in the form
\[
\sigma_{YWdo(X)}^{\tau_Z} = \sum_i \eta_{YF_i^L} \otimes \xi_{F_i^R}^{\tau_Z} 
\]
for appropriate positive operators $\eta_{YF_i^L}$ and $\xi_{F_i^R}^{\tau_Z}$, where we have adopted the same convention as previously and let the latter operators act as zero maps on all other subspaces $j \neq i$.
Consider a global intervention at $WX^{\text{out}}$, of the same form as that of \ref{QOS1}: that is, there is an additional locus $E$ such that for each node $N \in W\cup X$, one half of a maximally entangled state is fed into $N^{\text{out}}$, and the other half sent to $E$, and for each node $N\in W$, the system at $N^{\text{in}}$ is sent to $E$. The same arguments as in the proof of Prop.~\ref{Prop_Implication_QOS1a} imply that there exists a basis $\{\ket{i,f_i^L,f_i^R}\}$ of $E^{\text{in}}$, which corresponds to a basis $\{\ket{i,f_i^L}\ket{i,f_i^R}\}$ of $\bigoplus_i \mathcal{H}_{F_i^L} \otimes \mathcal{H}_{F_i^R}$. The agent at $E$ performs the associated von Neumann measurement. 

The probability of obtaining outcome $k_Y$ for an arbitrary intervention $\tau^{k_Y}_Y$ at $Y$, and outcome $k_E$, corresponding to $\ket{i,f_i^L,f_i^R}$, for the von Neumann measurement at $E$, is 
\begin{equation}
P(k_Y,k_E) = \Trace_Y \Big[ \bra{i,f_i^L}\eta_{YF_i^L} \ket{i,f_i^L}  \ \tau^{k_Y}_Y \Big] \ \bra{i,f_i^R}\xi_{F_i^R}^{\tau_Z} \ket{i,f_i^R} \ . 
\end{equation}
This product form implies that the probability of $k_Y$ conditional on obtaining any of the outcomes at $E$ is independent from $\tau_Z$. \hfill $\square$

\subsection{Proof of Theorem~\ref{Thm_Rule3_qu} \label{SubSec_Appendix_Proof_Rule3}} 

Let $G$ be a DAG with nodes $V = Y \cup Z \cup W \cup X \cup R$, for disjoint subsets $Y$, $Z$, $W$, $X$, and $R$, such that $(Y \bigCI Z | WX )_{G_{\overbar{X}\overbar{Z(W)}}}$. Then Lemma~\ref{Lem_DAGDSep} implies $SR(Y,Z ; WX)$ with respect to the mutilated DAG ${G_{\overbar{X}\overbar{Z(W)}}}$. 
There is therefore a partition of the set $W$ into $W_Y$ and $W_Z$, a partition of the set $R$ into $R_Y$, $R_Z$ and $R^c$, and a partition of the set $Z$ into $Z(W)$ and $Z' = Z \backslash Z(W)$, such that, with $X$ suppressed, the allowed parent-child relationships are as shown in Fig.~\ref{Fig_Rule3SubsetRelation_1}. 
Note in particular that it follows from the definition of $Z(W)$ that there is no arrow from $Z(W)$ to $W_Y$, $W_Z$ or $Z'$.

Let $\mathcal{R}:= \{ r \in R_Z : r \text{ is a descendant of a node in } Z(W) \}$ and define $\widetilde{R}_Z:=R_Z \setminus \mathcal{R}$ and $\widetilde{R}^c:=R^c  \cup \mathcal{R}$. The allowed parent-child relationships between the resulting sets are shown in Fig.~\ref{Fig_Rule3SubsetRelation_2}. Note in particular that there are no arrows from $\mathcal{R}$ to $W_Y$, $W_Z$ or $Z'$.\\
\begin{minipage}{16cm}
	\begin{minipage}{7.7cm}
		\begin{center}
			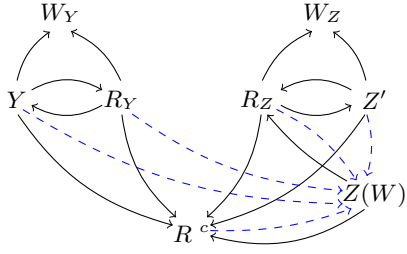
\begin{figure}[H]
				\begin{center}
					\small
					\renewcommand{\Scale}{2.3}
					\input{Figures/Fig_Rule3SubsetRelation_1.tex}
					\caption{\small{Allowed parent-child relations in $G$, with the set $X$, and arrows out of $W_Y$ and $W_Z$ suppressed. Blue dashed arrows show parent-child relationships that are allowed in $G$, but absent in $G_{\overbar{X}\overbar{Z(W)}}$.} \label{Fig_Rule3SubsetRelation_1}}
				\end{center}
			\end{figure}
		\end{center}
	\end{minipage}
	\hfill
	\begin{minipage}{7.5cm}
		\begin{center}
			\begin{figure}[H]
				\begin{center}
					\small
					\renewcommand{\Scale}{2.3}
					\input{Figures/Fig_Rule3SubsetRelation_2.tex}
					\caption{\small{Allowed parent-child relations in $G$, with re-defined sets $\widetilde{R}^c$ and $\widetilde{R}_Z$. The set $X$ and arrows out of $W_Y$ and $W_Z$ are suppressed. Blue dashed arrows show parent-child relationships that are allowed in $G$, but absent in $G_{\overbar{X}\overbar{Z(W)}}$.} \label{Fig_Rule3SubsetRelation_2}}
				\end{center}
			\end{figure}
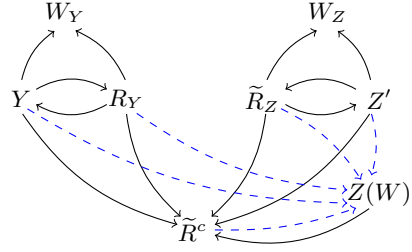
		\end{center}
	\end{minipage}
\end{minipage}
\\[0.2cm]
Let $\sigma_V$ be a process operator that is Markov for $G$.  
The constraints on allowed parent-child relations shown in Fig.~\ref{Fig_Rule3SubsetRelation_2} imply that for arbitrary local interventions $\tau_R = \tau_{R_Y}\otimes \tau_{\widetilde{R}_Z} \otimes \tau_{\widetilde{R}^c}$ and $\tau_Z = \tau_{Z(W)} \otimes \tau_{Z'}$,
\begin{eqnarray}
\sigma^{\tau_R , \tau_Z}_{YW do(X)} &=& 
\ \Trace_{Z'} \ \Trace_{R_Y} \ \Trace_{\widetilde{R}_Z} \Big[ \tau_{Z'} \ \tau_{R_Y} \ \tau_{\widetilde{R}_Z} \ \rho_{Y|Pa(Y)} \ \rho_{W_Y|Pa(W_Y)} \ \rho_{R_Y|Pa(R_Y)} \nonumber \\  
&&\hspace*{3cm} \rho_{Z'|Pa(Z')} \ \rho_{W_Z|Pa(W_Z)} \ \rho_{\widetilde{R}_Z|Pa(\widetilde{R}_Z)}  \nonumber \\
&&	\hspace{3cm} \Trace_{\widetilde{R}^c} \ \Trace_{Z(W)} \big[ \ \tau_{\widetilde{R}^c} \ \tau_{Z(W)} \ \rho_{Z(W)|Pa(Z(W))} \ \rho_{\widetilde{R}^c|Pa(\widetilde{R}^c)}\ \big]  \ \Big] \nonumber \\
&=& \  \Trace_{R_Y} \big[ \ \tau_{R_Y} \ \rho_{Y|Pa(Y)} \ \rho_{W_Y|Pa(W_Y)} \ \rho_{R_Y|Pa(R_Y)} \ \big] \nonumber \\  
&&\hspace*{1.0cm} \Trace_{Z'} \ \Trace_{\widetilde{R}_Z} \big[ \ \tau_{Z'} \ \tau_{\widetilde{R}_Z} \ \rho_{Z'|Pa(Z')} \ \rho_{W_Z|Pa(W_Z)} \ \rho_{\widetilde{R}_Z|Pa(\widetilde{R}_Z)} \big] \ , \nonumber 
\end{eqnarray}
where the second equality follows since
\[
\Trace_{\widetilde{R}^c} \ \Trace_{Z(W)} \big[ \ \tau_{\widetilde{R}^c} \ \tau_{Z(W)}\ \rho_{Z(W)|Pa(Z(W))} \ \rho_{\widetilde{R}^c|Pa(\widetilde{R}^c)}\ \big] = 1.
\]
Setting
\[
\eta_{YWX^{\text{out}}} =  \Trace_{R_Y} \big[ \ \tau_{R_Y} \ \rho_{Y|Pa(Y)} \ \rho_{W_Y|Pa(W_Y)} \ \rho_{R_Y|Pa(R_Y)} \ \big],
\]
and
\[
 \xi^{\tau_Z}_{WX^{\text{out}}} = \Trace_{Z'} \ \Trace_{\widetilde{R}_Z} \big[ \ \tau_{Z'} \ \tau_{\widetilde{R}_Z} \ \rho_{Z'|Pa(Z')} \ \rho_{W_Z|Pa(W_Z)} \ \rho_{\widetilde{R}_Z|Pa(\widetilde{R}_Z)} \big]
\]
(where the notation suppresses the dependence of these quantities on $\tau_R$), yields 
\[
\sigma^{\tau_R , \tau_Z}_{YW do(X)}  = \eta_{YWX^{\text{out}}} \ \xi^{\tau_Z}_{WX^{\text{out}}}. 
\]
Since the choices of $\tau_R$ and $\tau_Z$ were arbitrary, this gives that for all local interventions $\tau_R$, 
\[
(Y \bigCI Set(Z) | W do(X) )_{\sigma^{\tau_R}_{YZWX}},
\]
as required. \hfill $\square$ 

\end{document}

%% file: tikzstyles.tex
\usetikzlibrary{shapes.multipart}

\tikzstyle{arrowhead}=[regular polygon,regular polygon sides=3,draw,scale=0.2,inner sep=-0.15pt,minimum width=6mm,fill=black,regular polygon rotate=180]
\tikzstyle{trace}=[circuit ee IEC,thick,ground,rotate=0,scale=2]
\tikzstyle{wavy}=[decorate,decoration={snake, segment length=1mm, amplitude=0.3mm}]
\tikzstyle{mopoint}=[shape=semicircle, fill=white,draw=black,shape border rotate=180,scale =0.75]
\tikzstyle{mocopoint}=[shape=semicircle, fill=white,draw=black,minimum width = 0.9cm, scale =0.75, xscale=0.7]
\tikzstyle{cpoint}=[shape=semicircle, fill=white,draw=black,minimum width = 0.9cm, scale =0.75, xscale=1, yscale=0.7, shape border rotate = 90,font=\fontsize{14}{16}\selectfont]

\tikzstyle{smallcpoint}=[shape=semicircle, fill=white,draw=black,minimum width = 0.6cm, scale =0.75, xscale=1, yscale=1, shape border rotate = 180, font=\fontsize{14}{16}\selectfont]

\tikzstyle{cocpoint}=[shape=semicircle, fill=white,draw=black,minimum width = 0.9cm, scale =0.75, xscale=1, yscale=0.7, shape border rotate = 270,font=\fontsize{14}{16}\selectfont]
\tikzstyle{slit}=[line width=2]
\tikzstyle{block}=[line width=4,red,line cap=round]
\tikzstyle{screen}=[line width=4,black,line cap=round]
\tikzstyle{di}=[diamond,draw,inner sep=0.5pt,font=\small, minimum size = .5cm]
\tikzstyle{sbox}=[rectangle,draw]
\tikzstyle{mirror}=[line width=2,black]
\tikzstyle{traceState}=[circuit ee IEC,thick,ground,rotate=180,scale=2]
\tikzstyle{detEff}=[circuit ee IEC,thick,ground,rotate=180,scale=1.4]
\tikzstyle{maxMix}=[circuit ee IEC,thick,ground,scale=1.4]
\tikzstyle{particlePath}=[line width=2,gray!40, line cap =round]

\tikzstyle{bwSpider}=[
       rectangle split,
       rectangle split parts=2,
       rectangle split part fill={black,white},
 minimum size=3.6 mm, inner sep=-2mm, draw=black,scale=0.5,rounded corners=0.8 mm
       ]
 \tikzstyle{wbSpider}=[
       rectangle split,
       rectangle split parts=2,
       rectangle split part fill={white,black},
 minimum size=3.6 mm, inner sep=-2mm, draw=black,scale=0.5,rounded corners=0.8 mm
       ]
\tikzstyle{cWire}=[densely dotted, thick]
\tikzstyle{env}=[copoint,regular polygon rotate=0,minimum width=0.2cm, fill=black]

\tikzstyle{probs}=[shape=semicircle,fill=white,draw=black,shape border rotate=180,minimum width=1.2cm]

\tikzstyle{every picture}=[baseline=-0.25em,scale=0.5]
\tikzstyle{dotpic}=[] 
\tikzstyle{diredges}=[every to/.style={diredge}]
\tikzstyle{math matrix}=[matrix of math nodes,left delimiter=(,right delimiter=),inner sep=2pt,column sep=1em,row sep=0.5em,nodes={inner sep=0pt},text height=1.5ex, text depth=0.25ex]

\tikzstyle{inline text}=[text height=1.5ex, text depth=0.25ex,yshift=0.5mm]
\tikzstyle{label}=[font=\footnotesize,text height=1.5ex, text depth=0.25ex,yshift=0.5mm]
\tikzstyle{left label}=[label,anchor=east,xshift=1.5mm]
\tikzstyle{right label}=[label,anchor=west,xshift=-1mm]

\tikzstyle{braceedge}=[decorate,decoration={brace,amplitude=2mm,raise=-1mm}]
\tikzstyle{small braceedge}=[decorate,decoration={brace,amplitude=1mm,raise=-1mm}]

\tikzstyle{doubled}=[line width=1.6pt] 
\tikzstyle{boldedge}=[doubled,shorten <=-0.17mm,shorten >=-0.17mm]
\tikzstyle{boldedgegray}=[doubled,gray,shorten <=-0.17mm,shorten >=-0.17mm]
\tikzstyle{singleedgegray}=[gray]

\tikzstyle{semidoubled}=[line width=1.4pt] 
\tikzstyle{semiboldedgegray}=[semidoubled,gray,shorten <=-0.17mm,shorten >=-0.17mm]

\tikzstyle{boxedge}=[semiboldedgegray]

\tikzstyle{boldedgedashed}=[very thick,dashed,shorten <=-0.17mm,shorten >=-0.17mm]
\tikzstyle{vboldedgedashed}=[doubled,dashed,shorten <=-0.17mm,shorten >=-0.17mm]
\tikzstyle{left hook arrow}=[left hook-latex]
\tikzstyle{right hook arrow}=[right hook-latex]
\tikzstyle{sembracket}=[line width=0.5pt,shorten <=-0.07mm,shorten >=-0.07mm]

\tikzstyle{causal edge}=[->,thick,gray]
\tikzstyle{causal nondir}=[thick,gray]
\tikzstyle{timeline}=[thick,gray, dashed]

\tikzstyle{cedge}=[<->,thick,gray!70!white]

\tikzstyle{empty diagram}=[draw=gray!40!white,dashed,shape=rectangle,minimum width=1cm,minimum height=1cm]
\tikzstyle{empty diagram small}=[draw=gray!50!white,dashed,shape=rectangle,minimum width=0.6cm,minimum height=0.5cm]

\tikzstyle{dot}=[inner sep=0mm,minimum width=2mm,minimum height=2mm,draw,shape=circle]
\tikzstyle{preleak}=[trapezium, trapezium angle=67.5, draw, inner sep=0.1pt, outer sep=0pt, minimum height=2mm, minimum width=4pt,rotate=270]
\tikzstyle{leak}=[white dot, shape=regular polygon, minimum size=3.3 mm, regular polygon sides=3, outer sep=-0.2mm, regular polygon rotate=270]
\tikzstyle{proj}=[regular polygon,regular polygon sides=4,draw,scale=0.75,inner sep=-0.5pt,minimum width=6mm,fill=white]
\tikzstyle{projOut}=[regular polygon,regular polygon sides=3,draw,scale=0.75,inner sep=-0.5pt,minimum width=7.5mm,fill=white,regular polygon rotate=180]
\tikzstyle{projIn}=[regular polygon,regular polygon sides=3,draw,scale=0.75,inner sep=-0.5pt,minimum width=7.5mm,fill=white]
\tikzstyle{Vleak}=[white dot, shape=regular polygon, minimum size=3.3 mm, regular polygon sides=3, outer sep=-0.2mm, regular polygon rotate=90]
\tikzstyle{dleak}=[white dot, line width=1.6pt, shape=regular polygon, minimum size=3.3 mm, regular polygon sides=3, outer sep=-0.2mm, regular polygon rotate=270]

\tikzstyle{Wsquare}=[white dot, shape=regular polygon, rounded corners=0.8 mm, minimum size=3.3 mm, regular polygon sides=3, outer sep=-0.2mm]
\tikzstyle{Wsquareadj}=[white dot, shape=regular polygon, rounded corners=0.8 mm, minimum size=3.3 mm, regular polygon sides=3, outer sep=-0.2mm, regular polygon rotate=180]
\tikzstyle{ddot}=[inner sep=0mm, doubled, minimum width=2.5mm,minimum height=2.5mm,draw,shape=circle]

\tikzstyle{black dot}=[dot,fill=black]
\tikzstyle{white dot}=[dot,fill=white,,text depth=-0.2mm]
\tikzstyle{white Wsquare}=[Wsquare,fill=gray,,text depth=-0.2mm]
\tikzstyle{white Wsquareadj}=[Wsquareadj,fill=white,,text depth=-0.2mm]
\tikzstyle{green dot}=[white dot] 
\tikzstyle{gray dot}=[dot,fill=gray!40!white,,text depth=-0.2mm]
\tikzstyle{red dot}=[gray dot] 

\tikzstyle{black ddot}=[ddot,fill=black]
\tikzstyle{white ddot}=[ddot,fill=white]
\tikzstyle{gray ddot}=[ddot,fill=gray!40!white]

\tikzstyle{gray edge}=[gray!60!white]

\tikzstyle{small dot}=[inner sep=0.5mm,minimum width=0pt,minimum height=0pt,draw,shape=circle]

\tikzstyle{small black dot}=[small dot,fill=black]
\tikzstyle{small white dot}=[small dot,fill=white]
\tikzstyle{small gray dot}=[small dot,fill=gray!40!white]

\tikzstyle{causal dot}=[inner sep=0.4mm,minimum width=0pt,minimum height=0pt,draw=white,shape=circle,fill=gray!40!white]

\tikzstyle{phase dimensions}=[minimum size=5mm,font=\footnotesize,rectangle,rounded corners=2.5mm,inner sep=0.2mm,outer sep=-2mm]
\tikzstyle{dphase dimensions}=[minimum size=5mm,font=\footnotesize,rectangle,rounded corners=2.5mm,inner sep=0.2mm,outer sep=-2mm]

\tikzstyle{white phase dot}=[dot,fill=white,phase dimensions]
\tikzstyle{white phase ddot}=[ddot,fill=white,dphase dimensions]

\tikzstyle{white rect ddot}=[draw=black,fill=white,doubled,minimum size=5mm,font=\footnotesize,rectangle,rounded corners=2.5mm,inner sep=0.2mm]
\tikzstyle{gray rect ddot}=[draw=black,fill=gray!40!white,doubled,minimum size=6mm,font=\footnotesize,rectangle,rounded corners=3mm]

\tikzstyle{gray phase dot}=[dot,fill=gray!40!white,phase dimensions]
\tikzstyle{gray phase ddot}=[ddot,fill=gray!40!white,dphase dimensions]
\tikzstyle{grey phase dot}=[gray phase dot]
\tikzstyle{grey phase ddot}=[gray phase ddot]

\tikzstyle{small phase dimensions}=[minimum size=4mm,font=\tiny,rectangle,rounded corners=2mm,inner sep=0.2mm,outer sep=-2mm]
\tikzstyle{small dphase dimensions}=[minimum size=4mm,font=\tiny,rectangle,rounded corners=2mm,inner sep=0.2mm,outer sep=-2mm]

\tikzstyle{small gray phase dot}=[dot,fill=gray!40!white,small phase dimensions]
\tikzstyle{small gray phase ddot}=[ddot,fill=gray!40!white,small dphase dimensions]

\tikzstyle{small map}=[draw,shape=rectangle,minimum height=4mm,minimum width=4mm,fill=white]

\tikzstyle{cnot}=[fill=white,shape=circle,inner sep=-1.4pt]

\tikzstyle{asym hadamard}=[fill=white,draw,shape=NEbox,inner sep=0.6mm,font=\footnotesize,minimum height=4mm]
\tikzstyle{asym hadamard conj}=[fill=white,draw,shape=NWbox,inner sep=0.6mm,font=\footnotesize,minimum height=4mm]
\tikzstyle{asym hadamard dag}=[fill=white,draw,shape=SEbox,inner sep=0.6mm,font=\footnotesize,minimum height=4mm]

\tikzstyle{hadamard}=[fill=white,draw,inner sep=0.6mm,font=\footnotesize,minimum height=4mm,minimum width=4mm]
\tikzstyle{small hadamard}=[fill=white,draw,inner sep=0.6mm,minimum height=1.5mm,minimum width=1.5mm]
\tikzstyle{small hadamard rotate}=[small hadamard,rotate=45]
\tikzstyle{dhadamard}=[hadamard,doubled]
\tikzstyle{small dhadamard}=[small hadamard,doubled]
\tikzstyle{small dhadamard rotate}=[small hadamard rotate,doubled]
\tikzstyle{antipode}=[white dot,inner sep=0.3mm,font=\footnotesize]

\tikzstyle{scalar}=[diamond,draw,inner sep=0.5pt,font=\small]
\tikzstyle{dscalar}=[diamond,doubled, draw,inner sep=0.5pt,font=\small]

\tikzstyle{small box}=[rectangle,inline text,fill=white,draw,minimum height=5mm,yshift=-0.5mm,minimum width=5mm,font=\small]
\tikzstyle{small gray box}=[small box,fill=gray!30]
\tikzstyle{medium box}=[rectangle,inline text,fill=white,draw,minimum height=5mm,yshift=-0.5mm,minimum width=10mm,font=\small]
\tikzstyle{square box}=[small box] 
\tikzstyle{medium gray box}=[small box,fill=gray!30]
\tikzstyle{semilarge box}=[rectangle,inline text,fill=white,draw,minimum height=5mm,yshift=-0.5mm,minimum width=12.5mm,font=\small]
\tikzstyle{large box}=[rectangle,inline text,fill=white,draw,minimum height=5mm,yshift=-0.5mm,minimum width=15mm,font=\small]
\tikzstyle{large gray box}=[small box,fill=gray!30]

\tikzstyle{Bayes box}=[rectangle,fill=black,draw, minimum height=3mm, minimum width=3mm]

\tikzstyle{gray square point}=[small box,fill=gray!50]

\tikzstyle{dphase box white}=[dhadamard]
\tikzstyle{dphase box gray}=[dhadamard,fill=gray!50!white]
\tikzstyle{phase box white}=[hadamard]
\tikzstyle{phase box gray}=[hadamard,fill=gray!50!white]

\tikzstyle{point}=[regular polygon,regular polygon sides=3,draw,scale=0.75,inner sep=-0.5pt,minimum width=9mm,fill=white,regular polygon rotate=180]
\tikzstyle{point nosep}=[regular polygon,regular polygon sides=3,draw,scale=0.75,inner sep=-2pt,minimum width=9mm,fill=white,regular polygon rotate=180]
\tikzstyle{copoint}=[regular polygon,regular polygon sides=3,draw,scale=0.75,inner sep=-0.5pt,minimum width=9mm,fill=white]
\tikzstyle{dpoint}=[point,doubled]
\tikzstyle{dcopoint}=[copoint,doubled]

\tikzstyle{pointgrow}=[shape=cornerpoint,kpoint common,scale=0.75,inner sep=3pt]
\tikzstyle{pointgrow dag}=[shape=cornercopoint,kpoint common,scale=0.75,inner sep=3pt]

\tikzstyle{wide copoint}=[fill=white,draw,shape=isosceles triangle,shape border rotate=90,isosceles triangle stretches=true,inner sep=0pt,minimum width=1.5cm,minimum height=6.12mm]
\tikzstyle{wide point}=[fill=white,draw,shape=isosceles triangle,shape border rotate=-90,isosceles triangle stretches=true,inner sep=0pt,minimum width=1.5cm,minimum height=6.12mm,yshift=-0.0mm]
\tikzstyle{wide point plus}=[fill=white,draw,shape=isosceles triangle,shape border rotate=-90,isosceles triangle stretches=true,inner sep=0pt,minimum width=1.74cm,minimum height=7mm,yshift=-0.0mm]

\tikzstyle{wide dpoint}=[fill=white,doubled,draw,shape=isosceles triangle,shape border rotate=-90,isosceles triangle stretches=true,inner sep=0pt,minimum width=1.5cm,minimum height=6.12mm,yshift=-0.0mm]

\tikzstyle{tinypoint}=[regular polygon,regular polygon sides=3,draw,scale=0.55,inner sep=-0.15pt,minimum width=6mm,fill=white,regular polygon rotate=180]

\tikzstyle{white point}=[point]
\tikzstyle{white dpoint}=[dpoint]
\tikzstyle{green point}=[white point] 
\tikzstyle{white copoint}=[copoint]
\tikzstyle{gray point}=[point,fill=gray!40!white]
\tikzstyle{gray dpoint}=[gray point,doubled]
\tikzstyle{red point}=[gray point] 
\tikzstyle{gray copoint}=[copoint,fill=gray!40!white]
\tikzstyle{gray dcopoint}=[gray copoint,doubled]

\tikzstyle{white point guide}=[regular polygon,regular polygon sides=3,font=\scriptsize,draw,scale=0.65,inner sep=-0.5pt,minimum width=9mm,fill=white,regular polygon rotate=180]

\tikzstyle{black point}=[point,fill=black,font=\color{white}]
\tikzstyle{black copoint}=[copoint,fill=black,font=\color{white}]

\tikzstyle{tiny gray point}=[tinypoint,fill=gray!40!white]

\tikzstyle{diredge}=[->]
\tikzstyle{ddiredge}=[<->]
\tikzstyle{rdiredge}=[<-]
\tikzstyle{thickdiredge}=[->, very thick]
\tikzstyle{pointer edge}=[->,very thick,gray]
\tikzstyle{pointer edge part}=[very thick,gray]
\tikzstyle{dashed edge}=[dashed]
\tikzstyle{thick dashed edge}=[very thick,dashed]
\tikzstyle{thick gray dashed edge}=[thick dashed edge,gray!40]
\tikzstyle{thick map edge}=[very thick,|->]

\makeatletter
\newcommand{\boxshape}[3]{%
\pgfdeclareshape{#1}{
\inheritsavedanchors[from=rectangle] 
\inheritanchorborder[from=rectangle]
\inheritanchor[from=rectangle]{center}
\inheritanchor[from=rectangle]{north}
\inheritanchor[from=rectangle]{south}
\inheritanchor[from=rectangle]{west}
\inheritanchor[from=rectangle]{east}
\backgroundpath{
\southwest \pgf@xa=\pgf@x \pgf@ya=\pgf@y
\northeast \pgf@xb=\pgf@x \pgf@yb=\pgf@y

\@tempdima=#2
\@tempdimb=#3

\pgfpathmoveto{\pgfpoint{\pgf@xa - 5pt + \@tempdima}{\pgf@ya}}
\pgfpathlineto{\pgfpoint{\pgf@xa - 5pt - \@tempdima}{\pgf@yb}}
\pgfpathlineto{\pgfpoint{\pgf@xb + 5pt + \@tempdimb}{\pgf@yb}}
\pgfpathlineto{\pgfpoint{\pgf@xb + 5pt - \@tempdimb}{\pgf@ya}}
\pgfpathlineto{\pgfpoint{\pgf@xa - 5pt + \@tempdima}{\pgf@ya}}
\pgfpathclose
}
}}

\boxshape{NEbox}{0pt}{5pt}
\boxshape{SEbox}{0pt}{-5pt}
\boxshape{NWbox}{5pt}{0pt}
\boxshape{SWbox}{-5pt}{0pt}
\boxshape{EBox}{-3pt}{3pt}
\boxshape{WBox}{3pt}{-3pt}
\makeatother

\tikzstyle{cloud}=[shape=cloud,draw,minimum width=1.5cm,minimum height=1.5cm]

\tikzstyle{map}=[draw,shape=NEbox,inner sep=2pt,minimum height=6mm,fill=white]
\tikzstyle{dashedmap}=[draw,dashed,shape=NEbox,inner sep=2pt,minimum height=6mm,fill=white]
\tikzstyle{mapdag}=[draw,shape=SEbox,inner sep=2pt,minimum height=6mm,fill=white]
\tikzstyle{mapadj}=[draw,shape=SEbox,inner sep=2pt,minimum height=6mm,fill=white]
\tikzstyle{maptrans}=[draw,shape=SWbox,inner sep=2pt,minimum height=6mm,fill=white]
\tikzstyle{mapconj}=[draw,shape=NWbox,inner sep=2pt,minimum height=6mm,fill=white]

\tikzstyle{medium map}=[draw,shape=NEbox,inner sep=2pt,minimum height=6mm,fill=white,minimum width=7mm]
\tikzstyle{medium map dag}=[draw,shape=SEbox,inner sep=2pt,minimum height=6mm,fill=white,minimum width=7mm]
\tikzstyle{medium map adj}=[draw,shape=SEbox,inner sep=2pt,minimum height=6mm,fill=white,minimum width=7mm]
\tikzstyle{medium map trans}=[draw,shape=SWbox,inner sep=2pt,minimum height=6mm,fill=white,minimum width=7mm]
\tikzstyle{medium map conj}=[draw,shape=NWbox,inner sep=2pt,minimum height=6mm,fill=white,minimum width=7mm]
\tikzstyle{semilarge map}=[draw,shape=NEbox,inner sep=2pt,minimum height=6mm,fill=white,minimum width=9.5mm]
\tikzstyle{semilarge map trans}=[draw,shape=SWbox,inner sep=2pt,minimum height=6mm,fill=white,minimum width=9.5mm]
\tikzstyle{semilarge map adj}=[draw,shape=SEbox,inner sep=2pt,minimum height=6mm,fill=white,minimum width=9.5mm]
\tikzstyle{semilarge map dag}=[draw,shape=SEbox,inner sep=2pt,minimum height=6mm,fill=white,minimum width=9.5mm]
\tikzstyle{semilarge map conj}=[draw,shape=NWbox,inner sep=2pt,minimum height=6mm,fill=white,minimum width=9.5mm]
\tikzstyle{large map}=[draw,shape=NEbox,inner sep=2pt,minimum height=6mm,fill=white,minimum width=12mm]
\tikzstyle{large map conj}=[draw,shape=NWbox,inner sep=2pt,minimum height=6mm,fill=white,minimum width=12mm]
\tikzstyle{very large map}=[draw,shape=NEbox,inner sep=2pt,minimum height=6mm,fill=white,minimum width=17mm]

\tikzstyle{medium dmap}=[draw,doubled,shape=NEbox,inner sep=2pt,minimum height=6mm,fill=white,minimum width=7mm]
\tikzstyle{medium dmap dag}=[draw,doubled,shape=SEbox,inner sep=2pt,minimum height=6mm,fill=white,minimum width=7mm]
\tikzstyle{medium dmap adj}=[draw,doubled,shape=SEbox,inner sep=2pt,minimum height=6mm,fill=white,minimum width=7mm]
\tikzstyle{medium dmap trans}=[draw,doubled,shape=SWbox,inner sep=2pt,minimum height=6mm,fill=white,minimum width=7mm]
\tikzstyle{medium dmap conj}=[draw,doubled,shape=NWbox,inner sep=2pt,minimum height=6mm,fill=white,minimum width=7mm]
\tikzstyle{semilarge dmap}=[draw,doubled,shape=NEbox,inner sep=2pt,minimum height=6mm,fill=white,minimum width=9.5mm]
\tikzstyle{semilarge dmap trans}=[draw,doubled,shape=SWbox,inner sep=2pt,minimum height=6mm,fill=white,minimum width=9.5mm]
\tikzstyle{semilarge dmap adj}=[draw,doubled,shape=SEbox,inner sep=2pt,minimum height=6mm,fill=white,minimum width=9.5mm]
\tikzstyle{semilarge dmap dag}=[draw,doubled,shape=SEbox,inner sep=2pt,minimum height=6mm,fill=white,minimum width=9.5mm]
\tikzstyle{semilarge dmap conj}=[draw,doubled,shape=NWbox,inner sep=2pt,minimum height=6mm,fill=white,minimum width=9.5mm]
\tikzstyle{large dmap}=[draw,doubled,shape=NEbox,inner sep=2pt,minimum height=6mm,fill=white,minimum width=12mm]
\tikzstyle{large dmap conj}=[draw,doubled,shape=NWbox,inner sep=2pt,minimum height=6mm,fill=white,minimum width=12mm]
\tikzstyle{large dmap trans}=[draw,doubled,shape=SWbox,inner sep=2pt,minimum height=6mm,fill=white,minimum width=12mm]
\tikzstyle{large dmap adj}=[draw,doubled,shape=SEbox,inner sep=2pt,minimum height=6mm,fill=white,minimum width=12mm]
\tikzstyle{large dmap dag}=[draw,doubled,shape=SEbox,inner sep=2pt,minimum height=6mm,fill=white,minimum width=12mm]
\tikzstyle{very large dmap}=[draw,doubled,shape=NEbox,inner sep=2pt,minimum height=6mm,fill=white,minimum width=19.5mm]

\tikzstyle{muxbox}=[draw,shape=rectangle,minimum height=3mm,minimum width=3mm,fill=white]
\tikzstyle{dmuxbox}=[muxbox,doubled]

\tikzstyle{box}=[draw,shape=rectangle,inner sep=2pt,minimum height=6mm,minimum width=6mm,fill=white]
\tikzstyle{dbox}=[draw,doubled,shape=rectangle,inner sep=2pt,minimum height=6mm,minimum width=6mm,fill=white]
\tikzstyle{dmap}=[draw,doubled,shape=NEbox,inner sep=2pt,minimum height=6mm,fill=white]
\tikzstyle{dmapdag}=[draw,doubled,shape=SEbox,inner sep=2pt,minimum height=6mm,fill=white]
\tikzstyle{dmapadj}=[draw,doubled,shape=SEbox,inner sep=2pt,minimum height=6mm,fill=white]
\tikzstyle{dmaptrans}=[draw,doubled,shape=SWbox,inner sep=2pt,minimum height=6mm,fill=white]
\tikzstyle{dmapconj}=[draw,doubled,shape=NWbox,inner sep=2pt,minimum height=6mm,fill=white]

\tikzstyle{ddmap}=[draw,doubled,dashed,shape=NEbox,inner sep=2pt,minimum height=6mm,fill=white]
\tikzstyle{ddmapdag}=[draw,doubled,dashed,shape=SEbox,inner sep=2pt,minimum height=6mm,fill=white]
\tikzstyle{ddmapadj}=[draw,doubled,dashed,shape=SEbox,inner sep=2pt,minimum height=6mm,fill=white]
\tikzstyle{ddmaptrans}=[draw,doubled,dashed,shape=SWbox,inner sep=2pt,minimum height=6mm,fill=white]
\tikzstyle{ddmapconj}=[draw,doubled,dashed,shape=NWbox,inner sep=2pt,minimum height=6mm,fill=white]

\boxshape{sNEbox}{0pt}{3pt}
\boxshape{sSEbox}{0pt}{-3pt}
\boxshape{sNWbox}{3pt}{0pt}
\boxshape{sSWbox}{-3pt}{0pt}
\tikzstyle{smap}=[draw,shape=sNEbox,fill=white]
\tikzstyle{smapdag}=[draw,shape=sSEbox,fill=white]
\tikzstyle{smapadj}=[draw,shape=sSEbox,fill=white]
\tikzstyle{smaptrans}=[draw,shape=sSWbox,fill=white]
\tikzstyle{smapconj}=[draw,shape=sNWbox,fill=white]

\tikzstyle{dsmap}=[draw,dashed,shape=sNEbox,fill=white]
\tikzstyle{dsmapdag}=[draw,dashed,shape=sSEbox,fill=white]
\tikzstyle{dsmaptrans}=[draw,dashed,shape=sSWbox,fill=white]
\tikzstyle{dsmapconj}=[draw,dashed,shape=sNWbox,fill=white]

\boxshape{mNEbox}{0pt}{10pt}
\boxshape{mSEbox}{0pt}{-10pt}
\boxshape{mNWbox}{10pt}{0pt}
\boxshape{mSWbox}{-10pt}{0pt}
\tikzstyle{mmap}=[draw,shape=mNEbox]
\tikzstyle{mmapdag}=[draw,shape=mSEbox]
\tikzstyle{mmaptrans}=[draw,shape=mSWbox]
\tikzstyle{mmapconj}=[draw,shape=mNWbox]

\tikzstyle{mmapgray}=[draw,fill=gray!40!white,shape=mNEbox]
\tikzstyle{smapgray}=[draw,fill=gray!40!white,shape=sNEbox]

\makeatletter

\pgfdeclareshape{cornerpoint}{
\inheritsavedanchors[from=rectangle] 
\inheritanchorborder[from=rectangle]
\inheritanchor[from=rectangle]{center}
\inheritanchor[from=rectangle]{north}
\inheritanchor[from=rectangle]{south}
\inheritanchor[from=rectangle]{west}
\inheritanchor[from=rectangle]{east}
\backgroundpath{
\southwest \pgf@xa=\pgf@x \pgf@ya=\pgf@y
\northeast \pgf@xb=\pgf@x \pgf@yb=\pgf@y

\pgfmathsetmacro{\pgf@shorten@left}{\pgfkeysvalueof{/tikz/shorten left}}
\pgfmathsetmacro{\pgf@shorten@right}{\pgfkeysvalueof{/tikz/shorten right}}

\pgfpathmoveto{\pgfpoint{0.5 * (\pgf@xa + \pgf@xb)}{\pgf@ya - 5pt}}
\pgfpathlineto{\pgfpoint{\pgf@xa - 8pt + \pgf@shorten@left}{\pgf@yb - 1.5 * \pgf@shorten@left}}
\pgfpathlineto{\pgfpoint{\pgf@xa - 8pt + \pgf@shorten@left}{\pgf@yb}}
\pgfpathlineto{\pgfpoint{\pgf@xb + 8pt - \pgf@shorten@right}{\pgf@yb}}
\pgfpathlineto{\pgfpoint{\pgf@xb + 8pt - \pgf@shorten@right}{\pgf@yb - 1.5 * \pgf@shorten@right}}
\pgfpathclose
}
}

\pgfdeclareshape{cornercopoint}{
\inheritsavedanchors[from=rectangle] 
\inheritanchorborder[from=rectangle]
\inheritanchor[from=rectangle]{center}
\inheritanchor[from=rectangle]{north}
\inheritanchor[from=rectangle]{south}
\inheritanchor[from=rectangle]{west}
\inheritanchor[from=rectangle]{east}
\backgroundpath{
\southwest \pgf@xa=\pgf@x \pgf@ya=\pgf@y
\northeast \pgf@xb=\pgf@x \pgf@yb=\pgf@y

\pgfmathsetmacro{\pgf@shorten@left}{\pgfkeysvalueof{/tikz/shorten left}}
\pgfmathsetmacro{\pgf@shorten@right}{\pgfkeysvalueof{/tikz/shorten right}}

\pgfpathmoveto{\pgfpoint{0.5 * (\pgf@xa + \pgf@xb)}{\pgf@yb + 5pt}}
\pgfpathlineto{\pgfpoint{\pgf@xa - 8pt + \pgf@shorten@left}{\pgf@ya + 1.5 * \pgf@shorten@left}}
\pgfpathlineto{\pgfpoint{\pgf@xa - 8pt + \pgf@shorten@left}{\pgf@ya}}
\pgfpathlineto{\pgfpoint{\pgf@xb + 8pt - \pgf@shorten@right}{\pgf@ya}}
\pgfpathlineto{\pgfpoint{\pgf@xb + 8pt - \pgf@shorten@right}{\pgf@ya + 1.5 * \pgf@shorten@right}}
\pgfpathclose
}
}

\makeatother

\pgfkeyssetvalue{/tikz/shorten left}{0pt}
\pgfkeyssetvalue{/tikz/shorten right}{0pt}

\tikzstyle{kpoint common}=[draw,fill=white,inner sep=1pt,minimum height=4mm]
\tikzstyle{kpoint sc}=[shape=cornerpoint,kpoint common]
\tikzstyle{kpoint adjoint sc}=[shape=cornercopoint,kpoint common]
\tikzstyle{kpoint}=[shape=cornerpoint,shorten left=5pt,kpoint common]
\tikzstyle{kpoint adjoint}=[shape=cornercopoint,shorten left=5pt,kpoint common]
\tikzstyle{kpoint conjugate}=[shape=cornerpoint,shorten right=5pt,kpoint common]
\tikzstyle{kpoint transpose}=[shape=cornercopoint,shorten right=5pt,kpoint common]
\tikzstyle{kpoint symm}=[shape=cornerpoint,shorten left=5pt,shorten right=5pt,kpoint common]

\tikzstyle{wide kpoint sc}=[shape=cornerpoint,kpoint common, minimum width=1 cm]
\tikzstyle{wide kpointdag sc}=[shape=cornercopoint,kpoint common, minimum width=1 cm]

\tikzstyle{black kpoint}=[shape=cornerpoint,shorten left=5pt,kpoint common,fill=black,font=\color{white}]

\tikzstyle{black kpoint sm}=[shape=cornerpoint,shorten left=5pt,kpoint common,fill=black,font=\color{white},scale=0.75]

\tikzstyle{black kpoint adjoint}=[shape=cornercopoint,shorten left=5pt,kpoint common,fill=black,font=\color{white}]
\tikzstyle{black kpointadj}=[shape=cornercopoint,shorten left=5pt,kpoint common,fill=black,font=\color{white}]

\tikzstyle{black kpointadj sm}=[shape=cornercopoint,shorten left=5pt,kpoint common,fill=black,font=\color{white},scale=0.75]

\tikzstyle{black dkpoint}=[shape=cornerpoint,shorten left=5pt,kpoint common,fill=black, doubled,font=\color{white}]
\tikzstyle{black dkpoint adjoint}=[shape=cornercopoint,shorten left=5pt,kpoint common,fill=black, doubled,font=\color{white}]
\tikzstyle{black dkpointadj}=[shape=cornercopoint,shorten left=5pt,kpoint common,fill=black, doubled,font=\color{white}]

\tikzstyle{black dkpoint sm}=[shape=cornerpoint,shorten left=5pt,kpoint common,fill=black, doubled,font=\color{white},scale=0.75]
\tikzstyle{black dkpointadj sm}=[shape=cornercopoint,shorten left=5pt,kpoint common,fill=black, doubled,font=\color{white},scale=0.75]

\tikzstyle{kpointdag}=[kpoint adjoint]
\tikzstyle{kpointadj}=[kpoint adjoint]
\tikzstyle{kpointconj}=[kpoint conjugate]
\tikzstyle{kpointtrans}=[kpoint transpose]

\tikzstyle{big kpoint}=[kpoint, minimum width=1.2 cm, minimum height=8mm, inner sep=4pt, text depth=3mm]

\tikzstyle{wide kpoint}=[kpoint, minimum width=1 cm, inner sep=2pt]
\tikzstyle{wide kpointdag}=[kpointdag, minimum width=1 cm, inner sep=2pt]
\tikzstyle{wide kpointconj}=[kpointconj, minimum width=1 cm, inner sep=2pt]
\tikzstyle{wide kpointtrans}=[kpointtrans, minimum width=1 cm, inner sep=2pt]

\tikzstyle{wider kpoint}=[kpoint, minimum width=1.25 cm, inner sep=2pt]
\tikzstyle{wider kpointdag}=[kpointdag, minimum width=1.25 cm, inner sep=2pt]
\tikzstyle{wider kpointconj}=[kpointconj, minimum width=1.25 cm, inner sep=2pt]
\tikzstyle{wider kpointtrans}=[kpointtrans, minimum width=1.25 cm, inner sep=2pt]

\tikzstyle{gray kpoint}=[kpoint,fill=gray!50!white]
\tikzstyle{gray kpointdag}=[kpointdag,fill=gray!50!white]
\tikzstyle{gray kpointadj}=[kpointadj,fill=gray!50!white]
\tikzstyle{gray kpointconj}=[kpointconj,fill=gray!50!white]
\tikzstyle{gray kpointtrans}=[kpointtrans,fill=gray!50!white]

\tikzstyle{gray dkpoint}=[kpoint,fill=gray!50!white,doubled]
\tikzstyle{gray dkpointdag}=[kpointdag,fill=gray!50!white,doubled]
\tikzstyle{gray dkpointadj}=[kpointadj,fill=gray!50!white,doubled]
\tikzstyle{gray dkpointconj}=[kpointconj,fill=gray!50!white,doubled]
\tikzstyle{gray dkpointtrans}=[kpointtrans,fill=gray!50!white,doubled]

\tikzstyle{white label}=[draw,fill=white,rectangle,inner sep=0.7 mm]
\tikzstyle{gray label}=[draw,fill=gray!50!white,rectangle,inner sep=0.7 mm]
\tikzstyle{black label}=[draw,fill=black,rectangle,inner sep=0.7 mm]

\tikzstyle{dkpoint}=[kpoint,doubled]
\tikzstyle{wide dkpoint}=[wide kpoint,doubled]
\tikzstyle{dkpointdag}=[kpoint adjoint,doubled]
\tikzstyle{wide dkpointdag}=[wide kpointdag,doubled]
\tikzstyle{dkcopoint}=[kpoint adjoint,doubled]
\tikzstyle{dkpointadj}=[kpoint adjoint,doubled]
\tikzstyle{dkpointconj}=[kpoint conjugate,doubled]
\tikzstyle{dkpointtrans}=[kpoint transpose,doubled]

\tikzstyle{kscalar}=[kpoint common, shape=EBox, inner xsep=-1pt, inner ysep=3pt,font=\small]
\tikzstyle{kscalarconj}=[kpoint common, shape=WBox, inner xsep=-1pt, inner ysep=3pt,font=\small]

\tikzstyle{spekpoint}=[kpoint sc,minimum height=5mm,inner sep=3pt]
\tikzstyle{spekcopoint}=[kpoint adjoint sc,minimum height=5mm,inner sep=3pt]

\tikzstyle{dspekpoint}=[spekpoint,doubled]
\tikzstyle{dspekcopoint}=[spekcopoint,doubled]

 \tikzstyle{upground}=[circuit ee IEC,thick,ground,rotate=90,scale=2.5]
 \tikzstyle{downground}=[circuit ee IEC,thick,ground,rotate=-90,scale=2.5]
 \tikzstyle{bigground}=[regular polygon,regular polygon sides=3,draw=gray,scale=0.50,inner sep=-0.5pt,minimum width=10mm,fill=gray]

\tikzstyle{arrs}=[-latex,font=\small,auto]
\tikzstyle{arrow plain}=[arrs]
\tikzstyle{arrow dashed}=[dashed,arrs]
\tikzstyle{arrow bold}=[very thick,arrs]
\tikzstyle{arrow hide}=[draw=white!0,-]
\tikzstyle{arrow reverse}=[latex-]
\tikzstyle{cdnode}=[]

%% file: Figures/Fig_ExampleDAG.tex
\vspace*{-0.2cm}
\small
\begin{minipage}{9cm}
	\begin{minipage}{4cm}
		\centering
		\renewcommand{\Scale}{1.6}
		\begin{tikzpicture}
			\node [style=none](1)  at (0,0) {$X_1$};
			\node [style=none](1tl)  at (-0.1*\Scale,0.3*\Scale) {};
			\node [style=none](1tr)  at (0.0*\Scale,0.3*\Scale) {};
			\node [style=none](2)  at (-1.0*\Scale,1.2*\Scale) {$X_2$};
			\node [style=none](2b)  at (-1.0*\Scale,1.0*\Scale) {};
			\node [style=none](2t)  at (-0.9*\Scale,1.5*\Scale) {};
			\node [style=none](3)  at (1.0*\Scale,1.2*\Scale) {$X_3$};
			\node [style=none](3b)  at (0.9*\Scale,1.0*\Scale) {};
			\node [style=none](3br)  at (1.1*\Scale,1.0*\Scale) {};
			\node [style=none](3t)  at (0.9*\Scale,1.5*\Scale) {};
			\node [style=none](3tr)  at (1.0*\Scale,1.5*\Scale) {};
			\node [style=none](4)  at (0.0*\Scale,2.4*\Scale) {$X_4$};
			\node [style=none](4b)  at (0.0*\Scale,2.2*\Scale) {};
			\node [style=none](4t)  at (0.0*\Scale,2.7*\Scale) {};
			\node [style=none](5)  at (-1.9*\Scale,3.0*\Scale) {$X_5$};
			\node [style=none](5b)  at (-1.6*\Scale,2.9*\Scale) {};
			\draw [->] (1tl) to (2b);
			\draw [->] (1tr) to (3b);
			\draw [->] (2t) to (4b);
			\draw [->] (3t) to (4b);
			\draw [->] (4) to (5b);
			\draw [->] (1) to [bend left=55] (5);
		\end{tikzpicture}
	\end{minipage}
	\hfill
	\begin{minipage}{4cm}
		\begin{eqnarray}
			& P(X_1,X_2,X_3,X_4,X_5) \ =& \nonumber \\[0.1cm] 
			 & P(X_5|X_1,X_4) \ P(X_4|X_2,X_3) & \nonumber \\[0.1cm]  
				& P(X_2|X_1) \ P(X_3|X_1) & \nonumber \\[0.1cm]  
				& P(X_1) & \nonumber 
		\end{eqnarray}
		\vspace*{0.1cm}
	\end{minipage}
\end{minipage}

\vspace*{-0.3cm}

%% file: Figures/Fig_Examplefunctionalmodel.tex
\begin{minipage}{13cm}
	\begin{minipage}{6cm}
		\centering
		\renewcommand{\Scale}{2}
		\begin{tikzpicture}
			\node [style=none](1)  at (0,0) {$X_1$};
			\node [style=none](1tl)  at (-0.1*\Scale,0.2*\Scale) {};
			\node [style=none](1tr)  at (0.0*\Scale,0.2*\Scale) {};
			\node [style=none](1bl)  at (-0.25*\Scale,0.0*\Scale) {};
			\node [style=none](lam1)  at (0.7*\Scale,-0.3*\Scale) {\small{$\lambda_1$}};
			\node [style=none](2)  at (-1.0*\Scale,1.2*\Scale) {$X_2$};
			\node [style=none](2br)  at (-0.95*\Scale,1.0*\Scale) {};
			\node [style=none](2t)  at (-0.95*\Scale,1.45*\Scale) {};
			\node [style=none](lam2)  at (-1.4*\Scale,0.6*\Scale) {\small{$\lambda_2$}};
			\node [style=none](3)  at (1.0*\Scale,1.2*\Scale) {$X_3$};
			\node [style=none](3b)  at (0.9*\Scale,1.0*\Scale) {};
			\node [style=none](3br)  at (1.1*\Scale,1.0*\Scale) {};
			\node [style=none](3t)  at (0.95*\Scale,1.45*\Scale) {};
			\node [style=none](3tr)  at (1.0*\Scale,1.5*\Scale) {};
			\node [style=none](lam3)  at (1.55*\Scale,0.5*\Scale) {\small{$\lambda_3$}};
			\node [style=none](lam3t)  at (1.4*\Scale,0.6*\Scale) {};
			\node [style=none](4)  at (0.0*\Scale,2.4*\Scale) {$X_4$};
			\node [style=none](4bl)  at (-0.1*\Scale,2.2*\Scale) {};
			\node [style=none](4br)  at (0.1*\Scale,2.2*\Scale) {};
			\node [style=none](4bm)  at (0.0*\Scale,2.1*\Scale) {};
			\node [style=none](4t)  at (0.0*\Scale,2.7*\Scale) {};
			\node [style=none](lam4)  at (0*\Scale,1.6*\Scale) {\small{$\lambda_4$}};
			\node [style=none](5)  at (-1.9*\Scale,3.0*\Scale) {$X_5$};
			\node [style=none](5b)  at (-1.6*\Scale,2.9*\Scale) {};
			\node [style=none](5bm)  at (-1.9*\Scale,2.8*\Scale) {};
			\node [style=none](lam5)  at (-1.55*\Scale,2.35*\Scale) {\small{$\lambda_5$}};
			\node [style=none](lam5t)  at (-1.7*\Scale,2.45*\Scale) {};
			\draw [->] (1tl) to (2br);
			\draw [->] (1tr) to (3b);
			\draw [->] (lam1) to (1);
			\draw [->] (2t) to (4bl);
			\draw [->] (lam2) to (2);
			\draw [->] (3t) to (4br);
			\draw [->] (lam3t) to (3br);
			\draw [->] (4) to (5b);
			\draw [->] (lam4) to (4bm);
			\draw [->] (lam5t) to (5bm);
			\draw [->] (1bl) to [bend left=80] (5);
		\end{tikzpicture}
	\end{minipage}
	\hfill
	\begin{minipage}{6cm}
		\begin{eqnarray}
			X_5 &=&  f_5(X_1,X_4,\lambda_5) , \nonumber \\[0.1cm]
			X_4 &=& f_4(X_2,X_3,\lambda_4),  \nonumber \\[0.1cm]
			X_3 &=& f_3(X_1,\lambda_3) ,  \nonumber \\[0.1cm]
			X_2 &=& f_2(X_1,\lambda_2),  \nonumber \\[0.1cm]
			X_1 &=& f_1(\lambda_1),  \nonumber 
		\end{eqnarray}
		\\[-0.65cm]
		\hspace*{1.1cm}$ P(\lambda_1, \lambda_2,\lambda_3, \lambda_4, \lambda_5)= \prod\limits_{i=1}^{5} P(\lambda_i)$	.
		\vspace*{0.3cm}
	\end{minipage}
\end{minipage}
\vspace*{-0.3cm}

%% file: Figures/Fig_Exampledoconditional.tex
\vspace*{-0.5cm}

\begin{minipage}{12cm}
	\begin{minipage}{6cm}
		\centering
		\renewcommand{\Scale}{1.8}
		\begin{tikzpicture}
			\node [style=none](1)  at (0,0) {$X_1$};
			\node [style=none](1tl)  at (-0.1*\Scale,0.2*\Scale) {};
			\node [style=none](1tr)  at (0.0*\Scale,0.2*\Scale) {};
			\node [style=none](2)  at (-1.0*\Scale,1.2*\Scale) {$X_2$};
			\node [style=none](2b)  at (-1.0*\Scale,1.0*\Scale) {};
			\node [style=none](2t)  at (-0.9*\Scale,1.5*\Scale) {};
			\node [style=none](3)  at (1.0*\Scale,1.2*\Scale) {$X_3$};
			\node [style=none](3b)  at (0.9*\Scale,1.0*\Scale) {};
			\node [style=none](3br)  at (1.1*\Scale,1.0*\Scale) {};
			\node [style=none](3t)  at (0.9*\Scale,1.5*\Scale) {};
			\node [style=none](3tr)  at (1.0*\Scale,1.5*\Scale) {};
			\node [style=none](4)  at (0.0*\Scale,2.4*\Scale) {$X_4$};
			\node [style=none](4b)  at (0.0*\Scale,2.2*\Scale) {};
			\node [style=none](4t)  at (0.0*\Scale,2.7*\Scale) {};
			\node [style=none](5)  at (-1.9*\Scale,3.0*\Scale) {$X_5$};
			\node [style=none](5b)  at (-1.6*\Scale,2.9*\Scale) {};
			\draw [->] (1tl) to (2b);
			\draw [->] (1tr) to (3b);
			\draw [->] (4) to (5b);
			\draw [->] (1) to [bend left=55] (5);
		\end{tikzpicture}
	\end{minipage}
	\hfill
	\begin{minipage}{5cm}
		\begin{eqnarray}
			& P(X_1,X_2,X_3,X_5|do(X_4=x)) \ =& \nonumber \\[0.2cm] 
			 & P(X_5|X_1,X_4=x)&  \nonumber \\[0.2cm] 
				& P(X_2|X_1) \ P(X_3|X_1) & \nonumber \\[0.2cm]  
				& P(X_1) & \nonumber 
		\end{eqnarray}
		\vspace*{0.1cm}
	\end{minipage}
\end{minipage}
\vspace*{-0.5cm}

%% file: Figures/Fig_Examplefinetuning.tex
\vspace*{0.2cm}
\begin{center}
	\begin{minipage}{14.7cm}
		\begin{minipage}{3.3cm} 
			\textit{Functional model} \\[0.3cm]
			$Y= X + \lambda_Y \ \text{mod}\  2$ \\[0.1cm]
			$P(\lambda_Y)=1/2$ \\[0.1cm]
			$P(X)$
			\vspace*{0.4cm}
		\end{minipage}
		\hfill
		\begin{minipage}{2cm}
			\centering
			\renewcommand{\Scale}{2.0}
			\begin{tikzpicture}
				\node [style=none](1)  at (0.0*\Scale,0.0*\Scale) {};
				\node [style=none](2)  at (1.2*\Scale,0.0*\Scale) {};
				\node [style=none]  at (0.6*\Scale,0.3*\Scale) {\small{defines}};
				\draw [->] (1) to (2);
			\end{tikzpicture}
		\end{minipage}
		\hfill
		\begin{minipage}{2.8cm}
			\centering
			\textit{Causal structure} \\[0.55cm]
			\renewcommand{\Scale}{2.0}
			\hspace*{0.2cm}
			\begin{tikzpicture}
				\node [style=none](y)  at (0.05*\Scale,1.5*\Scale) {$Y$};
				\node [style=none](ybm)  at (0.0*\Scale,1.3*\Scale) {};
				\node [style=none](ybr)  at (0.1*\Scale,1.4*\Scale) {};
				\node [style=none](x)  at (0.0*\Scale,0.0*\Scale) {$X$};
				\node [style=none](xt)  at (0.0*\Scale,0.2*\Scale) {};
				\node [style=none](l)  at (0.8*\Scale,0.7*\Scale) {$\lambda_Y$};
				\node [style=none](lt)  at (0.6*\Scale,0.9*\Scale) {};
				\draw [->] (xt) to (ybm);
				\draw [->] (lt) to (ybr);
			\end{tikzpicture}
		\end{minipage}
		\hfill
		\begin{minipage}{2.5cm}
			\centering
			\vspace*{0.2cm}
			\renewcommand{\Scale}{2.0}
			\begin{tikzpicture}
				\node [style=none](1)  at (0.0*\Scale,0.0*\Scale) {};
				\node [style=none](2)  at (1.2*\Scale,0.0*\Scale) {};
				\node [style=none]  at (0.6*\Scale,0.3*\Scale) {\small{sum over}};			
				\node [style=none]  at (0.6*\Scale,-0.3*\Scale) {\small{$\lambda_Y$}};
				\draw [->] (1) to (2);
			\end{tikzpicture}
		\end{minipage}
		\hfill
		\begin{minipage}{3.5cm}
			\textit{No-signalling} \\[0.5cm]
			$P(Y|X)=P(Y)$ \\[0.2cm]
			$P(Y|do(X))=P(Y)$\\
			\vspace*{0.4cm}
		\end{minipage}
	\end{minipage}
\end{center}
\vspace*{-0.4cm}

%% file: Figures/Fig_Exampleqcm.tex
\vspace*{-0.5cm}

\begin{minipage}{9cm}
	\begin{minipage}{4cm}
		\centering
		\renewcommand{\Scale}{1.6}
		\begin{tikzpicture}
			\node [style=none](1)  at (0,0) {$A_1$};
			\node [style=none](1tl)  at (-0.1*\Scale,0.3*\Scale) {};
			\node [style=none](1tr)  at (0.0*\Scale,0.3*\Scale) {};
			\node [style=none](2)  at (-1.0*\Scale,1.2*\Scale) {$A_2$};
			\node [style=none](2b)  at (-1.0*\Scale,1.0*\Scale) {};
			\node [style=none](2t)  at (-0.9*\Scale,1.5*\Scale) {};
			\node [style=none](3)  at (1.0*\Scale,1.2*\Scale) {$A_3$};
			\node [style=none](3b)  at (0.9*\Scale,1.0*\Scale) {};
			\node [style=none](3br)  at (1.1*\Scale,1.0*\Scale) {};
			\node [style=none](3t)  at (0.9*\Scale,1.5*\Scale) {};
			\node [style=none](3tr)  at (1.0*\Scale,1.5*\Scale) {};
			\node [style=none](4)  at (0.0*\Scale,2.4*\Scale) {$A_4$};
			\node [style=none](4b)  at (0.0*\Scale,2.2*\Scale) {};
			\node [style=none](4t)  at (0.0*\Scale,2.7*\Scale) {};
			\node [style=none](5)  at (-1.9*\Scale,3.0*\Scale) {$A_5$};
			\node [style=none](5b)  at (-1.6*\Scale,2.9*\Scale) {};
			\draw [->] (1tl) to (2b);
			\draw [->] (1tr) to (3b);
			\draw [->] (2t) to (4b);
			\draw [->] (3t) to (4b);
			\draw [->] (4) to (5b);
			\draw [->] (1) to [bend left=55] (5);
		\end{tikzpicture}
	\end{minipage}
	\hfill
	\begin{minipage}{4cm}
		\begin{eqnarray}
			& \sigma_{A_1A_2A_3A_4A_5} \ =& \nonumber \\[0.1cm] 
			 & \rho_{A_5|A_1A_4} \ \rho_{A_4|A_2A_3} & \nonumber \\[0.1cm]  
				& \rho_{A_2|A_1} \ \rho_{A_3|A_1} & \nonumber \\[0.1cm]  
				& \rho_{A_1} & \nonumber 
		\end{eqnarray}
		\vspace*{0.3cm}
	\end{minipage}
\end{minipage}

\vspace*{-0.5cm}

%% file: Figures/Fig_Example_BrokenCiruit.tex
\renewcommand{\Scale}{1.8}
\newcommand{\FigFontSize}{9}
\newcommand{\FigFontSizeSkip}{12}
\begin{tikzpicture}

	\node at (0.75*\Scale,0.8*\Scale) {\fontsize{\FigFontSize}{\FigFontSizeSkip}$\mathcal{E}_1$};
	\draw (-0.5*\Scale, 0.5*\Scale) rectangle (2.0*\Scale, 1.1*\Scale);
	\node at (-1.25*\Scale,1.1*\Scale) {\fontsize{\FigFontSize}{\FigFontSizeSkip}$A$};
	\draw (-1.0*\Scale,0.0*\Scale) to (-1.0*\Scale, 0.9*\Scale);	
	\draw (-1.0*\Scale,1.3*\Scale) to (-1.0*\Scale, 2.1*\Scale);
	\node [style=smallcpoint] at (-1.0*\Scale,0.0*\Scale) {};
	\draw (0.3*\Scale,0) to (0.3*\Scale,0.5*\Scale);	
	\node [style=smallcpoint] at (0.3*\Scale,0.0*\Scale) {};
	\draw (1.2*\Scale,0) to (1.2*\Scale,0.5*\Scale);
	\node [style=smallcpoint] at (1.2*\Scale,0.0*\Scale) {};
	\draw (2.5*\Scale,0.0*\Scale) to (2.5*\Scale, 2.1*\Scale);
	\node [style=smallcpoint] at (2.5*\Scale,0.0*\Scale) {};
	
	\node at (-1.0*\Scale,2.4*\Scale) {\fontsize{\FigFontSize}{\FigFontSizeSkip}$\mathcal{E}_2$};
	\draw (-1.5*\Scale, 2.1*\Scale) rectangle (-0.5*\Scale, 2.7*\Scale);
	\node at (-1.25*\Scale,3.2*\Scale) {\fontsize{\FigFontSize}{\FigFontSizeSkip}$B$};
	\draw (-1.0*\Scale,2.7*\Scale) to (-1.0*\Scale, 3.0*\Scale);
	\draw (-1.0*\Scale,3.4*\Scale) to (-1.0*\Scale, 3.7*\Scale);
	
	\draw (0.0*\Scale,1.1*\Scale) to (0.0*\Scale,3.7*\Scale);
	
	\node at (1.65*\Scale,2.4*\Scale) {\fontsize{\FigFontSize}{\FigFontSizeSkip}$\mathcal{E}_3$};
	\draw (0.5*\Scale, 2.1*\Scale) rectangle (2.8*\Scale, 2.7*\Scale);
	\node at (1.25*\Scale,1.6*\Scale) {\fontsize{\FigFontSize}{\FigFontSizeSkip}$C$};
	\draw (1.5*\Scale,1.1*\Scale) to (1.5*\Scale, 1.4*\Scale);
	\draw (1.5*\Scale,1.8*\Scale) to (1.5*\Scale, 2.1*\Scale);
	\draw (0.8*\Scale,2.7*\Scale) to (0.8*\Scale,3.7*\Scale);	
	\node at (1.4*\Scale,3.2*\Scale) {\fontsize{\FigFontSize}{\FigFontSizeSkip}$D$};
	\draw (1.7*\Scale,2.7*\Scale) to (1.7*\Scale, 3.0*\Scale);
	\draw (1.7*\Scale,3.4*\Scale) to (1.7*\Scale, 3.7*\Scale);
	
	\node at (0.25*\Scale,4.0*\Scale) {\fontsize{\FigFontSize}{\FigFontSizeSkip}$\mathcal{E}_4$};
	\draw (-1.5*\Scale, 3.7*\Scale) rectangle (2.0*\Scale, 4.3*\Scale);

	\draw (-0.5*\Scale,4.3*\Scale) to (-0.5*\Scale, 4.8*\Scale);
	\node [style=upground] at (-0.5*\Scale,4.93*\Scale) {};
	\draw (1.0*\Scale,4.3*\Scale) to (1.0*\Scale, 4.8*\Scale);
	\node [style=upground] at (1.0*\Scale,4.93*\Scale) {};
	
	\node at (2.70*\Scale,3.7*\Scale) {\fontsize{\FigFontSize}{\FigFontSizeSkip}$E$};
	\draw (2.5*\Scale,2.7*\Scale) to (2.5*\Scale, 3.5*\Scale);
	\draw (2.5*\Scale,3.9*\Scale) to (2.5*\Scale, 4.8*\Scale);
	\node [style=upground] at (2.5*\Scale,4.93*\Scale) {};

\end{tikzpicture}

%% file: Figures/Fig_GenericBrokenUnitaryCiruit.tex
\renewcommand{\Scale}{2.4}
\begin{tikzpicture}

	\node [style=none] at (0.5*\Scale, -0.23*\Scale) {$\lambda_1$};
	\node [style=none] (a3) at (0.7*\Scale,-0.5*\Scale) {};
	\node [style=none] (a4) at (0.7*\Scale,-0.3*\Scale) {};
	\draw (a3) to (a4);
	\node [style=none] (ab3) at (0.7*\Scale,-0.2*\Scale) {};
	\node [style=none] (ab4) at (0.7*\Scale,0.025*\Scale) {};
	\draw (ab3) to (ab4);
	\node [style=smallcpoint] (a9) at (0.7*\Scale,-0.55*\Scale) {};
	\node [style=none] at (2.0*\Scale, -0.23*\Scale) {$\lambda_k$};
	\node [style=none] (a3) at (1.8*\Scale,-0.5*\Scale) {};
	\node [style=none] (a4) at (1.8*\Scale,-0.3*\Scale) {};
	\draw (a3) to (a4);
	\node [style=none] (ab3) at (1.8*\Scale,-0.2*\Scale) {};
	\node [style=none] (ab4) at (1.8*\Scale,0.025*\Scale) {};
	\draw (ab3) to (ab4);
	\node [style=smallcpoint] (a9) at (1.8*\Scale,-0.55*\Scale) {};
	\node [style=none] at (1.25*\Scale,-0.25*\Scale) {$\ldots$};
	\draw (0,0) rectangle (2.5*\Scale,0.6*\Scale);
	\node [style=none] at (1.25*\Scale, 0.3*\Scale) {$U_1$};
	
	\node [style=none] at (0.2*\Scale, 1.05*\Scale) {$A_1$};
	\node [style=none] (ba1) at (0.5*\Scale,1.15*\Scale) {};
	\node [style=none] (ba2) at (0.5*\Scale,1.525*\Scale) {};
	\draw (ba1) to (ba2);
	\node [style=none] (b1) at (0.5*\Scale,0.575*\Scale) {};
	\node [style=none] (b2) at (0.5*\Scale,0.95*\Scale) {};
	\draw (b1) to (b2);
	\node [style=none] at (2.3*\Scale, 1.05*\Scale) {$A_1'$};
	\node [style=none] (ba1) at (2.0*\Scale,1.15*\Scale) {};
	\node [style=none] (ba2) at (2.0*\Scale,1.525*\Scale) {};
	\node [style=none] (b1) at (2.0*\Scale,0.575*\Scale) {};
	\node [style=none] (b2) at (2.0*\Scale,0.95*\Scale) {};
	\draw (b1) to (ba2);
	\draw (0,1.5*\Scale) rectangle (2.5*\Scale,2.1*\Scale);
	\node [style=none] at (1.25*\Scale, 1.8*\Scale) {$U_2$};

	\node [style=none] at (0.2*\Scale, 2.4*\Scale) {$A_2$};
	\node [style=none] (c1) at (0.5*\Scale,2.075*\Scale) {};
	\node [style=none] (c2) at (0.5*\Scale,2.4*\Scale) {};
	\draw (c1) to (c2);
	\node [style=none] (ca1) at (0.5*\Scale,2.8*\Scale) {$\vdots$};
	\node [style=none] at 		(0.1*\Scale, 3.0*\Scale) {$A_{n-1}$};
	\node [style=none] (da1) at 	(0.5*\Scale,3.0*\Scale) {};
	\node [style=none] (da2) at 	(0.5*\Scale,3.325*\Scale) {};
	\draw (da1) to (da2);
	
	\node [style=none] at (2.3*\Scale, 2.4*\Scale) {$A_2'$};
	\node [style=none] (c5) at (2.0*\Scale,2.075*\Scale) {};
	\node [style=none] (c6) at (2.0*\Scale,2.4*\Scale) {};
	\draw (c5) to (c6);
	\node [style=none] (ca1) at 	(2.0*\Scale,2.8*\Scale) {$\vdots$};
	\node [style=none] at (2.4*\Scale, 3.0*\Scale) {$A_{n-1}'$};
	\node [style=none] (d13) at 	(2.0*\Scale,3.0*\Scale) {};
	\node [style=none] (d14) at 	(2.0*\Scale,3.325*\Scale) {};
	\draw (d13) to (d14);
	
	\draw (0,3.3*\Scale) rectangle (2.5*\Scale,3.9*\Scale);
	\node [style=none] at		(1.25*\Scale, 3.6*\Scale) {$U_{n}$};
	
	\node [style=none] (d3) at 	(0.5*\Scale,3.875*\Scale) {};
	\node [style=none] (d4) at 	(0.5*\Scale,4.25*\Scale) {};
	\draw (d3) to (d4);
	\node [style=none] at 		(0.2*\Scale, 4.35*\Scale) {$A_n$};
	\node [style=none] (d5) at 	(0.5*\Scale,4.45*\Scale) {};
	\node [style=none] (d6) at 	(0.5*\Scale,4.825*\Scale) {};
	\draw (d5) to (d6);
	\node [style=none] at 		(2.3*\Scale, 4.35*\Scale) {$A_n'$};
	\node [style=none] (d6) at 	(2.0*\Scale,3.875*\Scale) {};
	\node [style=none] (d7) at 	(2.0*\Scale,4.825*\Scale) {};
	\draw (d6) to (d7);
	\draw (0,4.8*\Scale) rectangle (2.5*\Scale,5.4*\Scale);
	\node [style=none] at		(1.25*\Scale, 5.1*\Scale) {$U_{n+1}$};
	
	\node [style=none] at 		(0.5*\Scale, 5.7*\Scale) {$F_1$};
	\node [style=none] (f1) at 	(0.7*\Scale,5.375*\Scale) {};
	\node [style=none] (f2) at 	(0.7*\Scale,5.6*\Scale) {};
	\draw (f1) to (f2);
	\node [style=none] (f3) at 	(0.7*\Scale,5.8*\Scale) {};
	\node [style=none] (f4) at 	(0.7*\Scale,6.02*\Scale) {};
	\draw (f3) to (f4);
	\node [style=upground] (d4) at 	(0.7*\Scale,6.1*\Scale) {};
	\node [style=none] at 		(2.0*\Scale, 5.7*\Scale) {$F_l$};
	\node [style=none] (f1) at 	(1.8*\Scale,5.375*\Scale) {};
	\node [style=none] (f2) at 	(1.8*\Scale,5.6*\Scale) {};
	\draw (f1) to (f2);
	\node [style=none] (f3) at 	(1.8*\Scale,5.8*\Scale) {};
	\node [style=none] (f4) at 	(1.8*\Scale,6.02*\Scale) {};
	\draw (f3) to (f4);
	\node [style=upground] (d4) at 	(1.8*\Scale,6.1*\Scale) {};
	\node [style=none] at (1.25*\Scale,5.7*\Scale) {$\ldots$};
\end{tikzpicture}

%% file: Figures/Fig_JustificationCircuit_NewNotation_2.tex
\renewcommand{\Scale}{2.4}
\begin{tikzpicture}

	\node [style=none] at (1.55*\Scale, -0.2*\Scale) {$\lambda_1$};
	\node [style=none] (a3) at (1.25*\Scale,-0.5*\Scale) {};
	\node [style=none] (a4) at (1.25*\Scale,-0.3*\Scale) {};
	\draw (a3) to (a4);
	\node [style=none] (ab3) at (1.25*\Scale,-0.2*\Scale) {};
	\node [style=none] (ab4) at (1.25*\Scale,0.025*\Scale) {};
	\draw (ab3) to (ab4);
	\node [style=smallcpoint] (a9) at (1.25*\Scale,-0.55*\Scale) {};
	\draw (0,0) rectangle (2.5*\Scale,0.6*\Scale);
	\node [style=none] at (1.25*\Scale, 0.3*\Scale) {$U_1$};
	
	\node [style=none] at (0.2*\Scale, 1.05*\Scale) {$A_1$};
	\node [style=none] (ba1) at (0.5*\Scale,1.15*\Scale) {};
	\node [style=none] (ba2) at (0.5*\Scale,1.525*\Scale) {};
	\draw (ba1) to (ba2);
	\node [style=none] (b1) at (0.5*\Scale,0.575*\Scale) {};
	\node [style=none] (b2) at (0.5*\Scale,0.95*\Scale) {};
	\draw (b1) to (b2);
	\node [style=none] at (1.55*\Scale, 1.3*\Scale) {$\lambda_2$};
	\node [style=none] (b3) at (1.25*\Scale,0.9*\Scale) {};
	\node [style=none] (b4) at (1.25*\Scale,1.2*\Scale) {};
	\draw (b3) to (b4);
	\node [style=none] (bb3) at (1.25*\Scale,1.3*\Scale) {};
	\node [style=none] (bb4) at (1.25*\Scale,1.525*\Scale) {};
	\draw (bb3) to (bb4);
	\node [style=smallcpoint] (b9) at (1.25*\Scale,0.95*\Scale) {};
	\node [style=none] at (2.3*\Scale, 1.05*\Scale) {$A_1'$};
	\node [style=none] (ba1) at (2.0*\Scale,1.15*\Scale) {};
	\node [style=none] (ba2) at (2.0*\Scale,1.525*\Scale) {};
	\node [style=none] (b1) at (2.0*\Scale,0.575*\Scale) {};
	\node [style=none] (b2) at (2.0*\Scale,0.95*\Scale) {};
	\draw (b1) to (ba2);
	\draw (0,1.5*\Scale) rectangle (2.5*\Scale,2.1*\Scale);
	\node [style=none] at (1.25*\Scale, 1.8*\Scale) {$U_2$};

	\node [style=none] at (0.2*\Scale, 2.4*\Scale) {$A_2$};
	\node [style=none] (c1) at (0.5*\Scale,2.075*\Scale) {};
	\node [style=none] (c2) at (0.5*\Scale,2.4*\Scale) {};
	\draw (c1) to (c2);
	\node [style=none] (ca1) at (0.5*\Scale,2.8*\Scale) {$\vdots$};
	\node [style=none] at 		(0.1*\Scale, 3.0*\Scale) {$A_{n-1}$};
	\node [style=none] (da1) at 	(0.5*\Scale,3.0*\Scale) {};
	\node [style=none] (da2) at 	(0.5*\Scale,3.325*\Scale) {};
	\draw (da1) to (da2);
	
	\node [style=none] at 		(1.55*\Scale, 3.1*\Scale) {$\lambda_n$};
	\node [style=none] (d7) at 	(1.25*\Scale,2.7*\Scale) {};
	\node [style=none] (d8) at 	(1.25*\Scale,3.0*\Scale) {};
	\draw (d7) to (d8);
	\node [style=none] (db7) at 	(1.25*\Scale,3.1*\Scale) {};
	\node [style=none] (db8) at 	(1.25*\Scale,3.325*\Scale) {};
	\draw (db7) to (db8);
	\node [style=smallcpoint] (d9) at (1.25*\Scale,2.75*\Scale) {};
	\node [style=none]  at 	(1.25*\Scale,2.5*\Scale) {$\vdots$};
	
	\node [style=none] at (2.3*\Scale, 2.4*\Scale) {$A_2'$};
	\node [style=none] (c5) at (2.0*\Scale,2.075*\Scale) {};
	\node [style=none] (c6) at (2.0*\Scale,2.4*\Scale) {};
	\draw (c5) to (c6);
	\node [style=none] (ca1) at 	(2.0*\Scale,2.8*\Scale) {$\vdots$};
	\node [style=none] at (2.4*\Scale, 3.0*\Scale) {$A_{n-1}'$};
	\node [style=none] (d13) at 	(2.0*\Scale,3.0*\Scale) {};
	\node [style=none] (d14) at 	(2.0*\Scale,3.325*\Scale) {};
	\draw (d13) to (d14);
	
	\draw (0,3.3*\Scale) rectangle (2.5*\Scale,3.9*\Scale);
	\node [style=none] at		(1.25*\Scale, 3.6*\Scale) {$U_{n}$};
	
	\node [style=none] (d3) at 	(0.5*\Scale,3.875*\Scale) {};
	\node [style=none] (d4) at 	(0.5*\Scale,4.25*\Scale) {};
	\draw (d3) to (d4);
	\node [style=none] at 		(0.2*\Scale, 4.35*\Scale) {$A_n$};
	\node [style=none] (d5) at 	(0.5*\Scale,4.45*\Scale) {};
	\node [style=none] (d6) at 	(0.5*\Scale,4.825*\Scale) {};
	\draw (d5) to (d6);
	\node [style=none] at 		(2.3*\Scale, 4.35*\Scale) {$A_n'$};
	\node [style=none] (d6) at 	(2.0*\Scale,3.875*\Scale) {};
	\node [style=none] (d7) at 	(2.0*\Scale,4.825*\Scale) {};
	\draw (d6) to (d7);
	\draw (0,4.8*\Scale) rectangle (2.5*\Scale,5.4*\Scale);
	\node [style=none] at		(1.25*\Scale, 5.1*\Scale) {$U_{n+1}$};
	
	\node [style=none] at 		(1.0*\Scale, 5.7*\Scale) {$F$};
	\node [style=none] (f1) at 	(1.25*\Scale,5.375*\Scale) {};
	\node [style=none] (f2) at 	(1.25*\Scale,5.6*\Scale) {};
	\draw (f1) to (f2);
	\node [style=none] (f3) at 	(1.25*\Scale,5.8*\Scale) {};
	\node [style=none] (f4) at 	(1.25*\Scale,6.02*\Scale) {};
	\draw (f3) to (f4);
	\node [style=upground] (d4) at 	(1.25*\Scale,6.1*\Scale) {};
\end{tikzpicture}
\renewcommand{\Scale}{2}

%% file: Figures/Fig_SchematicalOverviewCausalModels.tex
\begin{center}
	\begin{minipage}{16cm}
		\dbox{
		\begin{minipage}{3.6cm}
			\vspace*{0.1cm}
			\begin{center}
				\underline{QCM}
			\end{center}
			\small{
			\begin{itemize}[leftmargin=*]
				\item DAG $G$
				\item $i$th node: $\mathcal{H}_{A^{\text{in}}_i}$ , $\mathcal{H}_{A^{\text{out}}_i}$
				\item $\left\{ \rho_{A_i|\text{Pa}(A_i)} \right\}$	
			\end{itemize}
			}
			\vspace*{0.1cm}
		\end{minipage}
		}
		\hfill
		\begin{minipage}{2.0cm}
			\begin{center}
				\begin{tikzpicture}
					\node [style=none] at (2, 4.3) {
							\begin{minipage}{1.7cm}
								\begin{center}
									\small{induces in `all dia-\\ gonal' case}
								\end{center}
							\end{minipage}
								};
					\node [style=none] at (2, 6.4) {\small{$I_{\sigma\rightarrow\kappa}$}};
					\node [style=none] (1) at (0, 5.5) {};
					\node [style=none] (2) at (4, 5.5) {};
					\draw [->] (1) to [bend left=20] (2);
					\node [style=none] at (2, 1.5) {\small{induces}};
					\node [style=none] at (2, 0.2) {\small{$I_{\kappa\rightarrow\sigma}$}};
					\node [style=none] (1) at (0, 1.2) {};
					\node [style=none] (2) at (4, 1.2) {};
					\draw [->] (2) to [bend left=20] (1);
				\end{tikzpicture}
			\end{center}
		\end{minipage}
		\hfill
		\dbox{
		\begin{minipage}{3.4cm}
			\vspace*{0.1cm}
			\begin{center}
				\underline{CSM}
			\end{center}
			\small{
			\begin{itemize}[leftmargin=*]
				\item DAG $G$
				\item $i$th node: $X_i^{\text{in}}$ , $X_i^{\text{out}}$
				\item $\left\{ P(X_i^{\text{in}} |\text{Pa}(X_i)^{\text{out}}) \right\}$	
			\end{itemize}
			}
			\vspace*{0.1cm}
		\end{minipage}
		}
		\hfill
		\begin{minipage}{2.0cm}
			\begin{center}
				\begin{tikzpicture}
					\node [style=none] at (2, 4.3) {
							\begin{minipage}{1.7cm}
								\begin{center}
									\small{induces \\ for special intervention}
								\end{center}
							\end{minipage}
								};
					\node [style=none] at (2, 6.4) {\small{$I_{\kappa\rightarrow P}$}};
					\node [style=none] (1) at (0, 5.5) {};
					\node [style=none] (2) at (4, 5.5) {};
					\draw [->] (1) to [bend left=20] (2);
					\node [style=none] at (2, 1.5) {\small{induces}}; 
					\node [style=none] at (2, 0.2) {\small{$I_{CCM\rightarrow CSM}$}};
					\node [style=none] (1) at (0, 1.2) {};
					\node [style=none] (2) at (4, 1.2) {};
					\draw [->] (2) to [bend left=20] (1);
				\end{tikzpicture}
			\end{center}
		\end{minipage}
		\hfill
		\dbox{
		\begin{minipage}{2.8cm}
			\vspace*{0.1cm}
			\begin{center}
				\underline{CCM}
			\end{center}
			\small{
			\begin{itemize}[leftmargin=*]
				\item DAG $G$
				\item $i$th node: $X_i$ 
				\item $\left\{ P(X_i|\text{Pa}(X_i)) \right\}$	
			\end{itemize}
			}
			\vspace*{0.1cm}
		\end{minipage}
		}
	\end{minipage}
\end{center}

%% file: Figures/Fig_Example_DAGForRules.tex
\renewcommand{\Scale}{1.3}
\begin{tikzpicture}
	\node [style=none]  at (-3.0*\Scale,3.8*\Scale) {$G$};
	\node [style=none](1)  at (0.0*\Scale,0.5*\Scale) {$N_1$};
	\node [style=none](1tl)  at (-0.3*\Scale,0.8*\Scale) {};
	\node [style=none](1tr)  at (0.3*\Scale,0.8*\Scale) {};
	\node [style=none](2)  at (-2.0*\Scale,2.0*\Scale) {$N_2$};
	\node [style=none](2b)  at (-1.7*\Scale,1.6*\Scale) {};
	\node [style=none](2t)  at (-1.7*\Scale,2.4*\Scale) {};
	\node [style=none](2r)  at (-1.6*\Scale,2.0*\Scale) {};
	\node [style=none](3)  at (0.0*\Scale,2.0*\Scale) {$N_3$};
	\node [style=none](3l)  at (-0.5*\Scale,2.0*\Scale) {};
	\node [style=none](3r)  at (0.5*\Scale,2.0*\Scale) {};
	\node [style=none](4)  at (2.0*\Scale,2.0*\Scale) {$N_4$};
	\node [style=none](4b)  at (1.6*\Scale,1.6*\Scale) {};
	\node [style=none](4t)  at (1.6*\Scale,2.4*\Scale) {};
	\node [style=none](4l)  at (1.5*\Scale,2.0*\Scale) {};
	\node [style=none](5)  at (0.0*\Scale,3.5*\Scale) {$N_5$};
	\node [style=none](5bl)  at (-0.4*\Scale,3.2*\Scale) {};
	\node [style=none](5br)  at (0.4*\Scale,3.2*\Scale) {};
	\draw [->] (2b) to (1tl);
	\draw [->] (4b) to (1tr);
	\draw [->] (2t) to (5bl);
	\draw [->] (5br) to (4t);
	\draw [->] (3l) to (2r);
	\draw [->] (3r) to (4l);
\end{tikzpicture}

%% file: Figures/Fig_Example_Rule1_a.tex
\renewcommand{\Scale}{1.3}
\begin{tikzpicture}
	\node [style=none]  at (-3.0*\Scale,3.8*\Scale) {$G_{\overline{X}}$};
	\node [style=none](1)  at (0.0*\Scale,0.5*\Scale) {$N_1$};
	\node [style=none](1tl)  at (-0.3*\Scale,0.8*\Scale) {};
	\node [style=none](1tr)  at (0.3*\Scale,0.8*\Scale) {};
	\node [style=none](2)  at (-2.0*\Scale,2.0*\Scale) {$N_2$};
	\node [style=none](2b)  at (-1.7*\Scale,1.6*\Scale) {};
	\node [style=none](2t)  at (-1.7*\Scale,2.4*\Scale) {};
	\node [style=none](2r)  at (-1.6*\Scale,2.0*\Scale) {};
	\node [style=none](3)  at (0.0*\Scale,2.0*\Scale) {$N_3$};
	\node [style=none](3l)  at (-0.5*\Scale,2.0*\Scale) {};
	\node [style=none](3r)  at (0.5*\Scale,2.0*\Scale) {};
	\node [style=none](4)  at (2.0*\Scale,2.0*\Scale) {$N_4$};
	\node [style=none](4b)  at (1.6*\Scale,1.6*\Scale) {};
	\node [style=none](4t)  at (1.6*\Scale,2.4*\Scale) {};
	\node [style=none](4l)  at (1.5*\Scale,2.0*\Scale) {};
	\node [style=none](5)  at (0.0*\Scale,3.5*\Scale) {$N_5$};
	\node [style=none](5bl)  at (-0.4*\Scale,3.2*\Scale) {};
	\node [style=none](5br)  at (0.4*\Scale,3.2*\Scale) {};
	\draw [->] (2t) to (5bl);
	\draw [->] (5br) to (4t);
	\draw [->] (3l) to (2r);
	\draw [->] (3r) to (4l);
	\node [style=none]  at (-2.0*\Scale,-0.8*\Scale) {$Y=\{N_2\}$};
	\node [style=none]  at (-2.0*\Scale,-1.5*\Scale) {$Z=\{N_4\}$};
	\node [style=none]  at (2.0*\Scale,-0.8*\Scale) {$W=\{N_3,N_5\}$};
	\node [style=none]  at (1.5*\Scale,-1.5*\Scale) {$X=\{N_1\}$};
\end{tikzpicture}

	

%% file: Figures/Fig_Example_Rule1_b.tex
\renewcommand{\Scale}{1.3}
\begin{tikzpicture}
	\node [style=none]  at (-3.0*\Scale,3.8*\Scale) {$G_{\overline{X'}}$};
	\node [style=none](1)  at (0.0*\Scale,0.5*\Scale) {$N_1$};
	\node [style=none](1tl)  at (-0.3*\Scale,0.8*\Scale) {};
	\node [style=none](1tr)  at (0.3*\Scale,0.8*\Scale) {};
	\node [style=none](2)  at (-2.0*\Scale,2.0*\Scale) {$N_2$};
	\node [style=none](2b)  at (-1.7*\Scale,1.6*\Scale) {};
	\node [style=none](2t)  at (-1.7*\Scale,2.4*\Scale) {};
	\node [style=none](2r)  at (-1.6*\Scale,2.0*\Scale) {};
	\node [style=none](3)  at (0.0*\Scale,2.0*\Scale) {$N_3$};
	\node [style=none](3l)  at (-0.5*\Scale,2.0*\Scale) {};
	\node [style=none](3r)  at (0.5*\Scale,2.0*\Scale) {};
	\node [style=none](4)  at (2.0*\Scale,2.0*\Scale) {$N_4$};
	\node [style=none](4b)  at (1.6*\Scale,1.6*\Scale) {};
	\node [style=none](4t)  at (1.6*\Scale,2.4*\Scale) {};
	\node [style=none](4l)  at (1.5*\Scale,2.0*\Scale) {};
	\node [style=none](5)  at (0.0*\Scale,3.5*\Scale) {$N_5$};
	\node [style=none](5bl)  at (-0.4*\Scale,3.2*\Scale) {};
	\node [style=none](5br)  at (0.4*\Scale,3.2*\Scale) {};
	\draw [->] (2b) to (1tl);
	\draw [->] (4b) to (1tr);
	\draw [->] (5br) to (4t);
	\draw [->] (3l) to (2r);
	\draw [->] (3r) to (4l);
	\node [style=none]  at (-2.0*\Scale,-0.8*\Scale) {$Y=\{N_2\}$};
	\node [style=none]  at (-2.0*\Scale,-1.5*\Scale) {$Z=\{N_4\}$};
	\node [style=none]  at (2.0*\Scale,-0.8*\Scale) {$W'=\{N_1,N_3\}$};
	\node [style=none]  at (1.5*\Scale,-1.5*\Scale) {$X'=\{N_5\}$};;
\end{tikzpicture}

%% file: Figures/Fig_Example_Rule2_a.tex
\renewcommand{\Scale}{1.3}
\begin{tikzpicture}
	\node [style=none]  at (-3.0*\Scale,3.8*\Scale) {$G_{\overline{X},\underline{Z}}$};
	\node [style=none]  at (3.5*\Scale,3.8*\Scale) {};
	\node [style=none](1)  at (0.0*\Scale,0.5*\Scale) {$N_1$};
	\node [style=none](1tl)  at (-0.3*\Scale,0.8*\Scale) {};
	\node [style=none](1tr)  at (0.3*\Scale,0.8*\Scale) {};
	\node [style=none](2)  at (-2.0*\Scale,2.0*\Scale) {$N_2$};
	\node [style=none](2b)  at (-1.7*\Scale,1.6*\Scale) {};
	\node [style=none](2t)  at (-1.7*\Scale,2.4*\Scale) {};
	\node [style=none](2r)  at (-1.6*\Scale,2.0*\Scale) {};
	\node [style=none](3)  at (0.0*\Scale,2.0*\Scale) {$N_3$};
	\node [style=none](3l)  at (-0.5*\Scale,2.0*\Scale) {};
	\node [style=none](3r)  at (0.5*\Scale,2.0*\Scale) {};
	\node [style=none](4)  at (2.0*\Scale,2.0*\Scale) {$N_4$};
	\node [style=none](4b)  at (1.6*\Scale,1.6*\Scale) {};
	\node [style=none](4t)  at (1.6*\Scale,2.4*\Scale) {};
	\node [style=none](4l)  at (1.5*\Scale,2.0*\Scale) {};
	\node [style=none](5)  at (0.0*\Scale,3.5*\Scale) {$N_5$};
	\node [style=none](5bl)  at (-0.4*\Scale,3.2*\Scale) {};
	\node [style=none](5br)  at (0.4*\Scale,3.2*\Scale) {};
	\draw [->] (5br) to (4t);
	\draw [->] (3l) to (2r);
	\draw [->] (3r) to (4l);
\end{tikzpicture}

	

%% file: Figures/Fig_Example_Rule2_b.tex
\renewcommand{\Scale}{1.3}
\begin{tikzpicture}
	\node [style=none]  at (-3.0*\Scale,3.8*\Scale) {$G_{\overline{X},\underline{Y}}$};
	\node [style=none]  at (3.5*\Scale,3.8*\Scale) {};
	\node [style=none](1)  at (0.0*\Scale,0.5*\Scale) {$N_1$};
	\node [style=none](1tl)  at (-0.3*\Scale,0.8*\Scale) {};
	\node [style=none](1tr)  at (0.3*\Scale,0.8*\Scale) {};
	\node [style=none](2)  at (-2.0*\Scale,2.0*\Scale) {$N_2$};
	\node [style=none](2b)  at (-1.7*\Scale,1.6*\Scale) {};
	\node [style=none](2t)  at (-1.7*\Scale,2.4*\Scale) {};
	\node [style=none](2r)  at (-1.6*\Scale,2.0*\Scale) {};
	\node [style=none](3)  at (0.0*\Scale,2.0*\Scale) {$N_3$};
	\node [style=none](3l)  at (-0.5*\Scale,2.0*\Scale) {};
	\node [style=none](3r)  at (0.5*\Scale,2.0*\Scale) {};
	\node [style=none](4)  at (2.0*\Scale,2.0*\Scale) {$N_4$};
	\node [style=none](4b)  at (1.6*\Scale,1.6*\Scale) {};
	\node [style=none](4t)  at (1.6*\Scale,2.4*\Scale) {};
	\node [style=none](4l)  at (1.5*\Scale,2.0*\Scale) {};
	\node [style=none](5)  at (0.0*\Scale,3.5*\Scale) {$N_5$};
	\node [style=none](5bl)  at (-0.4*\Scale,3.2*\Scale) {};
	\node [style=none](5br)  at (0.4*\Scale,3.2*\Scale) {};
	\draw [->] (2t) to (5bl);
	\draw [->] (3l) to (2r);
	\draw [->] (3r) to (4l);
\end{tikzpicture}

%% file: Figures/Fig_Example_Rule3_a.tex
\renewcommand{\Scale}{1.3}
\begin{tikzpicture}
	\node [style=none]  at (-3.0*\Scale,3.8*\Scale) {$G_{\overline{X},\overline{Z(W)}}$};
	\node [style=none](1)  at (0.0*\Scale,0.5*\Scale) {$N_1$};
	\node [style=none](1tl)  at (-0.3*\Scale,0.8*\Scale) {};
	\node [style=none](1tr)  at (0.3*\Scale,0.8*\Scale) {};
	\node [style=none](2)  at (-2.0*\Scale,2.0*\Scale) {$N_2$};
	\node [style=none](2b)  at (-1.7*\Scale,1.6*\Scale) {};
	\node [style=none](2t)  at (-1.7*\Scale,2.4*\Scale) {};
	\node [style=none](2r)  at (-1.6*\Scale,2.0*\Scale) {};
	\node [style=none](3)  at (0.0*\Scale,2.0*\Scale) {$N_3$};
	\node [style=none](3l)  at (-0.5*\Scale,2.0*\Scale) {};
	\node [style=none](3r)  at (0.5*\Scale,2.0*\Scale) {};
	\node [style=none](4)  at (2.0*\Scale,2.0*\Scale) {$N_4$};
	\node [style=none](4b)  at (1.6*\Scale,1.6*\Scale) {};
	\node [style=none](4t)  at (1.6*\Scale,2.4*\Scale) {};
	\node [style=none](4l)  at (1.5*\Scale,2.0*\Scale) {};
	\node [style=none](5)  at (0.0*\Scale,3.5*\Scale) {$N_5$};
	\node [style=none](5bl)  at (-0.4*\Scale,3.2*\Scale) {};
	\node [style=none](5br)  at (0.4*\Scale,3.2*\Scale) {};
	\draw [->] (3l) to (2r);
	\node [style=none]  at (-2.0*\Scale,-0.8*\Scale) {$Z=\{N_4,N_5\}$};
	\node [style=none]  at (-2.5*\Scale,-1.5*\Scale) {$Y=\{N_2\}$};
	\node [style=none]  at (1.5*\Scale,-0.8*\Scale) {$W=\{N_3\}$};
	\node [style=none]  at (1.5*\Scale,-1.5*\Scale) {$X=\{N_1\}$};
\end{tikzpicture}

	

%% file: Figures/Fig_Example_Rule3_b.tex
\renewcommand{\Scale}{1.3}
\begin{tikzpicture}
	\node [style=none]  at (-3.0*\Scale,3.8*\Scale) {$G_{\overline{X},\overline{Z'(W')}}$};
	\node [style=none](1)  at (0.0*\Scale,0.5*\Scale) {$N_1$};
	\node [style=none](1tl)  at (-0.3*\Scale,0.8*\Scale) {};
	\node [style=none](1tr)  at (0.3*\Scale,0.8*\Scale) {};
	\node [style=none](2)  at (-2.0*\Scale,2.0*\Scale) {$N_2$};
	\node [style=none](2b)  at (-1.7*\Scale,1.6*\Scale) {};
	\node [style=none](2t)  at (-1.7*\Scale,2.4*\Scale) {};
	\node [style=none](2r)  at (-1.6*\Scale,2.0*\Scale) {};
	\node [style=none](3)  at (0.0*\Scale,2.0*\Scale) {$N_3$};
	\node [style=none](3l)  at (-0.5*\Scale,2.0*\Scale) {};
	\node [style=none](3r)  at (0.5*\Scale,2.0*\Scale) {};
	\node [style=none](4)  at (2.0*\Scale,2.0*\Scale) {$N_4$};
	\node [style=none](4b)  at (1.6*\Scale,1.6*\Scale) {};
	\node [style=none](4t)  at (1.6*\Scale,2.4*\Scale) {};
	\node [style=none](4l)  at (1.5*\Scale,2.0*\Scale) {};
	\node [style=none](5)  at (0.0*\Scale,3.5*\Scale) {$N_5$};
	\node [style=none](5bl)  at (-0.4*\Scale,3.2*\Scale) {};
	\node [style=none](5br)  at (0.4*\Scale,3.2*\Scale) {};
	\draw [->] (2t) to (5bl);
	\draw [->] (5br) to (4t);
	\draw [->] (3l) to (2r);
	\draw [->] (3r) to (4l);
	\node [style=none]  at (-2.0*\Scale,-0.8*\Scale) {$Z'=\{N_5\}$};
	\node [style=none]  at (-2.0*\Scale,-1.5*\Scale) {$Y=\{N_2\}$};
	\node [style=none]  at (2.0*\Scale,-0.8*\Scale) {$W'=\{N_3,N_4\}$};
	\node [style=none]  at (1.5*\Scale,-1.5*\Scale) {$X=\{N_1\}$};
\end{tikzpicture}

%% file: Figures/Fig_DAGProofJustification.tex
\renewcommand{\Scale}{1.4}
\begin{tikzpicture}
	\node [style=none]  at (0, 0) {$\lambda_1$};
	\node [style=none] (1) at (0, 0.15*\Scale) {};
	\node [style=none] (1c) at (-0.2*\Scale, 0.0*\Scale) {};
	\node [style=none] (2) at (0, \Scale) {$A_1$};
	\node [style=none] (2b) at (0, 1.2*\Scale) {};
	\node [style=none] (2c) at (-0.2*\Scale, 1.0*\Scale) {};
	\node [style=none] (2d) at (0.2*\Scale, 1.0*\Scale) {};
	
	\node [style=none] (3) at (0, 2*\Scale) {$\lambda_2$};
	\node [style=none] (3b) at (0, 2.15*\Scale) {};
	\node [style=none] (3c) at (-0.2*\Scale, 2.0*\Scale) {};
	\node [style=none] (4) at (0, 3*\Scale) {$A_2$};
	\node [style=none] (4b) at (0, 3.15*\Scale) {};
	\node [style=none] (4c) at (-0.2*\Scale, 3.0*\Scale) {};
	\node [style=none] (4d) at (0.2*\Scale, 3.0*\Scale) {};
	
	\node [style=none] at (0, 4.3*\Scale) {$\vdots$};
	
	\node [style=none] (5) at (0, 5.0*\Scale) {$\lambda_n$};
	\node [style=none] (5b) at (0, 5.15*\Scale) {};
	\node [style=none] (5c) at (-0.2*\Scale, 5.0*\Scale) {};
	\node [style=none] (6) at (0, 6.0*\Scale) {$A_n$};
	\node [style=none] (6b) at (0.1*\Scale, 5.8*\Scale) {};
	\node [style=none] (6c) at (-0.2*\Scale, 6.0*\Scale) {};
	
	\node [style=none] (7) at (0, 7.0*\Scale) {$F$};
	\node [style=none] (7b) at (0, 7.15*\Scale) {};
	
	\draw [->] (1) to (2);
	\draw [->] (3b) to (4);
	\draw [->] (5b) to (6);
	
	\draw [->] (2d) to [bend right=30] (4);
	\draw [->] (2d) to [bend right=45] (6b);
	\draw [->] (4d) to [bend right=30] (6b);
	\node [style=none, rotate=-45] at (0.7*\Scale, 2.8*\Scale) {$\vdots$};
	
	\draw [->] (6c) to [bend left=25] (7);
	\draw [->] (5c) to [bend left=35] (7);
	\draw [->] (4c) to [bend left=35] (7);
	\draw [->] (3c) to [bend left=35] (7);
	\draw [->] (2c) to [bend left=35] (7);
	\draw [->] (1c) to [bend left=35] (7);
	
\end{tikzpicture}

%% file: Figures/Fig_DSepSubSets_Statement.tex
\begin{tikzpicture}
	\node [style=none] (1)  at (0,0) {$R^c$};
	\node [style=none] (2)  at (-2*\Scale,1*\Scale) {$Y$};
	\node [style=none] (2b)  at (-2*\Scale,0.85*\Scale) {};
	\node [style=none] (2t)  at (-2*\Scale,1.2*\Scale) {};
	\node [style=none] (3)  at (-0.8*\Scale,1*\Scale) {$R_Y$};
	\node [style=none] (3b)  at (-0.8*\Scale,0.85*\Scale) {};
	\node [style=none] (3t)  at (-0.8*\Scale,1.2*\Scale) {};
	\node [style=none] (4)  at (0.8*\Scale,1*\Scale) {$R_Z$};
	\node [style=none] (4b)  at (0.8*\Scale,0.85*\Scale) {};
	\node [style=none] (4t)  at (0.8*\Scale,1.2*\Scale) {};
	\node [style=none] (5)  at (2*\Scale,1*\Scale) {$Z$};
	\node [style=none] (5b)  at (2*\Scale,0.85*\Scale) {};
	\node [style=none] (5t)  at (2*\Scale,1.2*\Scale) {};
	\node [style=none] (6)  at (-1.5*\Scale,2*\Scale) {$W_Y$};
	\node [style=none] (6bl)  at (-1.6*\Scale,1.85*\Scale) {};
	\node [style=none] (6br)  at (-1.4*\Scale,1.85*\Scale) {};
	\node [style=none] (7)  at (1.5*\Scale,2*\Scale) {$W_Z$};
	\node [style=none] (7bl)  at (1.4*\Scale,1.85*\Scale) {};
	\node [style=none] (7br)  at (1.6*\Scale,1.85*\Scale) {};
	\draw [->] (2b) to [bend right=20] (1);
	\draw [->] (3b) to [bend right=20] (1);
	\draw [->] (4b) to [bend left=20] (1);
	\draw [->] (5b) to [bend left=20] (1);
	\draw [->] (2) to [bend left=30] (3);
	\draw [->] (3) to [bend left=30] (2);
	\draw [->] (4) to [bend left=30] (5);
	\draw [->] (5) to [bend left=30] (4);
	\draw [->] (2t) to [bend left=20] (6bl);
	\draw [->] (3t) to [bend right=20] (6br);
	\draw [->] (4t) to [bend left=20] (7bl);
	\draw [->] (5t) to [bend right=20] (7br);
\end{tikzpicture}

%% file: Figures/Fig_DSepSubSets_Proof_LMC.tex
\begin{tikzpicture}
	\node [style=none] (1)  at (0,0) {$D$};
	\node [style=none] (2)  at (-2*\Scale,1*\Scale) {$X$};
	\node [style=none] (2b)  at (-2*\Scale,0.8*\Scale) {};
	\node [style=none] (2tr)  at (-1.9*\Scale,1.3*\Scale) {};
	\node [style=none] (5)  at (2*\Scale,1*\Scale) {$N$};
	\node [style=none] (5b)  at (2*\Scale,0.8*\Scale) {};
	\node [style=none] (5tl)  at (1.95*\Scale,1.3*\Scale) {};
	\node [style=none] (5tr)  at (2.05*\Scale,1.3*\Scale) {};
	\node [style=none] (7)  at (1.5*\Scale,2*\Scale) {$P$};
	\node [style=none] (7bl)  at (1.3*\Scale,2.0*\Scale) {};
	\node [style=none] (7br)  at (1.4*\Scale,1.8*\Scale) {};
	\draw [->] (2b) to [bend right=20] (1);
	\draw [->] (5b) to [bend left=20] (1);
	\draw [->] (7) to [bend left=20] (5tr);
	\draw [->] (5tl) to [bend left=20] (7);
	\draw [->] (7bl) to [bend right=20] (2tr);
	\draw [->] (7br) to [bend right=20] (1);
\end{tikzpicture}

%% file: Figures/Fig_DSepSubSets_Proof_Contraction_1.tex
\begin{tikzpicture}
	\node [style=none] (1)  at (0,0) {$R^c X^c $};
	\node [style=none] (1tll)  at (-0.4*\Scale,0.0*\Scale) {};
	\node [style=none] (1tl)  at (-0.3*\Scale,0.15*\Scale) {};
	\node [style=none] (1tr)  at (0.3*\Scale,0.15*\Scale) {};
	\node [style=none] (1trr)  at (0.45*\Scale,0.0*\Scale) {};
	\node [style=none] (2)  at (-2*\Scale,1*\Scale) {$Y$};
	\node [style=none] (2b)  at (-2*\Scale,0.85*\Scale) {};
	\node [style=none] (2t)  at (-2*\Scale,1.2*\Scale) {};
	\node [style=none] (2tr)  at (-1.85*\Scale,1.1*\Scale) {};
	\node [style=none] (2br)  at (-1.85*\Scale,0.95*\Scale) {};
	\node [style=none] (3)  at (-0.8*\Scale,1*\Scale) {\small{$R_Y X_Y$}};
	\node [style=none] (3b)  at (-0.8*\Scale,0.85*\Scale) {};
	\node [style=none] (3t)  at (-0.8*\Scale,1.2*\Scale) {};
	\node [style=none] (3bl)  at (-1.25*\Scale,0.95*\Scale) {};
	\node [style=none] (3tl)  at (-1.25*\Scale,1.1*\Scale) {};
	\node [style=none] (4)  at (0.8*\Scale,1*\Scale) {\small{$R_Z  X_Z$}};
	\node [style=none] (4b)  at (0.8*\Scale,0.85*\Scale) {};
	\node [style=none] (4t)  at (0.8*\Scale,1.2*\Scale) {};
	\node [style=none] (4tr)  at (1.85*\Scale,1.1*\Scale) {};
	\node [style=none] (4br)  at (1.85*\Scale,0.95*\Scale) {};
	\node [style=none] (5)  at (2*\Scale,1*\Scale) {$Z$};
	\node [style=none] (5b)  at (2*\Scale,0.85*\Scale) {};
	\node [style=none] (5t)  at (2*\Scale,1.2*\Scale) {};
	\node [style=none] (5bl)  at (1.25*\Scale,0.95*\Scale) {};
	\node [style=none] (5tl)  at (1.25*\Scale,1.1*\Scale) {};
	\node [style=none] (6)  at (-1.5*\Scale,2*\Scale) {$W_Y$};
	\node [style=none] (6bl)  at (-1.6*\Scale,1.8*\Scale) {};
	\node [style=none] (6br)  at (-1.4*\Scale,1.8*\Scale) {};
	\node [style=none] (7)  at (1.5*\Scale,2*\Scale) {$W_Z$};
	\node [style=none] (7bl)  at (1.4*\Scale,1.8*\Scale) {};
	\node [style=none] (7br)  at (1.6*\Scale,1.8*\Scale) {};
	\draw [->] (2b) to [bend right=20] (1tll);
	\draw [->] (3b) to [bend right=20] (1tl);
	\draw [->] (4b) to [bend left=20] (1tr);
	\draw [->] (5b) to [bend left=20] (1trr);
	\draw [->] (2tr) to [bend left=30] (3tl);
	\draw [->] (3bl) to [bend left=30] (2br);
	\draw [->] (4tr) to [bend right=30] (5tl);
	\draw [->] (5bl) to [bend right=30] (4br);
	\draw [->] (2t) to [bend left=20] (6bl);
	\draw [->] (3t) to [bend right=20] (6br);
	\draw [->] (4t) to [bend left=20] (7bl);
	\draw [->] (5t) to [bend right=20] (7br);
\end{tikzpicture}

%% file: Figures/Fig_DSepSubSets_Proof_Contraction_2.tex
\begin{tikzpicture}
	\node [style=none] (1)  at (0,0) {$R^{'c}$};
	\node [style=none] (2)  at (-2*\Scale,1*\Scale) {$Y$};
	\node [style=none] (2b)  at (-2*\Scale,0.85*\Scale) {};
	\node [style=none] (2t)  at (-2*\Scale,1.2*\Scale) {};
	\node [style=none] (2tr)  at (-1.85*\Scale,1.1*\Scale) {};
	\node [style=none] (2br)  at (-1.85*\Scale,0.95*\Scale) {};
	\node [style=none] (3)  at (-0.8*\Scale,1*\Scale) {\small{$R'_Y$}};
	\node [style=none] (3b)  at (-0.8*\Scale,0.85*\Scale) {};
	\node [style=none] (3t)  at (-0.8*\Scale,1.2*\Scale) {};
	\node [style=none] (3bl)  at (-1.0*\Scale,0.95*\Scale) {};
	\node [style=none] (3tl)  at (-1.0*\Scale,1.1*\Scale) {};
	\node [style=none] (4)  at (0.8*\Scale,1*\Scale) {\small{$R_X$}};
	\node [style=none] (4b)  at (0.8*\Scale,0.85*\Scale) {};
	\node [style=none] (4t)  at (0.8*\Scale,1.2*\Scale) {};
	\node [style=none] (4tr)  at (1.85*\Scale,1.1*\Scale) {};
	\node [style=none] (4br)  at (1.85*\Scale,0.95*\Scale) {};
	\node [style=none] (5)  at (2*\Scale,1*\Scale) {$X$};
	\node [style=none] (5b)  at (2*\Scale,0.85*\Scale) {};
	\node [style=none] (5t)  at (2*\Scale,1.2*\Scale) {};
	\node [style=none] (5bl)  at (1.0*\Scale,0.95*\Scale) {};
	\node [style=none] (5tl)  at (1.0*\Scale,1.1*\Scale) {};
	\node [style=none] (6)  at (-1.5*\Scale,2*\Scale) {$W'_Y Z_Y$};
	\node [style=none] (6bl)  at (-1.6*\Scale,1.8*\Scale) {};
	\node [style=none] (6br)  at (-1.4*\Scale,1.8*\Scale) {};
	\node [style=none] (7)  at (1.5*\Scale,2*\Scale) {$W_X Z_X$};
	\node [style=none] (7bl)  at (1.4*\Scale,1.8*\Scale) {};
	\node [style=none] (7br)  at (1.6*\Scale,1.8*\Scale) {};
	\draw [->] (2b) to [bend right=20] (1);
	\draw [->] (3b) to [bend right=20] (1);
	\draw [->] (4b) to [bend left=20] (1);
	\draw [->] (5b) to [bend left=20] (1);
	\draw [->] (2tr) to [bend left=30] (3tl);
	\draw [->] (3bl) to [bend left=30] (2br);
	\draw [->] (4tr) to [bend right=30] (5tl);
	\draw [->] (5bl) to [bend right=30] (4br);
	\draw [->] (2t) to [bend left=20] (6bl);
	\draw [->] (3t) to [bend right=20] (6br);
	\draw [->] (4t) to [bend left=20] (7bl);
	\draw [->] (5t) to [bend right=20] (7br);
\end{tikzpicture}

%% file: Figures/Fig_DSepSubSets_Proof_Contraction_3.tex
\begin{tikzpicture}
	\node [style=none] (1)  at (0,0) {$\widetilde{R}^c$};
	\node [style=none] (2)  at (-2*\Scale,1*\Scale) {$Y$};
	\node [style=none] (2b)  at (-2*\Scale,0.85*\Scale) {};
	\node [style=none] (2t)  at (-2*\Scale,1.2*\Scale) {};
	\node [style=none] (2tr)  at (-1.85*\Scale,1.1*\Scale) {};
	\node [style=none] (2br)  at (-1.85*\Scale,0.95*\Scale) {};
	\node [style=none] (3)  at (-0.8*\Scale,1*\Scale) {\small{$\widetilde{R}_Y$}};
	\node [style=none] (3b)  at (-0.8*\Scale,0.85*\Scale) {};
	\node [style=none] (3t)  at (-0.8*\Scale,1.2*\Scale) {};
	\node [style=none] (3bl)  at (-1.0*\Scale,0.95*\Scale) {};
	\node [style=none] (3tl)  at (-1.0*\Scale,1.1*\Scale) {};
	\node [style=none] (4)  at (0.8*\Scale,1*\Scale) {\small{$R_{XZ}$}};
	\node [style=none] (4b)  at (0.8*\Scale,0.85*\Scale) {};
	\node [style=none] (4t)  at (0.8*\Scale,1.2*\Scale) {};
	\node [style=none] (4tr)  at (1.75*\Scale,1.1*\Scale) {};
	\node [style=none] (4br)  at (1.75*\Scale,0.95*\Scale) {};
	\node [style=none] (5)  at (2*\Scale,1*\Scale) {$XZ$};
	\node [style=none] (5b)  at (2*\Scale,0.85*\Scale) {};
	\node [style=none] (5t)  at (2*\Scale,1.2*\Scale) {};
	\node [style=none] (5bl)  at (1.1*\Scale,0.95*\Scale) {};
	\node [style=none] (5tl)  at (1.1*\Scale,1.1*\Scale) {};		
	\node [style=none] (6)  at (-1.5*\Scale,2*\Scale) {$\widetilde{W}_Y$};
	\node [style=none] (6bl)  at (-1.6*\Scale,1.8*\Scale) {};
	\node [style=none] (6br)  at (-1.4*\Scale,1.8*\Scale) {};
	\node [style=none] (7)  at (1.5*\Scale,2*\Scale) {$W_{XZ}$};
	\node [style=none] (7bl)  at (1.4*\Scale,1.8*\Scale) {};
	\node [style=none] (7br)  at (1.6*\Scale,1.8*\Scale) {};
	\draw [->] (2b) to [bend right=20] (1);
	\draw [->] (3b) to [bend right=20] (1);
	\draw [->] (4b) to [bend left=20] (1);
	\draw [->] (5b) to [bend left=20] (1);
	\draw [->] (2tr) to [bend left=30] (3tl);
	\draw [->] (3bl) to [bend left=30] (2br);
	\draw [->] (4tr) to [bend right=30] (5tl);
	\draw [->] (5bl) to [bend right=30] (4br);
	\draw [->] (2t) to [bend left=20] (6bl);
	\draw [->] (3t) to [bend right=20] (6br);
	\draw [->] (4t) to [bend left=20] (7bl);
	\draw [->] (5t) to [bend right=20] (7br);
\end{tikzpicture}

%% file: Figures/Fig_Rule2SubsetRelation.tex
\begin{tikzpicture}
	\node [style=none] (1)  at (0,0) {$R^{c}$};
	\node [style=none] (2)  at (-2*\Scale,1*\Scale) {$Y$};
	\node [style=none] (2b)  at (-2*\Scale,0.85*\Scale) {};
	\node [style=none] (2t)  at (-2*\Scale,1.2*\Scale) {};
	\node [style=none] (2tr)  at (-1.85*\Scale,1.1*\Scale) {};
	\node [style=none] (2br)  at (-1.85*\Scale,0.95*\Scale) {};
	\node [style=none] (2bm)  at (-1.95*\Scale,0.95*\Scale) {};
	\node [style=none] (3)  at (-0.8*\Scale,1*\Scale) {\small{$R_Y$}};
	\node [style=none] (3b)  at (-0.8*\Scale,0.85*\Scale) {};
	\node [style=none] (3t)  at (-0.8*\Scale,1.2*\Scale) {};
	\node [style=none] (3bl)  at (-1.0*\Scale,0.95*\Scale) {};
	\node [style=none] (3tl)  at (-1.0*\Scale,1.1*\Scale) {};
	\node [style=none] (4)  at (0.8*\Scale,1*\Scale) {\small{$R_Z$}};
	\node [style=none] (4b)  at (0.8*\Scale,0.85*\Scale) {};
	\node [style=none] (4t)  at (0.8*\Scale,1.2*\Scale) {};
	\node [style=none] (4tr)  at (1.85*\Scale,1.1*\Scale) {};
	\node [style=none] (4br)  at (1.85*\Scale,0.95*\Scale) {};
	\node [style=none] (5)  at (2*\Scale,1*\Scale) {$Z$};
	\node [style=none] (5b)  at (2*\Scale,0.85*\Scale) {};
	\node [style=none] (5t)  at (2*\Scale,1.2*\Scale) {};
	\node [style=none] (5bl)  at (1.0*\Scale,0.95*\Scale) {};
	\node [style=none] (5tl)  at (1.0*\Scale,1.1*\Scale) {};
	\node [style=none] (6)  at (-1.5*\Scale,2*\Scale) {$W_Y$};
	\node [style=none] (6r)  at (-1.2*\Scale,2*\Scale) {};
	\node [style=none] (6bl)  at (-1.6*\Scale,1.8*\Scale) {};
	\node [style=none] (6br)  at (-1.4*\Scale,1.8*\Scale) {};
	\node [style=none] (7)  at (1.5*\Scale,2*\Scale) {$W_Z$};
	\node [style=none] (7bl)  at (1.4*\Scale,1.8*\Scale) {};
	\node [style=none] (7br)  at (1.6*\Scale,1.8*\Scale) {};
	\draw [->] (2b) to [bend right=20] (1);
	\draw [->] (3b) to [bend right=20] (1);
	\draw [->] (4b) to [bend left=20] (1);
	\draw [blue,dashed,->] (5b) to [bend left=20] (1);
	\draw [blue,dashed,->] (5b) to [bend left=30] (2bm);
	\draw [blue,dashed,->] (5t) to [bend right=20] (3t);
	\draw [blue,dashed,->] (5t) to [bend right=20] (6r);
	\draw [->] (2tr) to [bend left=30] (3tl);
	\draw [->] (3bl) to [bend left=30] (2br);
	\draw [blue,dashed,->] (4tr) to [bend right=30] (5tl);
	\draw [->] (5bl) to [bend right=30] (4br);
	\draw [->] (2t) to [bend left=20] (6bl);
	\draw [->] (3t) to [bend right=20] (6br);
	\draw [->] (4t) to [bend left=20] (7bl);
	\draw [blue,dashed,->] (5t) to [bend right=20] (7br);
\end{tikzpicture}

%% file: Figures/Fig_Rule3SubsetRelation_1.tex
\begin{tikzpicture}
	\node [style=none] (1)  at (0,-0.5*\Scale) {$R^{\phantom{'}c}$};
	\node [style=none] (2)  at (-2*\Scale,1*\Scale) {$Y$};
	\node [style=none] (2b)  at (-2*\Scale,0.85*\Scale) {};
	\node [style=none] (2t)  at (-2*\Scale,1.2*\Scale) {};
	\node [style=none] (2tr)  at (-1.85*\Scale,1.1*\Scale) {};
	\node [style=none] (2br)  at (-1.85*\Scale,0.95*\Scale) {};
	\node [style=none] (2brb)  at (-1.95*\Scale,0.9*\Scale) {};
	\node [style=none] (2bm)  at (-1.95*\Scale,0.95*\Scale) {};
	\node [style=none] (3)  at (-0.8*\Scale,1*\Scale) {\small{$R_Y$}};
	\node [style=none] (3b)  at (-0.8*\Scale,0.85*\Scale) {};
	\node [style=none] (3t)  at (-0.8*\Scale,1.2*\Scale) {};
	\node [style=none] (3bl)  at (-1.0*\Scale,0.95*\Scale) {};
	\node [style=none] (3br)  at (-0.7*\Scale,0.85*\Scale) {};
	\node [style=none] (3tl)  at (-1.0*\Scale,1.1*\Scale) {};
	\node [style=none] (4)  at (0.8*\Scale,1*\Scale) {\small{$R_Z \ $}};
	\node [style=none] (4b)  at (0.8*\Scale,0.85*\Scale) {};
	\node [style=none] (4t)  at (0.8*\Scale,1.2*\Scale) {};
	\node [style=none] (4tr)  at (1.85*\Scale,1.1*\Scale) {};
	\node [style=none] (4br)  at (1.85*\Scale,0.95*\Scale) {};
	\node [style=none] (5)  at (2*\Scale,1*\Scale) {$\ \ Z'$};
	\node [style=none] (5b)  at (2*\Scale,0.85*\Scale) {};
	\node [style=none] (5t)  at (2*\Scale,1.2*\Scale) {};
	\node [style=none] (5bl)  at (1.0*\Scale,0.95*\Scale) {};
	\node [style=none] (5tl)  at (1.0*\Scale,1.1*\Scale) {};
	\node [style=none] (5br)  at (0.95*\Scale,0.9*\Scale) {};
	\node [style=none] (5bm)  at (0.85*\Scale,0.85*\Scale) {};
	\node [style=none] (8)  at (2*\Scale,-0.1*\Scale) {$\ \ Z(W)$};
	\node [style=none] (8t)  at (2*\Scale,0.1*\Scale) {};
	\node [style=none] (8tl)  at (1.9*\Scale,0.05*\Scale) {};
	\node [style=none] (8tll)  at (1.80*\Scale,0.05*\Scale) {};
	\node [style=none] (8bl)  at (2.0*\Scale,-0.25*\Scale) {};
	\node [style=none] (8l)  at (1.75*\Scale,-0.05*\Scale) {};
	\node [style=none] (8lb)  at (1.75*\Scale,-0.2*\Scale) {};
	\node [style=none] (8lbr)  at (1.85*\Scale,-0.25*\Scale) {};
	\node [style=none] (6)  at (-1.5*\Scale,2*\Scale) {$W_Y$};
	\node [style=none] (6bl)  at (-1.6*\Scale,1.8*\Scale) {};
	\node [style=none] (6br)  at (-1.4*\Scale,1.8*\Scale) {};
	\node [style=none] (7)  at (1.5*\Scale,2*\Scale) {$W_Z$};
	\node [style=none] (7bl)  at (1.4*\Scale,1.8*\Scale) {};
	\node [style=none] (7br)  at (1.6*\Scale,1.8*\Scale) {};
	\draw [->] (2b) to [bend right=20] (1);
	\draw [->] (3b) to [bend right=20] (1);
	\draw [->] (4b) to [bend left=20] (1);
	\draw [->] (5b) to [bend left=20] (1);
	\draw [blue, dashed, ->] (5b) to [bend left=15] (8t);
	\draw [blue, dashed, ->] (5br) to [bend left=15] (8tl);
	\draw [blue, dashed, ->] (3br) to [bend right=15] (8l);
	\draw [blue, dashed, ->] (2brb) to [bend right=15] (8lb);
	\draw [blue, dashed, ->] (1) to [bend right=10] (8lbr);
	\draw [->] (8tll) to [bend left=10] (5bm);
	\draw [->] (8bl) to [bend left=25] (1);
	\draw [->] (2tr) to [bend left=30] (3tl);
	\draw [->] (3bl) to [bend left=30] (2br);
	\draw [->] (4tr) to [bend right=30] (5tl);
	\draw [->] (5bl) to [bend right=30] (4br);
	\draw [->] (2t) to [bend left=20] (6bl);
	\draw [->] (3t) to [bend right=20] (6br);
	\draw [->] (4t) to [bend left=20] (7bl);
	\draw [->] (5t) to [bend right=20] (7br);
\end{tikzpicture}

%% file: Figures/Fig_Rule3SubsetRelation_2.tex
\begin{tikzpicture}
	\node [style=none] (1)  at (0,-0.5*\Scale) {$\widetilde{R}^{c\phantom{'}}$};
	\node [style=none] (2)  at (-2*\Scale,1*\Scale) {$Y$};
	\node [style=none] (2b)  at (-2*\Scale,0.85*\Scale) {};
	\node [style=none] (2t)  at (-2*\Scale,1.2*\Scale) {};
	\node [style=none] (2tr)  at (-1.85*\Scale,1.1*\Scale) {};
	\node [style=none] (2br)  at (-1.85*\Scale,0.95*\Scale) {};
	\node [style=none] (2brb)  at (-1.95*\Scale,0.9*\Scale) {};
	\node [style=none] (2bm)  at (-1.95*\Scale,0.95*\Scale) {};
	\node [style=none] (3)  at (-0.8*\Scale,1*\Scale) {\small{$R_Y$}};
	\node [style=none] (3b)  at (-0.8*\Scale,0.85*\Scale) {};
	\node [style=none] (3t)  at (-0.8*\Scale,1.2*\Scale) {};
	\node [style=none] (3bl)  at (-1.0*\Scale,0.95*\Scale) {};
	\node [style=none] (3br)  at (-0.7*\Scale,0.85*\Scale) {};
	\node [style=none] (3tl)  at (-1.0*\Scale,1.1*\Scale) {};
	\node [style=none] (4)  at (0.8*\Scale,1*\Scale) {\small{$\widetilde{R}_Z \ $}};
	\node [style=none] (4b)  at (0.8*\Scale,0.85*\Scale) {};
	\node [style=none] (4t)  at (0.8*\Scale,1.2*\Scale) {};
	\node [style=none] (4tr)  at (1.85*\Scale,1.1*\Scale) {};
	\node [style=none] (4br)  at (1.85*\Scale,0.95*\Scale) {};
	\node [style=none] (5)  at (2*\Scale,1*\Scale) {$\ \ Z'$};
	\node [style=none] (5b)  at (2*\Scale,0.85*\Scale) {};
	\node [style=none] (5t)  at (2*\Scale,1.2*\Scale) {};
	\node [style=none] (5bl)  at (1.0*\Scale,0.95*\Scale) {};
	\node [style=none] (5tl)  at (1.0*\Scale,1.1*\Scale) {};
	\node [style=none] (5br)  at (0.95*\Scale,0.9*\Scale) {};
	\node [style=none] (5bm)  at (0.85*\Scale,0.85*\Scale) {};
	\node [style=none] (8)  at (2*\Scale,-0.1*\Scale) {$\ \ Z(W)$};
	\node [style=none] (8t)  at (2*\Scale,0.1*\Scale) {};
	\node [style=none] (8tl)  at (1.9*\Scale,0.05*\Scale) {};
	\node [style=none] (8tll)  at (1.80*\Scale,0.05*\Scale) {};
	\node [style=none] (8bl)  at (2.0*\Scale,-0.25*\Scale) {};
	\node [style=none] (8l)  at (1.75*\Scale,-0.05*\Scale) {};
	\node [style=none] (8lb)  at (1.75*\Scale,-0.2*\Scale) {};
	\node [style=none] (8lbr)  at (1.85*\Scale,-0.25*\Scale) {};
	\node [style=none] (6)  at (-1.5*\Scale,2*\Scale) {$W_Y$};
	\node [style=none] (6bl)  at (-1.6*\Scale,1.8*\Scale) {};
	\node [style=none] (6br)  at (-1.4*\Scale,1.8*\Scale) {};
	\node [style=none] (7)  at (1.5*\Scale,2*\Scale) {$W_Z$};
	\node [style=none] (7bl)  at (1.4*\Scale,1.8*\Scale) {};
	\node [style=none] (7br)  at (1.6*\Scale,1.8*\Scale) {};
	\draw [->] (2b) to [bend right=20] (1);
	\draw [->] (3b) to [bend right=20] (1);
	\draw [->] (4b) to [bend left=20] (1);
	\draw [->] (5b) to [bend left=20] (1);
	\draw [blue, dashed, ->] (5b) to [bend left=15] (8t);
	\draw [blue, dashed, ->] (5br) to [bend left=15] (8tl);
	\draw [blue, dashed, ->] (3br) to [bend right=15] (8l);
	\draw [blue, dashed, ->] (2brb) to [bend right=15] (8lb);
	\draw [blue, dashed, ->] (1) to [bend right=10] (8lbr);
	\draw [->] (8bl) to [bend left=25] (1);
	\draw [->] (2tr) to [bend left=30] (3tl);
	\draw [->] (3bl) to [bend left=30] (2br);
	\draw [->] (4tr) to [bend right=30] (5tl);
	\draw [->] (5bl) to [bend right=30] (4br);
	\draw [->] (2t) to [bend left=20] (6bl);
	\draw [->] (3t) to [bend right=20] (6br);
	\draw [->] (4t) to [bend left=20] (7bl);
	\draw [->] (5t) to [bend right=20] (7br);
\end{tikzpicture}